\documentstyle[11pt,epsf]{article}
 1
 1
 1

\def\lsim{\mathrel{\raise.3ex\hbox{$<$\kern-.75em\lower1ex\hbox{$\sim$}}}}
\def\gsim{\mathrel{\raise.3ex\hbox{$>$\kern-.75em\lower1ex\hbox{$\sim$}}}}
\def\Li2{{\rm Li}_2}
\def\Litri{{\rm Li}_3}
\catcode`\@=11
\def\slash{\mathpalette\make@slash}
\def\make@slash#1#2{\setbox\z@\hbox{$#1#2$}%
  \hbox to 0pt{\hss$#1/$\hss\kern-\wd0}\box0}
\catcode`\@=12 % @ signs are no longer letters
\begin{document}
\noindent
\thispagestyle{empty}
\renewcommand{\thefootnote}{\fnsymbol{footnote}}
\begin{flushright}
{\bf CERN-TH/2000-227}\\
{\bf hep-ph/0008102}\\
{\bf August 2000}\\
\end{flushright}
\vspace{.5cm}
\begin{center}
  \begin{Large}\bf
Bottom Quark Mass from $\Upsilon$ Mesons:\\[2mm]
 Charm Mass Effects
  \end{Large}
  \vspace{2.2cm}

\begin{large}
 A.~H.~Hoang
%$^{a}$ and
% M.~Melles$^{b}$ 
\end{large}

\vspace{.2cm}
\begin{center}
\begin{it}
%${}^a$ 
Theory Division, CERN,\\
   CH-1211 Geneva 23, Switzerland
%\\[.5cm]
%${}^b$  Paul Scherrer Institute (PSI),\\
%   CH-5232 Villigen, Switzerland
\end{it} 
\end{center}

  \vspace{3cm}
  {\bf Abstract}\\
\vspace{0.3cm}
%\setcounter{footnote}{0}
%\renewcommand{\thefootnote}{\arabic{footnote}}
%\addtocounter{footnote}{-1}
%
\noindent
\begin{minipage}{15.0cm}
\begin{small}
The effects of the finite charm quark mass on bottom quark mass
determinations from $\Upsilon$ sum rules are examined in detail. The
charm quark mass effects are calculated at next-to-next-to-leading
order in the non-relativistic power counting for the $\Upsilon$ sum
rules and at order $\alpha_s^3$ for the determination of the
bottom $\overline{\mbox{MS}}$ mass. For the bottom 1S mass, which is
extracted from the $\Upsilon$ sum rules directly, we obtain
$M_{\mbox{\tiny b}}^{\mbox{\tiny 1S}}=4.69\pm 0.03$~GeV with
negligible correlation  
to the value of the strong coupling. For the bottom
$\overline{\mbox{MS}}$ mass we obtain 
$\overline M_{\mbox{\tiny b}}(\overline M_{\mbox{\tiny b}}) = 
4.17 \pm 0.05$~GeV taking 
$\alpha^{(n_l=5)}_s(M_Z)=0.118\pm 0.003$ as an input. Compared with an
analysis where all quarks lighter than the bottom are treated as
massless, we find that the finite charm mass shifts the bottom 1S mass, 
$M_{\mbox{\tiny b}}^{\mbox{\tiny 1S}}$, by about $-20$~MeV and the
$\overline{\mbox{MS}}$ mass,
$\overline M_{\mbox{\tiny b}}(\overline M_{\mbox{\tiny b}})$, by
$-30$ to $-35$~MeV.  
\\[3mm]

\end{small}
\end{minipage}
\end{center}
\setcounter{footnote}{0}
\renewcommand{\thefootnote}{\arabic{footnote}}
\vspace{2.2cm}
{\bf CERN-TH/2000-227}\\
{\bf August 2000}
%
%
% text
%
\newpage
%\mbox{}
%\vspace{2cm}
\tableofcontents
\newpage
\noindent
\section{Introduction}
\label{sectionintroduction}
Precise and accurate determinations of the bottom quark mass
parameter are mainly motivated by current and future B physics
experiments, which aim at improved measurements of the
Cabibbo--Kobayashi--Maskawa (CKM) matrix elements. This will hopefully
get us closer to an understanding of the origin of flavour mixing,
mass generation and CP violation. A precise and accurate knowledge of
the bottom quark mass (in the scheme that is appropriate for the
particular application) will eliminate one source of uncertainties. In
particular, for the determination of $V_{ub}$, and also $V_{cb}$, from
the semileptonic B decays the bottom quark mass is important
in view of the strong dependence of the total rate and certain
distributions on the bottom mass parameter. Another field where a
precise bottom quark mass is desirable is represented by grand
unification models. Here, precise values of the $\overline{\mbox{MS}}$
running heavy quark masses that are determined at low energies and
that are as much as possible free of uncertainties from the strong
interaction at low scales can constrain the allowed parameter space
for a given scenario of flavour generation.

The bottom quark mass parameter can be obtained from properties of
hadrons containing one or more bottom quarks using perturbative
quantum chromodynamics (QCD) or lattice methods. Recently, lattice
calculations 
have been applied to bottom quark mass determinations from B and
$\Upsilon$ meson spectra~\cite{lattice1,lattice2}, whereas
perturbative methods using non-relativistic effective theories were 
applied to bottom quark mass determinations from the mass of the 
$\Upsilon(\mbox{1S})$~\cite{Pineda1}
and sum rules for the $\Upsilon$ masses and electronic
widths~\cite{Voloshin1,Kuhn1,Penin1,Hoang1,Melnikov1,Hoang2,Beneke1}.
Extractions of the bottom quark 
$\overline{\mbox{MS}}$ mass at the scale $M_Z$ have also been carried
out from analyses of LEP data on bottom--antibottom--gluon three-jet
events~\cite{Rodrigo1}. The results of the bottom $\overline{\mbox{MS}}$ mass
from these analyses have rather large uncertainties,
but they have established the evolution of the $\overline{\mbox{MS}}$
bottom quark mass at higher scales.

Using perturbative methods, non-relativistic sum rules for the masses
and electronic decay widths of $\Upsilon$ mesons, bottom--antibottom
quark bound states that have photonic quantum numbers and that can be
produced in $e^+e^-$ annihilations, are in principle the most reliable
tool to determine the bottom quark mass. Using causality and global
duality 
arguments, one can relate energy integrals (or ``moments'') over the
total cross section for the production of final states containing a
bottom--antibottom 
quark pair in $e^+e^-$ collisions to derivatives of the vacuum
polarization function of bottom quark currents at zero-momentum
transfer. For a particular range of numbers of derivatives the moments
are saturated by the very precise experimental data on the $\Upsilon$
mesons and, at the same time, can be calculated reliably using
perturbative QCD in the non-relativistic expansion. In contrast to the
bottom quark mass determination from the $\Upsilon(\mbox{1S})$ mass,
the sum rule analysis is practically free from model assumptions on
non-perturbative effects as long as the number of derivatives that is
employed is not too high.  Because the moments have, for dimensional
reasons, a strong dependence on the bottom quark mass, these sum rules
can be used to determine the bottom quark mass to high precision. 

In the most recent sum rule analyses at NNLO in the non-relativistic 
expansion~\cite{Melnikov1,Hoang2,Beneke1} 
special care was taken to consistently eliminate the
strong linear sensitivity to small momenta and the associated
(artificially) large perturbative corrections that were contained in
previous analyses that employed the bottom quark pole mass. This was
achieved by implementing bottom quark short-distance mass definitions
into the moments that were constructed, particularly for the situation
where the bottom quark is very close to its mass shell. For the
implementation of these masses the systematic cancellation of the
linear infrared-sensitive contributions and the compliance with the
non-relativistic power counting had to be ensured in each order of
perturbation theory. Melnikov--Yelkhovsky~\cite{Melnikov1} used the
kinetic mass, Hoang~\cite{Hoang2} the 1S mass and
Beneke--Signer~\cite{Beneke1} the PS 
mass. These ``low-virtuality short-distance masses'' can be used
directly for other 
processes where the bottom quark virtuality is small with respect to
its mass. In semileptonic B meson partial rates, for example, they
lead to a very well behaved perturbative expansion
(see e.g. Ref.~\cite{Hoang2,Melnikov2}). For situations 
where the bottom quark is very off-shell, on the other hand, the
$\overline{\mbox{MS}}$ mass definition is the appropriate
one. Starting from the low-virtuality short-distance bottom masses the
previously mentioned analyses obtained 
$\overline M_{\mbox{\tiny b}}(\overline M_{\mbox{\tiny b}})=
4.2\pm 0.1$~GeV~\cite{Melnikov1}, $4.20\pm 0.06$~GeV~\cite{Hoang2} and 
$4.25\pm 0.08$~GeV~\cite{Beneke1} for the bottom $\overline{\mbox{MS}}$
mass. These results represent determinations of
the bottom $\overline{\mbox{MS}}$ mass at order $\alpha_s^3$ in the
usual loop expansion of perturbative QCD. 
The results are compatible with each other and also with the
most recent lattice calculations (see Ref.~\cite{lattice2} for a convenient
compilation of results obtained after 1994). The uncertainties are
considerably smaller than $\Lambda_{\rm QCD}$ and amount to about 2\%. 
It will be the aim of future analyses to further reduce the
uncertainty, to maybe 1\% or less, i.e. to an amount of a few times
10~MeV. Thus, it is mandatory to account for effects that can lead to  
for shifts of several times 10~MeV. 
%In recent literature there are a number of analyses based on 
%spectrum calculations with perturbative QCD 
%where uncertainties at the level of a few $10$~MeV have already been
%quoted, see e.g. Refs.~\cite{XX}. These analyses, however, do have
%different standards in estimating systematic theoretical
%uncertainties.

One effect, which has been entirely neglected in all previous sum rule
analyses, is coming from the masses of those quarks that are lighter
than the bottom, most importantly the charm quark. The massless
approximation for the light quarks is a good one, provided that the
light quark mass is much smaller than any of the relevant dynamical
scales. In this case it is legitimate to simply
expand in the light quark mass. Because all linear sensitivity to
small momenta is eliminated by the use of a short-distance
bottom quark mass, the first non-vanishing correction to the massless
limit is of order 
$(\alpha_s/\pi)^2 m_{\rm light}^2/M_{\mbox{\tiny b}}$, where 
$m_{\rm light}$ is the light quark mass. Even for the charm quark this
would amount to a shift of only a few MeV in the bottom quark mass and
could be safely neglected. However, in order to judge whether the
massless approximation is really justified, one needs to compare the
light quark masses with the scales that are relevant to the
non-relativistic bottom--antibottom quark dynamics that is encoded in
the moments. There are two issues that make the effects of light quark
masses, and of the charm quark mass in particular, a subtle problem:
one is specific to the extraction of the bottom quark
$\overline{\mbox{MS}}$ mass from the non-relativistic sum rules and
the other exists for any process that is calculated at large orders of
perturbation theory. 

The issue that is specific to the dynamics of the non-relativistic
bottom--antibottom pair encoded in the moments arises from the fact that 
the inverse Bohr radius, $M_{\mbox{\tiny b}}\alpha_s$, is a relevant
dynamical scale. Because the inverse Bohr radius is about the same as
the mass of the charm quark, it is {\it a priori} not allowed to
expand in the 
charm quark mass. Rather, the effects of the charm quark mass have to
be treated exactly, leading to a non-trivial dependence on the ratio
of the charm quark mass and the inverse Bohr radius. In
Ref.~\cite{Hoang4} 
is has been shown that this leads to an incomplete cancellation of
linear charm quark mass terms $\propto \alpha_s^2 m_{\rm charm}$ in
the relation between the bottom $\overline{\mbox{MS}}$ and 1S masses.
The resulting shift in the $\overline{\mbox{MS}}$ mass $\overline
M_{\mbox{\tiny b}}(\overline M_{\mbox{\tiny b}})$ 
was estimated to be around $-20$~MeV, based on a
NLO calculation (which corresponds to an order $\alpha_s^2$ accuracy in
the $\overline{\mbox{MS}}$ mass). If an expansion in the
charm quark mass would be allowed, all linear charm quark mass terms
would cancel exactly. The large size of the shift caused by the
incomplete cancellation of the linear charm quark mass term 
comes from the fact that the linear charm quark mass term is enhanced.
Technically, this happens because linear light quark mass terms
represent non-analytic terms (in the square of the mass) that are
multiplied by a factor $\pi^2$, and because the scale of the strong
coupling governing them is the light quark mass and not the bottom
mass. Physically, this happens because light quark masses act as an
infrared cutoff for the gluon line in which the light quark loops are
inserted. A linear sensitivity to small momenta is, after expansion,
reflected by a non-analytic, linear dependence on the light quark
mass. Because this linear, non-analytic dependence can only arise from
momenta of the order of the light quark mass, the strong coupling
multiplying the linear terms is renormalized at the light quark
mass. Thus, the masses of the light quarks work in complete analogy to
the infinitesimally small 
fictitious gluon mass that is sometimes used in standard renormalon
analyses. The difference is that the light quark masses are not
fictitious, but finite numbers that are given by experimental
data. Whether we are allowed to expand in them (or not) depends on
their relation to the dynamical scales of the problem. 

The other, more general issue is associated with the fact that QCD
perturbation theory at large orders of perturbation theory becomes
dominated by exponentially decreasing momenta. This is the origin of
the divergent asymptotic behaviour of QCD perturbation theory
associated to the so-called ``infrared
renormalons''~\cite{renormalons1,renormalons2,renormalons3}. This 
means that the genuine effects from light quarks with finite masses
effectively decouple, as from a certain order, and that the large
order asymptotic 
behaviour is considerably affected by the masses of the light
quarks~\cite{Beneke3}. In fact, at large orders of perturbation theory, the
difference between the massless approximation for light quarks and a
calculation that includes the light quark masses becomes arbitrarily
large because the evolution of the strong coupling differs in the two
cases for momenta of order or below a specific light quark mass.
Even at intermediate (i.e. phenomenologically relevant and
calculable) orders, however, the effective decoupling of light quarks
can lead to considerable deviations from the massless approximation,
if the perturbative coefficients are already dominated by sufficiently
small momenta. The effective decoupling is therefore not just a purely
academic issue. 

From the issues mentioned above we can expect the following
qualitative effects coming from the finite charm quark mass in the
bottom $\overline{\mbox{MS}}$ mass obtained from the bottom 1S mass:
the largest correction arises at order 
$\alpha_s^2$ from the fact that one can expand in the charm quark mass
in the bottom pole--$\overline{\mbox{MS}}$ mass relation, but not in
the pole--1S mass relation~\cite{Hoang4}. The higher order charm quark mass
corrections should decrease because the importance of the dynamical
scales $M_{\mbox{\tiny b}}$, and $M_{\mbox{\tiny b}}\alpha_s$
becomes less prominent as the dominating momenta decrease
exponentially. This means that the charm mass corrections are
under control and can be calculated reliably\footnote{
We emphasize that this statement is not correct if one 
attempts a determination of the bottom quark pole mass. Its well-known
ambiguity of order $\Lambda_{\rm QCD}$ is directly reflected in  
an uncontrollable behaviour of the charm mass corrections. 
This is demonstrated in Sec.~\ref{subsectionmsbar1Smassexamination}.
}.  
From the light quarks other than the charm, we do not expect any
sizeable corrections because, at the accessible orders of perturbation
theory, their masses are simply too small to make any of the previous
considerations relevant. However, a quantitative statement about the
actual size and behaviour of the charm mass corrections can only be
made through an actual calculation. 
In addition, is it mandatory to assess how
the finite charm quark mass affects the extraction of the bottom 1S
mass from the $\Upsilon$ sum rules. For the bottom
$\overline{\mbox{MS}}$ mass the corrections might either add up or
cancel at the end. In the former case the charm quark mass effects for
the final value of 
$\overline M_{\mbox{\tiny b}}(\overline M_{\mbox{\tiny b}})$ might be
larger than the $20$~MeV estimate that has been obtained in the NLO
(or order $\alpha_s^2$) analysis of Ref.~\cite{Hoang4}.
 
In this paper we determine the charm mass corrections at order
$\alpha_s^3$ in the bottom $\overline{\mbox{MS}}$--1S mass relation as
well as for the bottom 1S mass determination from the $\Upsilon$
meson sum rules at NNLO in the non-relativistic expansion. This
includes the calculation of the NNLO light quark 
mass corrections to the perturbative contributions of the
$\Upsilon(\mbox{1S})$ mass in terms of the pole mass, and the order
$\alpha_s^3$ light 
quark mass corrections in the heavy quark pole--$\overline{\mbox{MS}}$
mass relation. The latter corrections are determined in the linear
light quark mass approximation because an expansion in the light quark
masses is allowed in that case. The determination of the linear
order $\alpha_s^3$ light quark mass corrections is based on the
conjecture 
that linear light quark mass terms are absent in the total static
energy of a heavy quark-antiquark pair. As far as the bottom
$\overline{\mbox{MS}}$--1S mass relation is concerned, we also attempt
to estimate the order $\alpha_s^4$ charm mass corrections based on a
calculation of the NNNLO (N$^3$LO) perturbative corrections in the
$\Upsilon(\mbox{1S})$ mass, in the large-$\beta_0$ approximation.
For the $\Upsilon$ sum rules we determine charm
mass corrections at NLO and NNLO in the non-relativistic expansion,
where we neglect the double-insertion contributions coming from
second order Rayleigh--Schr\"odinger time-independent perturbation
theory at NNLO. For all results we discuss in detail the behaviour of
the light quark mass  corrections. It is demonstrated that the
effects coming from the masses of up, down and strange quarks can, as
expected, be neglected. 
 
We note that, although the effects of the light quark masses can be
presented in a straightforward way in the framework of
effective theories for non-relativistic heavy quark--antiquark pairs,
we will not give such a presentation in this work. This is because
we feel the discussion of the many interesting
aspects of light quark mass effects, which do not depend at all on the
concepts of effective theories, would become cumbersome. We also
emphasize that all 
results presented in this work are given in the convention that the
evolution of the strong $\overline{\mbox{MS}}$ coupling includes the
light quarks with finite mass,
i.e. we use $\alpha_s^{(n_l=4)}$ for our results for the bottom quark
masses. We do this because all 
previous analyses, which neglected light quark masses, naturally used
this definition of $\alpha_s$. It is therefore easy to update the
previous analyses by simply adding the new corrections determined in
this work. In this context we also note that many conclusions and
statements made in this work about the behaviour and the size of the
charm mass corrections at the different orders are not adequate
if $\alpha_s^{(n_l=3)}$ is chosen as the definition of the
strong coupling, since this would lead to a reshuffling of
corrections. The final numbers for the bottom 1S and
$\overline{\mbox{MS}}$ masses, however, do not depend on this choice.  

The plan of this paper is as follows: in
Sec.~\ref{sectionpotential} we briefly review the ingredients needed
to describe the heavy-quark--antiquark dynamics at NNLO in the
non-relativistic expansion for massless light quarks and the result
for the NNLO light quark mass corrections to the static potential
obtained earlier by Melles~\cite{Melles1,Melles2}. The results by Melles are
rewritten in a more convenient dispersion relation representation. The
limit of vanishing light quark mass and the resulting linear light
quark mass terms are examined, and the prediction for the order
$\alpha_s^3$ linear light quark mass corrections in the heavy quark
pole--$\overline{\mbox{MS}}$ mass relation is made. The quality of the
linear light quark mass approximation in the  heavy quark
pole--$\overline{\mbox{MS}}$ mass relation is analysed in
Sec.~\ref{sectionpolemsbarmass}. In Sec.~\ref{sectionpole1Smass} the
NNLO light quark mass corrections to the relation between the heavy
quark 1S and pole masses are presented, and in
Sec.~\ref{sectionmsbar1Smass} the results from
Secs.~\ref{sectionpolemsbarmass} and \ref{sectionpole1Smass} are
combined to derive the order $\alpha_s^3$ (or NNLO) relation between
the heavy quark $\overline{\mbox{MS}}$ and 1S masses, where light
quark mass effects are taken into account. The upsilon
expansion~\cite{Hoang3}, a prescription for a modified perturbative
expansion that allows for a 
consistent derivation of the $\overline{\mbox{MS}}$--1S mass relation
in each order, is reviewed. The size of the charm mass corrections is
analysed and an estimate for the order $\alpha_s^4$ (or N$^3$LO) charm
mass corrections is given. In Sec.~\ref{sectionsumrules} we determine
the NNLO light quark mass corrections to the moments of the $\Upsilon$
sum rules, and in Sec.~\ref{sectionnumerical} we carry out the numerical
analysis to determine the bottom 1S mass from fitting theoretical and
experimental moments. The relation of our result for the bottom 1S
mass to the mass of the $\Upsilon(\mbox{1S})$ meson is
discussed, and the result for the bottom $\overline{\mbox{MS}}$ mass
is presented. All results are 
discussed focusing on the charm mass
effects. Section~\ref{sectionsummary} contains a summary. Attached to
the paper are five appendices: in App.~\ref{appendix1Smass} details on
the calculations of the NNLO light quark mass corrections to the mass
of the $J^{PC}=1^{--}$, ${}^3S_1$ heavy perturbative quarkonium ground
state are given. Appendix~\ref{appendix1Smasslargeb0}
contains the calculation of the N$^3$LO perturbative corrections to
the mass of the heavy quarkonium ground state in the large-$\beta_0$
approximation. In App.~\ref{appendixmsbar1Smasslargeb0} the results
for the order $\alpha_s^4$ (or N$^3$LO) perturbative
corrections to the heavy quark $\overline{\mbox{MS}}$--1S mass relation
are derived in the large-$\beta_0$ approximation. 
Appendix~\ref{appendixmsbarmassvacuum} contains exact analytic results
for 
the order $\alpha_s^3$ light quark mass corrections to the heavy quark
pole--$\overline{\mbox{MS}}$ mass relation coming from vacuum
polarization insertions, and App.~\ref{appendixmassless} gives some
formulae for the corrections to the moments of the $\Upsilon$ sum
rules from massless light quarks.  

\par
\vspace{0.5cm}
\section[Considerations on the Static Potential 
         including Light Quark Mass Effects]
       {Considerations on the Static Potential\\ 
         including Light Quark Mass Effects}
\label{sectionpotential}

The static potential represents the central quantity needed to
determine the effects of the masses of light quarks on the dynamics of
a perturbative non-relativistic heavy-quark--antiquark pair. In this
section we discuss the corrections to the static potential up to NNLO
in the non-relativistic expansion coming from the mass of a light
quark.   
\par
%\vspace{0.5cm}
%
\subsection{Massless Light Quarks}
\label{subsectionpotentialmassless}

At NNLO in the non-relativistic expansion, the dynamics of a
perturbative heavy-quark--antiquark pair\footnote{
In this context ``perturbative'' means that the heavy 
quark mass, $M_{\mbox{\tiny Q}}$,
is large enough, such  that the hierarchy 
$M_{\mbox{\tiny Q}}\gg M_{\mbox{\tiny Q}} v\gg 
M_{\mbox{\tiny Q}} v^2\gg\Lambda_{\rm QCD}$ holds, $v$ being
the heavy quark velocity.
}
represents a pure 2-body problem and can be described by a
textbook-like Schr\"odinger equation that contains the kinetic energy
term and a time-independent instantaneous potential. In configuration
space representation, the Schr\"odinger equation has the form
($E\equiv \sqrt{s}-2 M_{\mbox{\tiny Q}}^{\mbox{\tiny pole}}$):
\begin{eqnarray}
& &
\bigg(\,
-\frac{{\mbox{\boldmath $\nabla$}}^2}
{M^{\mbox{\tiny pole}}_{\mbox{\tiny Q}}} 
- \frac{{\mbox{\boldmath $\nabla$}}^4}
{4\,(M^{\mbox{\tiny pole}}_{\mbox{\tiny Q}})^3} 
+ \bigg[\,
  V_{\mbox{\tiny c}}^{\mbox{\tiny LO}}({\mbox{\boldmath $r$}}) +
  V_{\mbox{\tiny c}}^{\mbox{\tiny NLO}}({\mbox{\boldmath $r$}}) +
  V_{\mbox{\tiny c}}^{\mbox{\tiny NNLO}}({\mbox{\boldmath $r$}}) + 
  V_{\mbox{\tiny BF}}({\mbox{\boldmath $r$}}) + 
  V_{\mbox{\tiny NA}}({\mbox{\boldmath $r$}})
\,\bigg]  
- E
\,\bigg)
\nonumber\\[2mm] & &
\mbox{\hspace{9cm}}
\times\,G({\mbox{\boldmath $r$}},{\mbox{\boldmath $r$}}^\prime, E)
\,  = \, \delta^{(3)}({\mbox{\boldmath $r$}}-{\mbox{\boldmath $r$}}^\prime) 
\,,
\label{NNLOSchroedinger}
\end{eqnarray}
where $M^{\mbox{\tiny pole}}_{\mbox{\tiny Q}}$ is the heavy quark 
pole mass and $s$ the squared 
centre-of-mass energy. Although the use of the pole mass is a bad
choice for a phenomenological analysis, because of its strong
sensitivity to small momenta, we will use it for intermediate
calculations because it leads to the simple form of the Schr\"odinger
equation shown in Eq.~(\ref{NNLOSchroedinger}). The terms 
$V_{\mbox{\tiny c}}^{\mbox{\tiny LO}}$,
$V_{\mbox{\tiny c}}^{\mbox{\tiny NLO}}$ and
$V_{\mbox{\tiny c}}^{\mbox{\tiny NNLO}}$ represent the static
(Coulomb) potential at LO (order $v^2\sim\alpha_s^2$), 
NLO (order $v^3\sim\alpha_s^3$) and NNLO (order $v^4\sim\alpha_s^4$),
respectively, in the non-relativistic power counting, while
$V_{\mbox{\tiny BF}}$ is the Breit-Fermi potential known from
positronium and $V_{\mbox{\tiny NA}}$ is a non-Abelian potential
coming from a non-analytic gluonic correction to the exchange of a
Coulomb gluon; $V_{\mbox{\tiny BF}}$ and $V_{\mbox{\tiny NA}}$ are of
NNLO. The kinetic energy terms 
$-{\mbox{\boldmath $\nabla$}}^2/
M^{\mbox{\tiny pole}}_{\mbox{\tiny Q}}$ 
and $-{\mbox{\boldmath $\nabla$}}^4/
4(M^{\mbox{\tiny pole}}_{\mbox{\tiny Q}})^3$ are of LO
and NNLO, respectively.
Treating all $n_l$ quarks that a lighter than the heavy quark as
massless the individual potentials read 
($a_s\equiv\alpha_s^{(n_l)}(\mu)$,
$r\equiv|{\mbox{\boldmath $r$}}|$,
$\tilde\mu \, \equiv \, e^{\gamma_{\mbox{\tiny E}}}\,\mu$, 
$C_A=3$, $C_F=4/3$, $T=1/2$)
\begin{eqnarray}
V_{\mbox{\tiny c}}^{\mbox{\tiny LO}}({\mbox{\boldmath $r$}}) 
& = & -\,\frac{C_F\,a_s}{r}
\,,
\label{VcLOmassless}
\\[4mm] 
V_{\mbox{\tiny c,massless}}^{\mbox{\tiny NLO}}({\mbox{\boldmath $r$}})
& = &
V_{\mbox{\tiny c}}^{\mbox{\tiny LO}}({\mbox{\boldmath $r$}}) 
\, \bigg(\frac{a_s}{4\,\pi}\bigg)\,\bigg[\,
2\,\beta_0\,\ln(\tilde\mu\,r) + a_1
\,\bigg]
\,,
\label{VcNLOmassless}
\\[4mm]
V_{\mbox{\tiny c,massless}}^{\mbox{\tiny NNLO}}({\mbox{\boldmath $r$}})
& = &
V_{\mbox{\tiny c}}^{\mbox{\tiny LO}}({\mbox{\boldmath $r$}}) 
\, \bigg(\frac{a_s}{4\,\pi}\bigg)^2\,\bigg[\,
\beta_0^2\,\bigg(\,4\,\ln^2(\tilde\mu\,r) 
      + \frac{\pi^2}{3}\,\bigg) 
+ 2\,\Big(2\,\beta_0\,a_1 + \beta_1\Big)\,\ln(\tilde\mu\,r) 
+ a_2
\,\bigg]
\,,
\label{VcNNLOmassless}
\\[4mm]
V_{\mbox{\tiny BF}}({\mbox{\boldmath $r$}}) 
& = & 
\frac{C_F\,a_s\,\pi}{(M_{\mbox{\tiny Q}}^{\mbox{\tiny pole}})^2}\,
\bigg[\,
1 + \frac{8}{3}
\,{\mbox{\boldmath $S$}}_{\mbox{\tiny b}}\,{\mbox{\boldmath $S$}}_{\bar b}
\,\bigg]
\,\delta^{(3)}({\mbox{\boldmath $r$}}) 
+ \frac{C_F\,a_s}{2 (M_{\mbox{\tiny Q}}^{\mbox{\tiny pole}})^2}\,
\frac{1}{r}\bigg[\,
{\mbox{\boldmath $\nabla$}}^2 + 
\frac{1}{r^2}{\mbox{\boldmath $r$}}
  \, ({\mbox{\boldmath $r$}}\,{\mbox{\boldmath $\nabla$}}) 
{\mbox{\boldmath $\nabla$}}
\,\bigg]
\nonumber\\[2mm] & &
- \,\frac{3\,C_F\,a_s}
{(M_{\mbox{\tiny Q}}^{\mbox{\tiny pole}})^2}\,\frac{1}{r^3}
\bigg[\,
\frac{1}{3}
\,{\mbox{\boldmath $S$}}_b
\,{\mbox{\boldmath $S$}}_{\bar b}
- \frac{1}{r^2}
  \,\Big({\mbox{\boldmath $S$}}_b\,{\mbox{\boldmath $r$}}\,\Big)
  \,\Big({\mbox{\boldmath $S$}}_{\bar b}\,{\mbox{\boldmath $r$}}\,\Big)
\,\bigg]
+ \frac{3\,C_F\,a_s}
{2 (M_{\mbox{\tiny Q}}^{\mbox{\tiny pole}})^2}\,\frac{1}{r^3}
\,{\mbox{\boldmath $L$}}
\,({\mbox{\boldmath $S$}}_b+{\mbox{\boldmath $S$}}_{\bar b})
\,,
\label{VBF}
\\[4mm]
V_{\mbox{\tiny NA}}({\mbox{\boldmath $r$}})  & = & -\,
\frac{C_A\,C_F\,a_s^2}
{2 M_{\mbox{\tiny Q}}^{\mbox{\tiny pole}}}\,\frac{1}{r^2}
\,,
\label{VNA}
\end{eqnarray}
where
\begin{eqnarray}
\beta_0 & = & \frac{11}{3}\,C_A - \frac{4}{3}\,T\,n_l
\,,
\label{b0def}
\\[4mm]
\beta_1 & = & \frac{34}{3}\,C_A^2 
-\frac{20}{3}C_A\,T\,n_l
- 4\,C_F\,T\,n_l
\,,
\label{b1def}
\\[4mm]
a_1 & = &  \frac{31}{9}\,C_A - \frac{20}{9}\,T\,n_l
\,,
\label{a1def}
\\[4mm]
a_2 & = & 
\bigg(\,\frac{4343}{162}+4\,\pi^2-\frac{\pi^4}{4}
 +\frac{22}{3}\,\zeta_3\,\bigg)\,C_A^2 
-\bigg(\,\frac{1798}{81}+\frac{56}{3}\,\zeta_3\,\bigg)\,C_A\,T\,n_l
\nonumber\\[2mm] & &
-\bigg(\,\frac{55}{3}-16\,\zeta_3\,\bigg)\,C_F\,T\,n_l 
+\bigg(\,\frac{20}{9}\,T\,n_l\,\bigg)^2
\,,
\label{a2def}
\end{eqnarray}
${\mbox{\boldmath $S$}}_b$ and ${\mbox{\boldmath $S$}}_{\bar b}$ 
are the bottom and antibottom quark spin operators, 
${\mbox{\boldmath $L$}}$ is the
angular momentum operator, and 
$\gamma_{\mbox{\tiny E}} = 0.57721566\ldots$ is the
Euler--Mascheroni constant.
The constants $a_1$ and $a_2$ have been calculated in
Refs.~\cite{Fischler1,Billoire1} and Refs.~\cite{Schroeder1,Peter1},
respectively. 
For the determination of the corrections coming
from a finite light quark mass in physical heavy-quark--antiquark
properties at NNLO, the expressions for 
$V_{\mbox{\tiny c}}^{\mbox{\tiny LO}}$ and  
$V_{\mbox{\tiny c,massless}}^{\mbox{\tiny NLO}}$ will be needed
explicitly. The superscript ``$(n_l)$'' shown by the strong coupling
indicates that the results are
written in terms of the strong coupling that evolves with $n_l$ active
quark flavours.  If the light quark masses are much smaller than the
typical inverse distances of the heavy quarks, this convention is the
only sensible one, as it avoids the appearance of logarithms
of the light quark masses.  

\par
%\vspace{0.5cm}
%
\subsection{Effects of a Light Quark Mass}
\label{subsectionpotentialmassive}
From the explicit dependence of Eqs.~(\ref{VcLOmassless})--(\ref{VNA})
on the number of light quarks $n_l$, one can easily see that only the
static potential at NLO and NNLO acquire additional corrections from a
finite light quark mass. In the following we present the corrections
to the static potential at NLO and NNLO for the case that only one
light quark species has a finite mass $m$, whereas the $n_l-1$
other quark species are treated as massless. For convenience we adopt
the pole mass definition for the light quark mass $m$. The
generalization to other light quark mass definitions, and to the case
where more than one of the $n_l$ light quarks have a finite mass, is
straightforward. We note that all final results for the light quark
mass corrections presented in this work will be expressed in terms of
the light quark $\overline{\mbox{MS}}$ mass 
$\overline m(\overline m)$. 

\begin{figure}[t] % figVcNLOmassive
\begin{center}
\hspace{0.5cm}
\leavevmode
\epsfxsize=5cm
%\epsffile[220 410 420 540]{potential1.ps}
\epsffile[220 340 420 480]{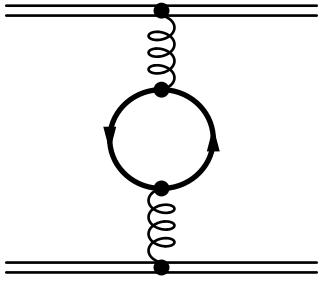}
\vskip  1.cm
 \caption{\label{figVcNLOmassive}
NLO contribution to the static potential coming from the insertion of
a one-loop vacuum polarization of a light quark with finite mass. 
}
 \end{center}
\end{figure}

At NLO the light quark mass corrections to the static potential arise
from the insertion of the light quark self energy into the gluon line
of the LO static potential (see Fig.~\ref{figVcNLOmassive}).
In momentum space representation the NLO static potential including
the light quark mass corrections reads
($a_s=\alpha_s^{(n_l)}(\mu)$)
\begin{eqnarray}
\tilde V_{\mbox{\tiny c}}^{\mbox{\tiny NLO}}({\mbox{\boldmath $p$}})
& = & 
\tilde V_{\mbox{\tiny c,massless}}^{\mbox{\tiny NLO}}({\mbox{\boldmath
$p$}})+
\delta\tilde V_{\mbox{\tiny c,m}}^{\mbox{\tiny NLO}}
({\mbox{\boldmath $p$}})
\,,
\\[4mm]
\tilde V_{\mbox{\tiny c,massless}}^{\mbox{\tiny NLO}}({\mbox{\boldmath
$p$}})
& = &
\tilde V_{\mbox{\tiny c}}^{\mbox{\tiny LO}}({\mbox{\boldmath $p$}})\,
\Big(\frac{a_s}{4\,\pi}\Big)\,\bigg[
-\beta_0\,\ln\Big(\frac{\mbox{\boldmath $p$}^2}{\mu^2}\Big) + a_1
\,\Bigg]
\,,
\\[4mm]
\delta\tilde V_{\mbox{\tiny c,m}}^{\mbox{\tiny NLO}}({\mbox{\boldmath
$p$}})
& = &
\tilde V_{\mbox{\tiny c}}^{\mbox{\tiny LO}}({\mbox{\boldmath $p$}})\,
\Big(\frac{a_s}{4\,\pi}\Big)\,
\frac{4}{3}\,T\,{\rm P}\Big(\frac{m^2}{{\mbox{\boldmath $p$}}^2}\Big)
\,,
\label{deltaVcNLOmomentumspace}
\\[4mm]
\tilde V_{\mbox{\tiny c}}^{\mbox{\tiny LO}}({\mbox{\boldmath $p$}})
& = & -\,\frac{4\, \pi\, C_F\, a_s}{{\mbox{\boldmath $p$}}^2}
\,,
\end{eqnarray}
where 
\begin{eqnarray}
{\rm P}\Big(\frac{m^2}{{\mbox{\boldmath $p$}}^2}\Big)
& = &
\bigg[\,
\Pi\Big(\frac{m^2}{{\mbox{\boldmath $p$}}^2}\Big)\,-\,
\bigg(\ln\frac{{\mbox{\boldmath $p$}}^2}{m^2}-\frac{5}{3}
\,\bigg)
\,\Bigg]
\,,
\label{Pdef}
\\[4mm]
\Pi\Big(\frac{m^2}{{\mbox{\boldmath $p$}}^2}\Big)
& = &
\frac{1}{3}\,-\,
(1-2\,z)\,\bigg[\,
2-\sqrt{1+4\,z}\,
\ln\Big(\frac{\sqrt{1+4\,z}+1}{\sqrt{1+4\,z}-1}\Big)
\,\bigg]\,\bigg|_{z=\frac{m^2}{{\mbox{\tiny\boldmath $p$}}^2}}
\,,
%\qquad 
%\bigg(\,z\,=\,\frac{m^2}{{\mbox{\boldmath $p$}}^2}\,\bigg)
%\,,
%\label{Pifull}
\nonumber
\\[2mm]
& = &
2\,{\mbox{\boldmath $p$}}^2\,
\int\limits_1^\infty\,\frac{dx\,f(x)}
{{\mbox{\boldmath $p$}}^2+4\,m^2\,x^2}
\,,
\label{Pidispersion}
\\[4mm]
f(x) & \equiv & \frac{1}{x^2}\,
\sqrt{x^2-1}\,\Big(\,1+\frac{1}{2\,x^2}\,\Big)
\,.
\label{fdef}
\end{eqnarray}
We note that ${\rm P}$ also contains the subtraction of the light
quark mass vacuum polarization in the limit $m\to 0$. 
Thus the light quark mass corrections 
$\delta\tilde V_{\mbox{\tiny c,m}}^{\mbox{\tiny NLO}}$ vanish for 
$m\to 0$ or ${\mbox{\boldmath $p$}}^2\to\infty$.
This is a consequence of our choice to include the light quark with
mass $m$ into the evolution of the strong coupling. 
The dispersion relation representation for the light quark vacuum
polarization function shown in Eq.~(\ref{Pidispersion}) is quite
useful because it allows for a quick determination of
the light quark mass corrections in configuration space
representation 
($\tilde m=e^{\gamma_{\mbox{\tiny E}}}\,m$):
\begin{eqnarray}
\delta V_{\mbox{\tiny c,m}}^{\mbox{\tiny NLO}}({\mbox{\boldmath
$r$}})
& = &
\int\!\frac{d^3{\mbox{\boldmath $p$}}}{(2\pi)^3}\,\,
\delta\tilde V_{\mbox{\tiny c,m}}^{\mbox{\tiny NLO}}
({\mbox{\boldmath $p$}})
\,\exp( i {\mbox{\boldmath $p$}} {\mbox{\boldmath $r$}} )
\nonumber\\[2mm]
& = &
V_{\mbox{\tiny c}}^{\mbox{\tiny LO}}({\mbox{\boldmath $r$}})\,
\Big(\frac{a_s}{3\,\pi}\Big)\,\bigg\{\,
\int\limits_1^\infty dx
%\frac{d x}{x^2}\,
%\sqrt{x^2-1}\,\Big(\,1+\frac{1}{x^2}\,\Big)\,e^{-2 m r x}\,
f(x)\,e^{-2 m r x}
+\bigg(\ln(\tilde m\,r)+\frac{5}{6}\,\bigg)
\,\bigg\}
\,.
\label{VcNLOmassiverspace}
\end{eqnarray}  
The expression in the second line of Eq.~(\ref{VcNLOmassiverspace})
(without the subtraction of the contributions for $m\to 0$)
has been derived a long time ago by Serber and Ueling~\cite{Landau1}
for the 
effects of the electron on the QED static potential of a charged
particle. The use of
the dispersion relation representation for the light quark mass
corrections to the static potential is advantageous for
the determination of higher order terms in Rayleigh--Schr\"odinger
perturbation theory due to the universality and simplicity of the
dependence on ${\mbox{\boldmath $r$}}$ (or ${\mbox{\boldmath
$p$}}$). The remaining dispersion integration can be carried out
numerically (or if possible analytically) at the very end. We believe
that this method is actually the only feasible one to determine the
light quark mass corrections, particularly for the light quark mass
corrections in the NNLO static potential and for multiple insertions
of the NLO static potential at higher order in Rayleigh--Schr\"odinger
perturbation theory. For the determination of the dispersion relation
representation, it is a useful fact that the corrections of the light
quark mass with the (pole) mass $m$ decouple (i.e. vanish for
${\mbox{\boldmath $p$}}^2\to 0$) if we exclude it from the evolution
of the strong coupling. Up to two loops the matching relation of the
strong coupling in the two schemes, where the light quark with mass
$m$ is included and excluded in the evolution, reads~\cite{alphasmatching}
\begin{eqnarray}
\alpha_s^{(n_l)}(\mu) 
& = &
\alpha_s^{(n_l-1)}(\mu)\,
\bigg\{\,
1\,+\,
\Big(\frac{\alpha_s^{(n_l-1)}(\mu)}{\pi}\Big)\,
 \bigg[\, \frac{1}{6}\,\ln\Big(\frac{\mu^2}{m^2}\Big)
   \,\bigg]
\nonumber
\\[2mm] & &\hspace{2.5cm}
+\,\Big(\frac{\alpha_s^{(n_l-1)}(\mu)}{\pi}\Big)^2\,
 \bigg[\, \frac{7}{24} + 
     \frac{19}{24}\,\ln\Big(\frac{\mu^2}{m^2}\Big) +
     \frac{1}{36}\,\ln^2\Big(\frac{\mu^2}{m^2}\Big)
   \,\bigg]
\,\bigg\}
\,. 
\label{alphasdecoupling}
\end{eqnarray}
Thus, after rewriting the static potential in terms of
$\alpha_s^{(n_l-1)}$, 
only the subtracted dispersion integral expression remains for the
corrections coming from the finite light quark mass $m$. We have used
this useful fact to determine the dispersion relation representation
for the light quark mass corrections to the static potential at NNLO.

At NNLO the light quark mass corrections to the static potential arise
from dressing the one-loop Feynman diagram displayed in
Fig.~\ref{figVcNLOmassive} with additional gluon, ghost and light
quark lines, where the heavy quark lines in the resulting loops and
the couplings of the 
heavy quarks to gluons correspond to those of static
quarks~\cite{Fischler1}
as they are used in Heavy-Quark Effective Theory. The 
corresponding diagrams were calculated numerically
by Melles~\cite{Melles1}. Fitted approximation formulae for the NNLO
static potential with the corrections of a light quark mass in
momentum as well as configuration space representation were given in
Ref.~\cite{Melles2}. There, the results were presented 
for $n_l=1$ and using $\alpha_s^{(n_l=1)}$ as the definitions for the
strong coupling. However, these results are
somewhat cumbersome to use for bound state calculations at higher
order in Rayleigh--Schr\"odinger perturbation theory. Using the
decoupling properties of the massive light quark corrections as
explained above, 
it is straightforward to derive the following dispersion relation
representation of the light quark mass corrections to the NNLO static
potential in momentum space 
representation\footnote{
I thank M.~Melles for checking Eqs.~(\ref{deltaVcmassiveNNLOqspace})
and (\ref{VcNNLOmassiverspace}).
}
($a_s=\alpha_s^{(n_l)}(\mu)$):  
\begin{eqnarray}
\tilde V_{\mbox{\tiny c}}^{\mbox{\tiny NNLO}}({\mbox{\boldmath $p$}})
& = & 
\tilde V_{\mbox{\tiny c,massless}}^{\mbox{\tiny NNLO}}({\mbox{\boldmath
$p$}})+
\delta\tilde V_{\mbox{\tiny c,m}}^{\mbox{\tiny NNLO}}
({\mbox{\boldmath $p$}})
\,,
\\[4mm]
\tilde V_{\mbox{\tiny c,massless}}^{\mbox{\tiny NNLO}}({\mbox{\boldmath
$p$}})
& = &
\tilde V_{\mbox{\tiny c}}^{\mbox{\tiny LO}}({\mbox{\boldmath $p$}})\,
\Big(\frac{a_s}{4\,\pi}\Big)^2\,\bigg[\,
\beta_0^2\,\ln^2\Big(\frac{\mbox{\boldmath $p$}^2}{\mu^2}\Big)  
- \Big(2\,\beta_0\,a_1 +
\beta_1\Big)\,\ln\Big(\frac{\mbox{\boldmath $p$}^2}{\mu^2}\Big) 
+ a_2
\,\Bigg]
\,,
\\[4mm]
\delta\tilde V_{\mbox{\tiny c,m}}^{\mbox{\tiny NNLO}}({\mbox{\boldmath
$p$}})
& = &
\tilde V_{\mbox{\tiny c}}^{\mbox{\tiny LO}}({\mbox{\boldmath $p$}})\,
\Big(\frac{a_s}{4\,\pi}\Big)^2\,
\bigg[\,
 \frac{8}{3}\,T\,{\rm P}\Big(\frac{m^2}{{\mbox{\boldmath $p$}}^2}\Big)\,
  \Big(-\beta_0\,\ln\Big(\frac{\mbox{\boldmath $p$}^2}{\mu^2}\Big) + a_1
  \,\Big) 
\nonumber
\\[2mm] & & \hspace{2.5cm}
+ \,\Big(\frac{4}{3}\,T\Big)^2
  \,{\rm P}^2\Big(\frac{m^2}{{\mbox{\boldmath $p$}}^2}\Big)
+ \frac{76}{3}\,T\,{\rm X}\Big(\frac{m^2}{{\mbox{\boldmath $p$}}^2}\Big)
\,\bigg]
\,,
\label{deltaVcmassiveNNLOqspace}
\end{eqnarray}
where 
\begin{eqnarray}
{\rm X}\Big(\frac{m^2}{{\mbox{\boldmath $p$}}^2}\Big)
& = &
\bigg[\,
\Xi\Big(\frac{m^2}{{\mbox{\boldmath $p$}}^2}\Big)\,-\,
\bigg(\ln\frac{{\mbox{\boldmath $p$}}^2}{m^2}-\frac{161}{114}
-\frac{26}{19}\,\zeta_3
\,\bigg)
\,\Bigg]
\,,
\label{Xdef}
\\[4mm]
\Xi\Big(\frac{m^2}{{\mbox{\boldmath $p$}}^2}\Big)
& = &
2\,c_1\,{\mbox{\boldmath $p$}}^2\,
\int\limits_{c_2}^\infty\,\frac{d x}{x}\,
\frac{1}{{\mbox{\boldmath $p$}}^2+4\,m^2\,x^2} +
2\,d_1\,{\mbox{\boldmath $p$}}^2\,
\int\limits_{d_2}^\infty\,\frac{d x}{x}\,
\frac{1}{{\mbox{\boldmath $p$}}^2+4\,m^2\,x^2}
\nonumber
\\[2mm]
& = &
c_1\,\ln\Big(1+\frac{{\mbox{\boldmath $p$}}^2}{4\,c_2^2\,m^2}\Big)+
d_1\,\ln\Big(1+\frac{{\mbox{\boldmath $p$}}^2}{4\,d_2^2\,m^2}\Big)
\,,
\label{Xidispersion}
\end{eqnarray}
and the definition of ${\rm P}$ has been given in Eq.~(\ref{Pdef}).
In contrast to the approximation formulae given in Ref.~\cite{Melles2},
the contributions involving the one-loop vacuum polarization
contributions coming from the light quark with mass $m$ are exact.
The remainder, which is parametrized in terms of the function $X$,
depends on the constants $c_1$, $c_2$, $d_1$ and $d_2$. Because 
$\Xi$ vanishes for ${\mbox{\boldmath $p$}}^2\to 0$ and 
${\rm X}$ for ${\mbox{\boldmath $p$}}^2\to\infty$, the constants satisfy the 
conditions $c_1+d_1=1$ and 
$c_1\ln(4c_2^2)+d_1\ln(4d_2^2)=\frac{161}{114}+\frac{26}{19}\zeta_3$.
A useful parametrization for the constants, which fulfils these
conditions and keeps $c_2$ and $d_2$ as free parameters, is
\begin{eqnarray}
c_1 & = & \frac{\,\ln\frac{A}{d_2}\,}{\ln\frac{c_2}{d_2}}
\,,
\\[4mm]
d_1 & = & \frac{\,\ln\frac{c_2}{A}\,}{\ln\frac{c_2}{d_2}}
\,,
\\[4mm]
A & = & \exp\bigg(
\,\frac{161}{228}+\frac{13}{19}\,\zeta_3-\ln 2
\bigg)
\,.
\end{eqnarray}
For the constants $c_2$ and $d_2$ we obtain
\begin{eqnarray}
c_2 & = & 0.470 \pm 0.005
\,,
\\[4mm]
d_2 & = & 1.120 \pm 0.010
\end{eqnarray}
from fitting to the results in Eq.~(30) of Ref.~\cite{Melles2}.
We assume a 1\% uncertainty for $c_2$ and $d_2$, but we note that this
estimate is not based on a statistical analysis. The uncertainties
have been fixed by hand in an ad hoc manner taking into account the
uncertainties quoted in Ref.~\cite{Melles2} and the fact
that the numerical result shown in Eq.~(30) of Ref.~\cite{Melles2}
does not decouple exactly. We note that the uncertainties in the charm
mass effects for the bottom mass extractions caused by the errors
in $c_2$ and $d_2$ are an order of magnitude smaller than those caused
by the other approximations made in this work. We therefore ignore the
errors in $c_2$ and $d_2$ for the rest of this paper.
Finally, the NNLO light quark mass corrections to
the static potential in configuration space representation read 
($a_s=\alpha_s^{(n_l)}(\mu)$)
\begin{eqnarray}
\lefteqn{
\delta V_{\mbox{\tiny c,m}}^{\mbox{\tiny NNLO}}({\mbox{\boldmath
$r$}})
%& = &
%\int\!\frac{d^3{\mbox{\boldmath $p$}}}{(2\pi)^3}\,\,
%\delta\tilde V_{\mbox{\tiny c,m}}^{\mbox{\tiny NNLO}}({\mbox{\boldmath
%$p$}})
%\,\exp( i {\mbox{\boldmath $p$}} {\mbox{\boldmath $r$}} )
%\nonumber\\[2mm]
\, = \,
V_{\mbox{\tiny c}}^{\mbox{\tiny LO}}({\mbox{\boldmath $r$}})\,
\Big(\frac{a_s}{3\,\pi}\Big)^2\,\bigg\{\,
}
\nonumber 
\\[2mm] & & \hspace{0.3cm}
\bigg[-\frac{3}{2}\,\int\limits_1^\infty dx\,f(x)\,e^{-2 m r x}\,
\bigg(\beta_0\,\Big(\ln\frac{4m^2x^2}{\mu^2}
 -{\rm Ei}(2\,m\,r\,x)-{\rm Ei}(-2\,m\,r\,x)\Big) - a_1
\bigg)
\nonumber 
\\[2mm] & & \hspace{0.6cm}
+\,3\,\bigg(\ln(\tilde m\,r)+\frac{5}{6}\,\bigg)\,
 \bigg(\beta_0\,\ln(\tilde \mu\,r)+\frac{a_1}{2}\bigg) 
+\beta_0\,\frac{\pi^2}{4}
\,\bigg]
\nonumber 
\\[2mm] & & 
-\bigg[\,
\int\limits_1^\infty dx\,f(x)\,e^{-2 m r x}\,
\bigg(\,\frac{5}{3}+\frac{1}{x^2}\bigg(
1+\frac{1}{2\,x}\,\sqrt{x^2-1}\,(1+2x^2)\,
\ln\Big(\frac{x-\sqrt{x^2-1}}{x+\sqrt{x^2-1}}\Big)
\bigg)
\,\bigg)
\nonumber
\\[2mm] & & \hspace{0.6cm}
+\int\limits_1^\infty dx\,f(x)\,e^{-2 m r x}\,
\bigg( \ln(4x^2)
 -{\rm Ei}(2\,m\,x\,r)-{\rm Ei}(-2\,m\,x\,r) - \frac{5}{3}
\bigg)
\nonumber
\\[2mm] & & \hspace{0.6cm}
+\bigg(\ln(\tilde m\,r)+\frac{5}{6}\,\bigg)^2 + \frac{\pi^2}{12}
\,\bigg]
\nonumber 
\\[2mm] & & 
+\bigg[\,\frac{57}{4}\,\bigg(\,
c_1\,\Gamma(0,2\,c_2\,m\,r) + d_1\,\Gamma(0,2\,d_2\,m\,r)
+ \ln(\tilde m\,r) + \frac{161}{228} + \frac{13}{19}\,\zeta_3
\,\bigg)
\,\bigg]
\,\bigg\}
\,,
\label{VcNNLOmassiverspace}
\end{eqnarray}  
where ${\rm Ei}$ is the exponential-integral function and
$\Gamma$ the incomplete gamma function, where
$\Gamma(0,c\,a) = \int_c^\infty e^{-a x}/x\,dx$. 
For the
exponential-integral function the relation 
${\rm Ei}(z)+{\rm Ei}(-z)=P\int_0^\infty 2t\,e^{(1+t)z}/(1-t^2)\,dt$,
where $P$ refers to the principal value prescription,
is quite useful. We note that of the three terms in the brackets on
the RHS of Eq.~(\ref{VcNNLOmassiverspace}), each one vanishes
individually for $m\to 0$. The first two terms incorporate all the
light quark mass corrections that can be obtained from insertions of
one-loop light quark vacuum polarizations.  

%It is instructive to consider the two limits $m\to 0$ and
%$m\to\infty$, which correspond to the situation where the distance
%$r$ between the heavy quarks is much smaller or larger than the
%inverse light quark mass. (For the latter case we ignore the issues of
%confinement and non-perturbative effects.) In the limit that $m$ is
%large the light quark mass corrections to the static potential in
%configuration space representation read
%($a_s=\alpha_s^{(n_l)}(\mu)$)
%\begin{eqnarray}
%\delta V_{\mbox{\tiny c,m}}^{\mbox{\tiny NLO}}({\mbox{\boldmath
%$r$}}) +
%\delta V_{\mbox{\tiny c,m}}^{\mbox{\tiny NNLO}}({\mbox{\boldmath
%$r$}}) 
%& \stackrel{m\to\infty}{\longrightarrow} & 
%\,.
%\end{eqnarray}
%The terms that are not exponentially suppressed contain large
%logarithms of $m\,r$ that indicate that in this limit it is more
%appropriate to exclude the massive ``light'' quark from the running of
%the strong coupling (i.e. to integrate it out). If the static
%potential is expressed in terms of $\alpha_s^{(n_l-1)}$ all terms that
%are not exponentially suppressed are absorbed into the massless static
%potential, rendering the decoupling of the light quark mass corrections
%explicit. The second case where $m$ is small is discussed in the next
%subsection. 

%
%
\par
%\vspace{0.5cm}
%
\subsection{Small-Distance Limit and Linear Light Quark Mass Terms}
\label{subsectionpotentialsmalldistance}
It is instructive to consider the light quark mass corrections
to the static potential for the case where the distance between the
heavy quark is much smaller then $1/m$, i.e. in the limit $m\ll 1/r$.
In this case the light quark mass corrections to the static potential
read
($a_s=\alpha_s^{(n_l)}(\mu)$):
\begin{eqnarray}
\lefteqn{
\delta V_{\mbox{\tiny c,m}}^{\mbox{\tiny NLO}}({\mbox{\boldmath
$r$}}) +
\delta V_{\mbox{\tiny c,m}}^{\mbox{\tiny NNLO}}({\mbox{\boldmath
$r$}}) 
%\, \stackrel{m\ll\frac{1}{r}}{\longrightarrow} \,
}
\nonumber
\\[2mm] & \stackrel{m\ll 1/r}{\longrightarrow} &
-\,2\,C_F\,\Big(\frac{a_s}{\pi}\Big)^2\,\frac{\pi^2}{8}\,m\,
\bigg\{
1+2\,\Big(\frac{a_s}{4\pi}\Big)\,
\bigg[\,a_1+
\beta_0\,\bigg(\ln\frac{\mu^2}{m^2}+3-4\ln 2
\bigg)
\nonumber
\\[2mm] & & \hspace{3cm}
+\,\frac{4}{45}\,\bigg(31-30\ln 2\bigg)
+\frac{76}{3\pi}\,\bigg(c_1\,c_2+d_1\,d_2\bigg)
\,\bigg]
\bigg\}
\, + \, {\cal{O}}(m^2\,r)
\nonumber
\\[2mm] & = &
-\,2\,C_F\,\Big(\frac{a_s}{\pi}\Big)^2\,\frac{\pi^2}{8}\,m\,
\bigg\{
1+2\,\Big(\frac{a_s}{4\pi}\Big)\,
\bigg[\,
\beta_0\,\bigg(\ln\frac{\mu^2}{m^2}-4\ln 2+\frac{14}{3}
\bigg)
\nonumber
\\[2mm] & & \hspace{3cm}
-\,\frac{4}{3}\,\bigg(\frac{59}{15}+2\ln 2\bigg)
+\frac{76}{3\pi}\,\bigg(c_1\,c_2+d_1\,d_2\bigg)
\,\bigg]
\bigg\}
\, + \, {\cal{O}}(m^2\,r)
%\nonumber
%\\[2mm] & = &
%-\,2\,C_F\,\Big(\frac{a_s}{\pi}\Big)^2\,\frac{\pi^2}{8}\,
%\overline m(\overline m)\,
%\bigg\{
%1+2\,\Big(\frac{a_s}{4\pi}\Big)\,
%\bigg[\,
%\beta_0\,\bigg(\ln\frac{\mu^2}{\overline m^2}-4\ln 2+\frac{14}{3}
%\bigg)
%\nonumber
%\\[2mm] & & \hspace{4cm}
%-\frac{4}{3}\,\bigg(\frac{29}{15}+2\ln 2\bigg)
%+\frac{76}{3\pi}\,\bigg(c_1\,c_2+d_1\,d_2\bigg)
%\,\bigg]
%\bigg\}
%\, + \, {\cal{O}}(\overline m^2\,r)
\,,
\label{deltaVcrspacesmallm}
\end{eqnarray}
where we have used the relation $a_1=\frac{5}{3}\beta_0-8$ in the
second line of Eq.~(\ref{deltaVcrspacesmallm}).
%In the third line we have written the result in terms of the
%light quark $\overline{\mbox{MS}}$ mass, 
%$m=\overline m(\overline m)[1+\frac{4}{3}(\frac{a_s}{\pi})]$.
The order $\alpha_s^3\beta_0$ term in the second line has already been
derived in Ref.~\cite{Hoang4}. 
The leading term in the expansion, which is linear in the light quark
mass\footnote{
In a non-relativistic effective field theory for heavy quarks, in
which the light quark is also integrated out, the
linear light quark mass term corresponds to a residual mass term 
operator~\cite{Hoang4}. This term accounts for the difference in the
analyticity structure at low energies of the effective theory in
which the light 
quark is integrated out and the theory in which the light quark is not
integrated out. This also means that the heavy quark pole masses in both
theories differ by the coefficient of the residual mass
operator~\cite{Wise1}, see also the section on quark masses in
Ref.~\cite{PDG}. 
}, has a 
number of remarkable features: it is independent of $r$, enhanced by a
factor of $\pi^2$, and its characteristic scale is the light quark
mass $m$ rather than the inverse of the distance $r$. We emphasize 
that the linear light quark mass term cannot be obtained by first
expanding Eqs.~(\ref{VcNLOmassiverspace}) and
(\ref{VcNNLOmassiverspace}) in the 
light quark mass and carrying out the dispersion 
integrations (over the variable $x$) afterwards, because it contains
contributions coming from large values of $x$, i.e. from momenta of
the order of the light quark mass or smaller (see
Eq.~(\ref{Pidispersion})). 
This means that the linear mass term is of
infrared origin, which explains that it represents a non-analytic
expression in $m^2$. In fact, the linear light quark mass term is
in complete analogy with the linear gluon mass
term that would arise if we were give the gluon a infinitesimally
small fictitious mass $\lambda$. The linear light quark mass term at
order $\alpha_s^n$ directly corresponds to the linear gluon 
mass term that would be obtained at order $\alpha_s^{n-1}$. (This can
be seen from Eq.~(\ref{Pidispersion}), using the change of variable
$\lambda^2=4\,m^2\,x^2$.) Just as for the linear gluon mass term,
the existence of the linear light quark mass term in
Eq.~(\ref{deltaVcrspacesmallm}) indicates that the static potential 
is, like the heavy quark pole mass, linearly sensitive to small
momenta. Thus, the static potential also contains an ($r$-independent)
ambiguity of order $\Lambda_{\rm QCD}$~\cite{Ligeti1}. 
This also explains why the perturbative series for the linear quark
mass term is completely out of control for all light
quarks, although the perturbative expansion of the static potential is
most trustworthy in the small distance limit. For the charm quark
the linear mass term amounts to $60$~MeV at order $\alpha_s^2$ and to
$86$~MeV at order $\alpha_s^3$ for $\overline m(\overline
m)=\mu=1.5$~GeV, $a_s=0.36$ and $n_l=4$ indicating a very bad higher
order behaviour\footnote{
This statement about the higher order behaviour of the linear charm
quark mass terms does not depend on which convention is used for the
strong coupling. 
}. (See Sec.~\ref{sectionpolemsbarmass} for a more
detailed discussion.) 

Fortunately, a linear sensitivity to small momenta and the
corresponding ambiguity of order $\Lambda_{\rm QCD}$, is not
contained in the total static energy~\cite{Hoang5,Beneke4,Uraltsev1}
\begin{eqnarray}
E_{\rm stat} & = &
2 M_{\mbox{\tiny Q}}^{\mbox{\tiny pole}} +  
V_{\mbox{\tiny c}}({\mbox{\boldmath
$r$}}) 
\,.
\label{totalstaticenergy}
\end{eqnarray}
In Refs.~\cite{Hoang5,Uraltsev1} this was proved at one loop. The
proof in Ref.~\cite{Beneke4} demonstrates this also at the two-loop
level, which is 
sufficient for the purpose of this work. A similar proof for inclusive
semileptonic B meson decays can be found in Ref.~\cite{Zakharov1}.
Equation~(\ref{totalstaticenergy}) means that---if we are actually
allowed to expand the static 
potential in the light quark mass---the linear light quark mass
corrections are cancelled, if a short-distance mass
definition is employed to describe the non-relativistic 
heavy-quark--antiquark dynamics. This remarkable feature allows us to 
determine the linear light quark mass corrections in the relation
between the heavy quark pole and a short-distance mass, such as the 
$\overline{\mbox{MS}}$ one, up to order $\alpha_s^3$:
\begin{eqnarray}
\bigg[\,
M_{\mbox{\tiny Q}}^{\mbox{\tiny pole}} -
 \overline M_{\mbox{\tiny Q}}(\overline M_{\mbox{\tiny Q}})
\,\bigg]_{\mbox{\tiny linear in $m$}}^
  {{\cal{O}}(\alpha_s^2,\alpha_s^3)} & = &
-\,\frac{1}{2}\,\bigg[\,
\delta V_{\mbox{\tiny c,m}}^{\mbox{\tiny NLO}}({\mbox{\boldmath
$r$}}) +
\delta V_{\mbox{\tiny c,m}}^{\mbox{\tiny NNLO}}({\mbox{\boldmath
$r$}}) 
\,\bigg]_{\mbox{\tiny linear in $m$}}
\,.
\label{linearmasspoleshortdist}
\end{eqnarray}
The order $\alpha_s^2$ linear light quark mass term of the static
potential shown in Eq.~(\ref{deltaVcrspacesmallm})
confirms the order $\alpha_s^2$ linear light quark mass term in the
heavy quark pole--$\overline{\mbox{MS}}$ mass relation  
calculated earlier in
Ref.~\cite{Broadhurst1}. The order $\alpha_s^3$ term that can be derived from
Eqs.~(\ref{deltaVcrspacesmallm}) and (\ref{linearmasspoleshortdist})
has to our knowledge not been obtained before.  
A detailed discussion on the linear light quark mass corrections in
the relation of the heavy quark pole and $\overline{\mbox{MS}}$ masses 
is given in Sec.~\ref{sectionpolemsbarmass}.

We are now in a position to substantiate the scenario for the light
quark mass corrections that we have pictured in
the introduction to this work for the extraction of bottom quark
masses from $\Upsilon$ mesons at intermediate orders of perturbation
theory. For the light quarks up, down and strange, the badly behaved
linear mass terms indeed cancel in an extraction of the bottom 
$\overline{\mbox{MS}}$ mass, because their masses are much smaller
than the dynamical scales 
$M_{\mbox{\tiny b}}$ and 
$\langle1/{\mbox{\boldmath $r$}}\rangle\approx 
M_{\mbox{\tiny b}}\alpha_s$ 
and one can expand in them everywhere, in the same way as in the
infinitesimally small gluon mass. For the charm quark, however, the 
expansion in its mass is allowed in the bottom quark
pole--$\overline{\mbox{MS}}$ mass relation (because it is only governed
by the scale $M_{\mbox{\tiny Q}}$), but not in quantities involving the
bottom--antibottom non-relativistic dynamics described 
through the Schr\"odinger equation~(\ref{NNLOSchroedinger}). This
means that, for an extraction of the bottom $\overline{\mbox{MS}}$ mass 
from $\Upsilon$ mesons, at least some part of the large linear charm mass
corrections remains uncancelled. Regarding
the very bad perturbative convergence for the linear charm mass terms in
Eq.~(\ref{linearmasspoleshortdist}), it is now natural to ask the
question of how the remaining linear charm mass corrections will behave
at higher orders. If they were to behave similarly to the linear charm
mass corrections in Eq.~(\ref{deltaVcrspacesmallm}), then the precision
of the bottom $\overline{\mbox{MS}}$ mass would be forever limited to
several tens of MeV. Fortunately, this is not the case because the
contributions from low momenta increase at higher orders of
perturbation theory. This means that the small momentum region, where
the linear mass terms originate, dominates over the larger
momenta. Thus, although it is formally impossible to expand in the
charm mass divided by the inverse Bohr radius, the coefficients of
higher powers of this ratio will decrease at higher orders, making the
linear term the dominant one. 
In Sec.~\ref{subsectionmsbar1Smassexamination} we will demonstrate
this interesting feature
explicitly. For the extraction of the bottom $\overline{\mbox{MS}}$
mass this means that the charm mass corrections are well under control
and calculable.  

\par
\vspace{0.5cm}
\section[Light Quark Mass Corrections in the 
         Heavy Quark Pole--$\overline{\mbox{MS}}$ Mass Relation]
        {Light Quark Mass Corrections in the \\
         Heavy Quark Pole--$\overline{\mbox{MS}}$ Mass Relation}
\label{sectionpolemsbarmass}
The perturbative relation between the pole and the
$\overline{\mbox{MS}}$ mass definition of a heavy quark is a main
ingredient in the determination of the bottom
$\overline{\mbox{MS}}$ mass from data on the $\Upsilon$ mesons. The
mass of a light quark leads to corrections in this relation starting
at order $\alpha_s^2$, originating from the insertion of a light quark
vacuum polarization into the one-loop gluon line. For a proper
determination of the bottom $\overline{\mbox{MS}}$ mass using
calculations for the non-relativistic bottom--antibottom quark dynamics
at NNLO (see Eq.~(\ref{NNLOSchroedinger})), we need the bottom
pole--$\overline{\mbox{MS}}$ mass relation at order
$\alpha_s^3$. Here, ``proper'' means that the artificially
large perturbative contributions coming from a linear sensitivity to
small momenta are eliminated consistently in each order\footnote{
The relation between perturbative orders (LO, NLO, etc.) in the
non-relativistic power-counting scheme and perturbative orders
($\alpha_s$, $\alpha_s^2$, etc., or one-loop, two-loop, etc.) in
quantities unrelated  
to non-relativistic quark--antiquark physics, such as the
pole--$\overline{\mbox{MS}}$ mass relation, can also be seen from
Eqs.~(\ref{deltaVcrspacesmallm}) and
(\ref{linearmasspoleshortdist}). The order $\alpha_s^2$ linear light 
quark mass term is obtained from the static potential at NLO, the
order $\alpha_s^3$ linear light quark mass term is obtained from the 
static potential at NNLO, etc.
}.
For massless light quarks the heavy quark pole--$\overline{\mbox{MS}}$
mass relation is known analytically at order
$\alpha_s^3$. The order $\alpha_s^2$ corrections have been determined
in Ref.~\cite{Broadhurst1} and the order $\alpha_s^3$ corrections in
Ref.~\cite{Melnikov2}. A numerical determination of the order $\alpha_s^3$
corrections can be found in Ref.~\cite{Steinhauser1}. In
Ref.~\cite{Broadhurst1} the order $\alpha_s^2$
corrections coming from the mass of a light quark have been determined
analytically for all values of the light quark mass. The light quark
mass corrections at order $\alpha_s^3$ 
are only known in the linear mass approximation and have been
determined in Sec.~\ref{subsectionpotentialsmalldistance} of this
work. At order $\alpha_s^3$ the relation between the pole and the
$\overline{\mbox{MS}}$ mass of a heavy quark for $n_l$ light quarks, of
which one has a finite pole mass $m$, reads 
($\bar a_s=\alpha_s^{(n_l)}(\overline M_{\mbox{\tiny Q}})$): 
\begin{eqnarray}
M_{\mbox{\tiny Q}}^{\mbox{\tiny pole}} & = & 
\overline M_{\mbox{\tiny Q}}(\overline M_{\mbox{\tiny Q}})\,
\bigg\{\, 1 
+ \epsilon\,\Big[\,\delta^{(1)}(\bar a_s)\,\Big]
+ \epsilon^2\,\Big[\,\delta^{(2)}_{\mbox{\tiny massless}}(\bar a_s)
       +\delta^{(2)}_{\mbox{\tiny massive}}
   (m,\overline M_{\mbox{\tiny Q}}(\overline M_{\mbox{\tiny Q}}),\bar a_s) 
   \,\Big]
\nonumber
\\[2mm]
&  & \hspace{2.0cm} 
+ \,\epsilon^3\,\Big[\,\delta^{(3)}_{\mbox{\tiny massless}}(\bar a_s)
       +\delta^{(3)}_{\mbox{\tiny massive}}
   (m,\overline M_{\mbox{\tiny Q}}(\overline M_{\mbox{\tiny Q}}),\bar a_s)
   \,\Big]
+ \ldots
\,\bigg\}
%\nonumber
%\\[2mm] & = & 
%\overline M_{\mbox{\tiny Q}}(\overline M_{\mbox{\tiny Q}})\,
%\bigg\{\, 1 
%+ \epsilon\,\Big[\,\delta^{(1)}(\bar a_s)\,\Big]
%+ \epsilon^2\,\Big[\,\delta^{(2)}_{\mbox{\tiny massless}}(\bar a_s)
%       +\delta^{(2)}_{\mbox{\tiny massive}}
%    (\overline m(\overline m),
%     \overline M_{\mbox{\tiny Q}}(\overline M_{\mbox{\tiny Q}}),\bar a_s)
%    \,\Big]
%\nonumber
%\\[2mm]
%&  & \hspace{0.5cm} 
%+ \,\epsilon^3\,\Big[\,\delta^{(3)}_{\mbox{\tiny massless}}(\bar a_s)
%       +\delta^{(3)}_{\mbox{\tiny massive}}
%    (\overline m(\overline m),
%       \overline M_{\mbox{\tiny Q}}(\overline M_{\mbox{\tiny Q}}),\bar a_s)
%\nonumber
%\\[2mm]
%&  & \hspace{0.5cm} 
%       +\,\delta^{(1)}(\bar a_s)\,
%        \delta^{(2)}_{\mbox{\tiny massive}}
%         (\overline m(\overline m),
%       \overline M_{\mbox{\tiny Q}}(\overline M_{\mbox{\tiny Q}}),\bar a_s)
%       \,\Big]
%+ \ldots
%\,\bigg\}
%\nonumber
%\\[2mm] & = &
%\overline M_{\mbox{\tiny Q}}(\overline M_{\mbox{\tiny Q}})\,
%\bigg\{\, 1 
%+ \epsilon\,\bar a_s\,\bigg[\,\frac{4}{3\,\pi} \,\bigg]  
%+ \epsilon^2\,\bar a_s^2\,\bigg[\, 1.362 - 0.1055\,n_l 
%+ \frac{1}{6}\,\overline m(\overline m)  \,\bigg]
%\nonumber
%\\[2mm] & & \hspace{0.5cm}
%+\, \epsilon^3\,\bar a_s^3\,\bigg[\,6.140 - 0.8597\,n_l + 0.02105\,n_l^2+
%\nonumber
%\\[2mm] & & \hspace{1cm}
%\overline m(\overline m)\,\bigg(\,
%0.4353 + 
%  0.2139\,(c_1\,c_2+d_1\,d_2) - 0.03349\,n_l +
%  \frac{\beta_0}{6\,\pi}\,\ln\Big(\frac{\mu}
%{\overline M_{\mbox{\tiny Q}}}\Big)
%  \,\bigg)
% \,\bigg]    
%\,\bigg\}
\,,
\label{polemsbarthreeloops}
\end{eqnarray}
where the corrections for massless light quarks read
\begin{eqnarray}
\delta^{(1)}
& = &
\frac{4}{3}\,\Big(\frac{\bar a_s}{\pi}\Big)
\,,
\label{delta1massless}
\\[4mm]
\delta^{(2)}_{\mbox{\tiny massless}}
& = &
\Big(\frac{\bar a_s}{\pi}\Big)^2\,
\Big(\,13.443 - 1.0414\, n_l\,\Big)
\,,
\label{delta2massless}
\\[4mm]
\delta^{(3)}_{\mbox{\tiny massless}}
& = & 
\Big(\frac{\bar a_s}{\pi}\Big)^3\,
\Big(\,190.4 -  26.66\,n_l + 0.6527\,n_l^2\,\Big)
\,.
\label{delta3massless}
\end{eqnarray}
The numbers displayed in Eq.~(\ref{delta3massless}) have been taken
from Ref.~\cite{Melnikov2}.
The light quark mass corrections in the linear approximation read
\begin{eqnarray}
\delta^{(2)}_{\mbox{\tiny massive}}
& = &
\frac{\bar a_s^2}{6}\,\frac{m}
 {\overline M_{\mbox{\tiny Q}}(\overline M_{\mbox{\tiny Q}})}
\,,
\label{delta2massive}
\\[4mm]
\delta^{(3)}_{\mbox{\tiny massive}}
& = &
\frac{\bar a_s^2}{12}\,\Big(\frac{\bar a_s}{\pi}\Big)\,
\frac{m}
{\overline M_{\mbox{\tiny Q}}}\,
\bigg[\,
\beta_0\,\bigg(\ln\frac{\overline M_{\mbox{\tiny Q}}^2}{m^2}
-4\ln 2+\frac{14}{3}
\bigg)
\nonumber
\\[2mm] & & \hspace{2cm}
-\frac{4}{3}\,\bigg(\frac{59}{15}+2\ln 2\bigg)
+\frac{76}{3\,\pi}\,\bigg(c_1\,c_2+d_1\,d_2\bigg)
\,\bigg]
\,.
\label{delta3massive}
\end{eqnarray}
If the $\overline{\mbox{MS}}$ definition is used for the
charm quark mass, 
$m=\overline m(\overline m)[1+\frac{4}{3}(\frac{a_s}{\pi})]$,
we have to apply the replacement
\begin{eqnarray}
\delta^{(2)}_{\mbox{\tiny massive}}
   (m,\overline M_{\mbox{\tiny Q}}(\overline M_{\mbox{\tiny Q}}),\bar a_s)
& \longrightarrow &
\delta^{(2)}_{\mbox{\tiny massive}}
    (\overline m(\overline m),
     \overline M_{\mbox{\tiny Q}}(\overline M_{\mbox{\tiny Q}}),\bar a_s)
+ \epsilon\,\delta^{(1)}(\bar a_s)\,
        \delta^{(2)}_{\mbox{\tiny massive}}
         (\overline m(\overline m),
       \overline M_{\mbox{\tiny Q}}(\overline M_{\mbox{\tiny Q}}),\bar a_s)
\nonumber
\\
\end{eqnarray}
in Eq.~(\ref{polemsbarthreeloops}).
The generalization to the case where more than one of the light quarks
has a finite mass 
is straightforward. The parameter $\epsilon$ ($\epsilon=1$) is an
auxiliary variable that facilitates the determination of the heavy
quark $\overline{\mbox{MS}}$--1S mass relation in
Sec.~\ref{sectionmsbar1Smass}, where the non-relativistic power
counting and the ordinary 
power counting via the number of loops have to be combined in such a
way that the linear sensitive 
contributions cancel systematically in each order. 
Equation~(\ref{polemsbarthreeloops}) is the basic equation that we
will use 
for the determination of the bottom $\overline{\mbox{MS}}$ mass in
this work. For the corrections coming from the charm quark mass, we
will only use the linear mass approximation. We show in this
section that this approximation should be sufficient to give a
description of the charm mass effects with a relative uncertainty of
10\%. We also demonstrate 
that the linear mass approximation becomes better at higher orders by
explicitly comparing the linear mass terms with known exact results. 

At order $\alpha_s^2$ the light quark mass corrections can be
represented as an integral over the one-loop gluon virtuality times
the subtracted light quark vacuum polarization 
${\rm P}$~\cite{Broadhurst1}, in analogy to the   
NLO light quark mass corrections to the static potential 
in Eq.~(\ref{deltaVcNLOmomentumspace})
($r=m/\overline M_{\mbox{\tiny Q}}$):  
\begin{eqnarray}
\delta^{(2),\mbox{\tiny full}}_{\mbox{\tiny massive}} 
& = &
\frac{4}{3}\,\Big(\frac{a_s}{\pi}\Big)^2\,
\frac{1}{24}\,
\int\limits_0^\infty\!
\frac{d q^2}{\overline M_{\mbox{\tiny Q}}^2}\,
\bigg[\,\frac{1}{2}\,
\frac{q^2}{\overline M_{\mbox{\tiny Q}}^2}
+\bigg(1-\frac{1}{2}\frac{q^2}{\overline M_{\mbox{\tiny Q}}^2}\bigg)\,
\bigg(1+4\,\frac{\overline M_{\mbox{\tiny Q}}^2}{q^2}\bigg)^{\frac{1}{2}}
\,\bigg]\,{\rm P}\Big(\frac{m^2}{q^2}\Big)
\nonumber
\\[2mm]
& = & \frac{4}{3}\,\Big(\frac{a_s}{\pi}\Big)^2\,
\frac{1}{4}\,
\bigg[\,
 \ln^2(r) + \frac{\pi^2}{6} - \bigg(\,\ln(r) + \frac{3}{2}\,\bigg)\,r^2 
\nonumber
\\[2mm] & & \hspace{1.5cm} +\,
(1 + r)\,(1 + r^3)\,\bigg(\Li2(-r) - \frac{1}{2}\,\ln^2(r) + \ln(r)\ln(1+r) + 
            \frac{\pi^2}{6}\bigg)
\nonumber
\\[2mm] & & \hspace{1.5cm} +\,
(1 - r)\,(1 - r^3)\,\bigg(\Li2(r) - \frac{1}{2}\,\ln^2(r) + \ln(r)\ln(1-r) - 
            \frac{\pi^2}{3}\bigg)
\,\bigg]
\nonumber
%\label{alphas2deltamanalytic}
\\[2mm]
& = & \frac{4}{3}\,\Big(\frac{a_s}{\pi}\Big)^2\,
\bigg[\,
\frac{\pi^2}{8}\,r - \frac{3}{4}\,r^2 + 
\frac{\pi^2}{8}\,r^3 - 
(\frac{1}{4}\,\ln^2(r)-\frac{13}{24}\,\ln(r) + 
 \frac{\pi^2}{24} + \frac{151}{288})\,r^4 
\nonumber
\\[2mm] & & \hspace{1.5cm} -\,
\sum\limits_{n=3}^\infty\Big(2\,{\rm F}(n)\,
\ln(r)+{\rm F}^\prime(n)\Big)\,r^{2n}
\,\bigg]
%\qquad \bigg(r\equiv\frac{m}{\overline M_{\mbox{\tiny Q}}}\bigg)
\,,
\label{alphas2deltam}
%\label{alphas2deltamanalyticexpanded}
\end{eqnarray} 
where $\rm P$ is defined in Eq.~(\ref{Pdef}) and
\begin{eqnarray}
{\rm F}(n) & \equiv & \frac{3\,(n-1)}{4\,n\,(n-2)\,(2n-1)\,(2n-3)}
\,,
\\[4mm]
{\rm F}^\prime(n) & = & \frac{d}{d n}\,F(n)
\,.
\end{eqnarray}
Equation~(\ref{alphas2deltam}) is ultraviolet-finite owing to the
subtraction of the massless vacuum polarization in the function P.
From the first line of Eq.~(\ref{alphas2deltam}) it is straightforward
to see that in 
the limit $m\ll M_{\mbox{\tiny Q}}$ two momentum regions 
contribute to the integral:
$q\sim m$ and $q\sim M_{\mbox{\tiny Q}}$. Thus, the expansion in 
terms of $m/M_{\mbox{\tiny Q}}$
cannot be obtained by naively Taylor-expanding the momentum integral
in the first line of Eq.~(\ref{alphas2deltam}),
since this would only extract the contributions of the high momentum
region $q\sim M_{\mbox{\tiny Q}}$. The linear light quark mass term 
and all the other
terms that are non-analytic in $m^2$, such as $m^3$ and 
$m^{2n}\,\ln m$, originate from the low momentum region $q\sim m$. To
obtain the linear light quark mass term directly, we therefore have to
carry out a non-relativistic expansion in $q^2/M_{\mbox{\tiny Q}}^2$:
\begin{eqnarray}
\delta^{(2)}_{\mbox{\tiny massive}}
& = &
\frac{4}{3}\,\Big(\frac{a_s}{\pi}\Big)^2\,
\frac{1}{24}\,
\int\limits_0^\infty\!\frac{d q^2}{\overline M_{\mbox{\tiny Q}}^2}\,
\bigg[\,2\,\frac{\overline M_{\mbox{\tiny Q}}}{q}
\,\bigg]\,{\rm P}\Big(\frac{m^2}{q^2}\Big)
\label{alphas2deltalinearm}
\\[2mm] 
& = & 
\frac{1}{6}\,a_s^2\,\frac{m}{\overline M_{\mbox{\tiny Q}}}
\label{alphas2deltalinarmanalytic}
\,.
\end{eqnarray}
We emphasize that the light quark vacuum polarization has to be fully
taken into account without any expansion. Comparing
Eq.~(\ref{alphas2deltalinearm}) with Eq.~(\ref{VcNLOmassiverspace}), we
also see the equality to $-1/2$ of the static potential at
zero distance, Eq.~(\ref{linearmasspoleshortdist}). This explicitly
demonstrates the cancellation of the linear light quark mass
terms in the total static energy~(\ref{totalstaticenergy}) at
order $\alpha_s^2$ for the case where the expansion in $m$ is allowed.
In Table~\ref{tabalphas2deltam} the size of the order $\alpha_s^2$
light quark mass corrections is displayed for the full result and the
linear mass approximation for $m/M_{\mbox{\tiny Q}}$ between $0.1$ and
$0.5$. The ratio of the light quark mass terms versus the full result
is also shown.
\begin{table}[tbp]  % tabalphas2deltam
\vskip 7mm
\begin{center}
\begin{tabular}{|c||c|c|c|c|c|} \hline
$m/\overline M_{\mbox{\tiny Q}}$ 
 & $0.1$ & $0.2$ & $0.3$ & $0.4$ &$0.5$ \\ \hline
$\delta^{(2),\mbox{\tiny full}}_{\mbox{\tiny massive}}/
[\frac{4}{3}(\frac{a_s}{\pi})^2]$
 & $0.117$ & $0.223$ & $0.320$ & $0.411$ &$0.496$ \\ \hline
$\delta^{(2)}_{\mbox{\tiny massive}}/
[\frac{4}{3}(\frac{a_s}{\pi})^2]$
 & $0.123$ & $0.247$ & $0.370$ & $0.493$ & $0.617$  \\ \hline
$\delta^{(2)}_{\mbox{\tiny massive}}/
\delta^{(2),\mbox{\tiny full}}_{\mbox{\tiny massive}}$
 & $1.06$ & $1.11$ & $1.16$ & $1.20$ & $1.24$ \\ \hline
\end{tabular}
\caption{\label{tabalphas2deltam}
Comparison of the full result (second line) and the linear mass
approximation (third line) for
the order $\alpha_s^2$ light quark mass corrections in the heavy quark
pole--$\overline{\mbox{MS}}$ 
mass relation for various values of the ratio 
$m/\overline M_{\mbox{\tiny Q}}$.
The ratio of linear mass approximation to full result is shown in the
fourth line.
}
\end{center}
\vskip 3mm
\end{table}
For $m/\overline M_{\mbox{\tiny Q}}=0.3$ ($0.5$) 
the linear mass terms account for  $116$\% ($124$\%)
of the full result. We see that the linear mass approximation is
remarkably good, even for rather large values of 
$m/\overline M_{\mbox{\tiny Q}}$, 
owing to the $\pi^2$ enhancement factor. 

As already mentioned before, the full form of the order $\alpha_s^3$
light quark mass corrections is not yet known. However, we can compare
the full result and the linear mass approximation for the corrections that
can be obtained from insertions of one-loop light quark vacuum
polarizations into the one-loop gluon line. For illustration,
in Eq.~(\ref{deltaVcmassiveNNLOqspace}) the light quark mass
corrections that can be obtained from insertions of one-loop light
quark vacuum polarizations correspond to the terms that involve
the function $\rm P$.
(In $\delta^{(3)}_{\mbox{\tiny massive}}$, Eq.~(\ref{delta3massive}),
the order $\alpha_s^3$ linear light quark mass corrections that can be 
obtained from one-loop light quark vacuum polarization insertions
correspond to the terms that remain after setting
$c_1=c_2=d_1=d_2=0$.) 
In the relation
between the heavy quark pole and the $\overline{\mbox{MS}}$ mass, the
light quark mass vacuum polarization corrections at order $\alpha_s^3$
read 
\begin{eqnarray}
\delta^{(3),\mbox{\tiny vac,full}}_{\mbox{\tiny massive}}
& = &
\frac{4}{3}\,\Big(\frac{a_s}{\pi}\Big)^3\,
\frac{1}{48}\,
\int\limits_0^\infty\!
\frac{d q^2}{\overline M_{\mbox{\tiny Q}}^2}\,
\bigg[\,\frac{1}{2}\,\frac{q^2}{\overline M_{\mbox{\tiny Q}}^2}
+\bigg(1-\frac{1}{2}\frac{q^2}{\overline M_{\mbox{\tiny Q}}^2}\bigg)\,
\bigg(1+4\,\frac{\overline M_{\mbox{\tiny Q}}^2}{q^2}\bigg)^{\frac{1}{2}}
\,\bigg]
\nonumber
\\[2mm]
& & \hspace{0.5cm} \times\,
\bigg[\,
{\rm P}\Big(\frac{m^2}{q^2}\Big)\,
\bigg(-\beta_0\,\ln\Big(\frac{q^2}{\mu^2}\Big) + a_1 \,\bigg)
+ \frac{1}{3}\,{\rm P}^2\Big(\frac{m^2}{q^2}\Big)
\,\bigg]
\nonumber
\\[2mm]
& = & 
\frac{4}{3}\,
\Big(\frac{a_s}{\pi}\Big)^3\,
\frac{1}{12}\,
\bigg[\,
\beta_0\,\bigg(
  \ln\Big(\frac{\mu^2}{\overline M_{\mbox{\tiny Q}}^2}\Big)\,
    {\rm f}_{\rm P}^{(0)}\Big(\frac{m}{\overline M_{\mbox{\tiny Q}}}\Big) - 
    {\rm f}_{\rm P}^{(1)}\Big(\frac{m}{\overline M_{\mbox{\tiny Q}}}\Big)
  \bigg) +
a_1\,{\rm f}_{\rm P}^{(0)}\Big(\frac{m}{\overline M_{\mbox{\tiny Q}}}\Big) +
\frac{1}{3}\,
  {\rm f}_{\rm PP}^{(0)}\Big(\frac{m}{\overline M_{\mbox{\tiny Q}}}\Big)
\,\bigg]
\,,
\nonumber
\\ & &
\label{Deltafull}
\end{eqnarray}
where
\begin{eqnarray}
{\rm f}_{\rm P}^{(n)}\Big(\frac{m}{M_{\mbox{\tiny Q}}}\Big)
& = &
\frac{1}{4}\, 
\int\limits_0^\infty\!\frac{d q^2}{M_{\mbox{\tiny Q}}^2}\,
\bigg[\,\frac{1}{2}\,\frac{q^2}{M_{\mbox{\tiny Q}}^2}
+\bigg(1-\frac{1}{2}\frac{q^2}{M_{\mbox{\tiny Q}}^2}\bigg)\,
\bigg(1+4\,\frac{M_{\mbox{\tiny Q}}^2}{q^2}\bigg)^{\frac{1}{2}}
\,\bigg]\,{\rm P}\Big(\frac{m^2}{q^2}\Big)
\,\bigg[\,\ln\Big(\frac{q^2}{M_{\mbox{\tiny Q}}^2}\Big)\,\bigg]^n
\label{fPlog}
\\[2mm]
{\rm f}_{\rm P}^{(n)}\Big(\frac{m}{M_{\mbox{\tiny Q}}}\Big)
& = & 
\frac{1}{4}\, 
\int\limits_0^\infty\!\frac{d q^2}{M_{\mbox{\tiny Q}}^2}\,
\bigg[\,\frac{1}{2}\,\frac{q^2}{M_{\mbox{\tiny Q}}^2}
+\bigg(1-\frac{1}{2}\frac{q^2}{M_{\mbox{\tiny Q}}^2}\bigg)\,
\bigg(1+4\,\frac{M_{\mbox{\tiny Q}}^2}{q^2}\bigg)^{\frac{1}{2}}
\,\bigg]\,\bigg[\,{\rm P}\Big(\frac{m^2}{q^2}\Big)\,\bigg]^2
\,\bigg[\,\ln\Big(\frac{q^2}{M_{\mbox{\tiny Q}}^2}\Big)\,\bigg]^n
\,,
\label{fPP}
\end{eqnarray}
and the function ${\rm P}$ is defined in Eq.~(\ref{Pdef}).
In the linear mass approximation, the functions 
${\rm f}_{\rm P}^{(0)}$, ${\rm f}_{\rm P}^{(1)}$ and
${\rm f}_{\rm PP}^{(0)}$ read  ($r=m/\overline M_{\mbox{\tiny Q}}$)
\begin{eqnarray} 
\bigg[\,{\rm f}_{\rm P}^{(0)}(r)\,\bigg]_{\mbox{\tiny linear in m}}
& = & 
\frac{3\,\pi^2}{4}\,r
\\[2mm]
\bigg[\,{\rm f}_{\rm P}^{(1)}(r)\,\bigg]_{\mbox{\tiny linear in m}}
& = &
\bigg(\,\frac{3}{2}\,\ln(r) + 3\,\ln(2)-\frac{9}{4}\,\bigg)\,\pi^2\,r
\\[2mm]
\bigg[\,{\rm f}_{\rm PP}^{(0)}(r)\,\bigg]_{\mbox{\tiny linear in m}}
& = &
\bigg(\,\frac{31}{5} - 6\,\ln(2)\,\bigg)\,\pi^2\,r 
\,.
\end{eqnarray}
Fully analytic results for ${\rm f}_{\rm P}^{(0)}$, 
${\rm f}_{\rm P}^{(1)}$ and  ${\rm f}_{\rm PP}^{(0)}$ can be found in
App.~\ref{appendixmsbarmassvacuum}.
\begin{table}[tbp]  % tabalphas3deltam
\vskip 7mm
\begin{center}
\begin{tabular}{|c||c|c|c|c|c|} \hline
$m/\overline M_{\mbox{\tiny Q}}$ 
 & $0.1$ & $0.2$ & $0.3$ & $0.4$ &$0.5$ \\ \hline
$\delta^{(3),\mbox{\tiny vac,full}}_{\mbox{\tiny massive}}/
[\frac{4}{3}(\frac{a_s}{\pi})^3]$
 & $0.689$ & $1.521$ & $2.419$ & $3.352$ & $4.305$ \\ \hline
$\delta^{(3),\mbox{\tiny vac}}_{\mbox{\tiny massive}}/
[\frac{4}{3}(\frac{a_s}{\pi})^3]$
 & $0.536$ & $1.072$ & $1.608$ & $2.144$ & $2.681$ \\ \hline
$\delta^{(3),\mbox{\tiny vac}}_{\mbox{\tiny massive}}/
\delta^{(3),\mbox{\tiny vac,full}}_{\mbox{\tiny massive}}$
 & $0.78$ & $0.71$ & $0.67$ & $0.64$ & $0.62$ \\ \hline
\end{tabular}
\caption{\label{tabalphas3deltam} 
Comparison of the full result (second line) and the linear mass
approximation (third line) for
the order $\alpha_s^3$ light quark mass corrections in the heavy quark
pole--$\overline{\mbox{MS}}$ 
mass relation for $\mu=m$, for various values of the ratio 
$m/\overline M_{\mbox{\tiny Q}}$.
The ratio of linear mass approximation to full result is shown in the
fourth line.
}
\end{center}
\vskip 3mm
\end{table}
In Table~\ref{tabalphas3deltam} the size of the order $\alpha_s^3$
light quark
vacuum polarization corrections is displayed for the full result and
for the linear mass approximation, for $m/\overline M_{\mbox{\tiny Q}}$ 
between $0.1$ and $0.5$ and $\mu=m$. The ratio of the linear
approximation to the full result is also shown. 
From the first line of Eq.~(\ref{Deltafull}), it is clear why one
expects the linear mass 
approximation to become better at higher orders: higher orders of
perturbation theory correspond to either larger powers of the
subtracted vacuum polarization function ${\rm P}$, or to larger powers
of $\ln q^2$. Both enhance the contributions coming from the low
momentum region
$q\sim m$ with respect to the contributions from 
$q\sim M_{\mbox{\tiny Q}}$.
We note that higher powers of ${\rm P}$ are not suppressed for small
$q$ due to the 
subtraction of the massless vacuum polarization, see Eq.~(\ref{Pdef}).
However, the numbers in Table~\ref{tabalphas3deltam} 
show that, in contrast to
this argument, the linear mass approximation for the order $\alpha_s^3$
vacuum polarization contributions is in fact much worse than for the
order $\alpha_s^2$ light quark mass corrections. For 
$m/\overline M_{\mbox{\tiny Q}}=0.3$
($0.5$) the linear mass terms only account for $67$\% ($62$\%) of the
complete result. Having a closer look at the second line of
Eq.~(\ref{Deltafull}), however, one finds that this behaviour is
caused by a cancellation between ${\rm f}_{\rm P}^{(0)}$ and 
${\rm f}_{\rm P}^{(1)}$, and the fact that the corrections to their 
linear mass 
approximations have a different sign. Individually, on the other
hand, the linear mass approximations are indeed improved for larger
powers of the logarithm and of P. This is demonstrated in
Table~\ref{tabffunctions}, where the ratios of linear mass
approximation to full result for ${\rm f}_{\rm P}^{(n)}$ and 
${\rm f}_{\rm PP}^{(n)}$ are shown for $m/\overline M_{\mbox{\tiny Q}}=0.3$ and
$n=0,\ldots,10$. 
\begin{table}[tbp]  % tabffunctions
\vskip 7mm
\begin{center}
\begin{tabular}{|c||c|c|c|c|c|c|c|c|c|c|c|} \hline
$n$ & $0$ & $1$ & $2$ & $3$ & $4$ & $5$ & $6$ & $7$ & $8$ & $9$ & $10$ \\ \hline
$[{\rm f}_{\rm P}^{(n)}]_{\mbox{\tiny lin}}/{\rm f}_{\rm P}^{(n)}$
 & $1.17$ & $0.88$ & $1.10$ & $0.92$ & $1.07$ & $0.94$ & $1.05$
 & $0.96$ & $1.04$ & $0.97$ & $1.03$ \\ \hline
$[{\rm f}_{\rm PP}^{(n)}]_{\mbox{\tiny lin}}/{\rm f}_{\rm PP}^{(n)}$
 & $1.016$ & $1.003$ & $1.001$ & $1.000$ & $1.000$ & $1.000$
 & $1.000$ & $1.000$ & $1.000$ & $1.000$ & $1.000$  \\ \hline
\end{tabular}
\caption{\label{tabffunctions}
The functions ${\rm f}_{\rm P}^{(n)}$ and ${\rm f}_{\rm PP}^{(n)}$,
which describe the corrections coming from light quark vacuum
polarization insertions into the one-loop gluon line of the heavy
quark pole--$\overline{\mbox{MS}}$ 
mass relation, for various values of $n$ and for
$m/\overline M_{\mbox{\tiny Q}}=0.3$.
}
\end{center}
\vskip 3mm
\end{table}
\begin{table}[tbp]  % tabrealistic
\vskip 7mm
\begin{center}
\begin{tabular}{|c||c|c|c|c|c|c|c|c|c|c|c|} \hline
$\mu/[\mbox{GeV}]$ & $1.0$ & $1.5$ & $2.0$ & $2.5$ & $3.0$ & $3.5$ & $4.0$ 
      & $4.5$ & $5.0$  \\ \hline
$\delta^{(2),\mbox{\tiny full}}_{\mbox{\tiny massive}}+
\delta^{(3),\mbox{\tiny vac,full}}_{\mbox{\tiny massive}}$
 & $0.0750$ & $0.0488$ & $0.0383$ & $0.0324$ & $0.0286$ & $0.0259$
 & $0.0239$ & $0.0223$ & $0.0210$ \\ \hline
$\delta^{(2)}_{\mbox{\tiny massive}}+
\delta^{(3),\mbox{\tiny vac}}_{\mbox{\tiny massive}}$
 & $0.0607$ & $0.0455$ & $0.0375$ & $0.0325$ & $0.0292$ & $0.0268$
 & $0.0249$ & $0.0234$ & $0.0222$ \\ \hline
$\mbox{linear app.}/\mbox{full}$
 & $0.81$ & $0.93$ & $0.98$ & $1.00$ & $1.02$ & $1.03$ & $1.04$ 
 & $1.05$ & $1.05$ \\ \hline
\end{tabular}
\caption{\label{tabrealistic} 
Comparison of the full result (second line) and the linear mass
approximation (third line) for
the sum of the order $\alpha_s^2$ and $\alpha_s^3$ light quark mass
corrections in the bottom quark pole--$\overline{\mbox{MS}}$  
mass relation for $m=1.5$~GeV,
$\overline M_{\mbox{\tiny Q}}=4.2$~GeV and
$\alpha_s^{(5)}(M_Z)=0.118$,
for various values of the renormalization scale $\mu$.
The ratio of linear mass approximation to full result is shown in the
fourth line.
}
\end{center}
\vskip 3mm
\end{table}
Despite the rather good approximation provided by the linear mass
terms at large orders, the fact remains that they only account for
about $70$\% of the order $\alpha_s^3$ light quark mass vacuum
polarization corrections, 
for values of $m/\overline M_{\mbox{\tiny Q}}$ that are relevant to
the charm  
mass effects in the bottom pole--$\overline{\mbox{MS}}$ mass relation. 
At this point it
is instructive to compare the sum of the order $\alpha_s^2$ light
quark mass corrections and the order $\alpha_s^3$ light quark mass
vacuum polarization corrections displayed in
Eqs.~(\ref{alphas2deltam}) and (\ref{Deltafull}),
$\delta^{(2),\mbox{\tiny full}}_{\mbox{\tiny massive}}+
\delta^{(3),\mbox{\tiny vac,full}}_{\mbox{\tiny massive}}$,
with the corresponding linear mass
approximation of the sum,
$\delta^{(2)}_{\mbox{\tiny massive}}+
\delta^{(3),\mbox{\tiny vac}}_{\mbox{\tiny massive}}$. 
Table~\ref{tabrealistic} shows the full
result (second line), the linear mass approximation (third line) and
the ratio (fourth line)
for $m=1.5$~GeV, $\overline M_{\mbox{\tiny b}}=4.2$, 
$\alpha_s^{(n_l=4)}(M_Z)=0.118$ and 
$\mu=1$--$5$~GeV, where $\mu$ is the scale used in the strong coupling.
For the strong coupling we used four-loop running~\cite{alphasrunning}
and three-loop matching conditions~\cite{alphasthreeloopmatching} 
at the five-four quark flavour threshold at 
$\mu=\overline M_{\mbox{\tiny b}}$.
It is an interesting observation that, although the linear
approximation differs from the full result 
by $+18$\% at order $\alpha_s^2$ and $-35$\%
at order $\alpha_s^3$ (at $\mu=m$), the difference amounts to only
$-7$\% for the sum. Therefore, instead of using the full result at
order $\alpha_s^2$ and the linear mass approximation at order
$\alpha_s^3$, it seems more advantageous to employ the linear mass
approximation at order $\alpha_s^2$ and $\alpha_s^3$. [We remind the
reader that the order $\alpha_s^3$ terms are as large as the order
$\alpha_s^2$ ones,  see Sec.~\ref{subsectionpotentialsmalldistance}.]
This is the reason why, in this work, we will employ only the linear
approximation for the charm quark mass corrections in 
the bottom pole--$\overline{\mbox{MS}}$ mass relation, as displayed in
Eq.~(\ref{polemsbarthreeloops}). We emphasize that this
decision is only based on the examination of the order $\alpha_s^3$
light quark mass corrections coming from vacuum polarization
insertions. They account for about $40$\% ($\mu=m$) to
$70$\% ($\mu=3 m$) of the full result in the linear mass approximation.
A calculation of the full order $\alpha_s^3$ light quark
mass corrections would be useful to remove this source of uncertainty
for the charm mass corrections determined in this work.  

\par
\vspace{0.5cm}
\section[Light Quark Mass Corrections in the 
         Heavy Quark Pole--1S Mass Relation]
        {Light Quark Mass Corrections in the \\
         Heavy Quark Pole--1S Mass Relation }
\label{sectionpole1Smass}

The heavy quark pole mass is quite a convenient parameter to carry out
bound state calculations because, in the pole mass scheme, the 
Schr\"odinger equation, which describes the non-relativistic dynamics
of a perturbative heavy-quark--antiquark pair at NNLO, has the simple
form shown in Eq.~(\ref{NNLOSchroedinger}). However, as has been
noted before, the heavy quark pole mass parameter should not be
employed in an actual analysis for the numerical
determination of the heavy quark mass. In practice, in the pole mass
scheme, the determination of the heavy quark mass parameter is spoiled
by a large correlation to the value of the strong coupling, large
perturbative corrections and large renormalization scale
variations. Thus it is advantageous to keep the pole mass
parameters for the actual bound state calculations, but to eliminate
it later in 
favour of a better defined mass with a reduced sensitivity to small
momenta and without an ambiguity of order $\Lambda_{\rm QCD}$. Such
mass definitions are called ``short-distance masses''. The
well-known $\overline{\mbox{MS}}$ mass definition is the prototype of
a short-distance mass.
However, for the description of the heavy-quark--antiquark bound state
dynamics through Eq.~(\ref{NNLOSchroedinger}), the
$\overline{\mbox{MS}}$ mass definition turns out to be a rather bad
choice because it breaks the non-relativistic power counting. This
breakdown can be visualized in Eq.~(\ref{NNLOSchroedinger}): after
eliminating the heavy quark pole mass in favour of the 
$\overline{\mbox{MS}}$ mass 
\begin{eqnarray}
M_{\mbox{\tiny Q}}^{\mbox{\tiny pole}} & \longrightarrow & 
\overline M_{\mbox{\tiny Q}}(\overline M_{\mbox{\tiny Q}})
 + \frac{4}{3}\,\Big(\frac{a_s}{\pi}\Big)
\,\overline M_{\mbox{\tiny Q}}(\overline M_{\mbox{\tiny Q}}) + \ldots
\,,
\end{eqnarray}
the additional correction terms coming from the external energy term
$E=\sqrt{s}-2M_{\mbox{\tiny Q}}^{\mbox{\tiny pole}}$ dominate all
other terms, which are proportional to $\alpha_s^2$ and higher powers
of $\alpha_s$. In practice, if we extract the 
heavy quark $\overline{\mbox{MS}}$ mass with this method, while taking
care of the proper cancellation of the contributions that are linearly
sensitive to small momenta, the results 
have a large correlation with the choice of the strong coupling
and a large renormalization scale dependence, which is
comparable to the pole mass case. 
(See Ref.~\cite{Hoang6} for a comparison of the heavy quark pole,
$\overline{\mbox{MS}}$ and 1S mass definition in the framework of
bottom and top mass determinations from bound state calculations.)
From the conceptual point of view the heavy quark
$\overline{\mbox{MS}}$ is not suited to the quark--antiquark bound
state problem, because it is a short-distance mass genuinely designed
for processes where the heavy quark is very off-shell. This means that
it is a good choice for high energy processes (such as heavy quark
production at high energies), or for processes where the
external energies are much smaller than the heavy quark mass (such as
virtual heavy quark corrections in low energy processes). A
short-distance mass definition that is suited to processes where the
heavy quark is very close to its mass-shell (i.e. for situations where
$q^2-M_{\mbox{\tiny Q}}^2\ll M_{\mbox{\tiny Q}}^2$) is the 1S mass. 
The 1S mass was introduced in
Refs.~\cite{Hoang2,Hoang7} to address the previously mentioned issues
for bottom 
and top quark mass extractions from the $\Upsilon$ sum rules and from
top--antitop quark pair production close to threshold at a future
Linear Collider. Other heavy quark mass definitions with similar
properties were proposed in Ref.~\cite{Czarnecki1} (kinetic mass) and
in Ref.~\cite{Beneke4} (PS mass). 
 
The heavy quark 1S mass is defined as half of the perturbative
contribution to the mass of a $J^{PC}=1^{--}$, ${}^3S_1$
quark--antiquark bound state, assuming that the heavy quark is
stable. In the case of the bottom quark, it corresponds to half of the
perturbative contributions of the mass of the $\Upsilon(1S)$ meson. We
emphasize 
that the 1S mass is a priori a purely formal parameter. Its relation
to the heavy quark pole mass, and any other mass definition that is
defined perturbatively, is determined using the non-relativistic power
counting for perturbative heavy-quark--antiquark systems (i.e. assuming
that the scale hierarchy $M_{\mbox{\tiny Q}}\gg M_{\mbox{\tiny Q}} v
\gg M_{\mbox{\tiny Q}} v^2\gg\Lambda_{\rm QCD}$ is
realized) regardless, whether this assumption is valid in reality or
not. For the top--antitop quark system it is certainly valid in most
cases, for the bottom--antibottom quark system it is debatable for low
radial excitations, whereas for the charm--anticharm system it is
definitely not valid at all. Nevertheless, the 1S mass could be
employed in 
all cases through its perturbative relation to other mass
definitions. However, the 1S mass is most useful for systems that are
truly perturbative, because for these systems it has a direct physical
interpretation as representing the perturbative contribution of 
the mass of the $J^{PC}=1^{--}$, ${}^3S_1$ heavy quarkonium ground
state. 

In this section we present the NNLO light quark mass corrections to
the relation between the heavy quark 1S and pole masses. As in the
previous section, we will assume that there are $n_l$ light quark
species, of which one has a finite (pole) mass $m$, and we adopt the
convention that the strong coupling evolves with $n_l$ quark
flavours. 
\par
%\vspace{0.5cm}
%
\subsection{Results}
\label{subsectionpole1Smassresults}
The relation between the heavy quark 1S and pole masses at NNLO in the
non-relativistic expansion including the light quark mass corrections
can be parametrized as ($a_s=\alpha_s^{(n_l)}(\mu)$):
\begin{eqnarray}
M_{\mbox{\tiny Q}}^{\mbox{\tiny 1S}} & = &
M_{\mbox{\tiny Q}}^{\mbox{\tiny pole}}\,\bigg\{\,
1 
\,- \,\epsilon\,\bigg[\,
     \Delta^{\mbox{\tiny LO}}(a_s)
  \,\bigg]
\,-\,\epsilon^2\,\bigg[\,
     \Delta^{\mbox{\tiny NLO}}_{\mbox{\tiny massless}}
  (M_{\mbox{\tiny Q}}^{\mbox{\tiny pole}},a_s,\mu) +
     \Delta^{\mbox{\tiny NLO}}_{\mbox{\tiny massive}}
  (m,M_{\mbox{\tiny Q}}^{\mbox{\tiny pole}},a_s) \,\bigg]
\nonumber
\\[2mm] & & \hspace{2cm}
\,-\,\epsilon^3\,\bigg[\,
     \Delta^{\mbox{\tiny NNLO}}_{\mbox{\tiny massless}}
  (M_{\mbox{\tiny Q}}^{\mbox{\tiny pole}},a_s,\mu) +
     \Delta^{\mbox{\tiny NNLO}}_{\mbox{\tiny massive}}
  (m,M_{\mbox{\tiny Q}}^{\mbox{\tiny pole}},a_s,\mu) \,\bigg]
\,\bigg\}
\,,
\label{1Spolegeneric}
\end{eqnarray}
where the subscript ``massless'' indicates the corrections for
massless light quarks and the subscript ``massive'' the corrections
from the light quark mass. The massless corrections are well known and
read 
\begin{eqnarray}
\Delta^{\mbox{\tiny LO}} & = &
 \frac{C_F^2\,a_s^2}{8}
\,,
\label{DeltaLOmassless}
\\[4mm]  
\Delta^{\mbox{\tiny NLO}}_{\mbox{\tiny massless}} & = &
\frac{C_F^2\,a_s^2}{8}\, 
\Big(\frac{a_s}{\pi}\Big)\,\bigg[\,
\beta_0\,\bigg( L + 1 \,\bigg) + \frac{a_1}{2} 
\,\bigg]
\,,
\label{DeltaNLOmassless}
\\[4mm] 
\Delta^{\mbox{\tiny NNLO}}_{\mbox{\tiny massless}} & = &
\frac{C_F^2\,a_s^2}{8}\, \Big(\frac{a_s}{\pi}\Big)^2\,
\bigg[\,
\beta_0^2\,\bigg(\, \frac{3}{4} L^2 +  L + 
                             \frac{\zeta_3}{2} + \frac{\pi^2}{24} +
                             \frac{1}{4} 
\,\bigg) + 
\beta_0\,\frac{a_1}{2}\,\bigg(\, \frac{3}{2}\,L + 1
\,\bigg)
\nonumber\\[3mm]
& & \hspace{1.5cm} +
\frac{\beta_1}{4}\,\bigg(\, L + 1
\,\bigg) +
\frac{a_1^2}{16} + \frac{a_2}{8} + 
\bigg(\, C_A - \frac{C_F}{48} \,\bigg)\, C_F \pi^2 
\,\bigg]
\,,
\label{DeltaNNLOmassless}
\end{eqnarray}
where
\begin{eqnarray}
L & \equiv & 
\ln\Big(\frac{\mu}{C_F\,a_s\,
M_{\mbox{\tiny Q}}^{\mbox{\tiny pole}}}\Big)
\,.
\end{eqnarray}
The one- and two-loop coefficients of the beta-function, 
$\beta_0$ and $\beta_1$, and the
constants $a_1$ and $a_2$ are given
in Eqs.~(\ref{b0def})--(\ref{a2def}). The NNLO corrections for massless
light quarks were first calculated in Ref.~\cite{Pineda1}. The contributions
at LO, NLO and NNLO have been labelled by powers of $\epsilon$,
$\epsilon^2$ and $\epsilon^3$, respectively, of the auxiliary
parameter $\epsilon=1$, that has already been employed for
Eq.~(\ref{polemsbarthreeloops}) and that will be relevant in
Sec.~\ref{sectionmsbar1Smass} when we determine the relation between 
the heavy quark $\overline{\mbox{MS}}$ and 1S masses.
The results at NLO and NNLO are obtained by starting from the known
S-wave ground state solution
($r=|{\mbox{\boldmath $r$}}|$),
\begin{eqnarray}
\phi_{1S}({\mbox{\boldmath $r$}}) & = & 
\langle\,{\mbox{\boldmath $r$}}\,|\,\mbox{1S}\,\rangle 
\, = \,
\langle\,\mbox{1S}\,|\,\mbox{\boldmath $r$}\,\rangle 
\, = \,
\pi^{-\frac{1}{2}}\,
\gamma^{\frac{3}{2}}\,
e^{-\gamma\,r}
\,,
\\[4mm]
\gamma & = & 
\frac{M_{\mbox{\tiny Q}}^{\mbox{\tiny pole}}\,C_F\,a_s}{2}
\,,
\end{eqnarray} 
of the LO Schr\"odinger equation,
\begin{eqnarray}
\bigg(
-\frac{{\mbox{\boldmath $\nabla$}}^2}{M^{\mbox{\tiny pole}}_{\mbox{\tiny Q}}}  
+   V_{\mbox{\tiny c}}^{\mbox{\tiny LO}}({\mbox{\boldmath $r$}})
- 2\,M_{\mbox{\tiny Q}}^{\mbox{\tiny pole}}\,\Delta^{\mbox{\tiny LO}}
\,\bigg)\,\phi_{1S}({\mbox{\boldmath $r$}}) & = & 0
\,,
\label{LOSchroedinger}
\end{eqnarray}
and by using Rayleigh--Schr\"odinger time-independent perturbation
theory to calculate the higher order corrections. The formal
results for the light quark mass corrections at NLO and NNLO can be
quickly written down using the Dirac notation:
\begin{eqnarray}
-\,2\,M_{\mbox{\tiny Q}}^{\mbox{\tiny pole}}\,
\Delta^{\mbox{\tiny NLO}}_{\mbox{\tiny massive}}
& = &
\langle\,1S\,|\,
\delta V_{\mbox{\tiny c,m}}^{\mbox{\tiny NLO}}
\,|\,1S\,\rangle
\,,
\label{DeltaNLOmassiveformal}
\\[4mm]
-\,2\,M_{\mbox{\tiny Q}}^{\mbox{\tiny pole}}\,
\Delta^{\mbox{\tiny NLO}}_{\mbox{\tiny massive}}
& = &
\langle\,1S\,|\,
\delta V_{\mbox{\tiny c,m}}^{\mbox{\tiny NNLO}}
\,|\,1S\,\rangle 
\, + \,
\sum\limits_{i\ne\mbox{\tiny 1S}}\hspace{-0.55cm}\int\,\,\,
\langle\,1S\,|\,
\delta V_{\mbox{\tiny c,m}}^{\mbox{\tiny NLO}}\,
\frac{|\,i\,\rangle\,\langle\,i\,|}{E_{\mbox{\tiny 1S}}-E_i}\,
\delta V_{\mbox{\tiny c,m}}^{\mbox{\tiny NLO}}\,
\,|\,1S\,\rangle 
\nonumber
\\[2mm] & & + \,
2\,\sum_{i\ne\mbox{\tiny 1S}}\hspace{-0.55cm}\int\,\,\,
\langle\,1S\,|\,
\delta V_{\mbox{\tiny c,m}}^{\mbox{\tiny NLO}}\,
\frac{|\,i\,\rangle\,\langle\,i\,|}{E_{\mbox{\tiny 1S}}-E_i}\,
V_{\mbox{\tiny c,massless}}^{\mbox{\tiny NLO}}\,
\,|\,1S\,\rangle 
\,,
\label{DeltaNNLOmassiveformal}
\end{eqnarray}
where
$V_{\mbox{\tiny c,massless}}^{\mbox{\tiny
NLO}}({\mbox{\boldmath $r$}})$, 
$\delta V_{\mbox{\tiny c,m}}^{\mbox{\tiny NLO}}({\mbox{\boldmath $r$}})$, 
and
$\delta V_{\mbox{\tiny c,m}}^{\mbox{\tiny NNLO}}({\mbox{\boldmath $r$}})$ 
are given in Eqs.~(\ref{VcNLOmassless}), (\ref{VcNLOmassiverspace})
and (\ref{VcNNLOmassiverspace}), respectively.
Details of the calculations required by
Eqs.~(\ref{DeltaNLOmassiveformal}) and  
(\ref{DeltaNNLOmassiveformal}) are presented in App.~\ref{appendix1Smass}.
The final results for 
$\Delta^{\mbox{\tiny NLO}}_{\mbox{\tiny massive}}$ and
$\Delta^{\mbox{\tiny NNLO}}_{\mbox{\tiny massive}}$ read
($a_s=\alpha_s^{(n_l)}(\mu)$)
\begin{eqnarray}
\Delta^{\mbox{\tiny NLO}}_{\mbox{\tiny massive}}
& = &
\frac{C_F^2\,a_s^2}{4}\,\Big(\frac{a_s}{3\,\pi}\Big)\,
\bigg\{\,
h_0(a) + \ln\Big(\frac{a}{2}\Big) + \frac{11}{6}
\,\bigg\}
\,,
\label{DeltaNLOmassiveexplicit}
\\[4mm]
\Delta^{\mbox{\tiny NNLO}}_{\mbox{\tiny massive}}
& = &
\frac{C_F^2\,a_s^2}{4}\,\Big(\frac{a_s}{3\,\pi}\Big)^2\,
\bigg\{\,
\nonumber
\\[2mm] 
& & \hspace{0.5cm}
  3\,\beta_0\,\bigg[\,
    \frac{3}{2}\,
    \bigg(\ln\Big(\frac{a}{2}\Big)+h_0(a)
          -\frac{1}{3}\,\bar h_0(a)+\frac{3}{2}\bigg)\,
    \bigg(\ln\Big(\frac{\mu}{2\,\gamma}\Big)+\frac{5}{6}\bigg)
\nonumber
\\[2mm] & & \hspace{1.5cm}
    +\,\ln\Big(\frac{a}{2}\Big)-\frac{1}{2}\,\bar h_0(a)-h_1(a)
    -h_7(a)+\zeta_3+\frac{\pi^2}{12}+\frac{4}{3}
  \,\bigg]
\nonumber
\\[2mm] 
& & \hspace{0.5cm}
  +\,\frac{3}{2}\,\ln\Big(\frac{a}{2}\Big)\,
    \bigg(\ln\Big(\frac{a}{2}\Big)+2\,h_0(a)
          -\frac{2}{3}\,\bar h_0(a)-9\bigg)
\nonumber
\\[2mm] 
& & \hspace{0.5cm}
  +\,\frac{1}{2}\,h_0(a)\,
    \bigg(h_0(a)-2\,\bar h_0(a)+4\,h_4(a)-31\bigg)
\nonumber
\\[2mm] 
& & \hspace{0.5cm}
  +\,\frac{25}{6}\,\bar h_0(a)-2\,h_1(a)-h_2(a)+h_5(a)
  -h_6(a)-2\,h_7(a)
  +\zeta_3+\frac{\pi^2}{12}-\frac{571}{24}
\nonumber
\\[2mm] 
& & \hspace{0.5cm}
  \frac{57}{4}\,
  \bigg(
    \frac{c_1\,c_2\,a}{1+c_2\,a} + \frac{d_1\,d_2\,a}{1+d_2\,a}
    + c_1\,\ln(1+c_2\,a) + d_1\,\ln(1+d_2\,a) 
  \bigg)
\,\bigg\}
\,,
\label{DeltaNNLOmassiveexplicit}
\end{eqnarray}
where
\begin{eqnarray}
a & \equiv & \frac{m}{\gamma} \, = \, \frac{2\,m}
{C_F\,a_s\,M_{\mbox{\tiny Q}}^{\mbox{\tiny pole}}}
\,,
\end{eqnarray}
and
\begin{eqnarray}
h_0(a) 
& \equiv &
\int\limits_1^\infty\,d x\,
\frac{f(x)}{(1+a\,x)^2}
\nonumber
\\[2mm] 
& = &
-\,\frac{11}{6} + \frac{3\,\pi}{4}\,a - 2\,a^2 + \pi\,a^3 
+ \frac{2-a^2-4\,a^4}{\sqrt{a^2-1}}\,
\arctan\Big(\frac{\sqrt{a-1}}{\sqrt{a+1}}\Big)
\,,
\label{h0func}
\\[2mm] 
\overline h_0(a)
& \equiv & a\,\frac{d}{d a}\,h(a)
\, = \,
2\,\Big(h_3(a)-h_0(a)\Big)
\nonumber
\\[2mm] 
& = &
\frac{2+7\,a^2-12\,a^4}{2\,(a^2-1)} 
+ \frac{3\,\pi}{4}\,(1+4\,a^2)\,a 
- \frac{3\,a^4\,(4\,a^2-5)}{(a^2-1)^{\frac{3}{2}}}\,
\arctan\Big(\frac{\sqrt{a-1}}{\sqrt{a+1}}\Big)
\,,
\label{h0barfunc}
\\[2mm] 
h_1(a)
& \equiv &
\int\limits_1^\infty\,d x\,
\frac{f(x)}{a^2\,x^2-1}\,\Big(
1-\frac{2\,a\,x}{a^2\,x^2-1}\,\ln(a\,x)
\Big)
\,,
\label{h1func}
\\[2mm] 
h_2(a)
& \equiv &
\int\limits_1^\infty\,d x\,
\frac{f(x)}{(1+a\,x)^2}\,
\bigg[\,\frac{5}{3}+\frac{1}{x^2}\bigg(\,
1+\frac{1}{2\,x}\,\sqrt{x^2-1}\,(1+2x^2)\,
\ln\Big(\frac{x-\sqrt{x^2-1}}{x+\sqrt{x^2-1}}\Big)
\,\bigg)
\,\bigg]
\,,
\label{h2func}
\\[2mm] 
h_3(a)
& \equiv &
\int\limits_1^\infty\,d x\,
\frac{f(x)}{(1+a\,x)^3}
\nonumber
\\[2mm] 
& = &
\frac{28+23\,a^2-60\,a^4}{12\,(a^2-1)} 
+ \frac{\pi}{8}\,(9+20\,a^2)\,a 
- \frac{4-6\,a^2-21\,a^4+20\,a^6}{2\,(a^2-1)^{\frac{3}{2}}}\,
\arctan\Big(\frac{\sqrt{a-1}}{\sqrt{a+1}}\Big)
\,,
\label{h3func}
\\[2mm] 
h_4(a)
& \equiv &
\int\limits_1^\infty\,d x\,
\frac{f(x)}{(1+a\,x)^2}\,\ln(1+a\,x)
\,,
\label{h4func}
\\[2mm] 
h_5(a)
& \equiv &
\int\limits_1^\infty\,d x\,
\int\limits_1^\infty\,d y\,
\frac{f(x)\,f(y)}{(1+a\,x)^2\,(1+a\,y)^2}\,
\frac{a^2\,x\,y}{1+a\,(x+y)}
\,,
\label{h5func}
\\[2mm] 
h_6(a)
& \equiv &
\int\limits_1^\infty\,d x\,
\int\limits_1^\infty\,d y\,
\frac{f(x)\,f(y)}{(1+a\,x)^2\,(1+a\,y)^2}\,
\ln\Big(1+a\,(x+y)\Big)
\,,
\label{h6func}
\\[2mm] 
h_7(a)
& \equiv &
\int\limits_1^\infty\,d x\,
\frac{f(x)}{(1+a\,x)^2}\,
\Li2\Big(\frac{a\,x}{1+a\,x}\Big)
\,.
\label{h7func}
\end{eqnarray}
The function $f$ is defined in Eq.~(\ref{fdef}).
%In Eq.~(\ref{DeltaNNLOmassiveexplicit}) we have rewritten the constant
%$a_1$ as $a_1=\frac{5}{3}\beta_0-8$.
It is instructive to consider the limit 
$m\ll M_{\mbox{\tiny Q}}\alpha_s$ of
Eqs.~(\ref{DeltaNLOmassiveexplicit}) and
(\ref{DeltaNNLOmassiveexplicit}),
%\footnote{
%At this point we note that the calculation of the
%charm mass effects in the bottom pole-1S mass relation at NLO carried
%out in Ref.~\cite{Yndurain1} is incorrect, as it treats the order 
%$m^2/(M_{\mbox{\tiny b}}\,a_s)$ term in
%Eq.~(\ref{DeltaNLOmassiveexplicitsmallm}) as the
%dominant contribution for the charm mass corrections and neglects all
%the other terms.
%},
\begin{eqnarray}
-\,2\,M_{\mbox{\tiny Q}}^{\mbox{\tiny pole}}\,
\Delta^{\mbox{\tiny NLO}}_{\mbox{\tiny massive}}
& \stackrel{m\ll M_{\mbox{\tiny Q}}\alpha_s}{\longrightarrow} &
-\,2\,C_F\,\Big(\frac{a_s}{\pi}\Big)^2\,\frac{\pi^2}{8}\,m\,
\, + \, 
2\,C_F\,\Big(\frac{a_s}{\pi}\Big)\,\frac{9}{16}\,
\frac{m^2}{M_{\mbox{\tiny Q}}^{\mbox{\tiny pole}}}
\, + \,
{\cal{O}}\Big(a_s^2\,\frac{m^3}{(M_{\mbox{\tiny Q}}\,a_s)^2}\Big)
\,,
\label{DeltaNLOmassiveexplicitsmallm}
\\[4mm]
-\,2\,M_{\mbox{\tiny Q}}^{\mbox{\tiny pole}}\,
\Delta^{\mbox{\tiny NNLO}}_{\mbox{\tiny massive}}
& \stackrel{m\ll M_{\mbox{\tiny Q}}\alpha_s}{\longrightarrow} &
-\,2\,C_F\,\Big(\frac{a_s}{\pi}\Big)^3\,\frac{\pi^2}{16}\,m\,
\bigg[\,
\beta_0\,\bigg(\ln\frac{\mu^2}{m^2}-4\ln 2+\frac{14}{3}
\bigg)
\nonumber
\\[2mm] & & \quad
-\,\frac{4}{3}\,\bigg(\frac{59}{15}+2\ln 2\bigg)
+\frac{76}{3\pi}\,\bigg(c_1\,c_2+d_1\,d_2\bigg)
\,\bigg]
\, + \, {\cal{O}}\Big(a_s^2\,\frac{m^2}{M_{\mbox{\tiny Q}}\,a_s}\Big)
\,.
\label{DeltaNNLOmassiveexplicitsmallm}
\end{eqnarray}
If $m\approx M_{\mbox{\tiny Q}}\alpha_s$, as is the case for the charm
mass effects in the bottom pole--1S mass relation, the expansion in the
charm quark mass is completely meaningless as each term in the
expansion is of equal size. 
From Eqs.~(\ref{DeltaNLOmassiveexplicitsmallm})
and (\ref{DeltaNNLOmassiveexplicitsmallm})
we see that the linear light quark mass terms coincide with those that
can be obtained from the corresponding expansion of the light quark
mass corrections to the static potential, 
Eq.~(\ref{deltaVcrspacesmallm}). This is expected, since the linear  
light quark mass terms in the static potential are 
${\mbox{\boldmath $r$}}$-independent and because all LO states form an
orthogonal set. Thus, multiple insertions of the
static potential at higher orders of Rayleigh--Schr\"odinger
perturbation theory do not lead to any linear light quark mass
terms. 
If the $\overline{\mbox{MS}}$ definition is used for the
charm quark mass, 
$m=\overline m(\overline m)[1+\frac{4}{3}(\frac{a_s}{\pi})]$,
we have to apply the replacement
\begin{eqnarray}
\Delta_{\mbox{\tiny massive}}^{\mbox{\tiny NLO}}
(m,M_{\mbox{\tiny Q}}^{\mbox{\tiny pole}},a_s)
& \longrightarrow &
\Delta_{\mbox{\tiny massive}}^{\mbox{\tiny NLO}}
(\overline m(\overline m),M_{\mbox{\tiny Q}}^{\mbox{\tiny pole}},a_s)
\, + \,\epsilon\, \frac{4}{3}\,\Big(\frac{a_s}{\pi}\Big)\,
\overline\Delta_{\mbox{\tiny massive}}^{\mbox{\tiny NLO}}
(\overline m(\overline m),M_{\mbox{\tiny Q}}^{\mbox{\tiny pole}},a_s)
\end{eqnarray}
in Eq.~(\ref{1Spolegeneric}), where
\begin{eqnarray}
\overline\Delta_{\mbox{\tiny massive}}^{\mbox{\tiny NLO}}
& = & a\,\frac{d}{d\,a}\,\Delta_{\mbox{\tiny massive}}^{\mbox{\tiny NLO}}
\, = \,
\frac{C_F^2\,a_s^2}{4}\,\Big(\frac{a_s}{3\,\pi}\Big)\,
\bigg\{\,
\bar h_0(a) + 1
\,\bigg\}
\,,
\label{DeltabarNNLOmassiveexplicit}
\end{eqnarray}
and $\bar h_0$ is defined in Eq.~(\ref{h0barfunc}).

In order to implement the 1S mass into the moments of the
bottom--antibottom cross section for the sum rule analysis we need the
inverse of Eq.~(\ref{1Spolegeneric}), using the non-relativistic power
counting, i.e. with respect to the non-relativistic expansion in
$\alpha_s$ (and not in $\epsilon$). The result reads
\begin{eqnarray}
M_{\mbox{\tiny Q}}^{\mbox{\tiny pole}} & = &
M_{\mbox{\tiny Q}}^{\mbox{\tiny 1S}}\,\bigg\{\,
1 \,+\,
\Delta^{\mbox{\tiny LO}}(a_s)
\,+\,
\bigg[\,
\Delta^{\mbox{\tiny NLO}}_{\mbox{\tiny massless}}
(M_{\mbox{\tiny Q}}^{1S},a_s,\mu)
\,+\,\Delta^{\mbox{\tiny NLO}}_{\mbox{\tiny massive}}
(m,M_{\mbox{\tiny Q}}^{1S},a_s)
\,\bigg]
\nonumber
\\[2mm] & &
\mbox{\hspace{1cm}} 
\,+\,
\bigg[\,
\Big(\Delta^{\mbox{\tiny LO}}(a_s)\Big)^2 \, + \,
\Delta^{\mbox{\tiny NNLO}}_{\mbox{\tiny massless}}
(M_{\mbox{\tiny Q}}^{\mbox{\tiny 1S}},a_s,\mu)
\,+\,
\Delta^{\mbox{\tiny NNLO}}_{\mbox{\tiny massive}}
(m,M_{\mbox{\tiny Q}}^{\mbox{\tiny 1S}},a_s,\mu)
\,\bigg]
\,\bigg\}
%\nonumber
%\\[2mm] & = &
%M_{\mbox{\tiny Q}}^{\mbox{\tiny 1S}}\,\bigg\{\,
%1 \,+\,
%\Delta^{\mbox{\tiny LO}}(a_s)
%\,+\,
%\bigg[\,
%\Delta^{\mbox{\tiny NLO}}_{\mbox{\tiny massless}}
%(M_{\mbox{\tiny Q}}^{1S},a_s,\mu)
%\,+\,\Delta^{\mbox{\tiny NLO}}_{\mbox{\tiny massive}}
%(\overline m(\overline m),M_{\mbox{\tiny Q}}^{1S},a_s)
%\,\bigg]
%\nonumber
%\\[2mm] & &
%\mbox{\hspace{1cm}} 
%\,+\,
%\bigg[\,
%\Big(\Delta^{\mbox{\tiny LO}}(a_s)\Big)^2 \, + \,
%\Delta^{\mbox{\tiny NNLO}}_{\mbox{\tiny massless}}
%(M_{\mbox{\tiny Q}}^{\mbox{\tiny 1S}},a_s,\mu)
%\nonumber
%\\[2mm] & & 
%\mbox{\hspace{1.5cm}} 
%\,+\,
%\frac{4}{3}\,\Big(\frac{a_s}{\pi}\Big)\,
%\overline \Delta^{\mbox{\tiny NLO}}_{\mbox{\tiny massive}}
%(\overline m(\overline m),M_{\mbox{\tiny Q}}^{\mbox{\tiny 1S}},a_s)
%\,+\,
%\Delta^{\mbox{\tiny NNLO}}_{\mbox{\tiny massive}}
%(\overline m(\overline m),M_{\mbox{\tiny Q}}^{\mbox{\tiny 1S}},a_s,\mu)
%\,\bigg]
%\,\bigg\}
\,.
\label{pole1Sgeneric}
\end{eqnarray}

\par
%\vspace{0.5cm}
%
\subsection{Double Insertion of the NLO Static Potential}
\label{subsectionpole1Smassdouble}

We already mentioned after Eq.~(\ref{DeltaNNLOmassiveexplicitsmallm})
that in the limit $m\to 0$ the linear light quark mass terms in 
$\Delta^{\mbox{\tiny NNLO}}_{\mbox{\tiny massive}}$ only arise from the
single insertion of the NNLO light quark mass corrections to the
static potential (first term on the RHS of
Eq.~(\ref{DeltaNNLOmassiveformal})). This means that---if one is
allowed to make the expansion in the light quark mass---the light
quark mass corrections coming from multiple insertions are subleading
with respect to those arising from single insertions. 
For the charm mass corrections, however, an expansion in the charm
mass is a priori not possible because $\langle 1/{\mbox{\boldmath
$r$}}\rangle\sim M_{\mbox{\tiny b}}\alpha_s\sim m_{\mbox{\tiny
charm}}$. Thus one can
not necessarily argue that charm mass corrections arising from multiple
insertions are smaller than single insertion ones, at least in low
orders of perturbation theory. In high orders of perturbation theory, on
the other hand, charm mass corrections coming from multiple insertions
are subleading because, as already mentioned, the contributions
from small momenta enhance the linear mass terms with
respect to higher powers of the charm mass.

\begin{figure}[t] % figdoublecompare
\begin{center}
\leavevmode
\epsfxsize=5.cm
\epsffile[230 580 440 710]{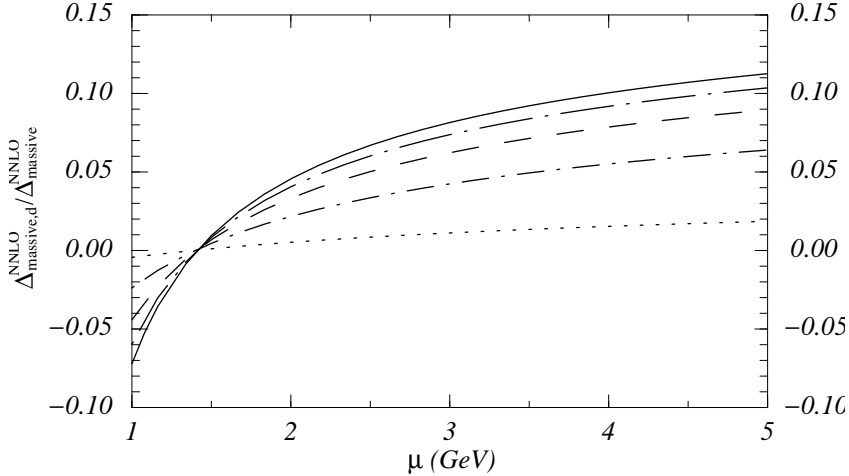}\\
\vskip  3.5cm
 \caption{\label{figdoublecompare} 
The ratio of the double-insertion contributions to the full result
for the NNLO light quark mass corrections in the bottom quark 1S--pole
mass relation, 
$\Delta^{\mbox{\tiny NNLO}}_{\mbox{\tiny massive,d}}/
\Delta^{\mbox{\tiny NNLO}}_{\mbox{\tiny massive}}$ for
$M_{\mbox{\tiny Q}}^{\mbox{\tiny pole}}=4.8$~GeV, 
$\alpha_s^{(5)}(M_Z)=0.118$ 
and $m=0.1$~GeV (dotted line), 
$0.5$~GeV (dash-dotted line),
$1.0$~GeV (dashed line),
$1.5$~GeV (long-dash-dotted line) and
$2.0$~GeV (solid line)
plotted over the renormalization scale $\mu$.
}
 \end{center}
\end{figure}

In this section we examine the size of the NNLO charm mass corrections
in the bottom 1S mass, 
$\Delta^{\mbox{\tiny NNLO}}_{\mbox{\tiny massive}}$,
coming from double insertions of the NLO static potential (second
and third terms on the RHS of Eq.~(\ref{DeltaNNLOmassiveformal})). This
will be important for the calculation of the NNLO charm mass
corrections to the sum rules in Sec.~\ref{sectionsumrules}, where we
neglect the double insertion contributions. From the results given in
App.~\ref{appendix1Smass}, the NNLO light quark mass corrections coming
from double insertions read
\begin{eqnarray}
\Delta^{\mbox{\tiny NNLO}}_{\mbox{\tiny massive,d}}
& = & -\,
\frac{C_F^2\,a_s^2}{4}\,\Big(\frac{a_s}{3\,\pi}\Big)^2\,
\bigg\{\,
\nonumber
\\[2mm] 
& & \hspace{0.5cm}
  \frac{3}{2}\,\beta_0\,\bigg[\,
    h_0(a)\,\bigg(3\,\ln\Big(\frac{2\,\gamma}{\mu}\Big)-2\bigg)
    +2\,h_3(a)\,\bigg(1-\ln\Big(\frac{2\,\gamma}{\mu}\Big)\bigg)
    +2\,h_7(a) 
\nonumber
\\[2mm] & & \hspace{1.5cm}
    +\,\bigg(\ln\Big(\frac{a}{2}\Big)+\frac{5}{6}\bigg)\,
    \ln\Big(\frac{2\,\gamma}{\mu}\Big)
    -2\,\zeta_3+\frac{\pi^2}{3}-1
  \,\bigg]
\nonumber
\\[2mm] 
& & \hspace{0.5cm}
  +\,\frac{3}{2}\,a_1\,\bigg[\,
    -\frac{3}{2}\,h_0(a)+h_3(a)
    -\frac{1}{2}\,\ln\Big(\frac{a}{2}\Big)-\frac{5}{12}
  \,\bigg]
\nonumber
\\[2mm] 
& & \hspace{0.5cm}
  +\,h_0(a)\,\bigg(-\frac{5}{2}\,h_0(a)+2\,h_3(a)-2\,h_4(a)\bigg)
  -h_5(a)+h_6(a)
\nonumber
\\[2mm] 
& & \hspace{0.5cm}
  -\,3\,h_0(a)\,\bigg(\ln\Big(\frac{a}{2}\Big)+\frac{3}{2}\bigg)
  +2\,h_7(a)
  +2\,h_3(a)\,\bigg(\ln\Big(\frac{a}{2}\Big)+\frac{11}{6}\bigg)
\nonumber
\\[2mm] 
& & \hspace{0.5cm}
  -\,\frac{1}{2}\,\ln^2\Big(\frac{a}{2}\Big)
  -\frac{5}{6}\,\ln\Big(\frac{a}{2}\Big)
  -\zeta_3+\frac{\pi^2}{6}-\frac{61}{72}
\,\bigg\}
\,.
\label{DeltaNNLOmassiveexplicitdouble}
\end{eqnarray} 
In Fig.~\ref{figdoublecompare} we have displayed the ratio
$\Delta^{\mbox{\tiny NNLO}}_{\mbox{\tiny massive,d}}/
\Delta^{\mbox{\tiny NNLO}}_{\mbox{\tiny massive}}$ for
$M_{\mbox{\tiny Q}}^{\mbox{\tiny pole}}=4.8$~GeV, 
$\alpha_s^{(5)}(M_Z)=0.118$ and 
$1$~GeV\,$\le\mu\le 5$~GeV for $m=0.1$~GeV (dotted line),  
$0.5$~GeV (dash-dotted line),
$1.0$~GeV (dashed line),
$1.5$~GeV (long-dash-dotted line) and
$2.0$~GeV (solid line). 
For the
charm case ($m\approx 1.5$~GeV) we see that the double-insertion
contributions amount to about 1\% for $\mu\approx 1.5$~GeV and
increase to about 10\% for $\mu\approx 5$~GeV. This shows that the
NNLO corrections already are considerably influenced by momenta
smaller than $\langle  1/{\mbox{\boldmath $r$}}
\rangle\sim M_{\mbox{\tiny b}}\alpha_s$ and that the single-insertion
contributions give the dominant contributions to the NNLO light quark
mass corrections even for the charm quark case. 
Thus, in view of the fact that the use of the linear mass
approximation for the charm mass corrections in the bottom 
pole--$\overline{\mbox{MS}}$ mass relation introduces a relative
uncertainty of around 10\% in the charm mass corrections, it would be
consistent to neglect all double insertions 
at the NNLO level. This is useful because the calculation of the
double insertions is considerably more time-consuming than that of the
single insertions. For this reason we believe that it is save, at the
level of 10\% accuracy for the charm mass corrections, to neglect
the double-insertion contributions for the NNLO charm mass corrections
to the moments of the $\Upsilon$ sum rules. 

\par
\vspace{0.5cm}
\section{Heavy Quark $\overline{\mbox{MS}}$-1S Mass Relation 
         and Upsilon Expansion}
\label{sectionmsbar1Smass}
In this work we use the bottom quark 1S mass as the mass definition
that is extracted directly from the $\Upsilon$ sum rule analysis. The
bottom quark $\overline{\mbox{MS}}$ mass is then determined in a
second step, using Eqs.~(\ref{polemsbarthreeloops}) and
(\ref{1Spolegeneric}). However, some care has to be taken in the
determination of the perturbative relation between the heavy quark
$\overline{\mbox{MS}}$ and 1S masses to ensure the proper cancellation
of the large corrections that are associated with the linear sensitivity
to small momenta of the pole mass in Eqs.~(\ref{polemsbarthreeloops})
and (\ref{1Spolegeneric}). Care is needed because
Eq.~(\ref{1Spolegeneric}), which gives the 1S mass in terms of the
pole mass, represents an expansion in the framework of the
non-relativistic power counting, where $\alpha_s\sim
v$. Equation~(\ref{polemsbarthreeloops}), on the other hand, which
gives the pole mass in terms of the $\overline{\mbox{MS}}$ mass,
represents a usual Feynman diagram expansion in the number of
loops, where the power of $\alpha_s$ corresponds to the number of 
strong coupling constants that occur in the corresponding diagrams. 
This means that Eqs.~(\ref{polemsbarthreeloops})
and (\ref{1Spolegeneric}) cannot simply be combined by using the
expansion in terms of $\alpha_s$. Rather, a modified perturbative
expansion, the ``upsilon expansion''~\cite{Hoang3}, has to be employed. 
The upsilon expansion gives the following prescription: the LO, NLO
and NNLO contributions in Eq.~(\ref{1Spolegeneric}) are of order
$\epsilon$, $\epsilon^2$ and $\epsilon^3$, respectively, of the
auxiliary parameter $\epsilon=1$. In Eq.~(\ref{polemsbarthreeloops}),
on the other hand, the terms of order $\alpha_s^n$ are of order
$\epsilon^n$ in the upsilon expansion. When the $\overline{\mbox{MS}}$
mass is then expressed in terms of the 1S mass one, must use the
expansion in $\epsilon$. This means that, in each order of $\epsilon$,
different orders in $\alpha_s$ are mixed. This also means
that a determination of the heavy quark 1S mass at NNLO in the
non-relativistic expansion corresponds to an order $\alpha_s^3$
determination 
of the heavy quark $\overline{\mbox{MS}}$ mass. If the mechanism
incorporated in the upsilon expansion is not used, the large
corrections mentioned previously can survive and lead to systematic
errors. The upsilon expansion prescription can be understood from the
asymptotic large order behaviour of the ``massless'' N$^{n}$LO
perturbative coefficients $c_n$ in Eq.~(\ref{1Spolegeneric}),
\begin{eqnarray}
\bigg[\,
c_n(a_s,M_{\mbox{\tiny Q}}^{\mbox{\tiny pole}},\mu)\,
M_{\mbox{\tiny Q}}^{\mbox{\tiny pole}}\,a_s^{n+2}
\,\bigg]_{n\gg 1}
& \sim &
-\,\mu\,(n+1)!\,(2\,\beta_0)^{n+1}\,a_s^{n+1}
\,,
\label{1Spoleasymptotic}
\end{eqnarray}
that arises from the contributions that are linearly sensitive to small
momenta and that dominate $c_n$ for large $n$. This relation comes
from the fact that terms that are linearly 
sensitive to small momenta in $c_n$ involve powers of the logarithmic term
$L=\ln(\mu/C_F\alpha_s M_{\mbox{\tiny Q}})$, which exponentiate at higher
orders, $\sum_{i=0}^{n} 
L^i/i!\approx\exp(L)=\mu/C_F\alpha_s M_{\mbox{\tiny Q}}$, and
effectively cancel one power of $\alpha_s$~\cite{Hoang3}. Another way
to visualize 
the mechanism behind the upsilon expansion is provided by the light
quark mass corrections themselves, as they provide a natural probe
for the linear sensitivity to small momenta: the linear light quark
mass terms in $\Delta^{\mbox{\tiny NLO}}_{\mbox{\tiny massive}}$ and 
$\Delta^{\mbox{\tiny NNLO}}_{\mbox{\tiny massive}}$ in 
Eqs.~(\ref{DeltaNLOmassiveexplicitsmallm})
and (\ref{DeltaNNLOmassiveexplicitsmallm}) are of order $\alpha_s^2$
and $\alpha_s^3$, respectively; 
$\Delta^{\mbox{\tiny NLO}}_{\mbox{\tiny massive}}$ and 
$\Delta^{\mbox{\tiny NNLO}}_{\mbox{\tiny massive}}$, on the other
hand, are of order $\alpha_s^3$ and $\alpha_s^4$, respectively, in the
non-relativistic power counting. The origin of this mismatch
is that the non-relativistic dynamics of a perturbative 
heavy-quark--antiquark system contains scales with powers of 
$\alpha_s$ such 
as the inverse Bohr radius $M_{\mbox{\tiny Q}}\alpha_s$ as dynamical
scales. These scales have to be accounted for by the non-relativistic
power counting. The question of sensitivity to small momenta and the
corresponding counting of orders, on the other hand, is only relevant
to scales that are much smaller than any dynamical one,
and therefore not tied to the non-relativistic power 
counting\footnote{
We emphasize that the issue of infrared sensitivity to small momenta, 
usually also referred to as the ``infrared renormalon problem'',
is a purely perturbative one. The typical hadronic scale
$\Lambda_{\rm QCD}$ does not arise as a relevant dynamical scale, but
only as a dimensionful parameter that parametrizes the renormalon
ambiguities in the perturbative series. 
}.

\par
%\vspace{0.5cm}
%
\subsection{Results}
\label{subsectionmsbar1Smassresults}
Combining Eqs.~(\ref{polemsbarthreeloops}) and (\ref{1Spolegeneric})
using the upsilon expansion up to order $\epsilon^3$ we arrive at the
following result for the relation between the heavy quark
$\overline{\mbox{MS}}$  and the 1S masses
($\tilde a_s\equiv\alpha_s^{(n_l)}(M_{\mbox{\tiny Q}}^{\mbox{\tiny 1S}})$),
\begin{eqnarray}
\overline M_{\mbox{\tiny Q}}(\overline M_{\mbox{\tiny Q}}) & = & 
M_{\mbox{\tiny Q}}^{\mbox{\tiny 1S}}\,
\bigg\{\, 1
+ \epsilon\,\bigg[\,
    \Delta^{(1)}(M_{\mbox{\tiny Q}}^{\mbox{\tiny 1S}},\tilde a_s)
  \,\bigg]
\nonumber
\\[2mm]
& & \hspace{1cm}
+ \, \epsilon^2\,\bigg[\,
\Delta^{(2)}
(M_{\mbox{\tiny Q}}^{\mbox{\tiny 1S}},\tilde a_s)+
\Delta^{(2)}_{\mbox{\tiny m}}
(\overline m(\overline m),M_{\mbox{\tiny Q}}^{\mbox{\tiny 1S}},\tilde a_s)
  \,\bigg]
\nonumber
\\[2mm]
& & \hspace{1cm}
+ \, \epsilon^3\,\bigg[\,
\Delta^{(3)}
(M_{\mbox{\tiny Q}}^{\mbox{\tiny 1S}},\tilde a_s)+
\Delta^{(3)}_{\mbox{\tiny m}}
(\overline m(\overline m),M_{\mbox{\tiny Q}}^{\mbox{\tiny 1S}},\tilde a_s)
  \,\bigg]
\,\bigg\} 
%\nonumber
%\\[2mm]
%& = & 
%M_{\mbox{\tiny Q}}^{\mbox{\tiny 1S}}\,
%\bigg\{\, 1
%+ \epsilon\,\bigg[\,
%    \Delta^{(1)}(M_{\mbox{\tiny Q}}^{\mbox{\tiny 1S}},\tilde a_s)
%  \,\bigg]
%\nonumber
%\\[2mm]
%& & \hspace{1cm}
%+ \, \epsilon^2\,\bigg[\,
%\Delta^{(2)}_{\mbox{\tiny massless}}(M_{\mbox{\tiny Q}}^{\mbox{\tiny 1S}},\tilde %a_s)+
%\Delta^{(2)}_{\mbox{\tiny m}}(\overline m(\overline m),M_{\mbox{\tiny Q}}^{\rm
%1S},\tilde a_s)
%  \,\bigg]
%\nonumber
%\\[2mm]
%& & \hspace{1cm}
%+ \, \epsilon^3\,\bigg[\,
%\Delta^{(3)}_{\mbox{\tiny massless}}(M_{\mbox{\tiny Q}}^{\mbox{\tiny 1S}},\tilde %a_s)+
%\overline \Delta^{(3)}_{\mbox{\tiny m}}
%(\overline m(\overline m),M_{\mbox{\tiny Q}}^{\mbox{\tiny 1S}},\tilde a_s) +
%\Delta^{(3)}_{\mbox{\tiny m}}
%(\overline m(\overline m),M_{\mbox{\tiny Q}}^{\mbox{\tiny 1S}},\tilde a_s)
%  \,\bigg]
%\,\bigg\}
\,,
\label{msbar1Sgeneric}
\end{eqnarray}
where we have separated the corrections for massless light quarks
(subscript ``massless'') and those coming from the light quark mass
(subscript ``massive''). Similar to what we did in the previous
sections, we consider 
$n_l$ light quarks, of which one has a $\overline{\mbox{MS}}$ mass
$\overline m(\overline m)$. The individual results for the
corrections for massless light quarks read
\begin{eqnarray}
\Delta^{(1)} & = &
\Delta^{\mbox{\tiny LO}}
-\delta^{(1)}
\,,
\label{Deltamassless1}
\\[4mm]
\Delta^{(2)} & = &
\Delta^{\mbox{\tiny NLO}} 
+ (\Delta^{\mbox{\tiny LO}})^2
- \delta^{(1)}\,\Delta^{\mbox{\tiny LO}}
- \delta^{(2)}
+ (\delta^{(1)})^2
\,,
\label{Deltamassless2}
\\[4mm]
\Delta^{(3)} & = &  
\Delta^{\mbox{\tiny NNLO}} 
+ 2\,\Delta^{\mbox{\tiny LO}}\,\Delta^{\mbox{\tiny NLO}} 
+ (\Delta^{\mbox{\tiny LO}})^3 
- \Big(\frac{a_s}{\pi}\Big)\,\beta_0\,(\Delta^{\mbox{\tiny LO}})^2
- \delta^{(1)}\,\Big[\Delta^{\mbox{\tiny NLO}} 
         + (\Delta^{\mbox{\tiny LO}})^2\Big] 
\nonumber
\\[2mm] & &
- \,\Delta^{\mbox{\tiny LO}}\,\Big[\delta^{(2)}-(\delta^{(1)})^2\Big]
- \delta^{(3)} 
+ 2\,\delta^{(1)}\,\delta^{(2)}
- (\delta^{(1)})^3 
- \Big(\frac{a_s}{2\pi}\Big)\,\beta_0\,\delta^{(1)}\,
     \Big[\delta^{(1)} - \Delta^{\mbox{\tiny LO}} \Big] 
\,,
\label{Deltamassless3}
\end{eqnarray}
where
\begin{eqnarray}
\Delta^{\mbox{\tiny LO}}
& \equiv &
\Delta^{\mbox{\tiny LO}}(\tilde a_s)
\,,
\label{firstmassless}
\\[4mm]
\Delta^{\mbox{\tiny NLO}}
& \equiv &
\Delta^{\mbox{\tiny NLO}}_{\mbox{\tiny massless}}
(M_{\mbox{\tiny Q}}^{\mbox{\tiny 1S}},\tilde a_s,
M_{\mbox{\tiny Q}}^{\mbox{\tiny 1S}})
\,,
\\[4mm]
\Delta^{\mbox{\tiny NNLO}}
& \equiv &
\Delta^{\mbox{\tiny NNLO}}_{\mbox{\tiny massless}}
(M_{\mbox{\tiny Q}}^{\mbox{\tiny 1S}},\tilde a_s,
M_{\mbox{\tiny Q}}^{\mbox{\tiny 1S}})
\,,
\\[4mm]
\delta^{(1)}
& \equiv &
\delta^{(1)}(\tilde a_s)
\,,
\\[4mm]
\delta^{(i)}
& \equiv &
\delta^{(i)}_{\mbox{\tiny massless}}(\tilde a_s)
\,,\qquad i=2,3
\,.
\label{lastmassless}
\end{eqnarray}
The formulae for the functions on the RHS of
Eqs.~(\ref{firstmassless})--(\ref{lastmassless}) can be found in 
Eqs.~(\ref{delta1massless})--(\ref{delta3massless}) and
(\ref{DeltaLOmassless})--(\ref{DeltaNNLOmassless}).
The individual results for the light quark mass corrections read
\begin{eqnarray}
\Delta^{(2)}_{\mbox{\tiny m}} 
& = &
\Delta^{\mbox{\tiny NLO}}_{\mbox{\tiny m}}
-\delta^{(2)}_{\mbox{\tiny m}}
\,,
\label{Deltamassive2}
\\[4mm]
\Delta^{(3)}_{\mbox{\tiny m}} 
& = &
\Delta^{\mbox{\tiny NNLO}}_{\mbox{\tiny m}}
+ \Delta^{\mbox{\tiny NLO}}_{\mbox{\tiny m}}\,
  \Big[2\,\Delta^{\mbox{\tiny LO}}-\delta^{(1)}\Big]
+ \overline\Delta^{\mbox{\tiny NLO}}_{\mbox{\tiny m}}
   \,\Big[\delta^{(1)}-\Delta^{\mbox{\tiny LO}}\Big]
-\delta^{(3)}_{\mbox{\tiny m}}
\,,
\label{Deltamassive3}
\end{eqnarray}
where
\begin{eqnarray}
\Delta^{\mbox{\tiny NLO}}_{\mbox{\tiny m}}
& \equiv &
\Delta^{\mbox{\tiny NLO}}_{\mbox{\tiny massive}}
(\overline m(\overline m),
M_{\mbox{\tiny Q}}^{\mbox{\tiny 1S}},\tilde a_s)
\,,
\label{firstmassive}
\\[4mm]
\overline\Delta^{\mbox{\tiny NLO}}_{\mbox{\tiny m}}
& \equiv &
\overline\Delta^{\mbox{\tiny NLO}}_{\mbox{\tiny massive}}
(\overline m(\overline m),
M_{\mbox{\tiny Q}}^{\mbox{\tiny 1S}},\tilde a_s)
\,,
\\[4mm]
\Delta^{\mbox{\tiny NNLO}}_{\mbox{\tiny m}}
& \equiv &
\Delta^{\mbox{\tiny NNLO}}_{\mbox{\tiny massive}}
(\overline m(\overline m),
M_{\mbox{\tiny Q}}^{\mbox{\tiny 1S}},\tilde a_s,
M_{\mbox{\tiny Q}}^{\mbox{\tiny 1S}})
\,,
\\[4mm]
\delta^{(2)}_{\mbox{\tiny m}}
& \equiv &
\delta^{(2)}_{\mbox{\tiny massive}}
(\overline m(\overline m),
M_{\mbox{\tiny Q}}^{\mbox{\tiny 1S}},\tilde a_s)
\,,
\\[4mm]
\delta^{(3)}_{\mbox{\tiny m}}
& \equiv &
\delta^{(3)}_{\mbox{\tiny massive}}
(\overline m(\overline m),
M_{\mbox{\tiny Q}}^{\mbox{\tiny 1S}},\tilde a_s)
\,.
\label{lastmassive}
\end{eqnarray}
The formulae for the functions on the RHS of
Eqs.~(\ref{firstmassive})--(\ref{lastmassive}) can be found in 
Eqs.~(\ref{delta2massive}), (\ref{delta3massive}), 
(\ref{DeltaNLOmassiveexplicit}), (\ref{DeltaNNLOmassiveexplicit}) 
and (\ref{DeltabarNNLOmassiveexplicit}).
The results for massless light quarks have already been explicitly
presented in Ref.~\cite{Hoang6}. It is worth mentioning that, in the
limit 
$\overline m(\overline m)\to 0$, $\Delta^{(3)}_{\mbox{\tiny m}}$
in Eq.~(\ref{Deltamassive3}) still contains a small linear light
quark mass term, $\overline m(\overline m)\alpha_s^4/27$. This term
arises because the heavy quark 1S mass receives, through the kinetic
energy term 
$-{\mbox{\boldmath $\nabla$}}^2/M^{\mbox{\tiny pole}}_{\mbox{\tiny Q}}$ 
in the Schr\"odinger equation~(\ref{NNLOSchroedinger}), an additional
dependence on the pole mass. This does not affect the cancellation of
linear light quark mass terms in the total static energy. The
associated ambiguity exists, regardless of which of the currently known 
short-distance mass definitions is employed in an extraction of the
bottom quark mass parameter from non-relativistic bottom--antibottom
observables. However, the ambiguity is suppressed by
$\Delta^{\mbox{\tiny LO}}=C_F^2\alpha_s^2/8=0.02$ with respect to the 
ambiguity of the pole mass definition, i.e. it is smaller than the
order $\Lambda_{\rm QCD}^2/M_{\mbox{\tiny Q}}$ ambiguity for the
bottom quark case. It can therefore be ignored for all
practical purposes.

For simplicity we have displayed
Eq.~(\ref{msbar1Sgeneric}) for the strong coupling at the scale
$\mu=M_{\mbox{\tiny Q}}^{\mbox{\tiny 1S}}$. 
Equation~(\ref{msbar1Sgeneric}) can be generalized to arbitrary scales
with the relation   
($\tilde a_s = \alpha_s^{(n_l)}(M_{\mbox{\tiny Q}}^{\mbox{\tiny 1S}})$, 
$a_s = \alpha_s^{(n_l)}(\mu)$):
\begin{eqnarray}
\tilde a_s & = &
a_s\,
\bigg\{\,
1 \, - \,\epsilon\,
    \Big(\frac{a_s}{2\,\pi}\Big)\,
    \beta_0\,\ln\Big(\frac{M_{\mbox{\tiny Q}}^{\mbox{\tiny 1S}}}{\mu}\Big)
\, +\,\epsilon^2\,
     \Big(\frac{a_s}{2\,\pi}\Big)^2\,
   \bigg[\,
 \beta_0^2\,\ln^2\Big(\frac{M_{\mbox{\tiny Q}}^{\mbox{\tiny 1S}}}{\mu}\Big)
 - \frac{1}{2}\,
 \beta_1\,\ln\Big(\frac{M_{\mbox{\tiny Q}}^{\mbox{\tiny 1S}}}{\mu}\Big)   
   \,\bigg]
\,\bigg\}
\,.
\label{runningalphas}
\end{eqnarray}
We note that, when we examine the scale dependence of the light quark
mass corrections in the following subsection, we leave the scales of
the quark masses unchanged. 
We also note that from now on we will designate to the terms of order
$\epsilon^n$ in Eq.~(\ref{msbar1Sgeneric}) by the expression
``order $\alpha_s^n$''. This accounts for 
the fact that a determination of the heavy quark 1S mass at NNLO in the
non-relativistic expansion corresponds to an order $\alpha_s^3$
determination of the heavy quark $\overline{\mbox{MS}}$ mass, as far as
the proper cancellation of the linear infrared-sensitive contributions
is concerned.

\par
%\vspace{0.5cm}
%
\subsection{Brief Examination}
\label{subsectionmsbar1Smassexamination}
\begin{figure}[t] % figmsbar1S
\begin{center}
\hspace{0.5cm}
\leavevmode
\epsfxsize=3.8cm
\epsffile[230 580 440 710]{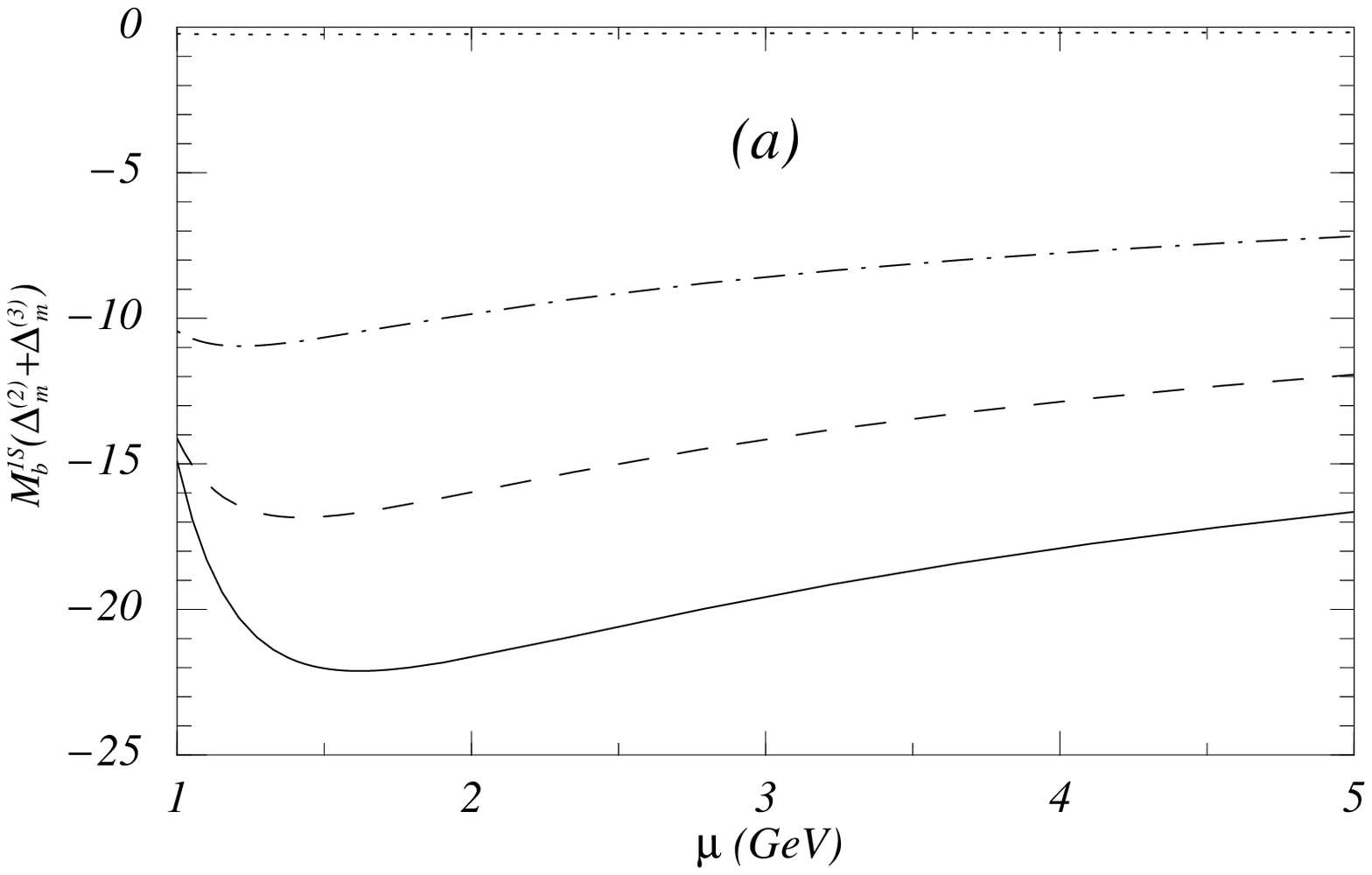}
\hspace{4.7cm}
\leavevmode
\epsfxsize=3.8cm
\epsffile[230 580 440 710]{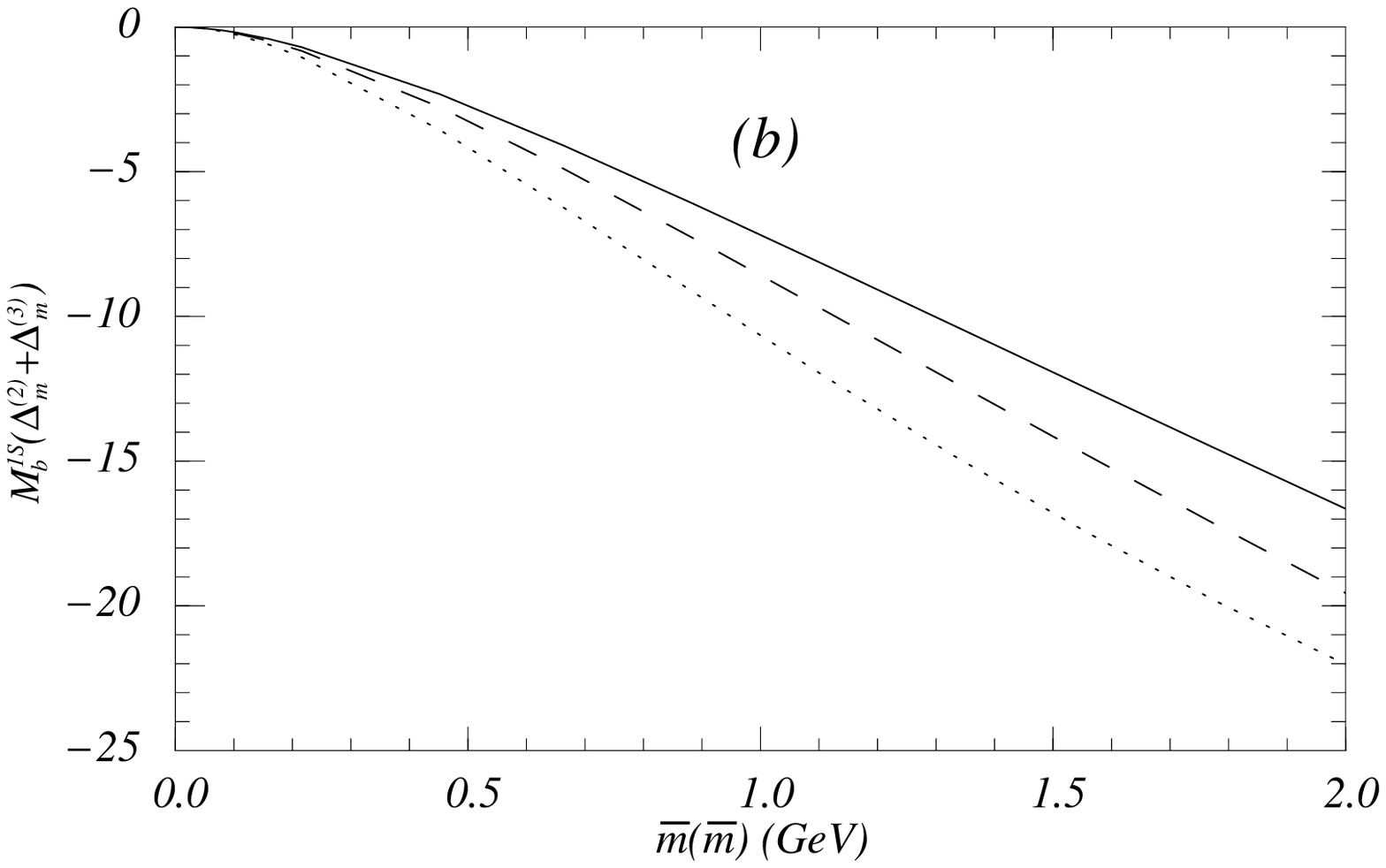}
\vskip  2.5cm
 \caption{\label{figmsbar1S} 
Charm mass corrections in the relation between the bottom 
$\overline{\mbox{MS}}$ mass 
$\overline M_{\mbox{\tiny b}}(\overline M_{\mbox{\tiny b}})$ 
and the bottom 1S mass.
Figure (a) displays the scale dependence for 
$M_{\mbox{\tiny b}}^{\mbox{\tiny 1S}}=4.7$~GeV and
$\alpha_s^{(5)}(M_Z)=0.118$ and using
$\overline m(\overline m)=0.1$~GeV (dotted line), 
$1.0$~GeV (dash-dotted line), $1.5$~GeV (dashed line) and
$2.0$~GeV (solid line) for the $\overline{\mbox{MS}}$ charm quark
mass.
Figure (b) displays the dependence on $\overline m(\overline m)$  
for $\mu=1.5$~GeV (dotted line), $3$~GeV (dashed line) and $5$~GeV
(solid line). The other parameters a chosen as in Fig.~(a).
}
 \end{center}
\end{figure}
\begin{table}[t]  % tabmsbar1Sscale
\vskip 7mm
\begin{center}
\begin{tabular}{|c||r|r|r|r|r|r|r|r|r|} \hline
$\mu [\mbox{GeV}]$ & $1.0$ & $1.5$ & $2.0$ & $2.5$ & $3.0$ & $3.5$ & $4.0$
      & $4.5$ & $5.0$ \\ \hline\hline
$M_{\mbox{\tiny b}}^{\mbox{\tiny 1S}}
\Delta^{(1)} [\mbox{MeV}]$
 & $-698$ & $-570$ & $-506$ & $-466$ & $-438$ & $-417$ & $-401$
 & $-387$ & $-376$ \\ \hline
$M_{\mbox{\tiny b}}^{\mbox{\tiny 1S}}
\Delta^{(2)} [\mbox{MeV}]$
 & $158$ & $33$ & $-22$ & $-52$ & $-70$ & $-82$ & $-90$ 
 & $-96$ & $-101$ \\ \hline
$M_{\mbox{\tiny b}}^{\mbox{\tiny 1S}}
\Delta^{(3)} [\mbox{MeV}]$
 & $240$ & $55$ & $19$ & $1$ & $-10$ & $-17$& $-23$ 
 & $-28$ & $-32$\\ \hline
$M_{\mbox{\tiny b}}^{\mbox{\tiny 1S}}
\sum_{i=1}^3\Delta^{(i)}/[\mbox{MeV}]$
 & $-300$ & $-482$ & $-509$ & $-517$ & $-518$ & $-516$ & $-514$
 & $-511$ & $-509$
\\ \hline\hline
$M_{\mbox{\tiny b}}^{\mbox{\tiny 1S}}
\Delta^{(2)}_{\mbox{\tiny m}} [\mbox{MeV}]$
 & $-24$ & $-16$ & $-12$ & $-11$ & $-9$ & $-9$ & $-8$
 & $-7$ & $-7$ \\ \hline
$M_{\mbox{\tiny b}}^{\mbox{\tiny 1S}}
\Delta^{(3)}_{\mbox{\tiny m}} [\mbox{MeV}]$
 & $10$ & $-1$ & $-4$ & $-4$ & $-5$ & $-5$ & $-5$
 & $-5$ & $-5$ \\ \hline
$M_{\mbox{\tiny b}}^{\mbox{\tiny 1S}}
\sum_{i=2}^3\Delta^{(i)}_{\mbox{\tiny m}} [\mbox{MeV}]$
 & $-14$ & $-17$ & $-16$ & $-15$ & $-14$ & $-14$ & $-13$
 & $-12$ & $-12$
\\ \hline\hline
$(\Delta^{(2)}+\Delta^{(2)}_{\mbox{\tiny m}})/\Delta^{(2)}$
 & $0.85$ & $0.52$ & $1.56$ & $1.21$ & $1.14$ & $1.10$ & $1.09$
 & $1.08$ & $1.07$ \\ \hline
$(\Delta^{(3)}+\Delta^{(3)}_{\mbox{\tiny m}})/\Delta^{(3)}$
 & $1.04$ & $0.98$ & $0.81$ & $-2.07$ & $1.50$ & $1.28$ & $1.21$
 & $1.18$ & $1.15$ \\ \hline\hline
$M_{\mbox{\tiny b}}^{\mbox{\tiny 1S}}
\Delta^{(1)}_{\beta_0} [\mbox{MeV}]$
% & $-577$ & $-502$ & $-459$ & $-431$ & $-410$ & $-393$ & $-380$
% & $-369$ & $-360$
 & $-698$ & $-570$ & $-506$ & $-466$ & $-438$ & $-417$ & $-401$
 & $-387$ & $-376$
\\ \hline
$M_{\mbox{\tiny b}}^{\mbox{\tiny 1S}}
\Delta^{(2)}_{\beta_0} [\mbox{MeV}]$
% & $-197$ & $-199$ & $-201$ & $-201$ & $-201$ & $-200$ & $-199$
% & $-198$ & $-197$
 & $-368$ & $-274$ & $-251$ & $-240$ & $-232$ & $-227$ & $-222$ 
 & $-219$ & $-215$
\\ \hline
$M_{\mbox{\tiny b}}^{\mbox{\tiny 1S}}
\Delta^{(3)}_{\beta_0} [\mbox{MeV}]$
% & $222$ & $92$ & $38$ & $8$ & $-11$ & $-25$ & $-35 $ 
% & $-42$ & $-49$
 & $512$ & $153$ & $58$ & $14$ & $-11$ & $-28$ & $-39$ 
 & $-48$ & $-54$
\\ \hline
$M_{\mbox{\tiny b}}^{\mbox{\tiny 1S}}
\Delta^{(4)}_{\beta_0} [\mbox{MeV}]$
% & $-265$ & $ -75$ & $-31$ & $-18$ & $-13$ & $-13$ & $-13$
% & $-15$ & $-16$
 & $-779$ & $-142$ & $-49$ & $-25$ & $-17$ & $-16$ & $-16$ 
 & $-17$ & $-19$
\\ \hline\hline
\end{tabular}
\caption{\label{tabmsbar1Sscale} 
The scale dependence of the ``massless'' (no subscript), ``massive''
(subscript ``m'')
and the large-$\beta_0$ (subscript ``$\beta_0$'') corrections 
in the bottom $\overline{\mbox{MS}}$--1S mass relation 
for $\overline m(\overline m)=1.5$~GeV, 
$M_{\mbox{\tiny b}}^{\mbox{\tiny 1S}}=4.7$~GeV and 
$\alpha_s^{(5)}(M_Z)=0.118$. The superscript ``($i$)'' corresponds to
corrections at order $\alpha_s^i$ (or $\epsilon^i$ in the upsilon
expansion). 
}
\end{center}
\vskip 3mm
\end{table}
It is instructive to examine the size of the light quark mass
corrections in Eq.~(\ref{msbar1Sgeneric}) for the case of the bottom
quark ($n_l=4$) taking into account the mass of the charm quark and
treating the other light quarks as massless. For 
$M_{\mbox{\tiny b}}^{\mbox{\tiny 1S}}=4.7$~GeV, 
$\overline m(\overline m)=1.5$~GeV and
$\alpha_s^{(4)}(\mu=4.7~\mbox{GeV})=0.216$ we obtain
\begin{eqnarray}
\overline M_{\mbox{\tiny b}}(\overline M_{\mbox{\tiny b}}) & = &
4.7 \, - \,
\epsilon\,\Big[\,
0.382
\,\Big] - \,
\epsilon^2\,\Big[\,
0.098 \, + \,0.0072_{\mbox{\tiny m}}
\,\Big] - \,
\epsilon^3\,\Big[\,
0.030 \, + \, 0.0049_{\mbox{\tiny m}}
\,\Big]
\,.
\label{msbar1Sfirstnumbers}
\end{eqnarray} 
The corrections coming from the finite charm quark mass are indicated
by the subscript ``m''. The shift in  
$\overline M_{\mbox{\tiny b}}(\overline M_{\mbox{\tiny b}})$ caused by
the charm mass corrections is $-7.2$~MeV  
at order $\alpha_s^2$ ($\epsilon^2$) and $-4.9$~MeV at order
$\alpha_s^3$ ($\epsilon^3$). If we neglect the NNLO double-insertion 
corrections in $\Delta^{\mbox{\tiny NNLO}}_{\mbox{\tiny massive}}$
in Eq.~(\ref{Deltamassless3}), we obtain $-6.1$~MeV at
order $\alpha_s^3$. It is quite instructive to confront 
the perturbative series of the bottom pole--$\overline{\mbox{MS}}$ and
the bottom 1S--pole mass relations in Eqs.~(\ref{polemsbarthreeloops}) and
(\ref{1Spolegeneric}) with the result for the
$\overline{\mbox{MS}}$--1S mass relation in
Eq.~(\ref{msbar1Sfirstnumbers}). For 
$M_{\mbox{\tiny b}}^{\mbox{\tiny pole}}=4.9$~GeV and 
$\overline M_{\mbox{\tiny b}}(\overline M_{\mbox{\tiny b}})=4.2$~GeV
and the same values for $\alpha_s$, $\mu$ and 
$\overline m(\overline m)$ as for Eq.~(\ref{msbar1Sfirstnumbers}), we
obtain:
\begin{eqnarray}
M_{\mbox{\tiny b}}^{\mbox{\tiny pole}} & = &
4.2 \, + \,
\epsilon\,\Big[\,
0.385
\,\Big] + \,
\epsilon^2\,\Big[\,
0.197 \, + \,0.0117_{\mbox{\tiny m}}
\,\Big] + \,
\epsilon^3\,\Big[\,
0.142 \, + \, 0.0176_{\mbox{\tiny m}}
\,\Big]
\,,
\label{msbarpoletest}
\\[4mm]
M_{\mbox{\tiny b}}^{\mbox{\tiny 1S}}
& = & 
4.9 \, - \,
\epsilon\,\Big[\,
0.051
\,\Big] - \,
\epsilon^2\,\Big[\,
0.074 \, + \,0.0045_{\mbox{\tiny m}}
\,\Big] - \,
\epsilon^3\,\Big[\,
0.099 \, + \, 0.0121_{\mbox{\tiny m}}
\,\Big]
\,.
\label{1Spoletest}
\end{eqnarray}
Clearly, the bottom $\overline{\mbox{MS}}$--1S mass relation has a much
better convergence for the corrections for massless light quarks as
well as for the charm mass corrections. 
As expected, we find that the cancellation of the 
linear charm mass term is more efficient at order $\alpha_s^3$
($\epsilon^3$) than at order $\alpha_s^2$ ($\epsilon^2$).   
In Fig.~\ref{figmsbar1S}a the scale dependence of the charm mass
corrections 
$M_{\mbox{\tiny b}}^{\mbox{\tiny 1S}}
(\Delta^{(2)}_{\mbox{\tiny m}}+\Delta^{(3)}_{\mbox{\tiny m}})$
is displayed for  $M_{\mbox{\tiny b}}^{\mbox{\tiny 1S}}=4.7$~GeV and
$\alpha_s^{(5)}(M_Z)=0.118$ for
$\overline m(\overline m)=0.1$~GeV (dotted line), 
$1.0$~GeV (dash-dotted line), $1.5$~GeV (dashed line) and
$2.0$~GeV (solid line). For the strong coupling we have employed
four-loop running and three-loop matching conditions at the five-four
quark flavour threshold. 
In Table~\ref{tabmsbar1Sscale} the scale dependence of the ``massless''
and ``massive'' corrections in Eq.~(\ref{msbar1Sgeneric}) are shown
for $\overline m(\overline m)=1.5$~GeV, 
$M_{\mbox{\tiny b}}^{\mbox{\tiny 1S}}=4.7$~GeV and 
$\alpha_s^{(5)}(M_Z)=0.118$, separately for each order. We see that the  
convergence of  the charm mass corrections is best for $\mu\approx
\overline m(\overline m)$, whereas for the massless corrections this
happens for $\mu\approx 2.5$~GeV. This reflects that the
characteristic scales for the two types of corrections are slightly
different. This is not unexpected as the relevant dynamical scales for
the charm mass corrections are the charm mass and the inverse Bohr
radius $M_{\mbox{\tiny b}}\alpha_s$; the massless corrections, on the
other hand, 
involve $M_{\mbox{\tiny b}}$ and $M_{\mbox{\tiny b}}\alpha_s$ as
relevant dynamical scales. 
The convergence of the charm mass corrections shown in 
Table~\ref{tabmsbar1Sscale} clearly shows that they are well under
control. For our final
determination of the bottom quark $\overline{\mbox{MS}}$ mass in
Sec.~\ref{sectionnumerical}, we will use renormalization scales between
$1.5$ and $7$~GeV. For the determination of the central value of the
bottom $\overline{\mbox{MS}}$ mass we will use 
$\mu=M_{\mbox{\tiny b}}^{\mbox{\tiny 1S}}$.

In Fig.~\ref{figmsbar1S}b, finally, the size of the charm
mass correction 
$M_{\mbox{\tiny b}}^{\mbox{\tiny 1S}}(\Delta^{(2)}_{\mbox{\tiny m}}+
\Delta^{(3)}_{\mbox{\tiny m}})$
is plotted over $\overline m(\overline m)$ for 
$M_{\mbox{\tiny b}}^{\mbox{\tiny 1S}}=4.7$~GeV,
$\alpha_s^{(5)}(M_Z)=0.118$
and $\mu=1.5$~GeV (dotted line), $3$~GeV (dashed line) and $5$~GeV
(solid line) to illustrate the impact of 
other light quark masses on 
$\overline M_{\mbox{\tiny b}}(\overline M_{\mbox{\tiny b}})$. 
Figures~\ref{figmsbar1S}a and b show that for 
$\overline m(\overline m)=0.1$~GeV the corrections amount to much less
than $1$~MeV. Thus, the effects of the masses from light quarks other than
the charm can be safely neglected. 

\par
%\vspace{0.5cm}
%
\subsection{Estimate of the Order $\alpha_s^4$ Charm Mass Effects}
\label{subsectionmsbar1Smassestimate}
The results shown in the 9th and the 10th row of
Table~\ref{tabmsbar1Sscale} reveal a quite interesting property of the
ratio between the full corrections and the massless ones at
order $\alpha_s^2$ and $\alpha_s^3$.
For $\mu$ around $4$ to $5$~GeV the ratios at order $\alpha_s^2$ and
$\alpha_s^3$ are approximately equal to 
$(\beta_0^{(n_l=3)}/\beta_0^{(n_l=4)})=1.08$ and 
$(\beta_0^{(n_l=3)}/\beta_0^{(n_l=4)})^2=1.17$, respectively.
We already mentioned that the inclusion of the charm mass leads
to an effective decoupling of the charm quark corrections at large
orders. Thus, because the order $\alpha_s^{n+1}$ term in the 
$\overline{\mbox{MS}}$--1S mass relation in Eq.~(\ref{msbar1Sgeneric})
contains $\beta_0^n$, the value
$(\beta_0^{(n_l=3)}/\beta_0^{(n_l=4)})^n$ for the ratio might not be
unexpected as the asymptotic behaviour for large orders of
perturbation theory.  
However, we emphasize that if ones employs $\alpha_s^{(n_l=4)}$
rather than $\alpha_s^{(n_l=3)}$, as we do in this work, the
decoupling is not explicit. (See the results for the light quark mass
corrections for the static potential discussed in
Sec.~\ref{sectionpotential}.) In addition, because the series in 
the bottom $\overline{\mbox{MS}}$--1S mass relation in
Eq.~(\ref{msbar1Sgeneric}) involves the cancellation of the leading
infrared-sensitive contributions, also subleading contributions,
which involve for example $\beta_1$, lower powers of $\beta_0$ or
higher order contributions in the 
non-relativistic expansion, will contribute to the dominant asymptotic
behaviour of the perturbative coefficients in
Eq.~(\ref{msbar1Sgeneric}). 
Thus the observation above is
certainly a coincidence, particularly the one at order
$\alpha_s^2$. Nevertheless, the ratio
$(\beta_0^{(n_l=3)}/\beta_0^{(n_l=4)})^n$ might be used as a rough 
estimate for the large order behaviour of the ratio of full to
massless corrections in the $\overline{\mbox{MS}}$--1S mass
relation. It should therefore be possible to estimate the size of the
order $\alpha_s^4$ charm mass 
corrections $\Delta^{(4)}_{\mbox{\tiny m}}$ from the order
$\alpha_s^4$ corrections for massless light quarks,
$\Delta^{(4)}$. At the present stage the full result for
$\Delta^{(4)}$ is not yet known. However, it can be rather easily
calculated in the large-$\beta_0$ approximation because for the
N$^3$LO corrections to the pole--1S mass relation this only 
amounts to the insertion of one-loop massless quark  vacuum
polarizations into the LO static potential. The order $\alpha_s^4$
large-$\beta_0$ corrections to the pole--$\overline{\mbox{MS}}$ mass
relation are already known from Ref.~\cite{Beneke5}, whereas 
the N$^3$LO corrections to the pole--1S mass relation in the
large-$\beta_0$ approximation are calculated in
App.~\ref{appendix1Smasslargeb0}. The formula for $\Delta^{(4)}$ in
the large-$\beta_0$ approximation is derived in
App.~\ref{appendixmsbar1Smasslargeb0} and reads 
($\tilde a_s=\alpha_s^{(4)}(M_{\mbox{\tiny b}}^{\mbox{\tiny 1S}})$)
\begin{eqnarray}
\Delta^{(4)}_{\beta_0}
& = &
\frac{C_F^2\,\tilde a_s^2}{8}\,
\Big(\frac{\tilde a_s}{\pi}\Big)^3\,\beta_0^3\,
\bigg[\,
\frac{1}{2}\,\tilde L^3 +
\frac{15}{8}\,\tilde L^2 + 
\bigg(\frac{\pi^2}{12} + \zeta_3 + \frac{25}{12}\bigg)\,\tilde L
\nonumber
\\[2mm] & & \hspace{0.5cm}
+ \frac{\pi^4}{1440}
+ \frac{19\,\pi^2}{144}
+ \frac{3}{2}\,\zeta_5
- \bigg(\frac{\pi^2}{8}-\frac{11}{6}\bigg)\,\zeta_3 
+ \frac{517}{864}
\,\bigg]
\nonumber
\\[2mm]  & & - \,
\Big(\frac{\tilde a_s}{\pi}\Big)^4\,\beta_0^3
\bigg(\frac{71\,\pi^4}{7680} + \frac{317}{768}\,\zeta_3
  + \frac{89\,\pi^2}{1152} + \frac{42979}{331776}
\bigg)
\,,
\label{Deltamassless4largeb0}
\end{eqnarray}
where $\tilde L=\ln(1/C_F \tilde a_s)$.
Equation~(\ref{Deltamassless4largeb0}) can be generalized to arbitrary 
renormalization scales in the strong coupling using the one-loop
running of the strong coupling, in analogy to
Eq.~(\ref{runningalphas}), which only involves 
corrections proportional to $\beta_0$.  
We have displayed the scale dependence of
$M_{\mbox{\tiny b}}^{\mbox{\tiny 1S}}\Delta^{(i)}_{\beta_0}$ ($i=1,2,3,4$)
for $\overline m(\overline m)=1.5$~GeV,
$M_{\mbox{\tiny b}}^{\mbox{\tiny 1S}}=4.7$~GeV and 
$\alpha_s^{(5)}(M_Z)=0.118$
in the 11th to the 14th rows of Table~\ref{tabmsbar1Sscale}.
The values for very 
small choices of $\mu$ are meaningless. As indicated in the
discussion about the ratio $(\beta_0^{(3)}/\beta_0^{(4)})^n$ at
the beginning of this subsection, we see that the large-$\beta_0$
approximation does certainly not provide a very good estimate of the
true corrections. At order $\alpha_s^2$ and $\alpha_s^3$ the
large-$\beta_0$ result overestimates the true correction. For
$\mu\gsim 3$~GeV 
the difference is between $50$ and $100$\%. This demonstrates
that the large-$\beta_0$ approximation does in general not give more than
an order of magnitude estimate if it is applied to a perturbative
expansion that does not contain a linear sensitivity to small
momenta, as is the case in the $\overline{\mbox{MS}}$--1S mass
relation. (In the pole-$\overline{\mbox{MS}}$ mass relation, which
contains a linear sensitivity to small momenta, the large-$\beta_0$
approximation is considerably better, see e.g. Ref.~\cite{Beneke5}.)
On the other hand, the large-$\beta_0$ approximation seems
to indicate quite a good convergence for larger values of $\mu$. 
Although we cannot prove that this property allows for any conclusion
about the size of the true 
expression for $M_{\mbox{\tiny b}}^{\mbox{\tiny 1S}}\Delta^{(4)}$,
we believe that this is likely to be the case\footnote{
We are aware of the fact that $\Delta^{(4)}$ contains ultrasoft
``Lamb-shift''-type corrections that involve 
$M_{\mbox{\tiny b}}\alpha_s^2$ as the
relevant dynamical scale and might lead to considerable additional
contributions. 
Calculations of the latter have been
carried out in Refs.~\cite{Brambilla1,Kniehl1}. However, no concrete
statement can be 
made before all corrections in $\Delta^{(4)}$ are determined. 
}.
If the relative size of the large-$\beta_0$ corrections roughly
reflects the relative size of the true corrections, one can expect
that $M_{\mbox{\tiny b}}^{\mbox{\tiny 1S}}\Delta^{(4)}$ amounts to
around $-10$ to $-15$~MeV for $\mu$ around $5$~GeV.
Using the value $(\beta_0^{(3)}/\beta_0^{(4)})^3=1.26$ as an estimate
for the ratio $(\Delta^{(4)}+\Delta^{(4)}_{\mbox{\tiny m}})/\Delta^{(4)}$,
this results in $\Delta^{(4)}_{\mbox{\tiny m}}=-3$ to $-4$~MeV for
$\mu$ around $5$~GeV. This estimate seems to be consistent with 
the sum $M_{\mbox{\tiny b}}^{\mbox{\tiny 1S}}
(\Delta^{(2)}_{\mbox{\tiny m}}+\Delta^{(3)}_{\mbox{\tiny m}})$ 
at the scale $\mu=1.5$~GeV, where we find the best convergence of the
charm mass corrections. 

To conclude the somewhat speculative discussion of this subsection, we
note that the main outcome is that the large-$\beta_0$ approximation
does not provide a particularly good estimate of higher order
corrections in the $\overline{\mbox{MS}}$--1S mass relation. This is
because subleading infrared-sensitive contributions would have to be
taken into account to determine the dominant asymptotic behaviour of
the perturbative relation of short-distance masses. As far as our estimate
for the order $\alpha_s^4$ charm mass corrections in the bottom 
$\overline{\mbox{MS}}$--1S mass relation is concerned, we believe that
it is more likely to be correct than the estimate for the order
$\alpha_s^4$ massless corrections, as the light quark mass corrections
at order $\alpha_s^4$ do not yet contain any ultrasoft contributions.

\par
\vspace{0.5cm}
\section{Light Quark Mass Effects in the $\Upsilon$ Sum Rules}
\label{sectionsumrules}
In this section we determine the corrections coming from the finite
charm quark mass to the large-$n$ moments of the bottom--antibottom
quark cross section in $e^+e^-$ annihilation at NNLO in the
non-relativistic expansion. In Sec.~\ref{subsectionsumrulesbasic} we
briefly review the basic concepts involved in the $\Upsilon$ sum
rules. More details can be found in Ref.~\cite{Hoang1}. In
Sec.~\ref{subsectionsumrulesmethod} we outline our method to calculate
the moments, and in Sec.~\ref{subsectionsumrulescalculation} we present
the calculations 
of the charm mass corrections. Results for the calculations for
massless light quarks are not given here; they have been given in
Ref.~\cite{Hoang1}. In Sec.~\ref{subsectionsumrulesexamination} the
results for the charm mass corrections are examined. 

\par
%\vspace{0.5cm}
%
\subsection{Basic Issues}
\label{subsectionsumrulesbasic}
The sum rules for the $\Upsilon$ mesons start from the correlator of
two electromagnetic currents of bottom quarks at momentum transfer
$q$,
\begin{eqnarray}
\Pi_{\mu\nu}(q) & = &
-\,i \int dx\,e^{i\,q.x}\,
   \langle\, 0\,|\,T\,j^b_\mu(x)\,j^b_\nu(0)\,|\,0\, \rangle
\nonumber
\\[2mm] 
& \equiv &
\mbox{Im}\,[\,-i\,
\langle\, 0\,| T\, \tilde j^b_\mu(q) \,
 \tilde j^{b,\mu}(-q)\, |\, 0\,\rangle]
\,,
\label{vacpoldef}
\end{eqnarray}
where
\begin{equation}
j^b_\mu(x) \, =  \bar b(x)\,\gamma_\mu\,b(x)
\,,
\end{equation}
and the symbol $b$ denotes the bottom quark Dirac field. The $n$-th
moment $P_n$ of the vacuum polarization function is defined as
\begin{equation}
P_n \, \equiv \,
\frac{4\,\pi^2\,Q_b^2}{n!\,q^2}\,
\bigg(\frac{d}{d q^2}\bigg)^n\,\Pi_\mu^{\,\,\,\mu}(q^2)\bigg|_{q^2=0}
\,,
\label{momentsdef1}
\end{equation}
where $Q_b=-1/3$ is the electric charge of the bottom quark.
Due to causality the $n$-th moment $P_n$ can also be written as a
dispersion integral
\begin{equation}
P_n \, = \,
\int\limits^\infty_{\sqrt{s}_{\rm min}} \frac{d s}{s^{n+1}}\,R(s)
\,,
\label{momentsdef2}
\end{equation}
where
\begin{equation}
R(s) \, = \, \frac{\sigma(e^+e^-\to\gamma^*\to \mbox{``$b\bar b$+X''})}
{\sigma_{pt}}
\, = \,
\frac{4\,\pi\,Q_b^2}{s}\,
\mbox{Im}
\,\Pi_\mu^{\,\,\,\mu}(s)
\label{Rdefinitioncovariant}
\end{equation}
is the total photon-mediated cross section of bottom quark--antiquark
production in $e^+e^-$ annihilation normalized to the Born cross
section for massless leptons, $\sigma_{pt}\equiv 4 \pi \alpha^2/3 s$,
and $s$ is the square of the centre-of-mass energy. The lower limit of
the integration in Eq.~(\ref{momentsdef2}) is set by the mass of the
lowest-lying resonance. Assuming global duality the moments $P_n$ can
be either calculated from experimental data on $R$ or theoretically
using perturbative QCD. 

The experimental moments $P_n^{\rm ex}$ are determined by using
latest data 
on the $\Upsilon$ meson masses, $M_{\rm kS}$, and electronic
decay widths, $\Gamma_{\rm kS}$, for $k=1,\ldots,6$. The formula
for the experimental moments used in this work reads
\begin{eqnarray}
P_n^{\rm ex} & = &  
\frac{9\,\pi}{\tilde\alpha^2_{\rm em}}\,\sum\limits_{k=1}^6\,
\frac{\Gamma_{\rm kS}}{M_{\rm kS}^{2n+1}} 
\, + \,
\int\limits_{(\sqrt{s})_{{\rm B}\bar{\rm B}}}^\infty
\frac{ds}{s^{n+1}}\,r_{\rm cont}(s)
\,,
\label{Pnexperiment}
\end{eqnarray}
and is based on the narrow width approximation for the known
$\Upsilon$ resonances; $\tilde\alpha_{\rm em}$ is the electromagnetic
coupling at the scale $10$~GeV. Because the difference in the
electromagnetic coupling for the different $\Upsilon$ masses is
negligible, we chose $10$~GeV as the scale of the electromagnetic
coupling for all resonances.
The continuum cross section above the ${\rm B}\bar{\rm B}$
threshold is approximated by the constant $r_c=1/3$, which is equal to
the Born cross section for $s\to\infty$, assuming a $50\%$
uncertainty,
\begin{equation}
r_{\rm cont}(s) = r_c\,(1 \pm 0.5)
\,.
\label{Rexperimentcontinuum}
\end{equation}
For $n\ge 4$ the continuum contribution is already suppressed 
sufficiently so that a more detailed description is not needed.
For a compilation of all experimental numbers used in this work, see
Table~\ref{tabdata}.

\begin{table}[t!]  % tabdata
\vskip 7mm
\begin{center}
\begin{tabular}{|c||r@{$.$}l|c|} \hline
 nS & \multicolumn{2}{|c||}{$M_{\rm nS}/[\mbox{GeV}]$} 
  & $\Gamma_{\rm nS}/[\mbox{keV}]$ 
\\ \hline\hline
 1S & $\hspace{0.6cm} 9$&$460$ & $1.32\pm 0.04\pm 0.03$
\\ \hline
 2S & $10$&$023$ & $0.52\pm 0.03\pm 0.01$
\\ \hline
 3S & $10$&$355$ & $0.48\pm 0.03\pm 0.03$
\\ \hline
 4S & $10$&$58$ & $0.25\pm 0.03\pm 0.01$ 
\\ \hline
 5S & $10$&$87$ & $0.31\pm 0.05\pm 0.07$
\\ \hline
 6S & $11$&$02$ & $0.13\pm 0.03\pm 0.03$
\\ \hline
\end{tabular}
\caption{\label{tabdata} 
The experimental numbers for the $\Upsilon$ masses and electronic decay
widths used for the calculation of the experimental moments
$P_n^{\rm ex}$. For the widths, the first error is statistical and the
second one systematic. The errors for the partial widths of
$\Upsilon(\mbox{1S})$ and $\Upsilon(\mbox{2S})$ are taken from
Ref.~\cite{Albrecht1}. All the other errors are estimated from the
numbers presented in Ref.~\cite{PDG}. For the electromagnetic coupling
at $10$~GeV and the ${\rm B}\bar{\rm B}$ threshold point we use
$\tilde \alpha_{\rm em}^{-1} = \alpha_{\rm em}^{-1}(10\,\mbox{GeV}) 
= 131.8(1\pm0.005)$ and
$(\sqrt{s})_{{\rm B}\bar{\rm B}} = 2\times 5.279$~GeV.
The small errors in the
$\Upsilon$ masses and the ${\rm B}\bar{\rm B}$ 
threshold $(\sqrt{s})_{{\rm B}\bar{\rm B}}$
are neglected.
}
\end{center}
\vskip 3mm
\end{table}

A reliable computation of the theoretical moments $P_n^{\rm th}$ based on 
perturbative QCD is only possible if the effective energy range
contributing to the integration in Eq.~(\ref{momentsdef2}) 
is sufficiently larger than
$\Lambda_{\rm QCD}\sim{\cal{O}}(200$--$300~\mbox{MeV})$~\cite{Poggio1}. 
For large values of $n$ it can be shown that the size of this
energy range is of order $M_{\mbox{\tiny b}}/n$. Qualitatively this 
dependence on
the moment parameter $n$ can be seen from Eq.~(\ref{momentsdef2}):
large values of $n$ enhance the low energy contributions and suppress
the high energy ones. This implies 
that $n$ should be chosen sufficiently smaller than $15$--$20$. This is
consistent with estimates of non-perturbative effects in the large-$n$
moments using the gluon condensate~\cite{Voloshin1,newcondensate}:
\begin{eqnarray}
P_n^{\mbox{\tiny non-pert}} 
& \approx & P_n^{\rm th}\,\bigg[\,1 \, - \,
\langle\,\alpha_s\,{\mbox{\boldmath $G$}}^2\,\rangle
\frac{n^3\,\pi}{72\,M_{\mbox{\tiny b}}^4}
\exp\Big(-0.4\,C_F\,\alpha_s\,\sqrt{n}\Big)
\,\bigg]
\,,
\label{momentsnonperturbative}
\end{eqnarray}
where the $P_n^{\rm th}$ represent the theoretical moments obtained with
pure perturbation theory. The RHS of Eq.~(\ref{momentsnonperturbative}) 
amounts to less than a per cent of the theoretical moments for
$n<20$. However, this result needs some careful interpretation
because the estimate of non-perturbative effects based on an expansion
in gluonic and light quark condensates is only reliable if
$M_{\mbox{\tiny b}} v^2\gg\Lambda_{\rm QCD}$. In addition, the values of the
dimension-$6$ (and higher) condensates as well as all perturbative
Wilson coefficients for the condensates are unknown, which introduces
additional systematic uncertainties. In order to suppress these sources of
uncertainties,
$n$ should be chosen as small as possible. However, it is
also desirable to choose $n$ as large as possible in order to suppress
the contribution from the $b\bar b$ continuum to
$R(s)$ above the ${\rm B}\bar{\rm B}$ threshold, which is rather
poorly known experimentally.
In other words, one has to choose $n$ large enough for the
bottom--antibottom quark dynamics encoded in the moments $P_n$ to be
non-relativistic. Because the size of the energy range
contributing to the $n$-th moment is of order $M_{\mbox{\tiny b}}/n$,
the mean centre-of-mass velocity of the bottom quarks in the $n$-th
moment is $v \, \sim \, 1/\sqrt{n}$.
This counting rule allows for a quantitative formulation
of the two requirements for the choice of $n$: theoretical reliability
demands $M_{\mbox{\tiny b}} v^2\gg\Lambda_{\rm QCD}$, and dominance of the
non-relativistic dynamics demands $v\ll 1$. If both requirements are
met, the bottom--antibottom quark dynamics encoded in the moments is
perturbative, i.e. the hierarchy $M_{\mbox{\tiny b}}\gg 
M_{\mbox{\tiny b}} v\gg M_{\mbox{\tiny b}} v^2 \gg
\Lambda_{\rm QCD}$ is valid. Only if this condition is satisfied, the
non-relativistic power counting described after
Eq.~(\ref{NNLOSchroedinger}) and used throughout this work is
truly justified. Only in this case can the non-relativistic
bottom--antibottom quark pair be considered as a Coulombic system
for which the perturbative treatment carried out in this work is
feasible and for which the power counting
\begin{equation}
v \, \sim \, \alpha_s \, \sim \,\frac{1}{\sqrt{n}}
\,,
\label{powercountinganv}
\end{equation} 
used for the calculation of the moments, can be applied.
In our analysis we choose
\begin{equation}
4\le n\le 10
\label{nrange}
\end{equation}
as the allowed range for the moments $P_n$. The upper bound was chosen
to avoid the problem of unknown systematic non-perturbative
uncertainties as much as possible. The gluon condensate contributions
mentioned above amounts to less than a per mille in the moments for
$n\le 10$. This is an order of magnitude less than the charm mass
corrections. Therefore, non-perturbative effects are not taken into
account in this work. The lower bound, on the other hand, is a
compromise born out of the requirement of the non-relativistic
dominance and of the fact that we would like to have a reasonable number
of moments that can be used for fitting. We also note that our
estimate of the size of non-perturbative effects is based on the
common faith that the expansion in gluon condensates really describes
the dominant source of non-perturbative contributions. This is not
necessarily the case, as has been pointed out for instance in
Refs.~\cite{Zakharov2}.
As we will discuss in Sec.~\ref{subsectionsumrulesexamination}, 
the perturbative behaviour of the moments does not seem to comply at
all the estimate of the smallness of non-perturbative effects 
based on the gluon condensate (presuming that perturbative convergence
has anything to do with the size non-perturbative effects). We believe
that it is probably the weakest point of the sum rule method.

\par
%\vspace{0.5cm}
%
\subsection{Method for the Calculation of the Moments}
\label{subsectionsumrulesmethod}
In this subsection we briefly review the method used in
Ref.~\cite{Hoang1} 
for the calculation of the NNLO moments for massless light quarks
assuming the pole definition for the bottom mass. This sets the stage
for the determination of the charm mass corrections carried out in
Sec.~\ref{subsectionsumrulescalculation}.

At NNLO in the non-relativistic expansion, we need to calculate
corrections up to order $\alpha_s^2$, $\alpha_s/\sqrt{n}$ and $1/n$ to
the moments $P_n$ with respect to the non-relativistic (LO)
limit. According to the power counting~(\ref{powercountinganv}) we
need to keep corrections of order 
$\alpha_s^2$, $\alpha_s v$ and $v^k\,n^l$ with $k-2\,l=2$ in the
non-relativistic expansion of Eq.~(\ref{momentsdef2}),
\begin{eqnarray}
P_n^{\rm th} \, = \,
\frac{1}{4^{n}\,(M_{\mbox{\tiny b}}^{\mbox{\tiny pole}})^{2n}}
\,\int\limits_{E_{\rm bind}}^\infty 
\frac{d E}{M_{\mbox{\tiny b}}^{\mbox{\tiny pole}}} \,
\exp\bigg(
-\frac{E}{M_{\mbox{\tiny b}}^{\mbox{\tiny pole}}}\,n
\,\bigg)\,\bigg(\,
1 - \frac{E}{2\,M_{\mbox{\tiny b}}^{\mbox{\tiny pole}}} + 
\frac{E^2}{4\,(M_{\mbox{\tiny b}}^{\mbox{\tiny pole}})^2}\,n
\,\bigg)\,R_{\mbox{\tiny NNLO}}^{\mbox{\tiny thr}}(E)
\,,
\label{Pnnonrelativistic}
\end{eqnarray} 
where $E=\sqrt{s}-2M_{\mbox{\tiny b}}^{\mbox{\tiny pole}}$ and 
$E_{\rm bind}$ is the perturbative
binding energy of the lowest-lying resonance, i.e. 
$E_{\rm bind}=2(M_{\mbox{\tiny b}}^{1S}-
M_{\mbox{\tiny b}}^{\mbox{\tiny pole}})$. The exponential
$\exp(-E/M_{\mbox{\tiny b}}^{\mbox{\tiny pole}}n)$ is the LO term in
the non-relativistic expansion of $ds/s^{n+1}$ and cannot be expanded
because $E/M_{\mbox{\tiny b}}^{\mbox{\tiny pole}}n$ is of order $1$
according to the power counting of Eq.~(\ref{powercountinganv}). 
We note that the non-relativistic expansion of $ds/s^{n+1}$ does not
contain any NLO corrections, as these would be non-analytic
functions of the energy $E$. Thus, because the light quark mass
corrections start at NLO, we only need to consider the LO term in the
non-relativistic expansion of $ds/s^{n+1}$ in the determination of the
light quark mass corrections up to NNLO. The term 
$R_{\mbox{\tiny NNLO}}^{\mbox{\tiny thr}}$ is the bottom--antibottom
quark cross section expanded at NNLO in the non-relativistic
expansion. It is obtained by expanding the relativistic
electromagnetic bottom quark currents in
Eq.~(\ref{Rdefinitioncovariant}) in terms of effective non-relativistic
${}^3S_1$ currents up to dimension $5$ ($i=1,2,3$):
\begin{eqnarray}
\tilde j^b_i(q) & = & c_1\,\Big({\tilde \psi}^\dagger \sigma_i 
\tilde \chi\Big)(q) -
\frac{c_2}{6 M_b^2}\,\Big({\tilde \psi}^\dagger \sigma_i
(\mbox{$-\frac{i}{2}$} 
\stackrel{\leftrightarrow}{\mbox{\boldmath $D$}})^2
 \tilde \chi\Big)(q) + \ldots
\,,
\label{currentexpansion1}
\\[4mm]
\tilde j^b_i(-q) & = & c_1\,\Big({\tilde \chi}^\dagger \sigma_i 
\tilde \psi\Big)(-q) -
\frac{c_2}{6 M_b^2}\,\Big({\tilde \chi}^\dagger \sigma_i
(\mbox{$-\frac{i}{2}$} 
\stackrel{\leftrightarrow}{\mbox{\boldmath $D$}})^2
 \tilde \psi\Big)(-q) + \ldots 
\,,
\label{currentexpansion2}
\end{eqnarray}
where $s=q^2$.
The non-relativistic currents are defined in the NRQCD factorization
scheme proposed by Lepage {\it et al.} in
Refs.~\cite{Caswell1,Bodwin1}, which 
separates contributions coming from momenta of order 
$M_{\mbox{\tiny b}}$ from non-relativistic momenta of order 
$M_{\mbox{\tiny b}}v$ and $M_{\mbox{\tiny b}}v^2$.
%and $\Lambda_{\rm QCD}$. 
The constants $c_1$ and $c_2$ are short-distance
Wilson coefficients that contain the contributions from momenta of
order $M_{\mbox{\tiny b}}$. At the Born level, $c_1=c_2=1$. At NNLO we
need $c_1$ at order $\alpha_s^2$; for $c_2$ we do not need to
calculate any corrections as the contributions of dimension-$5$
currents are already suppressed by two powers of $v$. From
Eqs.~(\ref{currentexpansion1}) and (\ref{currentexpansion2}) one
obtains 
\begin{eqnarray}    
R_{\mbox{\tiny NNLO}}^{\mbox{\tiny thr}}(E) & = &
\frac{\pi\,Q_b^2}{(M_{\mbox{\tiny b}}^{\mbox{\tiny pole}})^2}
\,C_1(\mu_{\rm hard},\mu_{\rm fac})\,
\mbox{Im}\Big[\,
{\cal{A}}_1(E,\mu_{\rm soft},\mu_{\rm fac})
\,\Big]
\nonumber\\[2mm]
& & - \,\frac{4 \, \pi\,Q_b^2}
{3 (M_{\mbox{\tiny b}}^{\mbox{\tiny pole}})^4}\,
\mbox{Im}\Big[\,{\cal{A}}_2(E,\mu_{\rm soft})
\,\Big]
+ \ldots
\,,
\label{Rthreshexpansion}
\end{eqnarray}
where $C_1=c_1^2$ and

\begin{eqnarray}
{\cal{A}}_1 & \equiv & i\,\langle \, 0 \, | 
\, ({\tilde\psi}^\dagger \vec\sigma \, \tilde \chi)\,
\, ({\tilde\chi}^\dagger \vec\sigma \, \tilde \psi)\,
| \, 0 \, \rangle
\,,
\label{A1definition}
\\[4mm]
{\cal{A}}_2 & \equiv & \mbox{$\frac{1}{2}$}\,i\,\langle \, 0 \, | 
\, ({\tilde\psi}^\dagger \vec\sigma \, \tilde \chi)\,
\, ({\tilde\chi}^\dagger \vec\sigma \, (\mbox{$-\frac{i}{2}$} 
\stackrel{\leftrightarrow}{\mbox{\boldmath $D$}})^2 \tilde \psi)\,
+ \mbox{h.c.}\,
| \, 0 \, \rangle
\,.
\label{A2definition}
\end{eqnarray}
The constant $Q_b=-1/3$ is the electric charge of the bottom quark.
Using the NRQCD equation of motion for the bottom and antibottom quark
fields, one can relate ${\cal{A}}_2$ to ${\cal{A}}_1$, 
${\cal{A}}_2=M_{\mbox{\tiny b}}^{\mbox{\tiny pole}} E {\cal{A}}_1$. In
Eq.~(\ref{Rthreshexpansion}) we have indicated the dependence of the
non-relativistic cross section on the renormalization scales used in
Refs.~\cite{Hoang1,Hoang2}. The scale $\mu_{\rm soft}$ is the
renormalization scale of 
the strong coupling governing the bottom--antibottom quark potential in
Eq.~(\ref{NNLOSchroedinger}) and $\mu_{\rm hard}$ the renormalizations
scale of the strong coupling in the short-distance coefficient
$C_1$. The scale $\mu_{\rm fac}$ is a factorization scale that
separates non-relativistic from hard momenta in the NRQCD
factorization scheme. The scales $\mu_{\rm hard}$, $\mu_{\rm fac}$ are
irrelevant to the charm quark mass corrections determined in this
work, since they contain no hard corrections up to NNLO. The
dimension-$6$ current correlator ${\cal{A}}_1$ is obtained at 
NNLO from the configuration space Green function of
Eq.~(\ref{NNLOSchroedinger}) for both spatial arguments evaluated at
zero distance:
\begin{eqnarray}
{\cal{A}}_1 & = & 6\,N_c\,G(0,0,E)
\,.
\label{A1Greenfunction}
\end{eqnarray}
(See Ref.~\cite{Hoang7} for details on the derivation of
Eq.~(\ref{A1Greenfunction}).) The short-distance coefficient $C_1$ is
obtained by matching the NNLO cross section in
Eq.~(\ref{Rthreshexpansion}) to the same cross section in full QCD in
the (formal) limit $\alpha_s\ll v\ll 1$ at the two-loop level and 
including terms in the velocity expansion up to NNLO. This method of
determining the short-distance coefficient $C_1$ at the level of the
cross section (rather than the amplitude) is called ``direct
matching''~\cite{Hoang8}.
The calculation of the NNLO zero-distance Green function of the
Schr\"odinger equation~(\ref{NNLOSchroedinger}) proceeds in analogy to
the determination of the NNLO corrections to the 1S mass carried out in
Sec.~\ref{sectionpole1Smass} via Rayleigh--Schr\"odinger
time-independent perturbation theory.
Parametrizing the zero-distance Green function as
\begin{eqnarray}
G(0,0,E) & = & 
G_c(0,0,E) + G^{\mbox{\tiny NLO}}(0,0,E)
+ G^{\mbox{\tiny NNLO}}(0,0,E)
\,,
\label{Greenfunctiongeneric}
\end{eqnarray}
where $G_c$ is the LO non-relativistic Coulomb Green function,
the formal expressions for
$G^{\mbox{\tiny NLO}}$ and $G^{\mbox{\tiny NNLO}}$ at zero distances
read  
($r\equiv|{\mbox{\boldmath $r$}}|$, 
$r^\prime\equiv|{\mbox{\boldmath $r$}^\prime}|$):
\begin{eqnarray}
G^{\mbox{\tiny NLO}}(0,0,E) 
& = &
-\,\int d^3{\mbox{\boldmath $r$}} \, 
G_c(0,r,E)
\,V_{\mbox{\tiny c,massless}}^{\mbox{\tiny NLO}}({\mbox{\boldmath $r$}})
\,G_c(r,0,E)
\,,
\label{GNLOformal}
\\[4mm] 
G^{\mbox{\tiny NNLO}}(0,0,E) 
& = &
-\,\int d^3{\mbox{\boldmath $r$}} \, 
G_c(0,r,E)
\,\bigg(\,
V_{\mbox{\tiny c,massless}}^{\mbox{\tiny NNLO}}
({\mbox{\boldmath $r$}})+ 
  V_{\mbox{\tiny BF}}({\mbox{\boldmath $r$}}) + 
  V_{\mbox{\tiny NA}}({\mbox{\boldmath $r$}}) +
  \delta H_{\mbox{\tiny kin}}\,\bigg)
\,G_c(r,0,E)
\nonumber
\\[2mm]  
\lefteqn{
+\,
\int d^3{\mbox{\boldmath $r$}}\,
\int d^3{\mbox{\boldmath $r$}^\prime}\,
G_c(0,r,E)\,
V_{\mbox{\tiny c,massless}}^{\mbox{\tiny NLO}}
({\mbox{\boldmath $r$}})\,
G_c^{S}(r,r^\prime,E)\,
V_{\mbox{\tiny c,massless}}^{\mbox{\tiny NLO}}
({\mbox{\boldmath $r$}^\prime})\,
G_c(r^\prime,0,E)
}
\label{GNNLOformal}
\end{eqnarray}
in configuration space representation,
where $\delta H_{\mbox{\tiny kin}}=
-{\mbox{\boldmath $\nabla$}}^4/4 M_{\mbox{\tiny b}}^
{\mbox{\tiny pole}}$,
\begin{eqnarray}
%\lefteqn{
G_c^{S}(r,r^\prime, E)
\,\Big|_{r^\prime \, < \, r} 
& \equiv &
\frac{1}{4 \pi}\,\int\! d\Omega \, 
G_c({\mbox{\boldmath $r$}},{\mbox{\boldmath $r$}}^\prime, E)
%}
\nonumber\\[2mm]
& = &
\frac{M_{\mbox{\tiny b}}^{\mbox{\tiny pole}}\,k}
{2\,\pi}\,\frac{\sin(\pi\,\lambda)}{\pi\,\lambda}\,
e^{-k(r+r^\prime)}\,
\nonumber
\\
& & \quad \times\,
\bigg[\,
\int\limits_0^\infty dt\,
e^{-2 k r t}\,
\bigg(\,\frac{1+t}{t}
\,\bigg)^{\lambda}
\,\bigg]\,
\bigg[\,
\int\limits_0^1 du\,
e^{2 k r^\prime u}\,
\bigg(\,\frac{1-u}{u}
\,\bigg)^{\lambda}
\,\bigg]
\,,
\label{GreenfunctionLOSwave}
\\[4mm]
G_c(0,r, E) & = & G_c(r,0, E)
 \, = \, 
\frac{M_{\mbox{\tiny b}}^{\mbox{\tiny pole}}\,k}
{2\,\pi}\,e^{- k\,r}\,
\int\limits_0^\infty\! dt \,
e^{-2\,k\,r\,t}\,
\bigg(\,\frac{1+t}{t}
\,\bigg)^{\lambda}
%\nonumber\\[2mm] & = & 
%-\,i\,\frac{M_{\mbox{\tiny b}}^{\mbox{\tiny pole}}\,p}
%{2\,\pi}\,e^{i\,p\,r}\,
%\Gamma(1-i\,\rho)\,U(1-i\,\rho,2,-2\,i\,p\,r)
%\nonumber\\[2mm] & = & 
%\frac{M_{\mbox{\tiny b}}^{\mbox{\tiny pole}}}
%{4\,\pi\,r}\,\Gamma(1-i\,\rho)\,
%   W_{i\,\rho\,,\frac{1}{2}}(-2\,i\,p\,r)
\,,
\label{GreenfunctionLOzeror}
\end{eqnarray}
and
\begin{eqnarray}
k 
& \equiv & 
\sqrt{-E\,M_{\mbox{\tiny b}}^{\mbox{\tiny pole}}}
\,,
\label{kdef}
\\[4mm]
\lambda
& \equiv &
\frac{C_F\,a_s\,M_{\mbox{\tiny b}}^{\mbox{\tiny pole}}}{2\,k}
\,.
\label{lambdadef}
\end{eqnarray}
The expressions for the Coulomb Green function in
Eqs.~(\ref{GreenfunctionLOSwave}) and (\ref{GreenfunctionLOzeror})
have been derived in Refs.~\cite{Wichmann1}.
For the NNLO contributions arising from second order
Rayleigh--Schr\"odinger perturbation theory we only need the S-wave
component of the Coulomb Green function, $G_c^{S}$, for the sum over
intermediate states because the static potential at NLO is spin- and
angle-independent. The case $r < r^\prime$ in 
Eq.~(\ref{GreenfunctionLOSwave}) is obtained by interchanging 
$r$ and $r^\prime$.

\begin{figure}[t] % figcontour
\begin{center}
\leavevmode
\epsfxsize=4.cm
\epsffile[220 420 420 550]{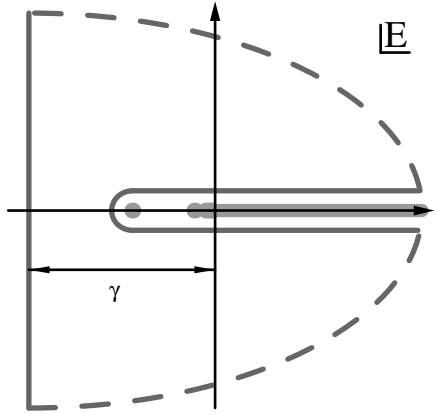}\\
\vskip  3.5cm
 \caption{\label{figcontour} 
Path of integration to calculate
expression~(\ref{Pnnonrelativisticdeformed}) for 
the theoretical moments. The dashed line closes the contour
at infinity and does not contribute to the integration. The free constant
$\gamma$ is chosen large enough to be safely away from the bound state
poles, which are indicated by the grey dots on the negative energy
axis. The thick grey line on the positive energy axis represents the
continuum. 
}
 \end{center}
\end{figure}

The calculation of the moments as shown in
Eq.~(\ref{Pnnonrelativistic}) is somewhat cumbersome, because it
requires the separate determination of bound state ($E<0$) and
continuum ($E>0$) contributions. In particular, for the bound state
contributions, the binding energies and the leptonic decay rates
(which are proportional to the modulus squared of the configuration
space wave functions at the origin) of all resonances would need to be
determined individually. A more economic way to calculate the moments
can be achieved by deforming the path of integration in
Eq.~(\ref{Pnnonrelativistic}) into the negative complex plane, as
shown in Fig.~\ref{figcontour}. 
Because the path that closes the
contour at infinity does not contribute, the result of the integration
in Eq.~(\ref{Pnnonrelativistic}) can be written as
\begin{eqnarray}
\lefteqn{
P_n^{\rm th} \, = \,
\frac{-2\, i\,Q_b^2\,\pi}{(4M_b^2)^{n+1}}
\int\limits_{-\gamma-i\infty}^{-\gamma+i\infty} \frac{d E}{M_b} 
\exp\bigg(-\frac{E}{M_b}\,\bigg)\,\bigg(
1 - \frac{E}{2M_b} + \frac{E^2}{4M_b^2} n
\bigg)\,\bigg[\,
C_1\,{\cal{A}}_1(E) -\frac{4}{3M_b^2}\,{\cal{A}}_2(E)
\,\bigg]
}
\nonumber
\\[2mm] & = &
\frac{4Q_b^2\pi^2}{(4M_b^2)^{n+1}}\,\frac{1}{2\pi i}\,
\int\limits_{\gamma-i\infty}^{\gamma+i\infty} \frac{d \tilde E}{M_b} \,
\exp\bigg(\frac{\tilde E}{M_b}n\bigg)\,\bigg(\,
1 + \frac{\tilde E}{2M_b} + \frac{\tilde E^2}{4M_b^2}\,n
\,\bigg)\bigg[\,
C_1 \,{\cal{A}}_1(-\tilde E) -\frac{4}{3M_b^2}\, 
{\cal{A}}_2(-\tilde E)
\,\bigg]
\,,
\nonumber\\&&
\label{Pnnonrelativisticdeformed}
\end{eqnarray}
where the free constant $\gamma$ is chosen much larger than the ground
state binding energy $E_{\rm bind}$. In the second line of 
Eq.~(\ref{Pnnonrelativisticdeformed}) the change of variable
$E\to-\tilde E$ has been performed. We note that, in contrast to
Eq.~(\ref{Pnnonrelativistic}), we also need the real part of the
correlators ${\cal{A}}_1$ and ${\cal{A}}_2$ in
Eq.~(\ref{Pnnonrelativisticdeformed}). The expression in
Eq.~(\ref{Pnnonrelativisticdeformed}) has three advantages: (i) the
bound state and continuum contributions are obtained at the same time,
(ii) the correlators, i.e. the Green functions shown in
Eq.~(\ref{Greenfunctiongeneric}), can be naively expanded in
$\alpha_s$ before the integration, and (iii) we can use a vast number
of tables for the calculations, because the complex energy integration
is nothing else than an inverse Laplace transform. This elegant way to
determine the moments has been used first in Ref.~\cite{Voloshin1}.  

The final result for the NNLO
theoretical moments for massless light quarks can be cast into the
form
\begin{eqnarray}
P_n^{\rm th} & = &
\frac{3\,N_c\,Q_b^2\,\sqrt{\pi}}{4^{n+1}\,
(M_{\mbox{\tiny b}}^{\mbox{\tiny pole}})^{2n} \, n^{3/2}}\,
\bigg\{\,
C_1\Big(\frac{\mu_{\rm hard}}
{M_{\mbox{\tiny b}}^{\mbox{\tiny pole}}},\frac{\mu_{\rm fac}}
{M_{\mbox{\tiny b}}^{\mbox{\tiny pole}}},
\alpha_s(\mu_{\rm hard})\Big)\,
\varrho_{n,1}\Big(\frac{\mu_{\rm soft}}
{M_{\mbox{\tiny b}}^{\mbox{\tiny pole}}},
    \frac{\mu_{\rm fac}}
{M_{\mbox{\tiny b}}^{\mbox{\tiny pole}}},\alpha_s(\mu_{\rm soft})\Big)
\nonumber\\[2mm] & &
\mbox{\hspace{4cm}}
 +\, \varrho_{n,2}\Big(\alpha_s(\mu_{\rm soft})\Big)
\,\bigg\}
\,,
\label{Pnnonrelativisticgeneric}
\end{eqnarray}
where
\begin{eqnarray}
\varrho_{n,1} & = &
\frac{8\,\pi^{3/2}\,n^{3/2}}
{(M_{\mbox{\tiny b}}^{\mbox{\tiny pole}})^2}\frac{1}{2\pi i}
\int\limits_{\gamma-i\infty}^{\gamma+i\infty} 
\frac{d \tilde E}{M_{\mbox{\tiny b}}^{\mbox{\tiny pole}}}\,
\exp\bigg(\frac{\tilde E}{M_b}\,n\bigg)\,
\bigg(
1 + \frac{\tilde E}{2M_{\mbox{\tiny b}}^{\mbox{\tiny pole}}} 
+ \frac{\tilde E^2}{4(M_{\mbox{\tiny b}}^{\mbox{\tiny pole}})^2}\,n
\bigg)\,
G(0,0,-\tilde E)
\,,
\nonumber
\\
\label{rho1massless}
\\[4mm]
\varrho_{n,2} & = &
\frac{8\,\pi^{3/2}\,n^{3/2}}
{(M_{\mbox{\tiny b}}^{\mbox{\tiny pole}})^2}\frac{1}{2\pi i}
\int\limits_{\gamma-i\infty}^{\gamma+i\infty} 
\frac{d \tilde E}{M_{\mbox{\tiny b}}^{\mbox{\tiny pole}}}\,
\exp\bigg(\frac{\tilde E}{M_{\mbox{\tiny b}}}\,n\bigg)\,
\bigg(\,\frac{4}{3}\,\frac{\tilde E}
{M_{\mbox{\tiny}}^{\mbox{\tiny pole}}}\,\bigg)\,
G_c(0,0,-\tilde E)
\,.
\label{rho2massless}
\end{eqnarray}
The explicit results for $\varrho_{n,1}$ and $\varrho_{n,2}$ can be
found in Ref.~\cite{Hoang1}. We note that $\varrho_{n,1}$ and
$\varrho_{n,2}$ are dimensionless and that $\varrho_{n,1}$ depends
only logarithmically 
on the bottom pole mass at NLO and NNLO through the dependence of the
strong coupling on the inverse Bohr radius.
We also note that only $\varrho_{n,1}$ receives
light quark mass corrections, since $\varrho_{n,2}$ contains only
NNLO contributions.

\par
%\vspace{0.5cm}
%
\subsection{Light Quark Mass Corrections to the Moments}
\label{subsectionsumrulescalculation}
The light quark mass corrections to the moments are determined
following the lines of the calculation for massless light quarks
described in the previous subsection. The insertions of the
NLO and NNLO static potentials that are needed are in complete analogy
to the calculation of the heavy quark 1S mass presented in
Sec.~\ref{sectionpole1Smass}. 

\par
%\vspace{0.5cm}
%
\subsubsection*{Light Quark Mass Corrections at NLO}
The NLO light quark mass corrections to the zero-distance Green
function read
\begin{eqnarray}
G^{\mbox{\tiny NLO}}_{\mbox{\tiny m}}(0,0,E) 
& = &
-\,\int d^3{\mbox{\boldmath $r$}} \, 
G_c(0,r,E)
\,\delta V_{\mbox{\tiny c,m}}^{\mbox{\tiny NLO}}({\mbox{\boldmath $r$}})
\,G_c(r,0,E)
\,,
\label{GVNLOmassiveGeneric}
\end{eqnarray}
where the formulae for 
$\delta V_{\mbox{\tiny c,m}}^{\mbox{\tiny NLO}}$ and $G_c$ are given in    
Eqs.~(\ref{VcNLOmassiverspace}) and (\ref{GreenfunctionLOzeror}),
respectively;
$\delta V_{\mbox{\tiny c,m}}^{\mbox{\tiny NLO}}
({\mbox{\boldmath $r$}})$ contains a contribution from the massive
light quark vacuum polarization function, called ``massive''
corrections in the following, and a contribution from the subtraction
of the massive light quark vacuum polarization function in the limit
$m\to 0$, called ``massless'' correction in the following. This
subtraction is a consequence of our convention that $n_l$ quark
species contribute to the evolution of the strong coupling. For the
massless corrections we can recycle the results for massless light
quarks given in Ref.~\cite{Hoang1}. The massive
corrections to the zero-distance Green function involve the term
($a_s=\alpha_s^{(4)}(\mu_{\rm soft})$):
\begin{eqnarray}
\lefteqn{
-\,\int d^3{\mbox{\boldmath $r$}} \, 
G_c(0,r,E)\,\bigg[\,
\bigg(-\frac{C_F\,a_s}{r}\,\bigg)\,
%\Big(\frac{a_s}{3\,\pi}\Big)\,
\int\limits_1^\infty d x\,f(x)\,e^{- 2 m r x}
\,\bigg]
\,G_c(r,0,E)
}
\nonumber
\\[2mm] 
& = &
C_F\,a_s\,%\Big(\frac{a_s}{3\,\pi}\Big)\,
\frac{(M_{\mbox{\tiny b}}^{\mbox{\tiny pole}})^2\,k^2}{\pi}\,
\int\limits_1^\infty d x\,f(x)\,
\int\limits_0^\infty d t\,
\int\limits_0^\infty d u\,
\int\limits_0^\infty d r\,
r\,e^{-2\,[k(1+t+u)+m x]\,r}\,
\Big(\frac{1+t}{t}\Big)^\lambda\,
\Big(\frac{1+u}{u}\Big)^\lambda
\nonumber
\\[2mm] 
%& = &
%C_F\,a_s\,%\Big(\frac{a_s}{3\,\pi}\Big)\,
%\frac{(M_{\mbox{\tiny b}}^{\mbox{\tiny pole}})^2\,k^2}{\pi}\,
%\int\limits_1^\infty d x\,f(x)\,
%\int\limits_0^\infty d t\,
%\int\limits_0^\infty d u\,
%\frac{1}{4\,[k(1+t+u)+m x]^2}\,
%\Big(\frac{1+t}{t}\Big)^\lambda\,
%\Big(\frac{1+u}{u}\Big)^\lambda
%\nonumber
%\\[2mm] 
& = &
C_F\,a_s\,%\Big(\frac{a_s}{3\,\pi}\Big)\,
\frac{(M_{\mbox{\tiny b}}^{\mbox{\tiny pole}})^2}{4\,\pi}\,
\sum\limits_{p=0}^\infty\,
\int\limits_1^\infty d x\,f(x)\,
\int\limits_0^\infty d t\,
\int\limits_0^\infty d u\,
\frac{\lambda^p}{p!}\,
\ln^p\Big(\frac{(1+t)(1+u)}{t\,u}\Big)\,
\frac{\tilde k^2}{(1+t+u)^2\,(\tilde k+\tilde a)^2}
\,,
\nonumber
\\
\label{GVNLOmassiveGexplicit}
\end{eqnarray}
where
\begin{eqnarray}
\tilde k 
& \equiv & 
\frac{k}{M_{\mbox{\tiny b}}^{\mbox{\tiny pole}}}
\,,
\\[2mm]
\tilde a 
& \equiv &
\frac{m\,x}{M_{\mbox{\tiny b}}^{\mbox{\tiny pole}}\,(1+t+u)}
\,,
\end{eqnarray}
and $k$ and $\lambda$ are defined in Eqs.~(\ref{kdef}) and
(\ref{lambdadef}), respectively.
In the last line of Eq.~(\ref{GVNLOmassiveGexplicit}) we have expanded
in $\alpha_s$ owing to the integration path in the negative
complex energy plane shown in Eq.~(\ref{Pnnonrelativisticdeformed}). 
To determine the contributions of Eq.~(\ref{GVNLOmassiveGexplicit}) to
the light quark mass corrections of the moments $\varrho_{n,1}$, we use
the following inverse Laplace transform~\cite{Prudnikov1}:
\begin{eqnarray}
\frac{1}{2\pi\,i}\,\int\limits_{\gamma-i\infty}^{\gamma+i\infty}\,
d z^2\,
\frac{z^{2-p}}{(z+A)^2}\,
e^{z^2n}
& = & n^{\frac{p}{2}-1}\,g_0(p,A^2n)
\,,
\label{Laplace1}
\end{eqnarray} 
where
\begin{eqnarray}
g_0(p,X) & \equiv &
\frac{1 + 2\,X}{\Gamma(\frac{p}{2})} 
- \frac{2\,\sqrt{X}\,(1 + X)}{\Gamma(\frac{1 + p}{2})} 
+ e^X\,X^{1 - \frac{p}{2}}\,(3 - p + 2\,X)\,
\bigg[\,
\frac{\Gamma(\frac{1 + p}{2},X)}{\Gamma(\frac{1 + p}{2})}
-\frac{\Gamma(\frac{p}{2},X)}{\Gamma(\frac{p}{2})} 
\,\bigg]
\,.
\nonumber
\\
\label{g0def}
\end{eqnarray}
The final result for the NLO light quark mass corrections to the
moments reads
\begin{eqnarray}
\delta\varrho_{n,1}^{\mbox{\tiny m,NLO}}
& = &
4\,\sqrt{\pi}\,\Big(\frac{a_s}{3\,\pi}\Big)\,\phi\,
\bigg\{\,
\bigg[\,
\frac{1}{2}\,\int\limits_1^\infty dx\,f(x)\,e^{n\,\eta^2\,x^2}\,
 \Big[1-{\rm erf}(\sqrt{n}\,\eta\,x)\Big]
%\nonumber
%\\[2mm]
%& & \hspace{1cm}
+\,\sum\limits_{p=1}^\infty\,\phi^p\,
\bar g_0\Big(n,p,\eta\Big)
\,\bigg]
\nonumber
\\[2mm]
& & 
+\,\bigg[\,
  \frac{1}{2}\,
  \ln\Big(\frac{\eta\,\sqrt{n}}{2}\Big)
   + \frac{\gamma_{\mbox{\tiny E}}}{4} + \frac{5}{12}
 + \sum\limits_{p=1}^\infty\,
   \phi^p\,\bigg(\,
   w_p^0\,
  \bigg(\frac{1}{2}\,\Psi\Big(\frac{p}{2}\Big)
   -\ln\Big(\frac{\eta\,\sqrt{n}}{2}\Big)
   -\frac{5}{6}\,\bigg) + 
   w_p^1\,\bigg)
\,\bigg]
\,\bigg\}
\,,
\nonumber
\\
\label{deltarhomassiveNLO}
\end{eqnarray}
where
\begin{eqnarray}
\phi 
& \equiv &
\frac{C_F\,a_s\,\sqrt{n}}{2}
\,,
\\[4mm]
\eta
& \equiv &
\frac{m}{M_{\mbox{\tiny b}}^{\mbox{\tiny pole}}}
\,,
\end{eqnarray}
and
\begin{eqnarray}
\bar g_0(n,p,\eta)
& = &
\int\limits_0^\infty\,dt\,
\int\limits_0^\infty\,du\,
\int\limits_1^\infty\,dx\,f(x)\,
\frac{1}{p!\,(1+t+u)^2}\,\ln^p\Big(\frac{(1+t)\,(1+u)}{t\,u}\Big)\,
g_0(p,X)\,\bigg|_{X=\frac{\eta^2\,x^2\,n}{(1+t+u)^2}}
\,.
\nonumber
\\
\label{g0bardef}
\end{eqnarray}
The first term in the curly brackets on the RHS of
Eq.~(\ref{deltarhomassiveNLO}) comes from the massive corrections and
the second from the massless ones.
The constants $w_p^0$ and $w_p^1$ have already been calculated in 
Ref.~\cite{Hoang1}; their expressions can be found in
App.~\ref{appendixmassless}.
The function $\bar g_0$ has been calculated numerically using
standard variable transforms to treat the logarithmic
singularities. We note that the numerical determination of $\bar g_0$
represents a considerable effort. 
For given values of $p$ and $n$ we have calculated $\bar g_0$ for
$r=0.01$, $0.2$, $0.3$, $0.4$ and $0.5$, and constructed an
interpolation function to determine the values for arbitrary values of
$r$. This was done in order to speed up the computations that arise
during the fitting procedure. We checked that the
relative deviation of the interpolation from the exact result
is at the per cent level. We also note that the terms of the sum over
$p$ 
%(which correspond to the expansion of the Green functions in
%terms of $\alpha_s$ that we have mentioned after
%Eq.~(\ref{Pnnonrelativisticdeformed})) 
only forms a convergent series
for $p$ larger than $10$. The larger the value of $n$, the more terms
in $p$ have to 
be taken into account. For $n\le 20$ it is sufficient to carry out the
sum up to $p=20$, which is what we have done is our numerical
evaluation of the moments. The issues just mentioned also apply to the
NNLO light quark mass corrections to $\varrho_{n,1}$.

\par
%\vspace{0.5cm}
%
\subsubsection*{Light Quark Mass Corrections at NNLO -- Single Insertions}
The NNLO light quark mass corrections to the zero-distance Green
function read
\begin{eqnarray}
\lefteqn{
\delta G^{\mbox{\tiny NNLO}}_{\mbox{\tiny m}}(0,0,E) 
\, = \,
-\,\int d^3{\mbox{\boldmath $r$}} \, 
G_c(0,r,E)\,
\delta V_{\mbox{\tiny c,m}}^{\mbox{\tiny NNLO}}
({\mbox{\boldmath $r$}})
\,G_c(r,0,E)
}
\nonumber
\\[2mm] & &
+\,
\int d^3{\mbox{\boldmath $r$}}\,
\int d^3{\mbox{\boldmath $r$}^\prime}\,
G_c(0,r,E)\,
\delta V_{\mbox{\tiny c,m}}^{\mbox{\tiny NLO}}
({\mbox{\boldmath $r$}})\,
G_c^{S}(r,r^\prime,E)\,
\delta V_{\mbox{\tiny c,m}}^{\mbox{\tiny NLO}}
({\mbox{\boldmath $r$}^\prime})\,
G_c(r^\prime,0,E)
\nonumber
\\[2mm] & &
+\,2\,
\int d^3{\mbox{\boldmath $r$}}\,
\int d^3{\mbox{\boldmath $r$}^\prime}\,
G_c(0,r,E)\,
\delta V_{\mbox{\tiny c,m}}^{\mbox{\tiny NLO}}
({\mbox{\boldmath $r$}})\,
G_c^{S}(r,r^\prime,E)\,
V_{\mbox{\tiny c,massless}}^{\mbox{\tiny NLO}}
({\mbox{\boldmath $r$}^\prime})\,
G_c(r^\prime,0,E)
\,.
\label{GVNNLOmassiveGgeneric}
\end{eqnarray}
Let us first consider the contributions from the single
insertion of the NNLO light quark mass corrections to the static
potential at first order Rayleigh--Schr\"odinger perturbation
theory. As for the NLO calculation, we distinguish between ``massive''
and ``massless'' corrections and only give details for the massive
contributions. Apart from expression~(\ref{GVNLOmassiveGexplicit})
there is one more term coming from the light quark mass corrections to 
the NNLO static potential that is relevant for the massive
corrections: 
\begin{eqnarray}
\lefteqn{
-\,\int d^3{\mbox{\boldmath $r$}} \, 
G_c(0,r,E)\,\bigg[\,
\frac{C_F\,a_s}{r}\,
\int\limits_1^\infty d x\,f(x)\,e^{- 2 m r x}\,
\bigg(\,
\ln(4\,x^2)-{\rm Ei}(2\,m\,x\,r)-{\rm Ei}(-2\,m\,x\,r)
\,\bigg)
\,\bigg]
\,G_c(r,0,E)
}
\nonumber
\\[2mm] 
& = &
-\,C_F\,a_s\,
\frac{(M_{\mbox{\tiny b}}^{\mbox{\tiny pole}})^2}{4\,\pi}\,
\sum\limits_{p=0}^\infty\,
\int\limits_1^\infty d x\,f(x)\,
\int\limits_0^\infty d t\,
\int\limits_0^\infty d u\,
\frac{\lambda^p}{p!}\,
\ln^p\Big(\frac{(1+t)(1+u)}{t\,u}\Big)\,
\frac{1}{(1+t+u)^2}
\hspace{2cm}
\nonumber
\\[2mm] & & \hspace{1cm} \times\,
\bigg\{\,
  \frac{\tilde k^2}{(\tilde k+\tilde a)^2}\,
    \ln\Big(\frac{4\,\tilde k^2\,x^2}{\tilde a^2}\Big)
 + 2\,\frac{\tilde k^2}{\tilde a^2-\tilde k^2}\,
    \bigg[\,
      1 - \frac{2\,\tilde k\,\tilde a}{\tilde a^2-\tilde k^2}\,
         \ln\Big(\frac{\tilde a}{\tilde k}\Big)
    \,\bigg]
\,\bigg\}
\,.
\label{GVNNLOmassiveGexplicit1}
\end{eqnarray}  
To determine the corrections to the moments $\varrho_{n,1}$ we need
the following inverse Laplace transforms~\cite{Prudnikov1}, in
addition to Eq.~(\ref{Laplace1}):
\begin{eqnarray}
\frac{1}{2\pi\,i}\,\int\limits_{\gamma-i\infty}^{\gamma+i\infty}\,
d z^2\,
\frac{2\,z^{2-p}}{A^2-z^2}\,
e^{z^2n}
& = & n^{\frac{p}{2}-1}\,g_1(p,A^2n)
\,,
\label{Laplace2}
\\[2mm]
\frac{1}{2\pi\,i}\,\int\limits_{\gamma-i\infty}^{\gamma+i\infty}\,
d z^2\,
\bigg[-\frac{4\,A\,z^{3-p}}{(A^2-z^2)^2}\,\bigg]\,
e^{z^2n}
& = & n^{\frac{p}{2}-1}\,g_2(p,A^2n)
\,,
\label{Laplace3}
\\[2mm]
\end{eqnarray} 
where
\begin{eqnarray}
g_1(p,X) & \equiv &
-\,2\,\bigg[\,
\frac{1}{\Gamma(\frac{p}{2})}
+ e^X\,X^{1 - \frac{p}{2}}\,
\bigg(\,
1 - \frac{\Gamma(\frac{p}{2},X)}{\Gamma(\frac{p}{2})}
\,\bigg)
\,\bigg]
\,,
\label{g1def}
\\[4mm]
g_2(p,X) & \equiv &
-\,2\,\bigg[\,
\frac{2\,\sqrt{X}\,(1 + X)}{\Gamma(\frac{1 + p}{2})}
+ e^X\,X^{1 - \frac{p}{2}}\,(3 - p + 2\,X)\,
\bigg(\,
1 - \frac{\Gamma(\frac{1 + p}{2},X)}{\Gamma(\frac{1 + p}{2})}
\,\bigg)
\,\bigg]
\,.
\label{g2def}
\end{eqnarray}
The final result for the single insertion NNLO light quark mass
corrections to the moments reads
\begin{eqnarray}
\lefteqn{
\Big[\,\delta\varrho_{n,1}^{\mbox{\tiny m,NNLO}}\,\Big]_{\mbox{\tiny single}} 
\, = \,
}
\nonumber
\\[2mm]
%
% quark-gluon contributions
%
& = &
4\,\sqrt{\pi}\,\Big(\frac{a_s}{3\,\pi}\Big)^2\,\phi\,
\bigg\{\,
\bigg[\,
-\,\frac{3}{4}\int\limits_{1}^\infty dx\,f(x)\,
\bigg(\,h(x,n,\eta)\,\Big(\beta_0\ln\Big(\frac{4\,m^2\,x^2}{\mu^2}\Big)-a_1\Big)
+\beta_0\,e^{\eta^2 x^2 n}\,\Gamma(0,\eta^2\,x^2\,n)
\,\bigg) 
\nonumber
\\[2mm]
& & \qquad
+ \,\frac{3}{2}\,\sum\limits_{p=1}^\infty\,\phi^p\,
\bigg(
\beta_0\,\Big(\,
  \ln\Big(\frac{\mu^2}{m^2}\Big)\,\bar g_0(n,p,\eta)
  + \bar g_2(n,p,\eta)\,\Big)
  +a_1\,\bar g_0(n,p,\eta)
\bigg)
\nonumber
\\[2mm]
& & \qquad + \,
3\,\beta_0\,\bigg(\,
\frac{1}{2}\,
% \bigg(\,
 \ln^2\Big(\frac{c\,\eta\,\sqrt{n}}{2}\Big)
%  +\frac{\gamma_{\mbox{\tiny E}}}{2}+\frac{5}{6}\,\bigg)^2 
 -\frac{5}{12}\,
%\bigg(
  \ln\Big(\frac{c\,\eta\,\sqrt{n}}{2}\Big)
%   +\frac{\gamma_{\mbox{\tiny E}}}{2}+\frac{5}{6}
%  \bigg)
 +\frac{5\,\pi^2}{48}
\,\bigg)
%\nonumber
%\\[2mm]
%& & \qquad
+ \frac{3}{4}\,
\bigg(\beta_0\,\ln\Big(\frac{\mu^2}{m^2}\Big)+a_1\bigg)
%\bigg(
  \ln\Big(\frac{c\,\eta\,\sqrt{n}}{2}\Big)
% +\frac{\gamma_{\mbox{\tiny E}}}{2}+\frac{5}{6}\bigg)
\nonumber
\\[2mm]
& & \qquad
- \,3\,\beta_0\,\sum\limits_{p=1}^\infty\,\phi^p\,
\bigg(\,
   w_p^0\,\Big(
  {\rm cln2}(n,p,\eta)
%   \Big(\frac{1}{2}\Psi\Big(\frac{p}{2}\Big)
%       -\ln\Big(\frac{\eta\,\sqrt{n}}{2}\Big)-\frac{5}{6}\Big)^2
   +\frac{5}{6}\,{\rm cln}(n,p,\eta)
%   +\frac{5}{6}\,\Big(\frac{1}{2}\Psi\Big(\frac{p}{2}\Big)
%       -\ln\Big(\frac{\eta\,\sqrt{n}}{2}\Big)
%   -\frac{5}{6}\Big) 
%\nonumber
%\\[2mm]
%& & \hspace{2cm}
%   - \,\frac{1}{4}\,\Psi^\prime\Big(\frac{p}{2}\Big) 
   + \frac{\pi^2}{12}\Big)
 +  2\,w_p^1\,\Big(\,
   {\rm cln}(n,p,\eta)+\frac{5}{12}
%     \frac{1}{2}\Psi\Big(\frac{p}{2}\Big) -
%     \ln\Big(\frac{\eta\,\sqrt{n}}{2}\Big) - \frac{5}{12}
     \,\Big) 
 - w_p^2
\,\bigg) 
\nonumber
\\[2mm]
& & \qquad
+\,\frac{3}{2}\,\bigg(\beta_0\,\ln\Big(\frac{\mu^2}{m^2}\Big)+a_1\bigg)\,
\sum\limits_{p=1}^\infty\,\phi^p\,
\bigg(\,
  w_p^0\,
%\bigg(
  {\rm cln}(n,p,\eta)
%    \frac{1}{2}\Psi\Big(\frac{p}{2}\Big)
%   -\ln\Big(\frac{\eta\,\sqrt{n}}{2}\Big)-\frac{5}{6}
%  \bigg)
+ w_p^1
\,\bigg) 
\,\bigg]
\nonumber 
\\[2mm]
%
% quark-quark contributions
%
& &
+\,
\bigg[\,
-\,\frac{1}{2}\int\limits_{1}^\infty dx\,f(x)\,
\bigg(\,h(x,n,\eta)\,\Big(g(x)+\ln(4\,x^2)-\frac{5}{3}\Big)
+e^{\eta^2 x^2 n}\,\Gamma(0,\eta^2\,x^2\,n)
\,\bigg) 
\nonumber
\\[2mm]
& & \qquad
+ \,\sum\limits_{p=1}^\infty\,\phi^p\,
\bigg(
\bar g_1(n,p,\eta) 
+ \bar g_2(n,p,\eta) 
+ \frac{5}{3}\,\bar g_0(n,p,\eta)
\bigg)
\nonumber
\\[2mm]
& & \qquad
+\, \frac{1}{2}\,
 \ln^2\Big(\frac{c\,\eta\,\sqrt{n}}{2}\Big) 
 +\frac{5\,\pi^2}{48}
%\nonumber
%\\[2mm]
%& & \qquad
- \,\sum\limits_{p=1}^\infty\,\phi^p\,
\bigg(\,
   w_p^0\,\Big(\,
  {\rm cln2}(n,p,\eta)
%   \Big(\frac{1}{2}\Psi\Big(\frac{p}{2}\Big)
%       -\ln\Big(\frac{\eta\,\sqrt{n}}{2}\Big)-\frac{5}{6}\Big)^2
%   - \frac{1}{4}\,\Psi^\prime\Big(\frac{p}{2}\Big) 
   + \frac{\pi^2}{12}\,\Big) 
%\nonumber
%\\[2mm]
%& & \hspace{2cm}
 + \,2\,w_p^1\,
%\bigg(\,
  {\rm cln}(n,p,\eta)
%     \frac{1}{2}\Psi\Big(\frac{p}{2}\Big) -
%     \ln\Big(\frac{\eta\,\sqrt{n}}{2}\Big) - \frac{5}{6}
%     \,\bigg) 
 - w_p^2
\,\bigg) 
\,\bigg]
\nonumber
\\[2mm]
%
% Melles contribution 
%
& &
+\,
\frac{57}{4}\,\bigg[
\frac{c_1}{2}\int\limits_{c_2}^\infty \frac{dx}{x}\,
  h(x,n,\eta)
% e^{\eta^2 x^2 n}\,\Big[1-{\rm erf}(\eta\,x\,\sqrt{n})\Big]
+\frac{d_1}{2}\int\limits_{d_2}^\infty \frac{dx}{x}\,
  h(x,n,\eta)
% e^{\eta^2 x^2 n}\,\Big[1-{\rm erf}(\eta\,x\,\sqrt{n})\Big]
+ \sum\limits_{p=1}^\infty\,\phi^p\,
\bar g_3(n,p,\eta)
\nonumber
\\[2mm]
& & \qquad +\,
 \frac{1}{2}\,\bigg(\,\ln\Big(\frac{\eta\sqrt{n}}{2}\Big)
   + \frac{\gamma_{\mbox{\tiny E}}}{2} +
      \frac{161}{228} + \frac{13}{19}\,\zeta_3\,\bigg)  
\nonumber
\\[2mm]
& & \qquad 
   +  \,\sum\limits_{p=1}^\infty
   \bigg(\,
   w_p^0\,
   \bigg(\frac{1}{2}\,\Psi\Big(\frac{p}{2}\Big)
   -\ln\Big(\frac{\eta\,\sqrt{n}}
      {2}\Big)
   -\frac{161}{228} - \frac{13}{19}\,\zeta_3 \,\bigg)
   + w_p^1 
  \bigg)\,
\,\bigg]
\,\bigg\}
\,,
\label{deltarhomassiveNNLOsingle}
\end{eqnarray}
where
\begin{eqnarray}
c 
& \equiv &
\exp\Big(\frac{\gamma_{E}}{2}+\frac{5}{6}\Big)
\,,
\\[4mm]
g(x)
& \equiv &
\frac{5}{3}+\frac{1}{x^2}\bigg(
1+\frac{1}{2\,x}\,\sqrt{x^2-1}\,(1+2x^2)\,
\ln\Big(\frac{x-\sqrt{x^2-1}}{x+\sqrt{x^2-1}}\Big)
\bigg)
\,,
\\[4mm]
h(x,n,\eta)
& \equiv &
e^{\eta^2 x^2 n}\,\Big[1-{\rm erf}(\eta\,x\,\sqrt{n})\Big]
\, = \, 
\frac{1}{\sqrt{\pi}}\,
e^{\eta^2 x^2 n}\,\Gamma\Big(\frac{1}{2},\eta^2\,x^2\,n\Big)
\,,
\\[4mm]
{\rm cln}(n,p,\eta)
& \equiv &
 \frac{1}{2}\Psi\Big(\frac{p}{2}\Big) -
     \ln\Big(\frac{\eta\,\sqrt{n}}{2}\Big) - \frac{5}{6}
\,,
\\[4mm]
{\rm cln2}(n,p,\eta)
& \equiv &
\bigg(\,\frac{1}{2}\Psi\Big(\frac{p}{2}\Big)
    -\ln\Big(\frac{\eta\,\sqrt{n}}{2}\Big)-\frac{5}{6}\,\bigg)^2
   - \frac{1}{4}\,\Psi^\prime\Big(\frac{p}{2}\Big) 
\,,
\end{eqnarray}
and
\begin{eqnarray}
\bar g_1(n,p,\eta)
& = &
-\,
\int\limits_0^\infty\,dt\,
\int\limits_0^\infty\,du\,
\int\limits_1^\infty\,dx\,f(x)\,
\bigg[\,\frac{5}{3}+\frac{1}{x^2}\bigg(
 1+\frac{1}{2\,x}\,\sqrt{x^2-1}\,(1+2x^2)\,
  \ln\Big(\frac{x-\sqrt{x^2-1}}{x+\sqrt{x^2-1}}\Big)
 \,\bigg)
\,\bigg]
\nonumber
\\[2mm]
& & \hspace{.2cm} \times\,
\frac{1}{p!\,(1+t+u)^2}\,\ln^p\Big(\frac{(1+t)\,(1+u)}{t\,u}\Big)\,
g_0(p,X)\,\bigg|_{X=\frac{\eta^2\,x^2\,n}{(1+t+u)^2}}
\,,
\label{g1bardef}
\\[2mm]
\bar g_2(n,p,\eta)
& = &
-\,
\int\limits_0^\infty\,dt\,
\int\limits_0^\infty\,du\,
\int\limits_1^\infty\,dx\,f(x)\,
\frac{1}{p!\,(1+t+u)^2}\,\ln^p\Big(\frac{(1+t)\,(1+u)}{t\,u}\Big)\,
\nonumber
\\[2mm]
& & \hspace{0.2cm} \times\,
\bigg[\,
 \bigg(\ln\Big(\frac{4\,x^2}{X}\Big)-2\frac{\partial}{\partial p}\bigg)
  g_0(p,X) + g_1(p,X)
  + \bigg(\frac{1}{2}\,\ln(X)+\frac{\partial}{\partial p}\bigg)
  g_2(p,X)
\,\bigg]\,
\bigg|_{X=\frac{\eta^2\,x^2\,n}{(1+t+u)^2}}
\,,
\nonumber\\
\label{g2bardef}
\\[2mm]
\bar g_3(n,p,\eta)
& = &
c_1\,
\int\limits_0^\infty\,dt\,
\int\limits_0^\infty\,du\,
\int\limits_{c_2}^\infty\,\frac{dx}{x}\,
\frac{1}{p!\,(1+t+u)^2}\,\ln^p\Big(\frac{(1+t)\,(1+u)}{t\,u}\Big)\,
g_0(p,X)\,\bigg|_{X=\frac{\eta^2\,x^2\,n}{(1+t+u)^2}}
\nonumber
\\[2mm]
& & \hspace{0.2cm} 
+ \,
d_1\,
\int\limits_0^\infty\,dt\,
\int\limits_0^\infty\,du\,
\int\limits_{d_2}^\infty\,\frac{dx}{x}\,
\frac{1}{p!\,(1+t+u)^2}\,\ln^p\Big(\frac{(1+t)\,(1+u)}{t\,u}\Big)\,
g_0(p,X)\,\bigg|_{X=\frac{\eta^2\,x^2\,n}{(1+t+u)^2}}
\,.
\nonumber\\
\label{g3bardef}
\end{eqnarray}
The constants $w_p^0$, $w_p^1$ and $w_p^2$ have already been
calculated in Ref.~\cite{Hoang1}; their expressions can be found in
App.~\ref{appendixmassless}.
The three terms in the curly brackets on the RHS of
Eq.~(\ref{deltarhomassiveNNLOsingle}) originate from the
corresponding three
terms given in Eq.~(\ref{VcNNLOmassiverspace}), which all vanish
individually for $m\to 0$. 
If the $\overline{\mbox{MS}}$ definition is used for the
charm quark mass, 
$m=\overline m(\overline m)[1+\frac{4}{3}(\frac{a_s}{\pi})]$,
we get an additional contribution to the moments at NNLO through the
replacement 
\begin{eqnarray}
\lefteqn{
\delta\varrho_{n,1}^{\mbox{\tiny m,NLO}}
(n,\eta,a_s)
\, \longrightarrow \,
\delta\varrho_{n,1}^{\mbox{\tiny m,NLO}}
(n,\bar\eta,a_s)\, +
}
\nonumber
\\[2mm]
&  & + \,
32\,\sqrt{\pi}\,\Big(\frac{a_s}{3\,\pi}\Big)^2\,\phi\,
\bigg\{\,
\bigg[\,
-\frac{1}{2}\,\int\limits_1^\infty dx\,f(x)\,
\bigg(\,
  \frac{\bar\eta\,x\,\sqrt{n}}{\sqrt{\pi}}
  - \bar\eta^2\,x^2\,n\,\,h(x,n,\bar\eta)
\,\bigg)
%\nonumber
%\\[2mm]
%& & \hspace{1cm}
+\,\sum\limits_{p=1}^\infty\,\phi^p\,
\bar g_4\Big(n,p,\bar\eta\Big)
\,\bigg]
\nonumber
\\[2mm]
& & \hspace{3cm}
+\,\bigg[\, \frac{1}{4}
 -\frac{1}{2}\, \sum\limits_{p=1}^\infty\,
   \phi^p\, w_p^0
\,\bigg]
\,\bigg\}
\,,
\label{deltarhomassiveNLOmsbar}
\end{eqnarray}
where
\begin{eqnarray}
\bar\eta
& = &
\frac{\overline m(\overline m)}{M_{\mbox{\tiny b}}^{\mbox{\tiny pole}}}
\,,
\end{eqnarray}
and
\begin{eqnarray}
\bar g_4(n,p,\eta)
& = &
\int\limits_0^\infty\,dt\,
\int\limits_0^\infty\,du\,
\int\limits_1^\infty\,dx\,f(x)\,
\frac{1}{p!(1+t+u)^2}\ln^p\Big(\frac{(1+t)(1+u)}{t\,u}\Big)
\bigg[
X\,\frac{\partial}{\partial X}g_0(p,X)\bigg]
\bigg|_{X=\frac{\eta^2\,x^2\,n}{(1+t+u)^2}}
\,.
\nonumber
\\
\label{g4bardef}
\end{eqnarray}

\par
%\vspace{0.5cm}
%
\subsubsection*{Light Quark Mass Corrections at NNLO -- Double Insertions}

The second and third terms on the RHS of
Eq.~(\ref{GVNNLOmassiveGgeneric}) come from double insertions of the
NLO static potential. The calculation of these 
contributions is much more involved than that of the ones coming from
single insertions and will not be presented in this
work. On the other hand, it is not necessary to include them in the
present analysis. We have shown that the NNLO double-insertion
contributions in the heavy quark pole--1S mass relation are suppressed 
with respect to the single-insertion ones; for the NNLO charm mass 
corrections, the double-insertion contributions amount to about 10\% of
the full NNLO result. As was pointed out in
Sec.~\ref{subsectionpole1Smassdouble}, this happens because at NNLO
the corrections are dominated by momenta low enough for the
linear (and $r$-independent) charm quark mass terms to dominate in the
static potential. The same is expected for the double-insertion
contributions of the NNLO charm quark mass corrections to the moments.
Thus, the double-insertion contributions are of the same order as the
terms $\propto(m_{\mbox{\tiny charm}}/M_{\mbox{\tiny b}})^n$,
$n\ge 2$, in the bottom quark pole--$\overline{\mbox{MS}}$ mass
relation, which are already neglected in this work. For the reasons 
mentioned here, we neglect the double-insertion contributions. 

\par
%\vspace{0.5cm}
%
\subsubsection*{Implementation of the 1S Mass}
The moments displayed in Eq.~(\ref{Pnnonrelativisticgeneric}), including
the charm quark mass corrections to
$\varrho_{n,1}$ in Eqs.~(\ref{deltarhomassiveNLO}) and
(\ref{deltarhomassiveNNLOsingle}), still depend on the bottom quark
pole mass through 
the global factor $(M_{\mbox{\tiny b}}^{\mbox{\tiny pole}})^{-2n}$
and through logarithms of ratios with the renormalization scales. The
ratios of the pole mass with the renormalization scales originate from
the running of the strong coupling and from anomalous dimensions of
the ${}^3S_1$ NRQCD bottom quark currents. These ratios only arise at
NLO and NNLO in $\varrho_{n,1}$ and $C_1$. They do not lead to any
modifications in the 1S mass scheme, because the difference between the
bottom 1S and the pole masses is of order $v^2\sim\alpha_s^2\sim 1/n$
in the non-relativistic power counting. The only relevant modification
that arises from the implementation of the bottom 1S mass comes from
the global factor 
$(M_{\mbox{\tiny b}}^{\mbox{\tiny pole}})^{-2n}$. Using 
Eq.~(\ref{pole1Sgeneric}) and the scaling $\alpha_s^2\sim 1/n$, the
global factor has the following form in the 1S mass scheme at NNLO in
the non-relativistic expansion 
($a_s=\alpha_s^{(4)}(\mu_{\rm soft})$):
\begin{eqnarray}
\lefteqn{
\frac{1}{(M_{\mbox{\tiny b}}^{\mbox{\tiny pole}})^{2n}} \, = \,
\frac{1}{(M_{\mbox{\tiny b}}^{\mbox{\tiny 1S}})^{2n}}\,
\exp\Big(-2\,n\,
\Delta^{\mbox{\tiny LO}}(a_s)\Big)
}
\nonumber
\\[2mm] & & \times\,
\bigg\{\,
1 - 2\,n\,\bigg[\,
  \Delta^{\mbox{\tiny NLO}}_{\mbox{\tiny massless}}
    (M_{\mbox{\tiny b}}^{\mbox{\tiny 1S}},a_s,\mu_{\rm soft}) + 
  \Delta^{\mbox{\tiny NLO}}_{\mbox{\tiny massive}}
    (\overline m(\overline m),M_{\mbox{\tiny b}}^{\mbox{\tiny 1S}},a_s) 
\,\bigg]
\nonumber
\\[2mm] & & \quad
+\,n\bigg[\,
\Big(\Delta^{\mbox{\tiny LO}}(a_s)\Big)^2   +
2\,n\,\Big(\,\Delta^{\mbox{\tiny NLO}}_{\mbox{\tiny massless}}
(M_{\mbox{\tiny b}}^{\mbox{\tiny 1S}},a_s,\mu_{\rm soft})
+\Delta^{\mbox{\tiny NLO}}_{\mbox{\tiny massive}}
(\overline m(\overline m),M_{\mbox{\tiny b}}^{\mbox{\tiny 1S}},a_s)\,\Big)^2
\nonumber
\\[2mm] & & \qquad
 -\, 2\,\Big(\,
 \Delta^{\mbox{\tiny NNLO}}_{\mbox{\tiny massless}}
(M_{\mbox{\tiny b}}^{\mbox{\tiny 1S}},a_s,\mu_{\rm soft})
 + \Delta^{\mbox{\tiny NNLO}}_{\mbox{\tiny massive}}
(\overline m(\overline m),
   M_{\mbox{\tiny b}}^{\mbox{\tiny 1S}},a_s,\mu_{\rm soft})\,
\Big)
\nonumber
\\[2mm] & & \qquad
 - \, 2\, \delta^{(1)}(a_s)\,
 \overline \Delta^{\mbox{\tiny NLO}}_{\mbox{\tiny m}}
    (\overline m(\overline m),M_{\mbox{\tiny b}}^{\mbox{\tiny
        1S}},a_s)\,
\bigg]\,
\bigg\}
\,,
\label{globalfactorpole1S}
\end{eqnarray}
where the $\Delta$'s have been given in
Eqs.~(\ref{delta1massless}),
(\ref{DeltaLOmassless})--(\ref{DeltaNNLOmassless}), 
(\ref{DeltaNLOmassiveexplicit}), (\ref{DeltaNNLOmassiveexplicit})
and (\ref{DeltabarNNLOmassiveexplicit}). The second and third terms
in the curly brackets on the RHS of Eq.~(\ref{globalfactorpole1S}) are
the NLO and NNLO contributions, respectively. We emphasize that
using Eq.~(\ref{globalfactorpole1S}) for the global factor 
$1/(M_{\mbox{\tiny b}}^{\mbox{\tiny pole}})^{2n}$ in
Eq.~(\ref{Pnnonrelativisticgeneric}) is equivalent to
implementing the bottom 1S mass via Eq.~(\ref{pole1Sgeneric}) into the
Schr\"odinger 
equation~(\ref{NNLOSchroedinger}) before any calculation is carried
out. The renormalization scale on the RHS of Eq.~(\ref{globalfactorpole1S})
is $\mu_{\rm soft}$, which governs the strong coupling in the static
potential. This is because the corrections on the RHS of
Eq.~(\ref{globalfactorpole1S}) only involve non-relativistic
momenta. From the technical point of view,  they are supposed to
compensate for the bad large order behaviour of the static
potential. For the same reason we have to consistently expand the LO,
NLO and NNLO corrections in Eq.~(\ref{globalfactorpole1S}) with the
LO, NLO and NNLO corrections in $\varrho_{n,1}$. Aside from the issue
of the cancellation of the large high order corrections that are associated
with the bottom quark pole mass definition, it is easy to understand
how the 1S mass definition reduces the correlation of the moments (and
also of the mass determination) to the strong coupling and the
dependence on $\mu_{\rm soft}$: in the 1S scheme the exponential
energy-enhancement of the ${}^3S_1$ bottom--antibottom ground state
that dominates the moments for large values of $n$, is eliminated. A
detailed discussion of this effect has been given in Ref.~\cite{Hoang2}.

\par
%\vspace{0.5cm}
%
\subsubsection*{Light Quark Mass Corrections to the 
                Short-Distance Coefficient $C_1$}
The short-distance coefficient $C_1$ contains the effects of the
bottom--antibottom quark production process that arise from momenta of
order $M_{\mbox{\tiny b}}$. This has two consequences: the first is
that the light quark mass corrections can be naively expanded in $m^2$
without the emergence of any terms that are non-analytic in $m^2$. Thus,
$C_1$ does not contain any linear dependence on a light quark
mass. The second consequence is that one can expand in the light quark
mass because $m/M_{\mbox{\tiny b}}$ is small. Therefore, the dominant
light quark mass effect in $C_1$ is proportional to 
$(\alpha_s/\pi)^2(m/M_{\mbox{\tiny b}})^2$, where the scale in the
strong coupling is of order $M_{\mbox{\tiny b}}$. For the charm quark
this correction is at the per mille level and subleading. The charm
mass corrections to $C_1$ are of the same order as the 
$m^2/M_{\mbox{\tiny b}}$ corrections in the
pole--$\overline{\mbox{MS}}$ mass relation that have been
neglected in this 
work. The light quark mass corrections to $C_1$ are therefore also
neglected in this work. We note that the absence of any linear light
quark mass corrections in $C_1$ has been proved by explicit
calculation in Ref.~\cite{Hoang9}. It was shown, for any energy at
order $\alpha_s^2$, that the total cross section $R$, 
Eq.~(\ref{Rdefinitioncovariant}), does not contain any linear light
quark mass corrections if a short-distance definition is employed for
the bottom quark mass. Thus, because $C_1=1$ (and is mass independent)
at the Born approximation, it cannot contain linear light quark mass
corrections at order $\alpha_s^2$.   

\par
%\vspace{0.5cm}
%
\subsection{Brief Examination}
\label{subsectionsumrulesexamination}
Before turning to the statistical analysis it is useful to examine the
size and behaviour of the light quark mass corrections to the moments
for various choices of the light quark mass. This will be helpful for
the interpretation of the result that is obtained for the bottom 1S
mass from the sum rule analysis carried out in the next
section. Detailed discussions on the behaviour of the moments in the
pole and the 1S mass schemes for the case of massless light quarks can
be found in Refs.~\cite{Hoang1,Hoang2}. The observations for the
behaviour of the moments in the 1S mass scheme made in the analysis of 
Ref.~\cite{Hoang2} remain qualitatively true if the effects of light quark
masses are included. Therefore we do not repeat any of the detailed
discussions of Ref.~\cite{Hoang2}. 
\begin{table}[t] % tabmoments
\vskip 7mm
\begin{center}
\begin{tabular}{|r@{$ $}l||c|c|c|c|c|} \hline
$\overline m(\overline m)$&$[\mbox{GeV}]$ 
 & $0.0$ & $0.1$ & $1.0$ & $1.5$ & $2.0$ \\ \hline\hline 
$P_4^{\mbox{\tiny th,LO}}$&$[10^{8}\,\mbox{GeV}^{8}]$ 
 & $0.301$ & $0.301$ & $0.301$ & $0.301$ & $0.301$ \\ \hline
$P_4^{\mbox{\tiny th,NLO}}$&$[10^{8}\,\mbox{GeV}^{8}]$ 
 & $0.157$ & $0.157$ & $0.158$ & $0.159$ & $0.159$ \\ \hline
$P_4^{\mbox{\tiny th,NNLO}}$&$[10^{8}\,\mbox{GeV}^{8}]$ 
 & $0.232$ & $0.232$ & $0.231$ & $0.231$ & $0.231$ \\ \hline\hline 
$P_6^{\mbox{\tiny th,LO}}$&$[10^{12}\,\mbox{GeV}^{12}]$ 
 & $0.268$ & $0.268$ & $0.268$ & $0.268$ & $0.268$ \\ \hline
$P_6^{\mbox{\tiny th,NLO}}$&$[10^{12}\,\mbox{GeV}^{12}]$ 
 & $0.145$ & $0.145$ & $0.146$ & $0.147$ & $0.148$ \\ \hline
$P_6^{\mbox{\tiny th,NNLO}}$&$[10^{12}\,\mbox{GeV}^{12}]$ 
 & $0.223$ & $0.223$ & $0.221$ & $0.220$ & $0.220$ \\ \hline\hline 
$P_8^{\mbox{\tiny th,LO}}$&$[10^{16}\,\mbox{GeV}^{16}]$ 
 & $0.273$ & $0.273$ & $0.273$ & $0.273$ & $0.273$ \\ \hline
$P_8^{\mbox{\tiny th,NLO}}$&$[10^{16}\,\mbox{GeV}^{16}]$ 
 & $0.152$ & $0.152$ & $0.154$ & $0.155$ & $0.156$ \\ \hline
$P_8^{\mbox{\tiny th,NNLO}}$&$[10^{16}\,\mbox{GeV}^{16}]$ 
 & $0.239$ & $0.238$ & $0.235$ & $0.234$ & $0.233$ \\ \hline\hline 
$P_{10}^{\mbox{\tiny th,LO}}$&$[10^{20}\,\mbox{GeV}^{20}]$ 
 & $0.297$ & $0.297$ & $0.297$ & $0.297$ & $0.297$ \\ \hline 
$P_{10}^{\mbox{\tiny th,NLO}}$&$[10^{20}\,\mbox{GeV}^{20}]$ 
 & $0.170$ & $0.170$ & $0.173$ & $0.174$ & $0.176$ \\ \hline 
$P_{10}^{\mbox{\tiny th,NNLO}}$&$[10^{20}\,\mbox{GeV}^{20}]$ 
 & $0.272$ & $0.270$ & $0.264$ & $0.263$ & $0.261$ \\ \hline\hline 
$P_{20}^{\mbox{\tiny th,LO}}$&$[10^{40}\,\mbox{GeV}^{40}]$ 
 & $0.679$ & $0.679$ & $0.679$ & $0.679$ & $0.679$ \\ \hline 
$P_{20}^{\mbox{\tiny th,NLO}}$&$[10^{40}\,\mbox{GeV}^{40}]$ 
 & $0.437$ & $0.470$ & $0.743$ & $0.865$ & $0.963$ \\ \hline 
$P_{20}^{\mbox{\tiny th,NNLO}}$&$[10^{40}\,\mbox{GeV}^{40}]$ 
 & $0.713$ & $0.698$ & $0.643$ & $0.628$ & $0.617$ \\ \hline\hline
\end{tabular}
\caption{\label{tabmoments} 
The theoretical moments $P_n^{\rm th}$ at LO, NLO and NNLO for
$M_{\mbox{\tiny b}}^{\mbox{\tiny 1S}}=4.7$~GeV,
$\alpha_s^{(4)}(M_Z)=0.118$,
$\mu_{\rm soft}=2.5$~GeV and $\mu_{\rm hard}=\mu_{\rm fac}=5$~GeV,
for various values of $\overline m(\overline m)$ and $n=4,6,8,10,20$.  
The values for $P_n^{\rm th}$ for $\overline m(\overline m)=0.0$~GeV 
are slightly different from the numbers shown in
Ref.~\cite{Hoang2} as we used four-loop running and three-loop matching
conditions at the five-four flavour threshold for the determination of
strong coupling at the lower scales. In Ref.~\cite{Hoang2} two-loop
running was employed.
 }
\end{center}
\vskip 3mm
\end{table}

In Table~\ref{tabmoments} the values of $P_n^{\rm th}$ at LO,
NLO and NNLO have been displayed for
$M_{\mbox{\tiny b}}^{\mbox{\tiny 1S}}=4.7$~GeV,
$\alpha_s^{(4)}(M_Z)=0.118$,
$\mu_{\rm soft}=2.5$~GeV and $\mu_{\rm hard}=\mu_{\rm fac}=5$~GeV.
for various values of $\overline m(\overline m)$ and $n=4,6,8,10,20$.

Let us first discuss the results for $n\le 10$.
The numbers show that the light quark mass corrections are positive 
at NLO and negative at NNLO. (The observation that the NLO light quark
mass corrections to the moments are positive is consistent with the
NLO results for the charm quark mass effects in the perturbative  
bottom--antibottom 1S wave function at the origin that were
determined in Ref.~\cite{Eiras1}.) 
The NNLO light quark mass corrections
are actually larger than the NLO ones and lead to an overall negative
shift in the moments at NNLO. For $n=4$ the overall shift is
around $-1$\% and for $n=10$ around $-5$\%, for 
$\overline m(\overline m)\approx 1.5$~GeV. This means that the
bottom 1S mass that is extracted from the $\Upsilon$ sum rule analysis 
receives a negative shift of the order of $15$~MeV from the
charm mass effects. This can be estimated from the overall mass
dependence of the moments, see for instance Eqs.~(\ref{Pnexperiment})
and (\ref{Pnnonrelativistic}). 
We will see in Sec.~\ref{sectionnumerical} that
this is indeed the case. Interestingly, for the determination of the
bottom $\overline{\mbox{MS}}$ mass 
$\overline M_{\mbox{\tiny b}}(\overline M_{\mbox{\tiny b}})$, this means
that the charm mass corrections in the moments of the $\Upsilon$ sum
rules and in the $\overline{\mbox{MS}}$--1S mass relation are additive.
The fact that the NNLO light quark corrections are larger than the NLO
ones and lead to an overcompensation is not necessarily a point of
concern, because the light quark mass corrections should be viewed as
a part of the full corrections. Their behaviour is essentially not
affected by that due to the small overall size of the light quark mass
corrections.      
Nevertheless, one can and should ask why the behaviour of the light
quark mass corrections in the moments is so much worse than in
the bottom $\overline{\mbox{MS}}$--1S mass relation discussed in
Sec.~\ref{sectionmsbar1Smass}. To judge the situation properly,
however, one also has to take into account that the convergence of the
massless corrections is also much worse in the moments than in
the bottom $\overline{\mbox{MS}}$--1S mass relation.
(Compare the numbers shown in Table~\ref{tabmsbar1Sscale} with those
in Table~\ref{tabmoments}.) The difference in the behaviour
of the perturbative series comes from the fact that the moments also
depend on the square of the bottom--antibottom wave function at the
origin (through their dependence on the bottom--antibottom decay and
production rate).
It is well known that the non-relativistic expansion
of the bound state wave function at the origin is much worse than for
the binding energy. A complete theoretical understanding of the
physical origin of this behaviour, which could subsequently lead to an
improvement of the situation, has not been achieved yet. However, it
seems obvious 
that the problem arises from the infrared region\footnote{
Voloshin~\cite{Voloshin2} (see also Ref.~\cite{Leutwyler1}) made the
observation that 
the relative gluon condensate correction is about three times larger 
for the square of the wave functions at the origin than for the 
energy levels. 
}.
Paradoxically, the wave function at the origin is not affected by the
question of the quark mass definition, which is generally argued to
represent the dominant issue as far as infrared sensitivity is
concerned; whether we use the bottom 1S mass or the bottom pole
mass does essentially not affect corrections to the wave function
as shifts to the energy in the Schr\"odinger
equation~(\ref{NNLOSchroedinger}) leave the wave function unchanged. 
The behaviour of the light quark mass corrections in the
moments therefore seems to be just a reflection of the present
situation, in particular because the light quark mass corrections are,
as mentioned several times earlier, very sensitive to moments smaller
than the light quark mass. 
The present situation is that the moments of the $\Upsilon$ sum rules 
have a perturbative expansion that is much worse than the perturbative
expansion of the bottom $\overline{\mbox{MS}}$--1S mass relation,
although general arguments based on global 
duality tell that the infrared sensitivity of the moments should be
smaller~\cite{Poggio1}, and 
although the gluon condensate contributions to the moments are
negligibly small compared to the ones in the $\Upsilon(\mbox{1S})$ 
mass. For the fits that we carry out in Sec.~\ref{sectionnumerical} to
determine the bottom 1S mass, we will disregard this problem as
was done in previous $\Upsilon$ sum rule analyses.       

Let us now discuss the results for $n=20$. We emphasize that moments
$P_n$ for $n>10$ are not employed in our analysis. 
The light quark mass corrections for $\overline m(\overline m)\approx
1.5$~GeV at NLO and NNLO are scaringly large individually (of the
order of $50$\% of the full corrections in each order), but lead to a
relatively moderate overall correction at NNLO (of the order of $15$\%).    
We cannot exclude that, for very large values of $n$, the
neglected double-insertion contributions at NNLO are very strongly
enhanced for some reason, but we believe that this is very unlikely.
The behaviour of the light quark mass corrections to the moments at
very large values of $n$ therefore seem to reflect the increased
infrared sensitivity of the moments if $n$ is considerably larger than
$10$. This is in agreement with the general considerations made by
Poggio, 
Quinn and Weinberg~\cite{Poggio1} about the required minimal size of
the energy interval in the integration that defines the moments in
Eq.~(\ref{momentsdef2}). 
Thus the use of moments with $n$ much larger than $10$ could
potentially lead to large systematic uncertainties and should be
avoided. 

\par
\vspace{0.5cm}
\section{Determination of the Bottom Mass from the $\Upsilon$ Sum Rules}
\label{sectionnumerical}
In this section we carry out the numerical analysis for the
determination of the bottom quark 1S and $\overline{\mbox{MS}}$ masses
using the $\Upsilon$ sum rules. In the analysis we put the main focus
on the effects coming from the finite charm quark mass. In the
previous sections the effects of the up, down and strange quark masses
have have shown to be negligible. They are not included in the
analysis. In Sec.~\ref{subsectionnumericalchi} we will briefly review
the statistical procedure that is used in the sum rule analysis, and in
Sec.~\ref{subsectionnumericalsumrule} we show the results for the
bottom quark 1S mass. In Sec.~\ref{subsectionnumericalupsilon} the
results for the bottom quark 1S mass are confronted with the mass of
the $\Upsilon(\mbox{1S})$ and the implications for non-perturbative
corrections in the $\Upsilon(\mbox{1S})$ mass are examined. In
Sec.~\ref{subsectionnumericalmsbar} the bottom $\overline{\mbox{MS}}$
mass is derived from the 1S mass result.

\par
%\vspace{0.5cm}
%
\subsection{Statistical Procedure}
\label{subsectionnumericalchi}
To obtain numerical results for the 1S mass we use a statistical
procedure
%first used in Ref.~\cite{Hoang1}, which is 
based on the 
$\chi^2$-function 
\begin{equation}
\chi^2\Big(M_b^{1S},\alpha^{(5)}_s(M_Z),\mu_{\rm soft},\mu_{\rm
  hard},\mu_{\rm fac}\Big) \, = \,
\sum\limits_{\{n\},\{m\}}\,
\Big(\,P_n^{\rm th}-P_n^{\rm ex}\,\Big)\,(S^{-1})_{n m}\,
\Big(\,P_m^{\rm th}-P_m^{\rm ex}\,\Big)
\,,
\label{x2general}
\end{equation}
where
$\{n\}$ represents the set of $n$'s for which the fit is carried out
and $S^{-1}$ is the inverse covariance matrix describing the
experimental errors and the correlation between the experimental
moments. (See Ref.~\cite{Hoang1} for a detailed description of the
covariance matrix.)
We emphasize that the renormalization scale dependence of the
theoretical moments implemented in $\chi^2$ is exactly as displayed in
the formulae in Secs.~\ref{subsectionsumrulesmethod} and
\ref{subsectionsumrulescalculation} and that no optimization procedure
has been carried out.
The covariance matrix contains the errors in the $\Upsilon$
electronic decay widths, the electromagnetic coupling
$\tilde\alpha_{\rm em}$, and the continuum cross section 
$r_{\rm cont}$. The small errors in the $\Upsilon$ 
masses are neglected. The correlations between the individual
measurements of the electronic decay widths are estimated to be equal
to the product of the respective systematic errors given in
Table~\ref{tabdata}. In our two previous analyses in
Refs.~\cite{Hoang1,Hoang2} we 
have varied the correlation between this product and zero, and we
found that the results are insensitive to this choice. In order to
estimate the theoretical uncertainties in the mass extraction, the
renormalization scales $\mu_{\rm soft}$, $\mu_{\rm hard}$ and 
$\mu_{\rm fac}$ are varied randomly in the
ranges
\begin{eqnarray}
1.5\,\mbox{GeV} \, \le & \mu_{\rm soft} & \le \, 3.5\,\mbox{GeV}
\,, 
\nonumber\\[2mm]
2.5\,\mbox{GeV} \, \le & \mu_{\rm hard} & \le \, 10\,\mbox{GeV} 
\,,
\nonumber\\[2mm]
2.5\,\mbox{GeV} \, \le & \mu_{\rm fac} & \le \, 10\,\mbox{GeV} 
\,,
\label{scaleranges}
\end{eqnarray}
and the following sets of $n$'s
\begin{equation}
\{n\} \, = \,
\{4,5,6,7\}\,,\{7,8,9,10\}\,,\{4,6,8,10\}
\label{nsets}
\end{equation}
are employed. Because the strong coupling cannot be extracted from the
sum rules based on the $\chi^2$ function in Eq.~(\ref{x2general})
with relative uncertainties smaller than 10\% (see the
corresponding results of Refs.~\cite{Hoang1,Hoang2}) we fit 
$M_{\mbox{\tiny b}}^{\mbox{\tiny 1S}}$, taking $\alpha_s^{(5)}(M_Z)$ as
an input\footnote{
For this work we have also carried out a simultaneous fit of 
$M_{\mbox{\tiny b}}^{\mbox{\tiny 1S}}$ and $\alpha_s^{(5)}(M_Z)$.
The results are not discussed in this work.
As in the previous analyses in Refs.~\cite{Hoang1,Hoang2} we found a relative
uncertainty of 10\% in $\alpha_s^{(5)}(M_Z)$. The results for 
$M_{\mbox{\tiny b}}^{\mbox{\tiny 1S}}$ were found to be equivalent to
those obtained when $\alpha_s^{(5)}(M_Z)$ is taken as an input.
}. 
For each choice of the renormalization scales $\mu_{\rm soft}$, 
$\mu_{\rm hard}$, $\mu_{\rm fac}$ in the ranges~(\ref{scaleranges}),
each set of $n$'s and each given value of $\alpha_s^{(5)}(M_Z)$,  
$M_{\mbox{\tiny b}}^{\mbox{\tiny 1S}}$ is obtained for which
$\chi^2$ is minimal.

We note that the dependence of the theoretical moments on variations of the
renormalization scales, and in particular of $\mu_{\rm soft}$, is
rather large, see Table~3 of Ref.~\cite{Hoang2}. This strong dependence is a
reflection of the bad convergence properties of the moments
caused by their dependence on the square of the bottom--antibottom wave
functions at the origin, which we have already discussed in
Sec.~\ref{subsectionsumrulesexamination}. 
It turns out that the $\chi^2$-function in Eq.~(\ref{x2general}) has a 
smaller renormalization scale dependence than the individual moments,
because it puts higher weight on linear combinations of the moments 
$(P_n^{\rm th}-P_n^{\rm ex})$ for
which the overall dependence on $\mu_{\rm soft}$ in the wave functions
partially cancels. This feature arises from the fact that the scale
sensitivity of the squared wave functions for low radial
bottom--antibottom excitations roughly corresponds to the size and
correlation of the experimental uncertainties of the respective
electronic $\Upsilon$ decay rates. Since       
the inverse covariance matrix puts higher weight on those linear
combinations of the moments $(P_n^{\rm th}-P_n^{\rm ex})$
for which the experimental uncertainties
coming from the electronic decay widths cancel, this consequently
leads to partial cancellation of the scale dependence.
This means that the $\chi^2$-function essentially fits the relative
size of the moments (as a function of $n$ for a given value of
$\mu_{\rm soft}$) and not so much their overall normalization. 
In particular, this means that the resulting estimate for the
theoretical uncertainty in the bottom 1S mass based on the scale
dependence of the $\chi^2$-function in Eq.~(\ref{x2general}) is smaller
than an estimate that is only based on a single moment. The estimate
based on the $\chi^2$-function is more realistic if
the true theoretical uncertainties of the moments (and of the squared
wave functions at the origin for the low radial excitations in
particular), which are
arising from truncating the perturbative series, are
correlated according to their scale dependence. The stability of the
estimate for the bottom 1S mass based on the $\chi^2$-function in
Eq.~(\ref{x2general}) under 
inclusion of higher order perturbative corrections (see
Figs.~\ref{fig1SNLO} and \ref{fig1SNNLO} in
Sec.~\ref{subsectionnumericalsumrule}) seems to support
that this is the case.  

\par
%\vspace{0.5cm}
%
\subsection{Results for the Bottom 1S Mass}
\label{subsectionnumericalsumrule}

\begin{figure}[t!] %fig1SNLO
\begin{center}
\hspace{0.5cm}
\leavevmode
\epsfxsize=3.8cm
\epsffile[230 580 440 710]{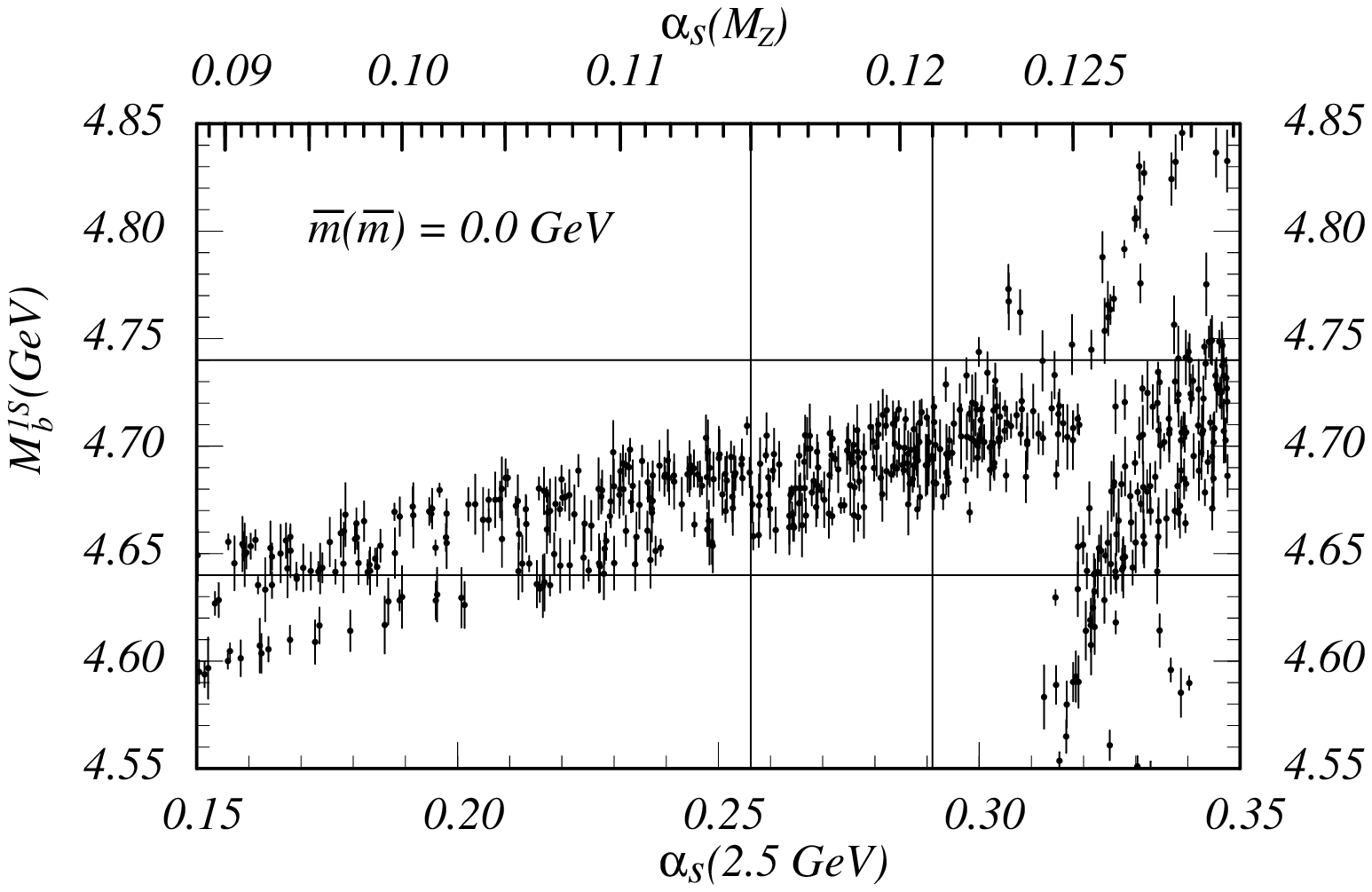}
\hspace{4.7cm}
\leavevmode
\epsfxsize=3.8cm
\epsffile[230 580 440 710]{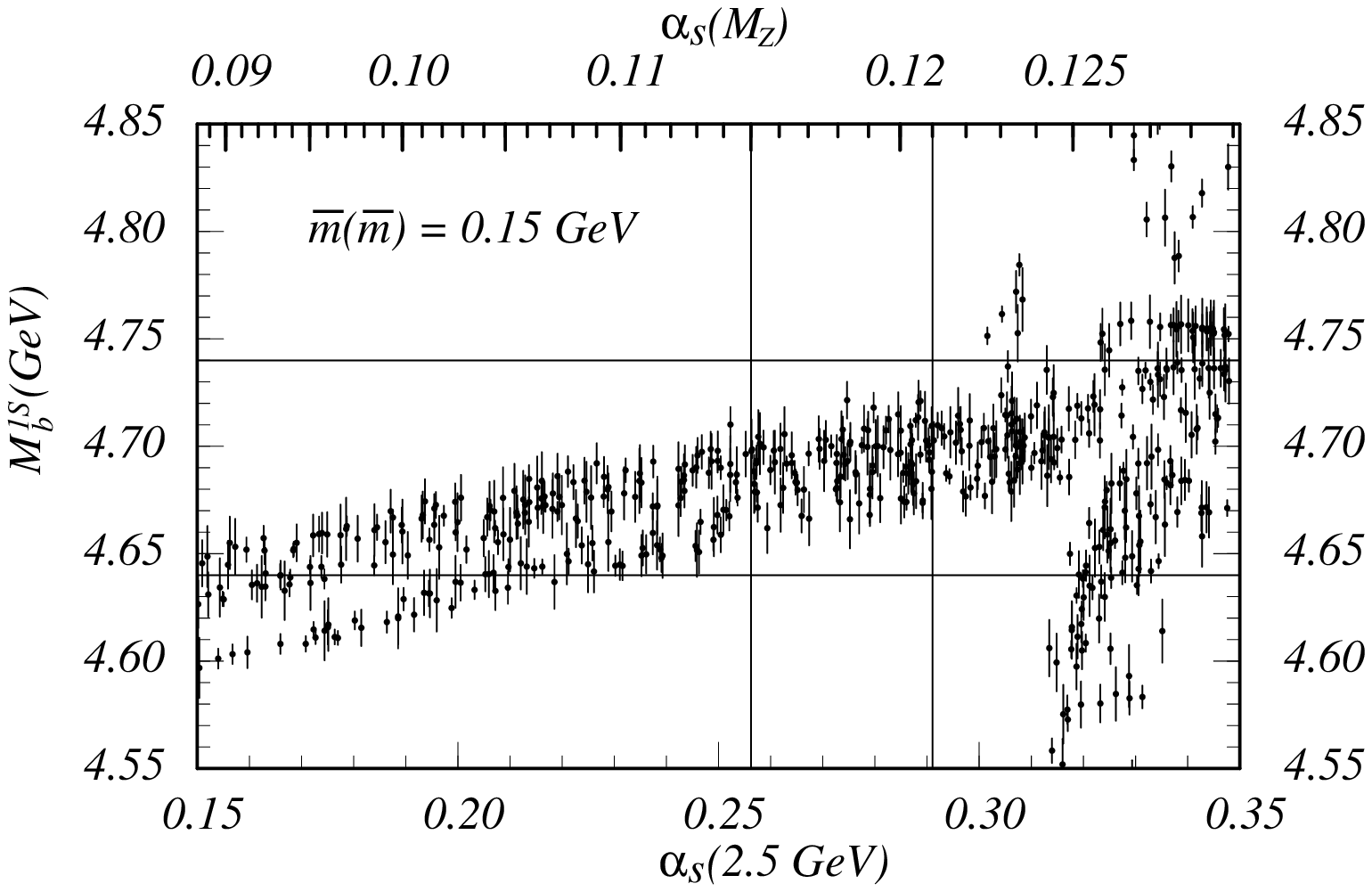}\\
\vskip  3.0cm
\hspace{0.5cm}
\leavevmode
\epsfxsize=3.8cm
\epsffile[230 580 440 710]{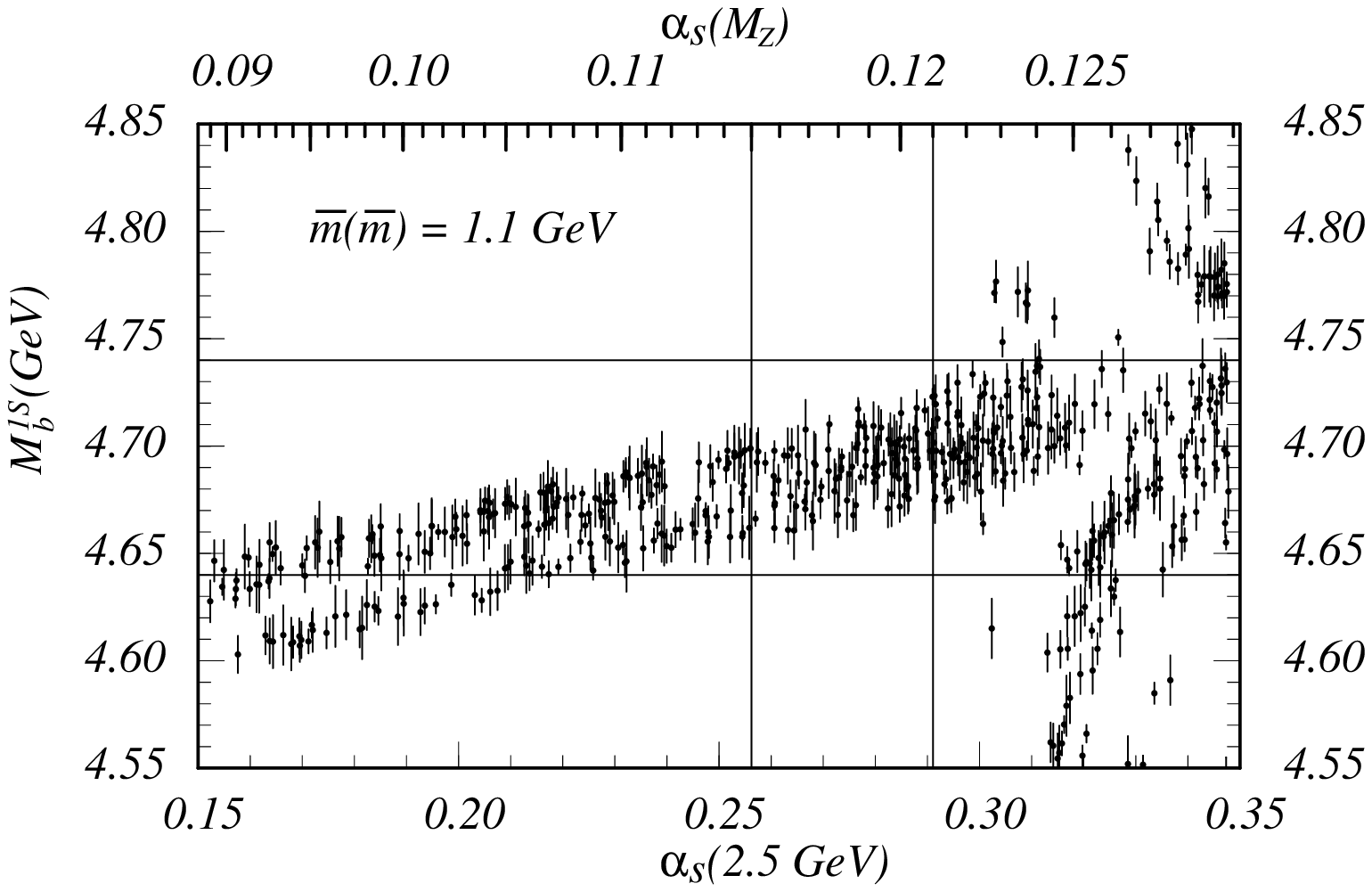}
\hspace{4.7cm}
\leavevmode
\epsfxsize=3.8cm
\epsffile[230 580 440 710]{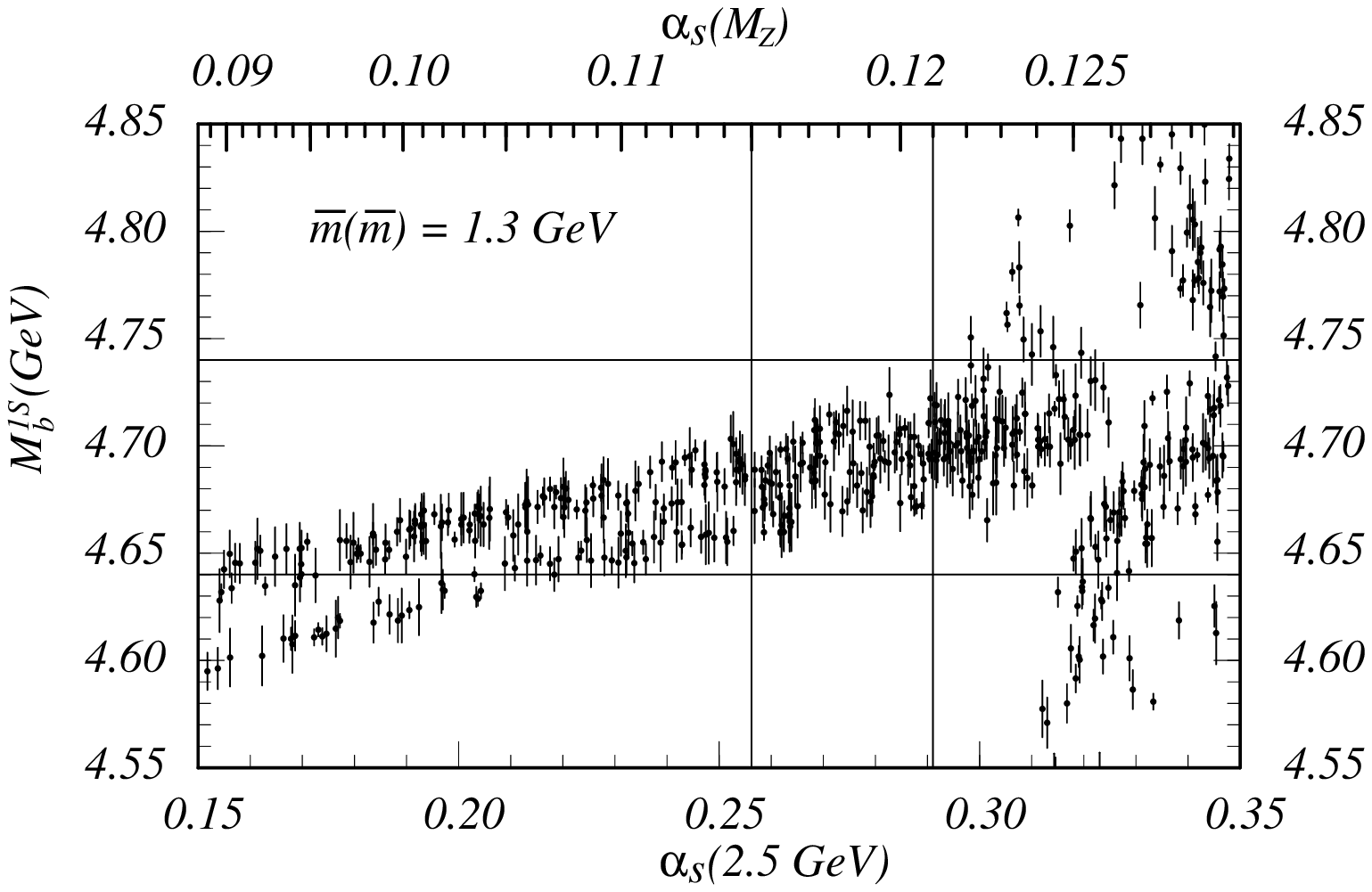}\\
\vskip  3.0cm
\hspace{0.5cm}
\leavevmode
\epsfxsize=3.8cm
\epsffile[230 580 440 710]{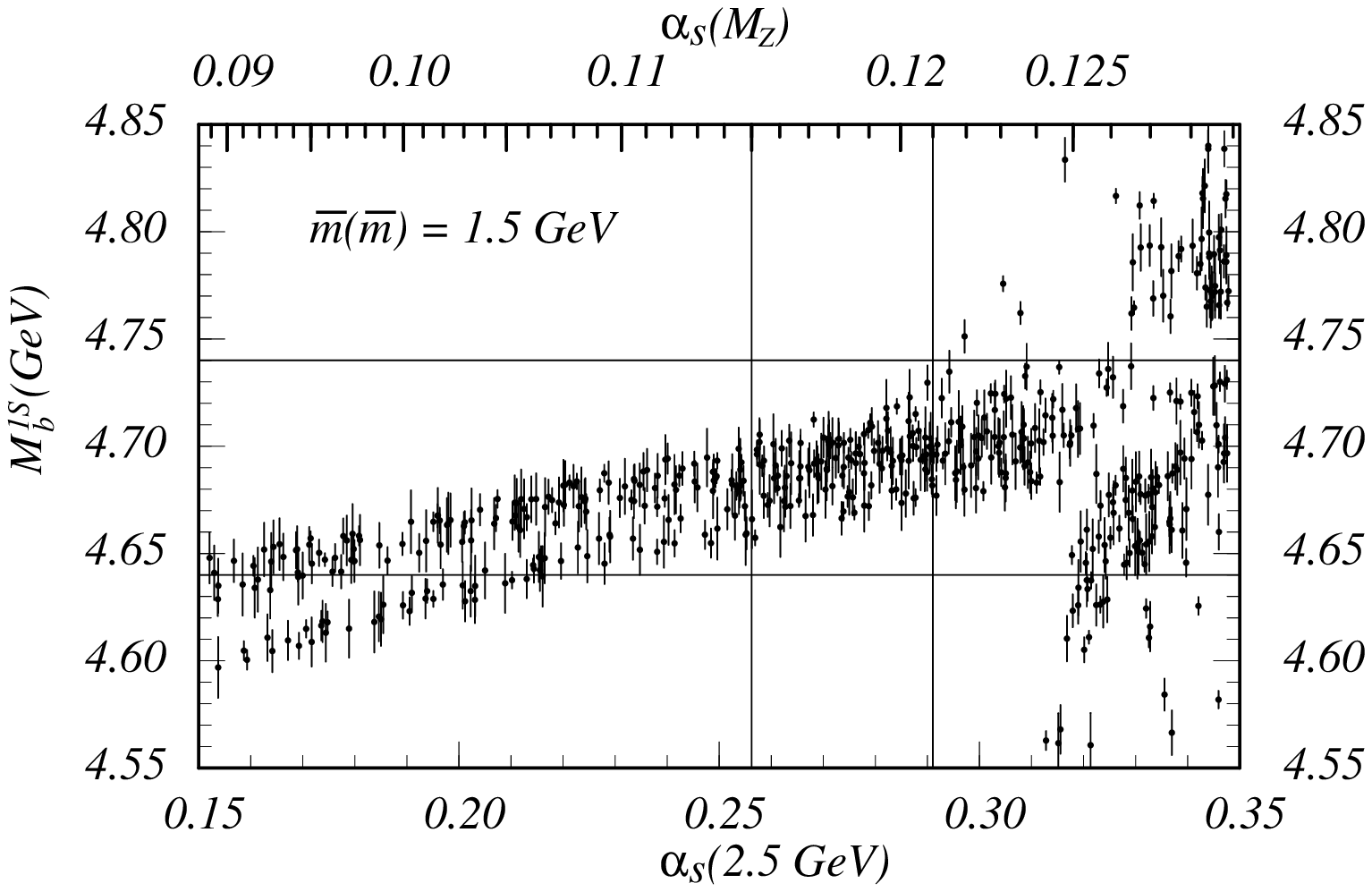}
\hspace{4.7cm}
\leavevmode
\epsfxsize=3.8cm
\epsffile[230 580 440 710]{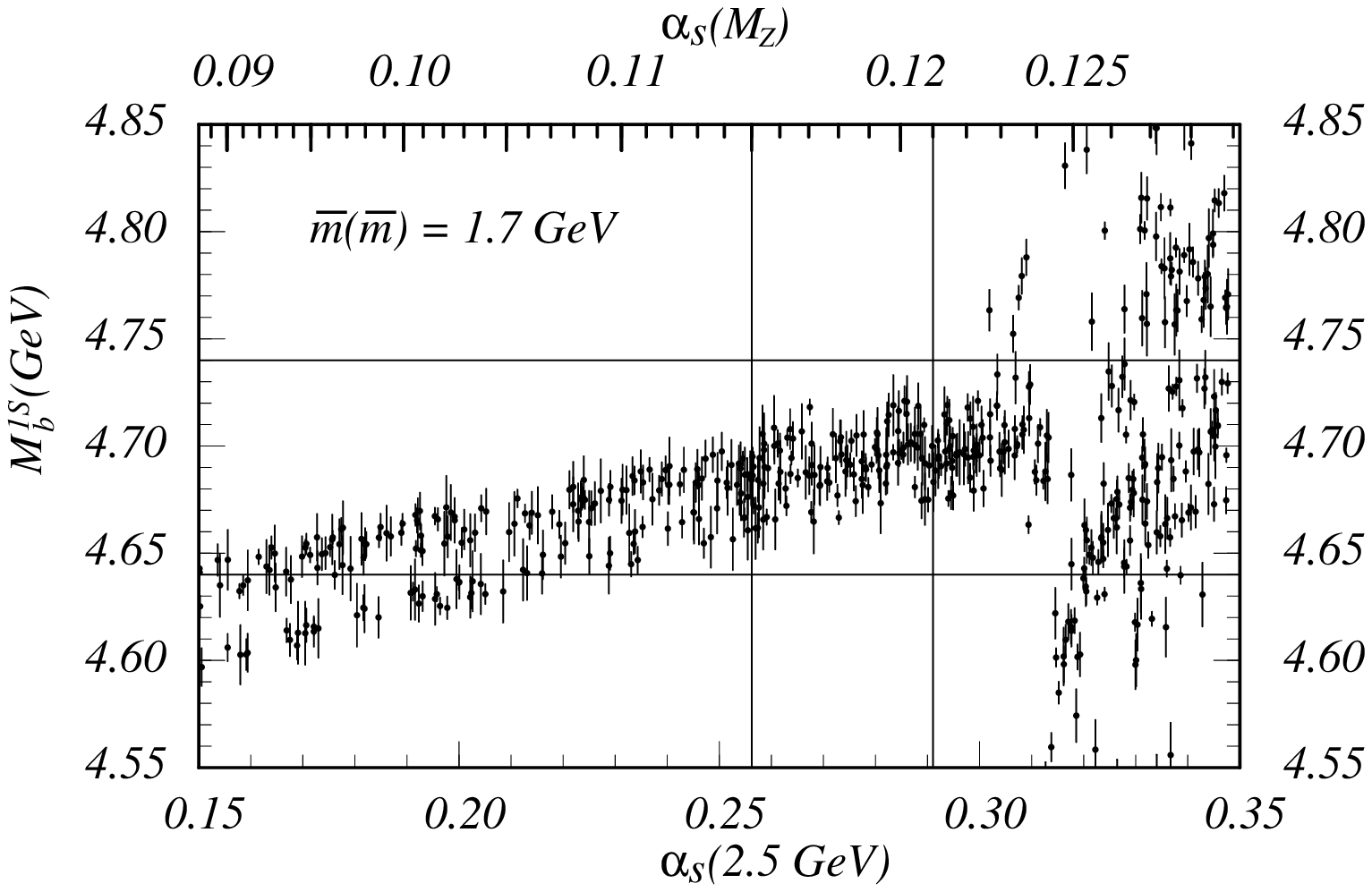}
%
%\centerline{\epsfig{file=got.ps,height=3.5in,width=3.5in}}
%\vspace{10pt}
%
\vskip  2.5cm
 \caption{\label{fig1SNLO} 
Results for the allowed range of $M_b^{\mbox{\tiny 1S}}$ for given
values of $\alpha_s^{(5)}(M_Z)$ 
(and the corresponding values of $\alpha_s^{(4)}(2.5\,\mbox{GeV})$) 
at NLO for different choices of the
$\overline{\mbox{MS}}$ charm quark mass. The dots represent points of
minimal $\chi^2$ for a large number of random choices within the
ranges~(\ref{scaleranges}) and the sets~(\ref{nsets}), and
randomly chosen values of the strong coupling.
Experimental errors at $95\%$ CL are displayed as vertical
lines. It is illustrated how the allowed range for 
$M_b^{\mbox{\tiny 1S}}$ is
obtained if $0.115\le\alpha_s^{(5)}(M_Z)\le 0.121$ is taken as an
input. 
}
%\label{fig1SNLO}
 \end{center}
\end{figure}

\begin{figure}[t!] %fig1SNNLO
\begin{center}
\hspace{0.5cm}
\leavevmode
\epsfxsize=3.8cm
\epsffile[230 580 440 710]{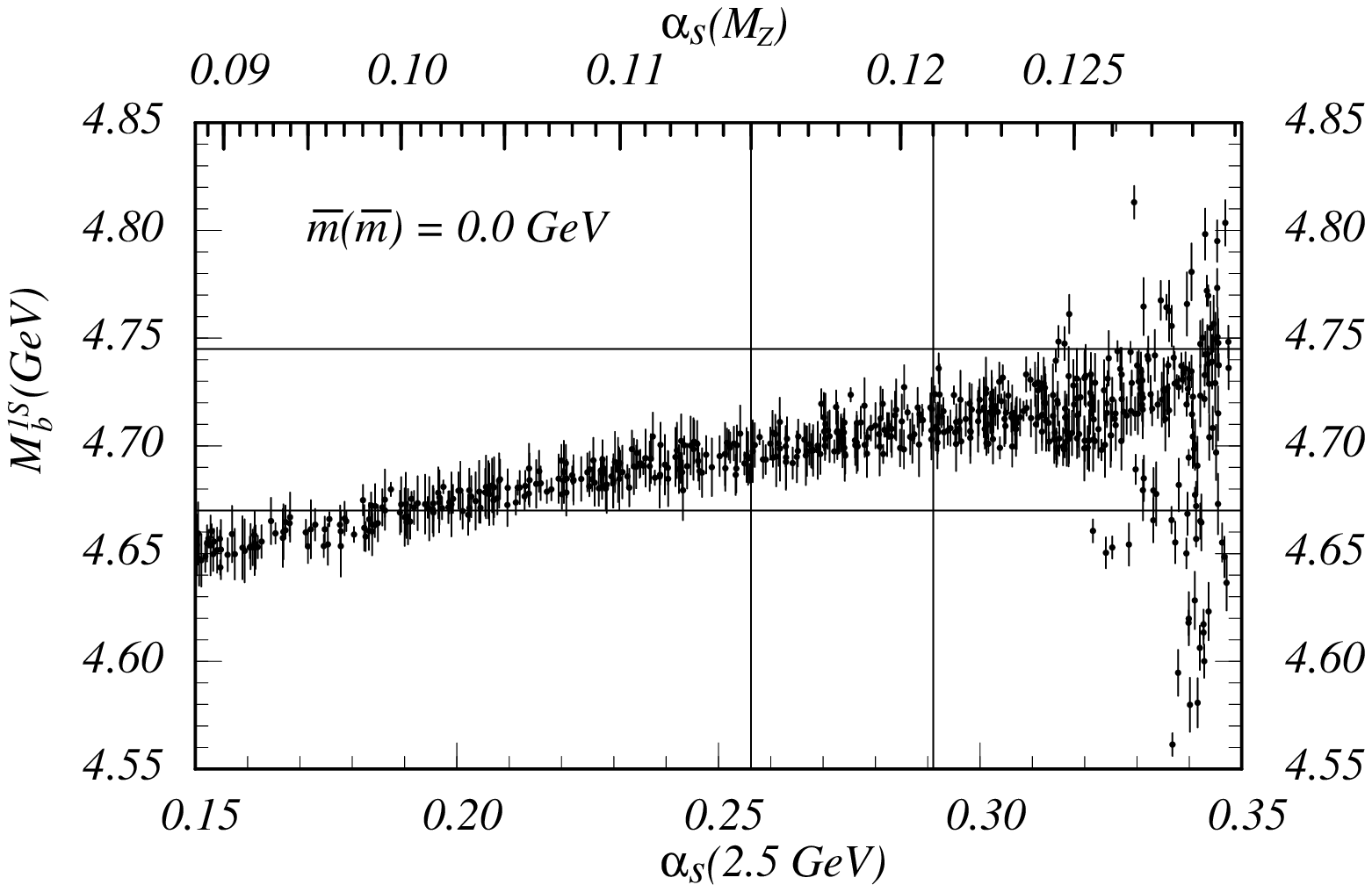}
\hspace{4.7cm}
\leavevmode
\epsfxsize=3.8cm
\epsffile[230 580 440 710]{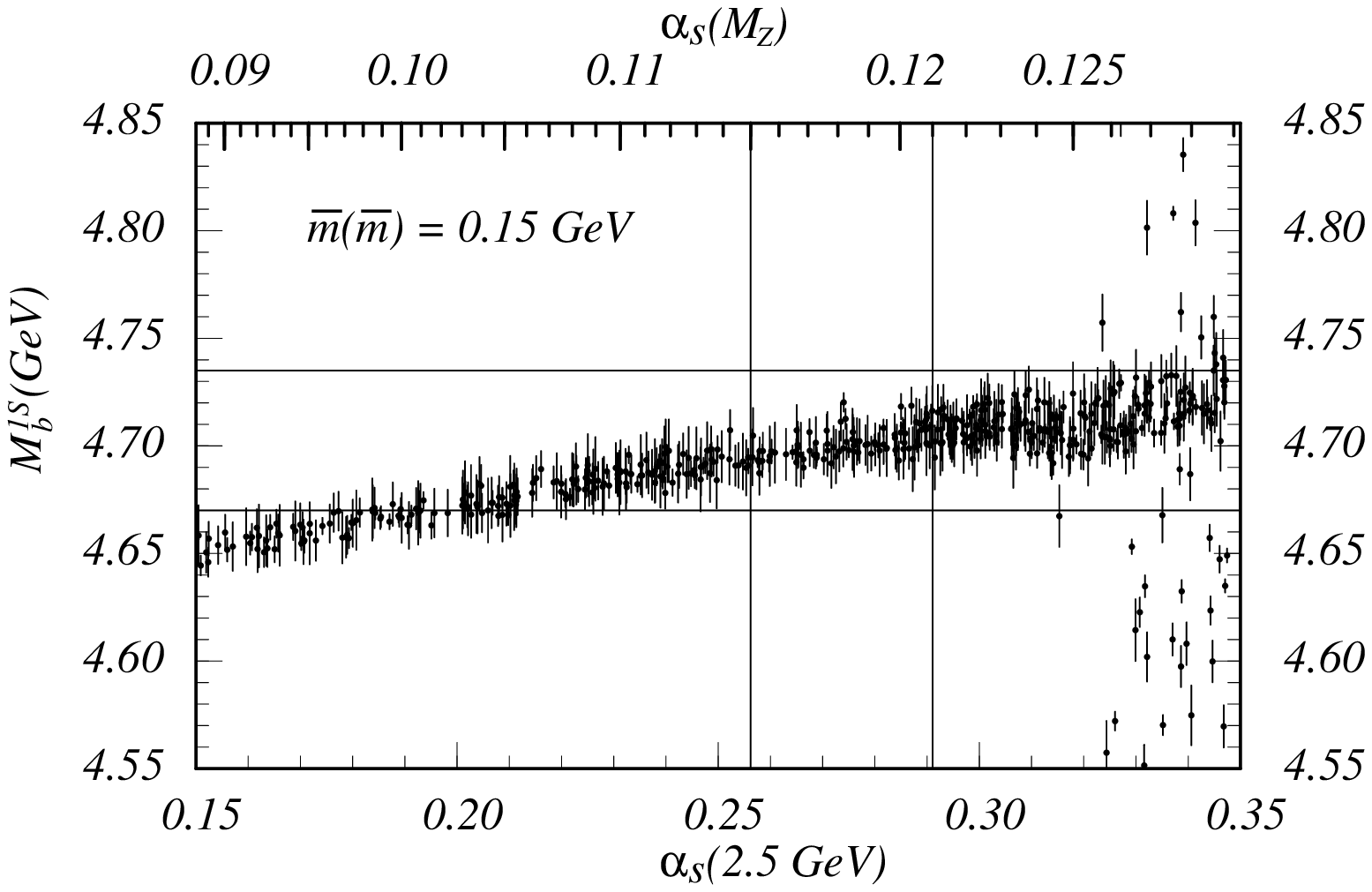}\\
\vskip  3.0cm
\hspace{0.5cm}
\leavevmode
\epsfxsize=3.8cm
\epsffile[230 580 440 710]{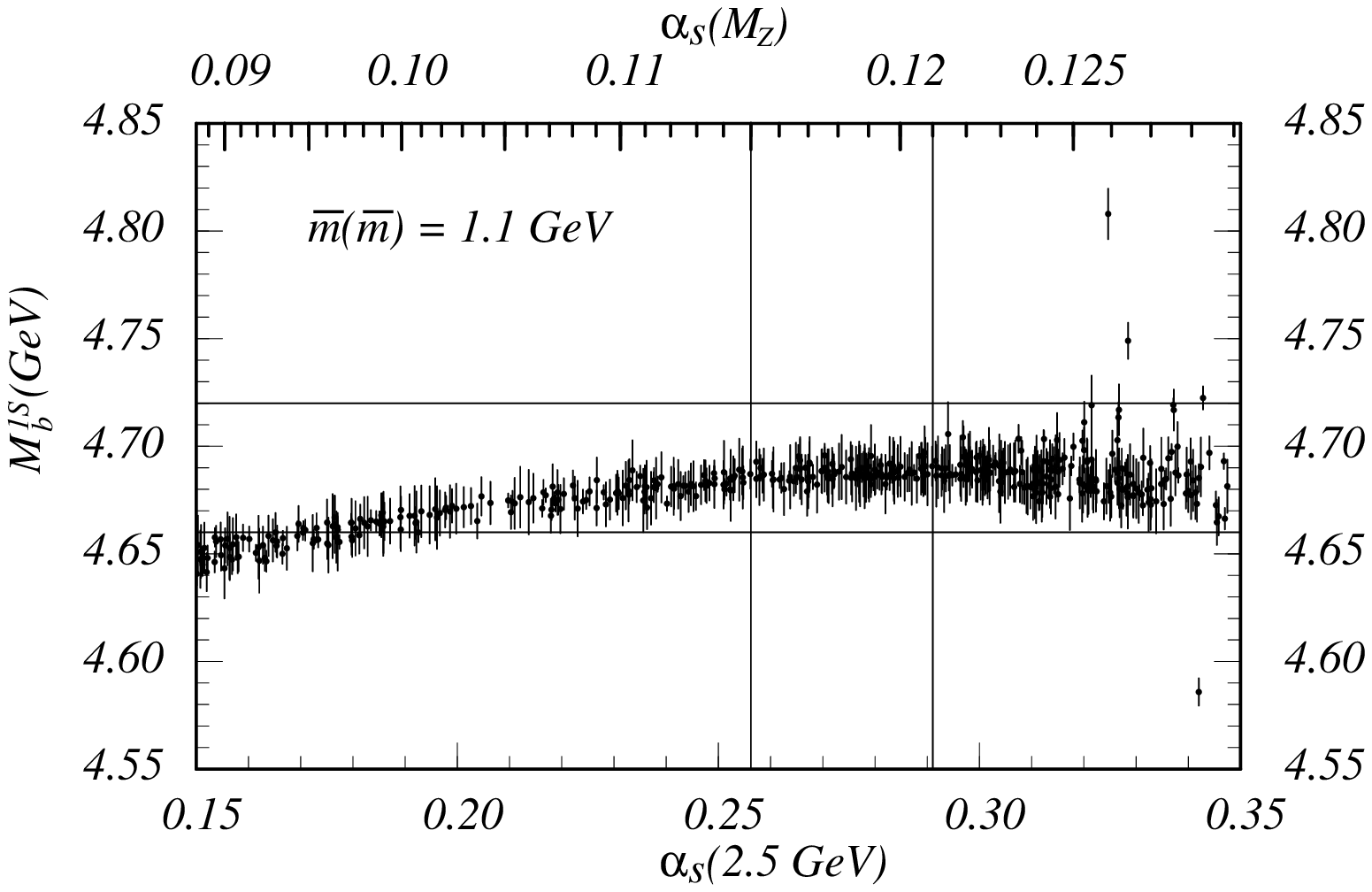}
\hspace{4.7cm}
\leavevmode
\epsfxsize=3.8cm
\epsffile[230 580 440 710]{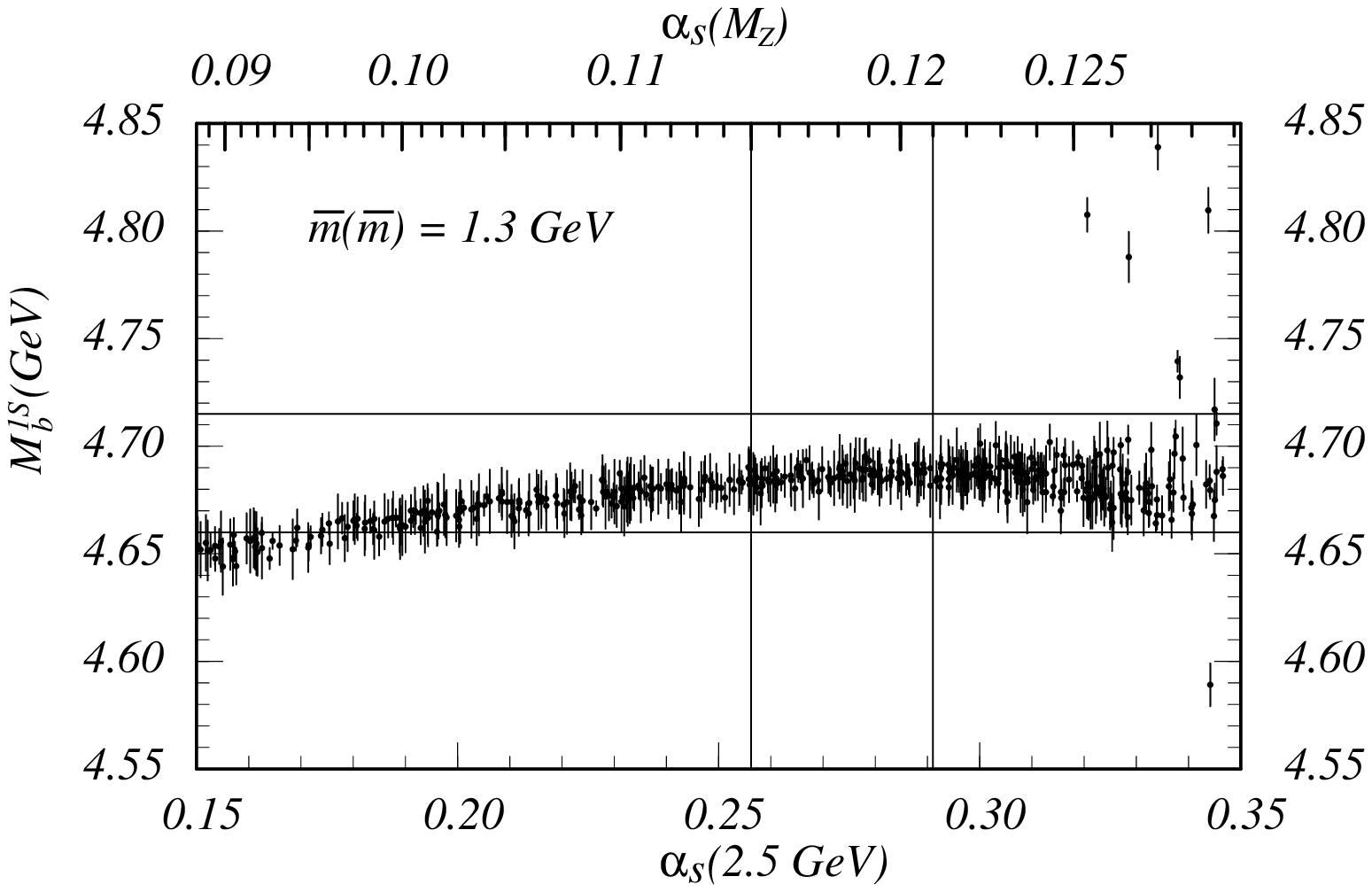}\\
\vskip  3.0cm
\hspace{0.5cm}
\leavevmode
\epsfxsize=3.8cm
\epsffile[230 580 440 710]{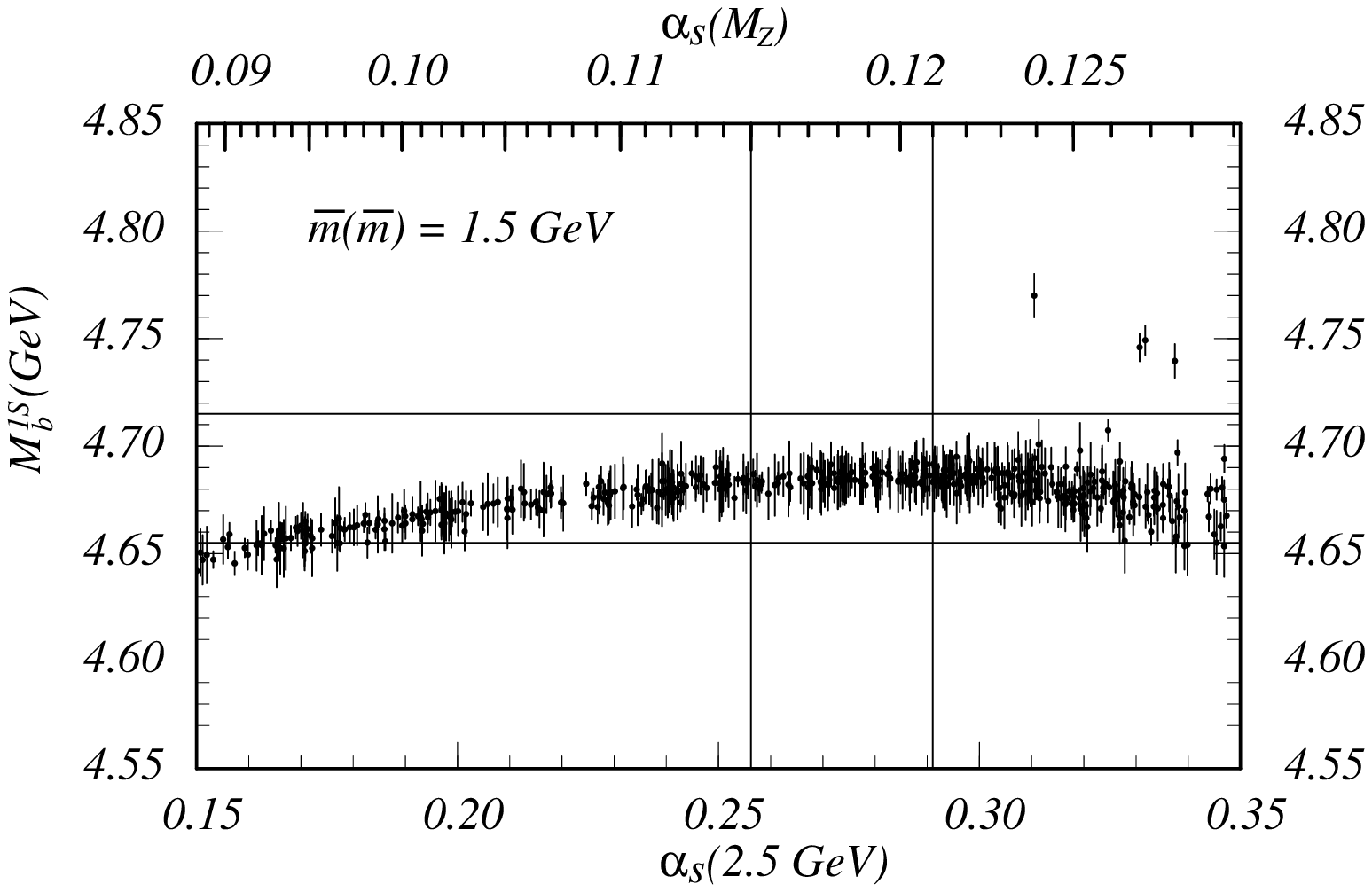}
\hspace{4.7cm}
\leavevmode
\epsfxsize=3.8cm
\epsffile[230 580 440 710]{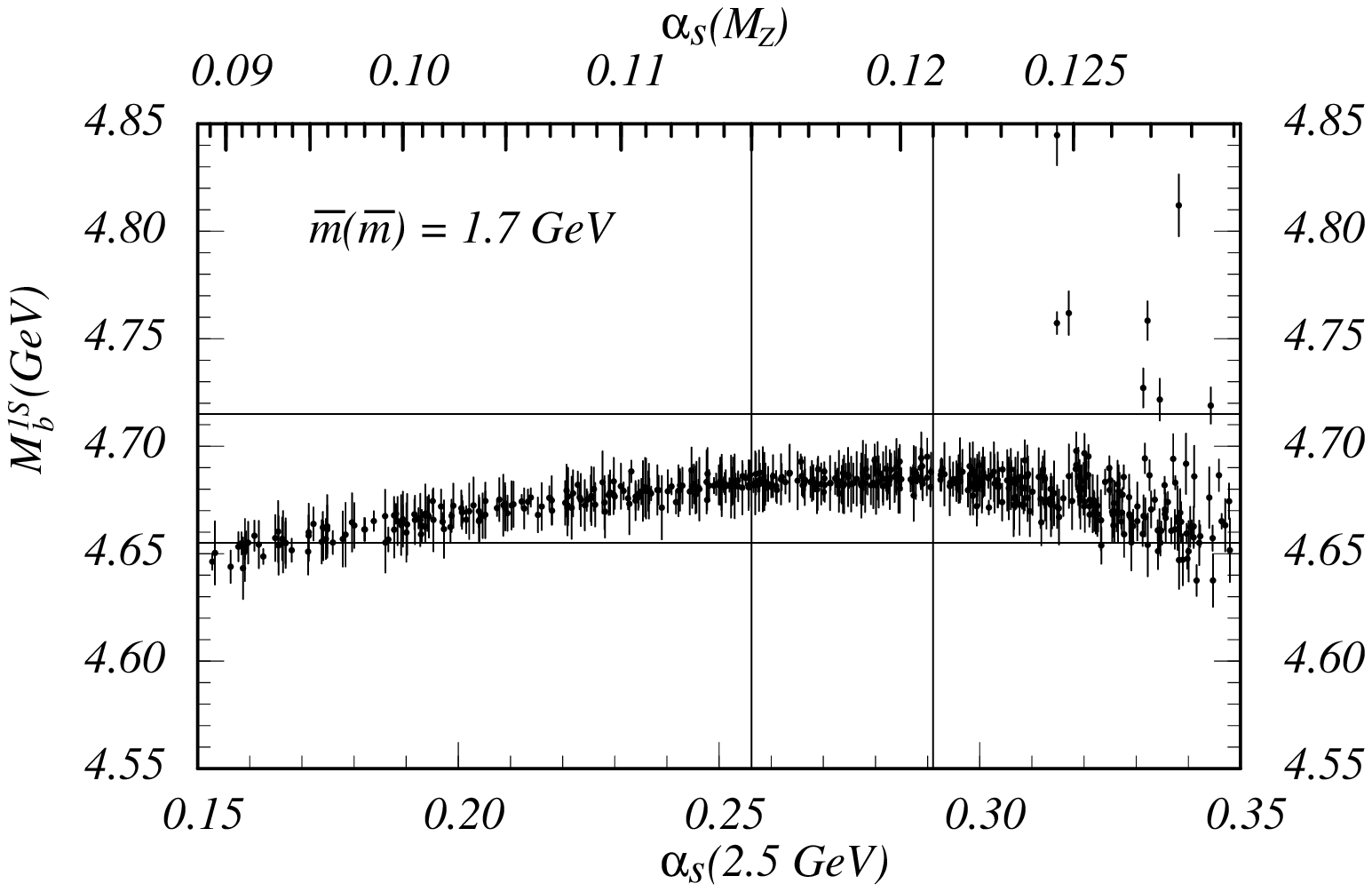}
%
%\centerline{\epsfig{file=got.ps,height=3.5in,width=3.5in}}
%\vspace{10pt}
%
\vskip  2.5cm
 \caption{\label{fig1SNNLO} 
Results for the allowed range of $M_b^{\mbox{\tiny 1S}}$ for given
values of $\alpha_s^{(5)}(M_Z)$ 
(and the corresponding values of $\alpha_s^{(4)}(2.5\,\mbox{GeV})$) 
at NNLO for different choices of the
$\overline{\mbox{MS}}$ charm quark mass. The dots represent points of
minimal $\chi^2$ for a large number of random choices within the
ranges~(\ref{scaleranges}) and the sets~(\ref{nsets}), and
randomly chosen values of the strong coupling.
Experimental errors at $95\%$ CL are displayed as vertical
lines. It is illustrated how the allowed range for 
$M_b^{\mbox{\tiny 1S}}$ is obtained if 
$0.115\le\alpha_s^{(5)}(M_Z)\le 0.121$ is taken as an input. 
}
%\label{fig1SNNLO}
 \end{center}
\end{figure}

In Figs.~\ref{fig1SNLO} and \ref{fig1SNNLO}, the allowed range for the
bottom 1S mass is displayed as a function of the strong coupling based
on the NLO and NNLO theoretical moments, respectively, for the charm 
$\overline{\mbox{MS}}$ masses $\overline m(\overline m)=1.1$, $1.3$,
$1.5$ and $1.7$~GeV. As a comparison, the results for $\overline
m(\overline m)=0.0$ and $0.15$~GeV are also shown. Each dot represents
the bottom 1S mass for which the function $\chi^2$ is minimal for 
given (random) values of $\alpha_s$ and the renormalization
scales and a given set of $n$'s. For the running of the strong 
coupling we employed the four-loop beta function and three-loop
matching conditions at the five-four flavour threshold. 
The experimental and statistical errors that are indicated by the
vertical lines correspond to 95\% CL. They are below $15$~MeV for all
dots displayed in Figs.~\ref{fig1SNLO} and \ref{fig1SNNLO}. We note
that the NNLO results do not depend on whether the NNLO
double-insertion charm mass corrections in the relation between the
bottom pole and 1S masses in Eq.~(\ref{globalfactorpole1S}) are taken
into account or not.

Figures~\ref{fig1SNLO} show that at NLO the inclusion of the finite
charm mass does not at all affect the results for the bottom 1S
mass. For all choices of the $\overline{\mbox{MS}}$ charm quark mass
between $1.1$ and $1.7$~GeV the allowed range for 
$M_{\mbox{\tiny b}}^{\mbox{\tiny 1S}}$ is practically identical to the
analysis where the charm mass is neglected. Assuming 
\begin{eqnarray}
\alpha_s^{(5)}(M_Z) & = & 0.118 \, \pm \, 0.003
\label{alphasworldaverage}
\end{eqnarray}
for the strong coupling, which is motivated by the compilations given
in Refs.~\cite{Bethke1,Manohar1}, we arrive at
$M_{\mbox{\tiny b}}^{\mbox{\tiny 1S}}=4.70\pm 0.06$~GeV from the fits
to the NLO theoretical moments. Although we assume a smaller
uncertainty for $\alpha_s$ than in our previous analysis of
Ref.~\cite{Hoang2}, where we used 
$\alpha_s^{(5)}(M_Z) = 0.118\pm 0.004$, the result for the bottom 1S
mass is equivalent to the one obtained at NLO in
Ref.~\cite{Hoang2}. This 
is because the correlation of the allowed range for the 1S mass to the
value of the strong coupling is rather weak (and because we round the
results to $10$~MeV precision). In contrast to the NLO results shown
in Ref.~\cite{Hoang2}, Fig.~\ref{fig1SNLO} exhibits an instability in the
allowed 1S mass range for input vales of $\alpha_s^{(5)}(M_Z)$ larger
than $0.125$. This is caused by the use of the four-loop running for
the strong coupling, which leads to larger values of $\alpha_s$ at
lower scales than the two-loop running that has been used in
Ref.~\cite{Hoang2}. For the extraction of the bottom 1S mass using
Eq.~(\ref{alphasworldaverage}) as an input, this instability is
irrelevant. 

Figures~\ref{fig1SNNLO} show the results for the allowed bottom 1S
mass range obtained from the NNLO theoretical moments. The results
have a number of remarkable features. Most prominently, the already
weak correlation of the allowed bottom 1S mass range with the strong 
coupling visible in the NLO analysis
has completely vanished in the relevant strong coupling range 
$\alpha_s^{(5)}(M_Z)\approx 0.118$ if the effects of the charm quark
mass are included. This leads to a variation of the
bottom 1S mass slightly smaller than the analysis with zero charm
quark mass. We 
also find that the central value for the bottom 1S mass is about
$20$~MeV lower if the charm quark mass is taken into account. Taking
Eq.~(\ref{alphasworldaverage}) as an input we arrive at
\begin{eqnarray}
M_{\mbox{\tiny b}}^{\mbox{\tiny 1S}}=4.69 \, \pm \, 0.03\,\,\mbox{GeV}
\label{M1Sfinalresult}
\end{eqnarray}
for the bottom 1S mass from the NNLO theoretical moments, taking into
account the charm $\overline{\mbox{MS}}$ mass 
$\overline m(\overline m)=1.4\pm 0.3$~GeV. 
We emphasize that the uncertainty of $30$~MeV in
Eq.~(\ref{M1Sfinalresult}) does not depend at all
on the uncertainty of the strong coupling. It is also interesting that
the allowed range for the bottom 1S mass is stabilized for 
$\alpha_s^{(5)}(M_Z)$ larger than $0.125$ compared to the NLO
analysis. 
We note that the fact that the bottom 1S mass gets a negative shift
from the finite charm quark mass is expected, because
a light quark mass gives negative corrections to the NNLO theoretical
moments (see the discussion in
Sec.~\ref{subsectionsumrulesexamination}). Neglecting the charm quark
mass, the result reads 
$M_{\mbox{\tiny b}}^{\mbox{\tiny 1S}}=4.71\pm 0.04$~GeV, where the
variation is slightly larger than in Ref.~\cite{Hoang2} owing to the
use of the four-loop beta function for $\alpha_s$. 

\par
%\vspace{0.5cm}
%
\subsection{Comparison with the Mass of the $\Upsilon(\mbox{1S})$ Meson}
\label{subsectionnumericalupsilon}
It is a quite interesting feature of our NNLO result for the bottom 1S
mass, $M_{\mbox{\tiny b}}^{\mbox{\tiny 1S}}=4.69\pm 0.03$~GeV, that it
could be used to constrain the non-perturbative effects in the mass of
the $\Upsilon(\mbox{1S})$ meson, 
$M(\Upsilon(\mbox{1S}))=9460.37\pm 0.21$~MeV. 
We remind the reader that, because we only used moments $P_n$ with
$n\le 10$, the bottom--antibottom quark dynamics in the sum rules can
be treated as perturbative (i.e. that the hierarchy
``$M_{\mbox{\tiny b}}\gg M_{\mbox{\tiny b}}v\gg 
M_{\mbox{\tiny b}}v^2\gg\Lambda_{\rm QCD}$'' is satisfied). Thus,
non-perturbative corrections to the 
moments can be reliably determined in an expansion in gluonic and
light quark condensates of increasing dimensions.\footnote{
We assume for now that the expansion in terms of
condensates is valid for the moments if $n$ is not too large, and
disregard the discussions in Secs.~\ref{subsectionsumrulesbasic}
and \ref{subsectionsumrulesexamination}. 
}
For the moments
$P_n$ with $n\le 10$ the dominant non-perturbative corrections from
the dimension-$4$ gluon condensate corrections are below the per mille
level (see Eq.~(\ref{momentsnonperturbative})). We therefore assume 
that non-perturbative effects in general can be safely neglected in
our sum rule analysis. This means that our result for the bottom 1S
mass in Eq.~(\ref{M1Sfinalresult}) is not affected at all by
non-perturbative effects. On the other 
hand, the bottom quark 1S mass is just half of the perturbative
contribution of the mass of the $\Upsilon(\mbox{1S})$ meson, assuming
that the latter is also a perturbative system. 

If we assume that the $\Upsilon(\mbox{1S})$ meson can be treated as a
perturbative system (at least as far as the calculation of its mass is
concerned), non-perturbative corrections in $M(\Upsilon(\mbox{1S}))$
can also be reliably determined 
in an expansion in the condensates. However, the corrections
caused by the condensates are much larger than for the sum rules and
cannot be neglected as in the sum rule analysis. A precise and
accurate determination of the bottom 1S mass from the $\Upsilon$ sum
rules could therefore be used to seriously test the hypothesis that
the $\Upsilon(\mbox{1S})$ is perturbative and, eventually, to extract
more precise values for the condensates. From the NNLO sum rule result
for the bottom 1S mass in Eq.~(\ref{M1Sfinalresult}), we find
\begin{equation}
\Big[\,M(\Upsilon(\mbox{1S}))\,\Big]^{\mbox{\tiny non-pert}}
\, = \, 
M(\Upsilon(\mbox{1S}))-2\, M_{\mbox{\tiny b}}^{\mbox{\tiny 1S}}
\, = \, 
80\,\pm\,60\,\mbox{MeV}
\label{upsilon1Snonperturbative}
\end{equation}
for the non-perturbative contributions in the $\Upsilon(\mbox{1S})$
mass. We emphasize again that the result in
Eq.~(\ref{upsilon1Snonperturbative}) can only be interpreted as ``the
non-perturbative contributions in the $\Upsilon(\mbox{1S})$ mass''
based on the hypothesis that the $\Upsilon(\mbox{1S})$ is a
perturbative 
system. It is interesting that Eq.~(\ref{upsilon1Snonperturbative}) 
seems indeed consistent with the leading dimension-four gluon
condensate contribution obtained by Voloshin and
Leutwyler~\cite{Voloshin2,Leutwyler1}:
\begin{eqnarray}
\Big[\,M(\Upsilon(\mbox{1S}))\,\Big]^{\mbox{\tiny non-pert}}
& = &
\frac{1872}{1275}\,\frac{M_{\mbox{\tiny b}}\,n^6\,\pi}
{(M_{\mbox{\tiny b}}\,C_F\,\alpha_s)^4}\,
\langle\,\alpha_s\,
{\mbox{\boldmath
$G$}}^2\,\rangle
\, + \, \ldots
\,.
\label{upsilon1Scondensate}
\end{eqnarray}
Equation~(\ref{upsilon1Scondensate}) obtains corrections from 
higher-dimension condensates as well as higher order
perturbative corrections to the Wilson coefficients multiplying each
of the condensates. The subleading dimension-$6$ condensate
contributions have been calculated in Ref.~\cite{Pineda2}.
The unknown perturbative corrections induce a very
large scale uncertainty of the dimension-$4$ condensate contribution
shown in Eq.~(\ref{upsilon1Scondensate}). Assuming the standard
literature range 
$\langle\,\alpha_s\,{\mbox{\boldmath $G$}}^2\,\rangle=
0.05\pm 0.03\,\mbox{GeV}^4$
the leading gluon condensate correction can range from $10$ up to
$300$~MeV (for $0.2<\alpha_s<0.5$), where large numbers
paradoxically correspond to smaller values of the strong coupling. As
said before, this range is consistent with
Eq.~(\ref{upsilon1Snonperturbative}), but at the present stage it does
not provide any meaningful determination of
$\langle\,\alpha_s\,{\mbox{\boldmath $G$}}^2\,\rangle$, let alone a
test of the hypothesis that the $\Upsilon(\mbox{1S})$ is
perturbative. The calculation of higher order perturbative corrections
in Eq.~(\ref{upsilon1Scondensate}) would be very helpful to clarify the
situation.

\par
%\vspace{0.5cm}
%
\subsection{Results for the Bottom $\overline{\mbox{MS}}$ Mass}
\label{subsectionnumericalmsbar}
The heavy quark 1S mass is, by design, a short-distance mass adapted
to systems where the heavy quark is very close to its mass shell, or,
in other word, where the heavy quark virtuality is much smaller than
its mass, 
$q^2-M_{\mbox{\tiny Q}}^2\ll M_{\mbox{\tiny Q}}^2$. 
Apart from heavy-quark--antiquark systems the 1S mass definition has
also been applied to B mesons~\cite{Hoang3}, where the bottom quark
virtuality is of order $\Lambda_{\rm QCD} M_{\mbox{\tiny Q}}$. For
high-energy or low-energy processes, however, where the heavy
quark virtuality is of order 
$M_{\mbox{\tiny Q}}^2$ or larger, the $\overline{\mbox{MS}}$ mass
definition is 
the preferred one. In this section we determine the bottom
$\overline{\mbox{MS}}$ mass from the NNLO result for the 1S mass in
Eq.~(\ref{M1Sfinalresult}) using formula~(\ref{msbar1Sgeneric}).

\begin{figure}[t!] %figmsbardependence
\begin{center}
\hspace{0.5cm}
\leavevmode
\epsfxsize=3.8cm
\epsffile[230 580 440 710]{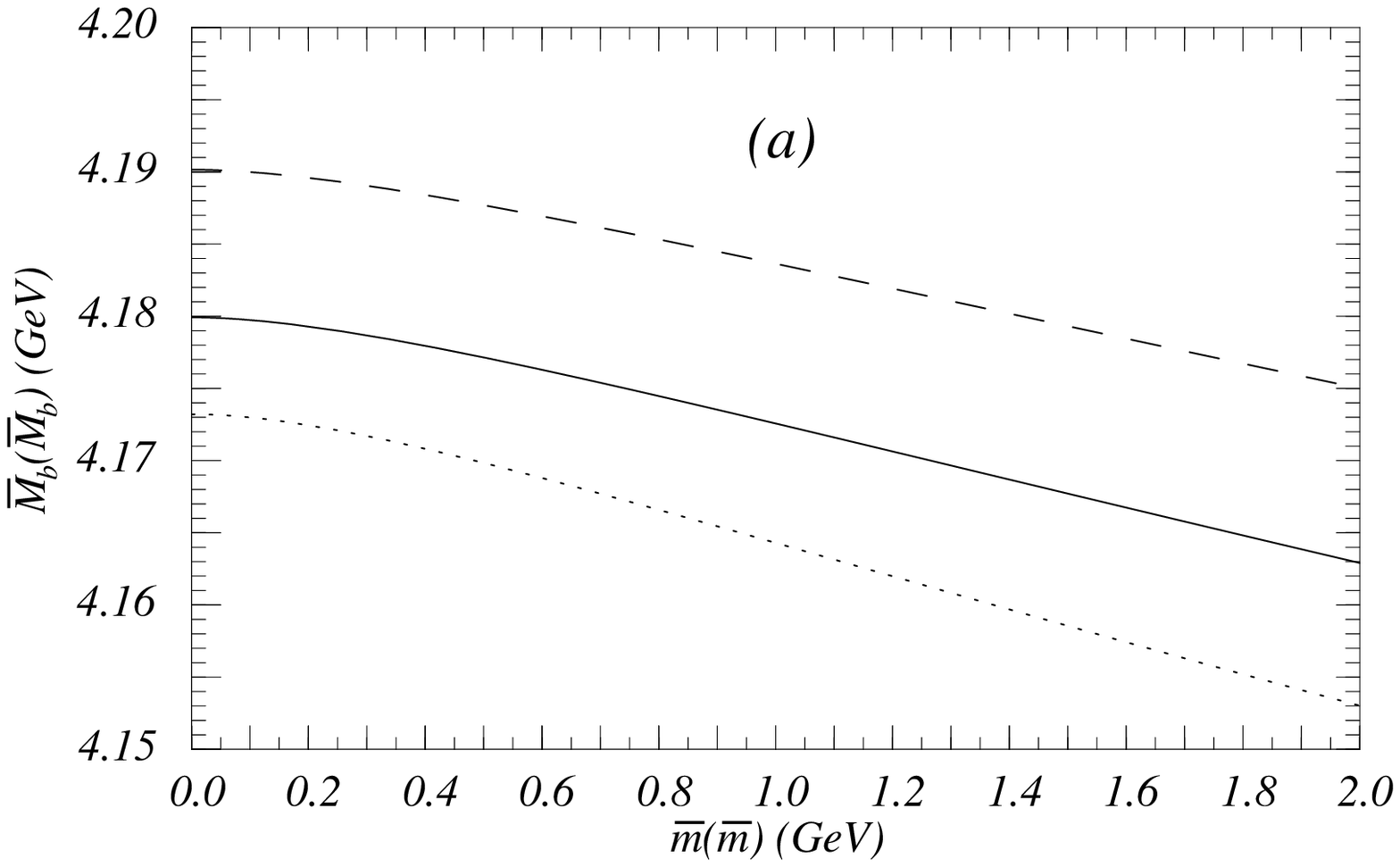}
\hspace{4.7cm}
\leavevmode
\epsfxsize=3.8cm
\epsffile[230 580 440 710]{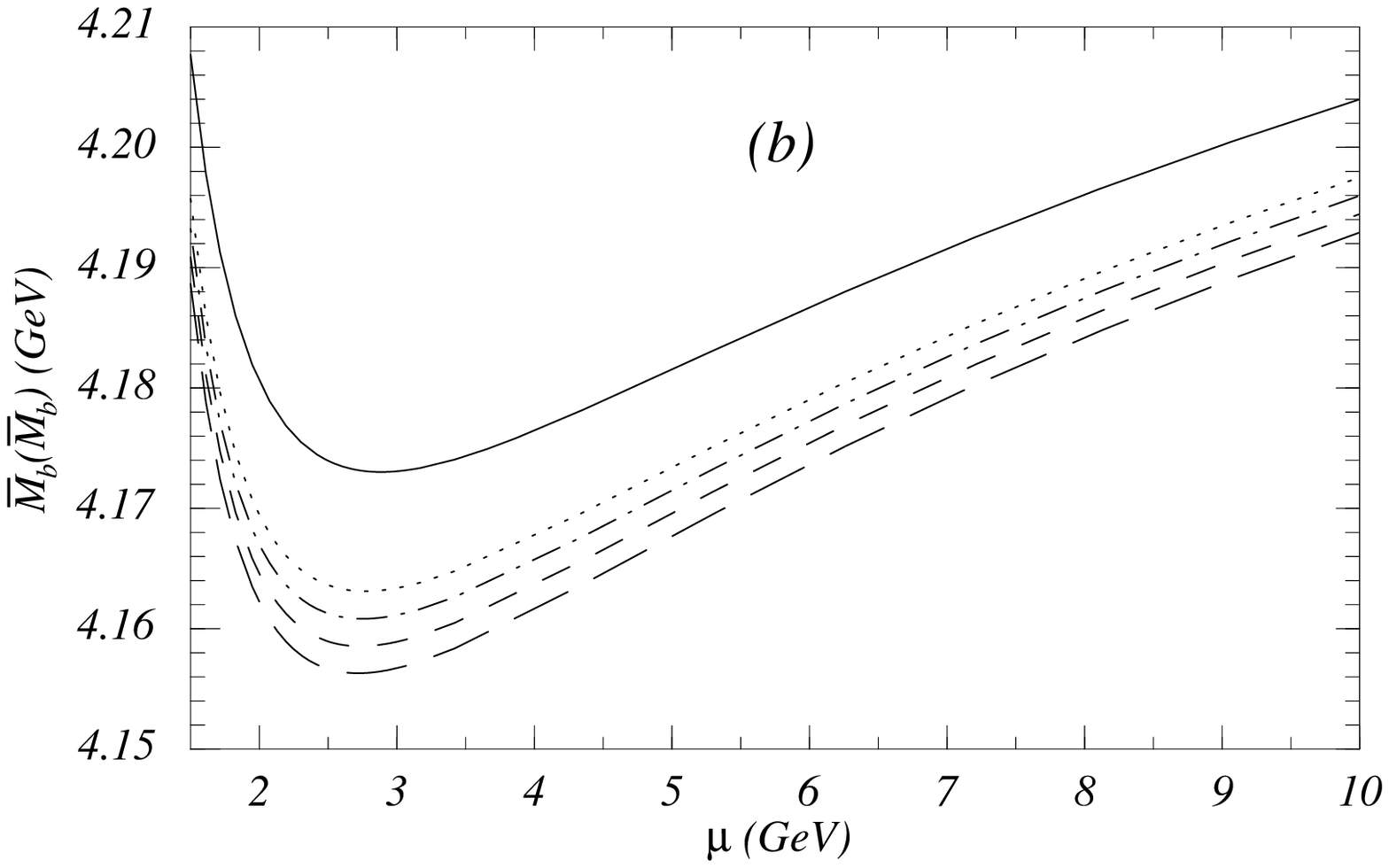}\\
\vskip  3.0cm
\hspace{0.5cm}
\leavevmode
\epsfxsize=3.8cm
\epsffile[230 580 440 710]{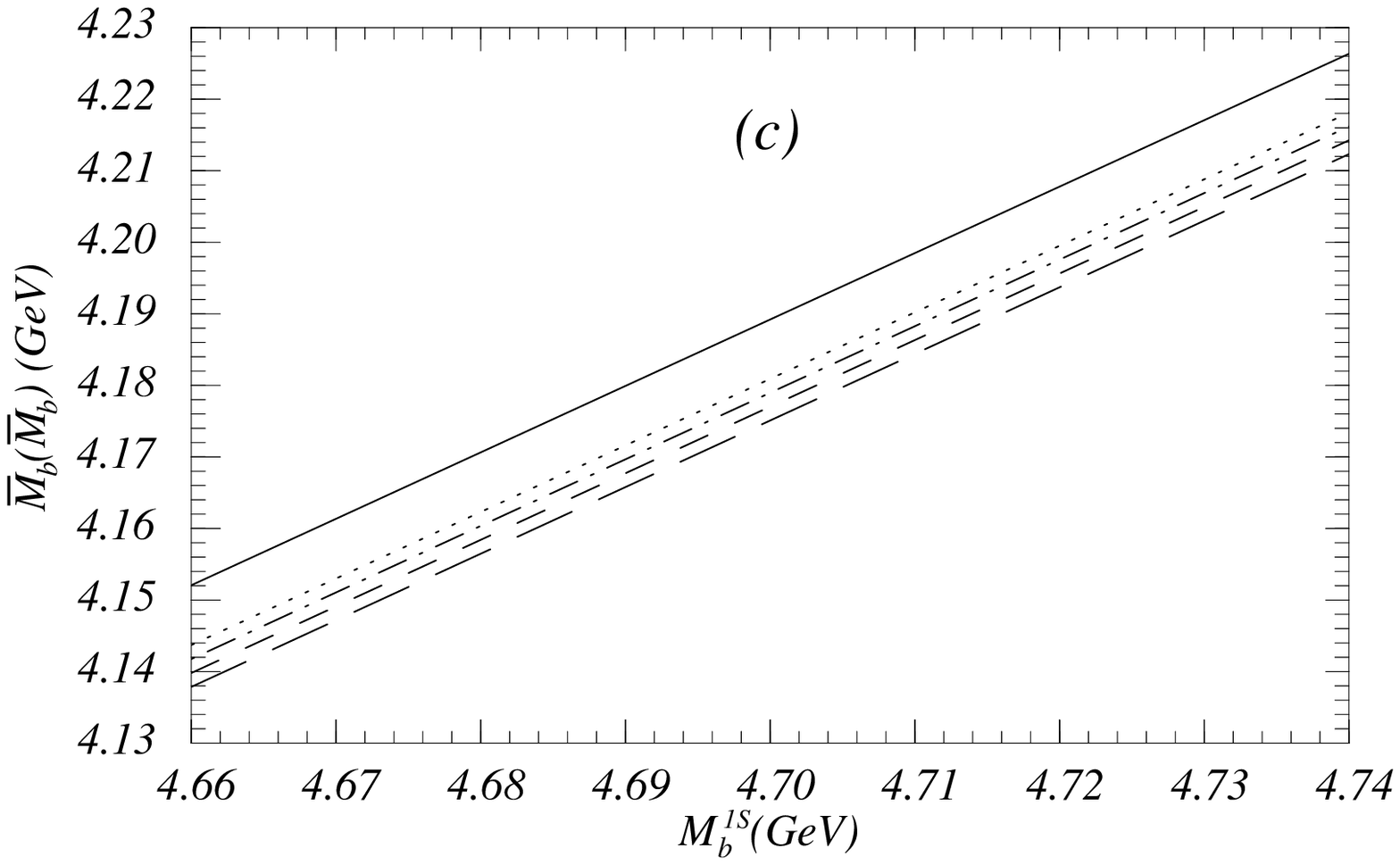}
\hspace{4.7cm}
\leavevmode
\epsfxsize=3.8cm
\epsffile[230 580 440 710]{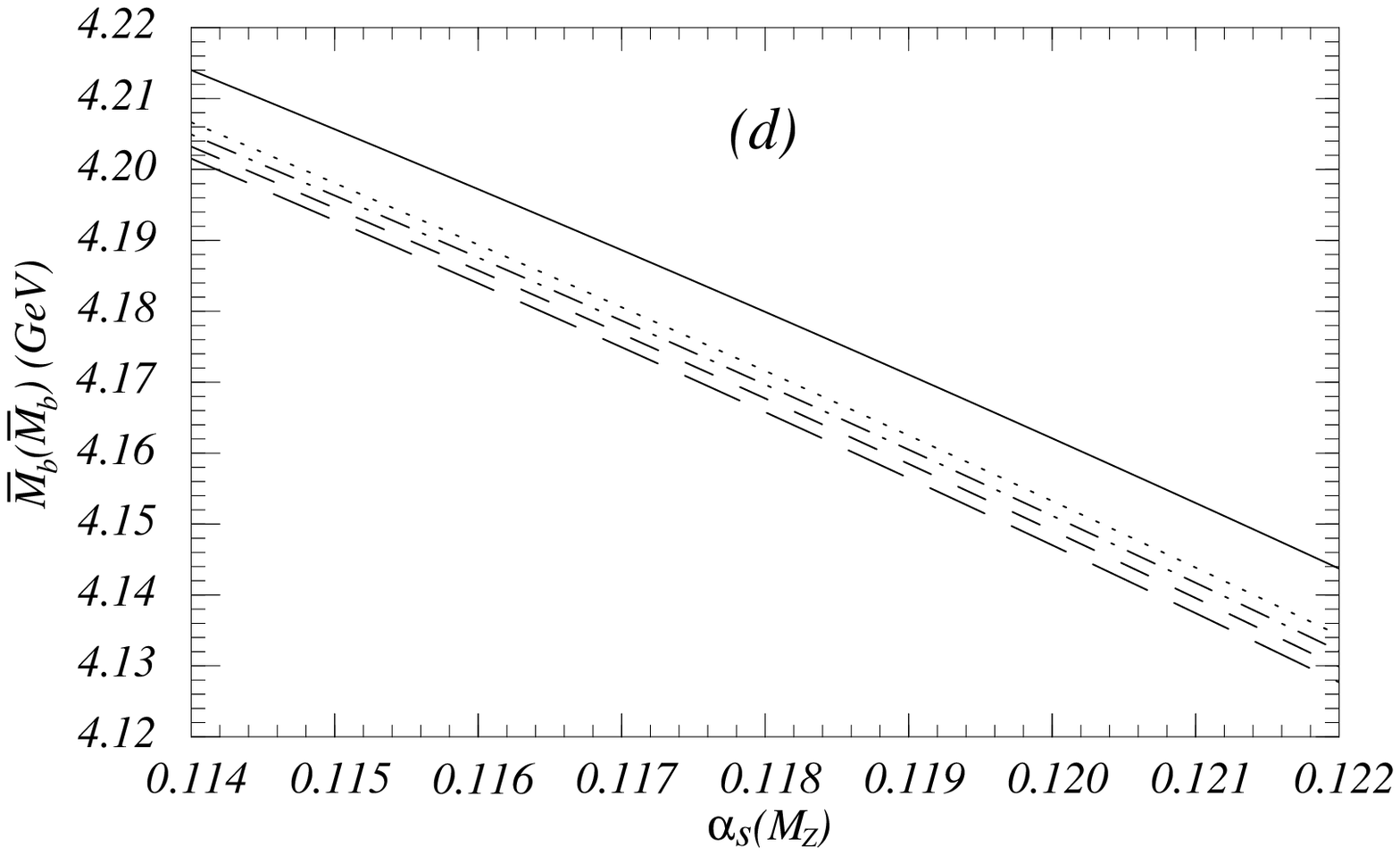}
%
%\centerline{\epsfig{file=got.ps,height=3.5in,width=3.5in}}
%\vspace{10pt}
%
\vskip  2.5cm
 \caption{\label{figmsbardependence}
The dependence of the bottom 
$\overline{\mbox{MS}}$ mass
$\overline M_{\mbox{\tiny b}}(\overline M_{\mbox{\tiny b}})$,
determined from the bottom 1S mass, 
on the $\overline{\mbox{MS}}$ charm quark mass (a), 
the renormalization scale $\mu$ (b),
the bottom 1S mass $M_{\mbox{\tiny b}}^{\mbox{\tiny 1S}}$ (c) and
the strong coupling $\alpha_s^{(5)}(M_Z)$ (d).
In each plot the fixed parameters 
have been chosen as $\alpha_s^{(5)}(M_Z)=0.118$, 
$M_{\mbox{\tiny b}}^{\mbox{\tiny 1S}}=4.69$~GeV and $\mu=4.69$~GeV,
if not stated otherwise. 
In (a) the result is displayed for $\mu=6.7$ (dashed line), 
$4.69$ (solid line) and $2.7$~GeV (dotted line).
In (b,c,d) the result is displayed for 
$\overline m(\overline m)=0.0$~GeV (solid lines),
$1.1$~GeV (dotted lines), $1.3$~GeV (dash-dotted lines), $1.5$~GeV
(dashed lines) and $1.7$~GeV (long-dashed lines).
}
%\label{figmsbardependence}
 \end{center}
\end{figure}

In Figs.~\ref{figmsbardependence} the dependence of the bottom 
$\overline{\mbox{MS}}$ mass
$\overline M_{\mbox{\tiny b}}(\overline M_{\mbox{\tiny b}})$ 
on the charm quark $\overline{\mbox{MS}}$ mass 
$\overline m(\overline m)$, the strong coupling 
$\alpha_s^{(5)}(M_Z)$, the renormalization scale $\mu$ (of the
strong coupling entering the relation between the bottom
$\overline{\mbox{MS}}$ and 1S masses)
and the bottom 1S mass $M_{\mbox{\tiny b}}^{\mbox{\tiny 1S}}$
is displayed. In each plot the respective 
fixed parameters have been chosen as $\alpha_s^{(5)}(M_Z)=0.118$, 
$M_{\mbox{\tiny b}}^{\mbox{\tiny 1S}}=4.69$~GeV and $\mu=4.69$~GeV, 
if not stated otherwise.
Figure~\ref{figmsbardependence}a shows 
$\overline M_{\mbox{\tiny b}}(\overline M_{\mbox{\tiny b}})$ as a
function of the charm mass for $\mu=6.7$ (dashed line), $4.69$ (solid
line) and $2.7$~GeV (dotted line). 
Figures~\ref{figmsbardependence}b,c,d show 
$\overline M_{\mbox{\tiny b}}(\overline M_{\mbox{\tiny b}})$ as a
function of $\mu$, $M_{\mbox{\tiny b}}^{\mbox{\tiny 1S}}$ and
$\alpha_s^{(5)}(M_Z)$ for $\overline m(\overline m)=0.0$~GeV (solid lines),
$1.1$~GeV (dotted lines), $1.3$~GeV (dash-dotted lines), $1.5$~GeV
(dashed lines) and $1.7$~GeV (long-dashed lines). It is conspicuous
that for   
$\overline m(\overline m)>0.4$~GeV and $\mu\gsim 2.5$~GeV the
dependence of 
$\overline M_{\mbox{\tiny b}}(\overline M_{\mbox{\tiny b}})$ on all
four parameters is approximately linear. This
welcome feature allows for the derivation of the following handy
approximation formula:
\begin{eqnarray}
\overline M_{\mbox{\tiny b}}(\overline M_{\mbox{\tiny b}})
& = &
\bigg[\,
4.169\,\mbox{GeV} 
\, - \, 0.01\,\Big(\,
  \overline m(\overline m)-1.4\,\mbox{GeV}\,\Big)
\, + \, 0.925\,\Big(\,
  M_{\mbox{\tiny b}}^{\mbox{\tiny 1S}}
   -4.69\,\mbox{GeV}\,\Big)
\nonumber
\\[2mm] & & \hspace{2cm}
\, -\, 9.1\,\Big(\,
  \alpha_s^{(5)}(M_Z) - 0.118
\,\Big)\,\mbox{GeV}
\, + \, 0.0057\,\Big(\,
   \mu - 4.69
\,\Big)\,\mbox{GeV}
\,\bigg]
\,.
\label{msbarapproximation}
\end{eqnarray}
For $\overline m(\overline m)>0.4$~GeV and $\mu>2.5$~GeV the
difference between this approximation formula and the exact result is
less than $3$~MeV. Figures~\ref{figmsbardependence} and
Eq.~(\ref{msbarapproximation}) show that at present the dominant
uncertainties arise from the errors in the strong coupling and the
bottom quark 1S mass. The error in 
$\overline M_{\mbox{\tiny b}}(\overline M_{\mbox{\tiny b}})$ induced
by $M_{\mbox{\tiny b}}^{\mbox{\tiny 1S}}$ is roughly equal to the error
in $M_{\mbox{\tiny b}}^{\mbox{\tiny 1S}}$, and the one induced by
$\alpha_s$ roughly amounts to $x$ times $10$~MeV for an uncertainty of
$x$ times $0.001$ in $\alpha_s^{(5)}(M_Z)$. At the present stage 
$M_{\mbox{\tiny b}}^{\mbox{\tiny 1S}}$ and $\alpha_s^{(5)}(M_Z)$ each
contribute about $30$~MeV to the uncertainty of 
$\overline M_{\mbox{\tiny b}}(\overline M_{\mbox{\tiny b}})$. The
uncertainty coming from the scale variation and from the error in the
charm mass corrections, which arises from the uncertainty in 
$\overline m(\overline m)$ and the neglected order
$\overline m^2/M_{\mbox{}\tiny b}$ corrections in the bottom
pole--$\overline{\mbox{MS}}$ mass relation, is much smaller. Varying
the renormalization scale by $1$~GeV shifts 
$\overline M_{\mbox{\tiny b}}(\overline M_{\mbox{\tiny b}})$ by about
$5$~MeV. This variation reflects the remaining perturbative
uncertainty contained in the order $\alpha_s^3$ relation between the
bottom $\overline{\mbox{MS}}$ and 1S masses. We assign a perturbative
uncertainty based on a $3$~GeV variation of the renormalization scale,
which gives $15$~MeV. We estimate the error in the charm mass
corrections as $5$~MeV, which includes the sources of
uncertainties in the charm mass corrections mentioned before and mass
corrections coming from the other light quarks. Adding all
uncertainties quadratically, we obtain 
\begin{eqnarray}
\overline M_{\mbox{\tiny b}}(\overline M_{\mbox{\tiny b}})
& = &
4.17 \, \pm \, 0.05\,\mbox{GeV}
\,,
\label{msbarfinalresults}
\end{eqnarray}
where the central value has been determined for
$M_{\mbox{\tiny b}}^{\mbox{\tiny 1S}}=\mu=4.69$~GeV,
$\alpha_s^{(5)}(M_Z)=0.118$ and $\overline m(\overline m)=1.4$~GeV.
The result is rounded to units of $10$~MeV.

It is instructive to compare the result with the one obtained if the
charm quark mass were neglected entirely: starting from 
$4.71\pm 0.04$~GeV for the bottom 1S mass we would arrive at
$\overline M_{\mbox{\tiny b}}(\overline M_{\mbox{\tiny b}})=
4.20\,\mbox{GeV}
\pm 30\,\mbox{MeV} [\delta\alpha_s]
\pm 40\,\mbox{MeV} [\delta M_{\mbox{\tiny b}}^{\mbox{\tiny 1S}}]
\pm 15\,\mbox{MeV} [\mu]
=4.20\pm 0.05\,\mbox{GeV}$, where the central value is obtained for
$M_{\mbox{\tiny b}}^{\mbox{\tiny 1S}}=\mu=4.71$~GeV and
$\alpha_s^{(5)}(M_Z)=0.118$. Thus, after rounding to units of
$10$~MeV, the overall effect of the finite
charm quark mass on the bottom $\overline{\mbox{MS}}$ mass is a shift
of $-30$~MeV.

Let us briefly comment on the results for
$\overline M_{\mbox{\tiny b}}(\overline M_{\mbox{\tiny b}})$ that we 
gave in Refs.~\cite{Hoang2,Hoang6}, where the bottom 1S mass was also
used as an intermediate step to determine 
$\overline M_{\mbox{\tiny b}}(\overline M_{\mbox{\tiny b}})$, but
where charm mass effects were neglected.
In Ref.~\cite{Hoang2} we have obtained
$\overline M_{\mbox{\tiny b}}(\overline M_{\mbox{\tiny b}})
=4.20\pm 0.06$~GeV. The central value was obtained for
$M_{\mbox{\tiny b}}^{\mbox{\tiny 1S}}=\mu=4.71$~GeV and
$\alpha_s^{(5)}(M_Z)=0.118$, not using
formula~(\ref{msbar1Sgeneric}) for $\overline m(\overline m)=0$, but
solving Eqs.~(\ref{polemsbarthreeloops}) and (\ref{1Spolegeneric})
numerically for a given value of 
$M_{\mbox{\tiny b}}^{\mbox{\tiny 1S}}$. The order $\alpha_s^3$ terms
were not included because at that time the order $\alpha_s^3$
corrections in the pole--$\overline{\mbox{MS}}$ mass 
relation~\cite{Melnikov2,Steinhauser1} where not
yet known. The agreement of the central value with the one mentioned
in the previous paragraph is a coincidence. The $60$~MeV uncertainty
was estimated by assuming a $40$~MeV perturbative error, a $40$~MeV
error caused by the strong coupling (assuming
$\alpha_s^{(5)}(M_Z)=0.118\pm 0.004$) and a $30$~MeV error for 
$M_{\mbox{\tiny b}}^{\mbox{\tiny 1S}}$. The result 
$\overline M_{\mbox{\tiny b}}(\overline M_{\mbox{\tiny b}})
=4.21\pm 0.07$~GeV in the proceedings quoted in Ref.~\cite{Hoang6} 
was obtained from formula~(\ref{msbar1Sgeneric}) for $\overline
m(\overline m)=0$, assuming  
$M_{\mbox{\tiny b}}^{\mbox{\tiny 1S}}=4.73\pm 0.05$~GeV,
$\mu=4.73$~GeV and
$\alpha_s^{(5)}(M_Z)=0.118\pm 0.004$. 
The value for the bottom 1S mass
was determined from 
$M_{\mbox{\tiny b}}^{\mbox{\tiny 1S}}=
1/2(M_{\Upsilon(1S)}\pm\Delta_{\Upsilon(1S)}^{\mbox{\tiny non-pert}})$
assuming $\Delta_{\Upsilon(1S)}\approx 100$~MeV for the
non-perturbative contribution in the $\Upsilon(1S)$ mass.
After rounding to units of $10$~MeV, the error was $10$~MeV larger
than the one obtained 
from formula~(\ref{msbar1Sgeneric}) (for $\overline m(\overline m)=0$)
used in this work because in Ref.~\cite{Hoang6} the corresponding formula
was based on the numerical order $\alpha_s^3$ corrections to the
pole-$\overline{\mbox{MS}}$ mass relation from Ref.~\cite{Steinhauser1}. 
Equation~(\ref{msbar1Sgeneric}), on the other hand, is based on the
analytic result of Ref.~\cite{Melnikov2}. 

\par
\vspace{0.5cm}
\section{Summary}
\label{sectionsummary}
In this work we have examined, for the first time, the effects of
light quark masses on bottom quark mass determinations based on sum
rules for the  
$\Upsilon$ mesons at NNLO in the non-relativistic expansion. 
The effects coming from the charm quark mass are of
particular interest, because one of the relevant physical scales
governing the dynamics of the bottom--antibottom quark pair described
in the sum rules is the inverse Bohr radius 
$M_{\mbox{\tiny b}}\alpha_s$, which is as large as the charm
quark mass. Thus, a naive expansion in the charm quark mass is 
{\it a priori} 
not possible in the calculations for the sum rules and the
extraction of the bottom quark masses. Based on a NNLO analysis our 
result is that the charm mass effects amount to a shift of $-20$~MeV
in the bottom 1S mass, the mass parameter that is extracted directly
from the sum rules. Our final result for the bottom 1S mass reads
\begin{eqnarray}
M_{\mbox{\tiny b}}^{\mbox{\tiny 1S}}=4.69 \, \pm \, 0.03\,\,\mbox{GeV}
\,,
\label{M1Sfinalsummary}
\end{eqnarray}
where the error arises from renormalization scale variations;
the correlation to the value of the strong coupling is negligible.
The bottom 1S mass is a short-distance mass, designed to be used in
systems where the bottom quark virtuality is small with respect to its
mass. Except for non-relativistic bottom--antibottom quark systems, it
can also be applied directly to B meson decays. For inclusive
semileptonic rates, it leads to very well behaved perturbative
expansions~\cite{Hoang3}. Starting from the result for the bottom 1S
mass in Eq.~(\ref{M1Sfinalsummary}), we have examined the light quark
mass effects in the determination of the bottom $\overline{\mbox{MS}}$
mass. Here, after rounding to $10$~MeV precision, we find another
shift of $-10$~MeV coming from the finite charm quark mass.
Our final result for the bottom $\overline{\mbox{MS}}$ mass reads
\begin{eqnarray}
\overline M_{\mbox{\tiny b}}(\overline M_{\mbox{\tiny b}})
& = &
4.17 \, \pm \, 0.05\,\mbox{GeV}
\,,
\label{msbarfinalresultsummary}
\end{eqnarray}
where the $50$~MeV error includes the error in 
$M_{\mbox{\tiny b}}^{\mbox{\tiny 1S}}$, in the strong coupling
($\alpha_s^{(5)}(M_Z)=0.118\pm 0.003$), the
perturbative error estimated from variations of the renormalization
scale $\mu$ and the uncertainties in the charm mass
corrections. Equation~(\ref{msbarfinalresultsummary}) represents a 
determination of the bottom $\overline{\mbox{MS}}$ mass at order
$\alpha_s^3$.
A handy approximation formula that shows the dependence
of $\overline M_{\mbox{\tiny b}}(\overline M_{\mbox{\tiny b}})$ on
the previously mentioned parameters can be found in
Eq.~(\ref{msbarapproximation}). We 
note that the charm quark mass effects in the bottom 1S mass obtained
from the sum rule analysis and those obtained in the relation
between the 
bottom $\overline{\mbox{MS}}$ and 1S masses have the same sign and
lead to an overall shift of about $-30$~MeV in 
$\overline M_{\mbox{\tiny b}}(\overline M_{\mbox{\tiny b}})$. 
We emphasize that, for all practical applications, the values for the
bottom 1S and $\overline{\mbox{MS}}$ masses should, in contrast to the
pole mass, not be considered as order-dependent numbers, since they
do not contain an ambiguity of order $\Lambda_{\rm QCD}$.   
Therefore, the results shown in Eqs.~(\ref{M1Sfinalsummary})
and (\ref{msbarfinalresultsummary}) can be used as an input for
calculations at any order of perturbation theory.

The order of magnitude of the charm mass corrections in the 
bottom $\overline{\mbox{MS}}$ mass can be anticipated from general
considerations because the fact that one cannot expand in
$m_{\mbox{\tiny charm}}/(M_{\mbox{\tiny b}}\alpha_s)$ in the bottom
1S--pole mass relation, whereas it is allowed to expand in
$m_{\mbox{\tiny charm}}$ in the bottom
pole--$\overline{\mbox{MS}}$ mass relation, leads to are non-analytic
correction $\propto\alpha_s^2 m_{\mbox{\tiny charm}}$ that is
$\pi^2$-enhanced and governed by the renormalization scale
$m_{\mbox{\tiny charm}}$ rather than $M_{\mbox{\tiny b}}$. The
enhancement of the linear charm mass term can be understood from the
fact that terms that a non-analytic in the square of the light quark
masses are sensitive to small (infrared) momenta, in analogy to the
non-analytic fictitious gluon mass terms that are often used in
standard renormalon analyses. 
We have explicitly checked that the charm quark
mass effects are under control in the bottom
$\overline{\mbox{MS}}$--1S mass relation, i.e. they do not lead to an
uncontrollable higher order uncertainty that would be relevant at 
present (or in the foreseeable future). This is because at higher
orders of perturbation theory the charm quark (with finite mass)
effectively decouples. It was one of the main motivations of this work
to study the subtle interplay of the two mechanisms just mentioned 
at NNLO in the non-relativistic expansion. 
  
In this work we have determined the complete light quark mass
corrections for the bottom 1S mass obtained from the $\Upsilon$ sum
rules at NNLO in the non-relativistic expansion, and for the bottom
$\overline{\mbox{MS}}$ at order $\alpha_s^3$, with the following
exceptions: in the moments for the sum rules, we neglected the NNLO
double-insertion corrections from the NLO static potential at second
order Rayleigh--Schr\"odinger perturbation theory, since they are
suppressed with respect to the 
single-insertion contributions coming from the NNLO static potential
at first order Rayleigh--Schr\"odinger perturbation theory. For the
charm mass corrections in the bottom
pole--$\overline{\mbox{MS}}$ mass relation we have only taken into
account the linear charm quark mass terms 
$\propto m_{\mbox{\tiny charm}}$ 
and have neglected higher order terms
$(m_{\mbox{\tiny charm}}/M_{\mbox{\tiny b}})^n
m_{\mbox{\tiny charm}}$ for $n\ge 1$. This is feasible because the
bottom pole--$\overline{\mbox{MS}}$ mass relation does not involve 
the inverse Bohr radius as a dynamical scale and because the
subleading terms are less sensitive to small momenta and, therefore,
not enhanced. We demonstrated that
the linear charm mass approximation deviates from the full result for
the charm mass corrections by
only 10\% for the cases where the full result is known. 
Finally, we have neglected the order $\alpha_s^2$ light quark mass
corrections 
$\propto\alpha_s^2(m_{\mbox{\tiny charm}}/M_{\mbox{\tiny b}})^n$ for
$n\ge 2$ in the short-distance Wilson coefficient of the NRQCD bottom
quark currents that describe bottom--antibottom production and
annihilation in the non-relativistic regime. The neglected terms 
are small because the short-distance coefficient only contains
contributions from momenta of order $M_{\mbox{\tiny b}}$, which means
that an expansion in the charm quark mass is justified.
We estimate that all charm mass corrections that have been neglected
in this work are an order of magnitude smaller than the ones that have
been taken into account.
We estimate the effects of the neglected terms and the uncertainties
coming from the error in value of the charm quark mass to be at most
at the level of $5$~MeV. 

{\it Note added:} After this paper was completed we received
Ref.~\cite{Sumino1}, where the N$^3$LO large-$\beta_0$ corrections in
the heavy quark pole--1S mass relation where presented. The results
agree with ours in App.~\ref{appendix1Smasslargeb0}.
%The conclusions about the quality of the large-$\beta_0$ approximation
%in the bottom 1S--$\overline{\mbox{MS}}$ mass relation drawn in
%Sec.~\ref{subsectionmsbar1Smassestimate} are different from 
%Ref.~\cite{Sumino1} because in Ref.~\cite{Sumino1} some terms that are 
%subleading in large-$\beta_0$ limit have been taken into account.

%
%
\par
\vspace{.5cm}
\section*{Acknowledgements}
\addcontentsline{toc}{section}{Acknowledgements}
I am grateful to M.~Melles for discussions and his
collaboration during the initial stages of this work. I thank 
M.~Beneke, G.~Buchalla, M.~Mangano and A.~V.~Manohar for helpful
discussions.  
This work is supported in part by the EU Fourth Framework Program
``Training and Mobility of Researchers'', Network ``Quantum
Chromodynamics and Deep Structure of Elementary Particles'', contract
FMRX-CT98-0194 (DG12-MIHT). 

\par
\vspace{1.5cm}

\begin{appendix}
\section[NNLO Light Quark Mass Corrections to the S-Wave 
  $Q\bar Q$ Ground State Mass]
  {NNLO Light Quark Mass Corrections to the S-Wave Heavy
  Quark--Antiquark Ground State Mass} 
\label{appendix1Smass}
In this appendix we present details of the calculation of the light
quark mass corrections to the ground
state mass of the heavy-quark--antiquark $J^{PC}=1^{--}$, ${}^3S_1$
bound state at NNLO in the non-relativistic expansion. The result is
presented using the pole mass definition for the heavy and the light
quark masses. We assume that the
dynamics of the heavy-quark--antiquark pair is perturbative, i.e. that
the scales $M_{\mbox{\tiny Q}}$, $M_{\mbox{\tiny Q}}v$ and
$M_{\mbox{\tiny Q}}v^2$ are assumed to be larger than 
$\Lambda_{\rm QCD}$. We also assume that there are $n_l$ light quarks,
of which one quark species has a finite pole mass $m$. We adopt the
definition of the strong coupling where all $n_l$ light quarks species
contribute to the running. 

The formal expressions for the light quark mass corrections up to NNLO
have been presented in Eqs.~(\ref{DeltaNLOmassiveformal}) 
and (\ref{DeltaNNLOmassiveformal}). In
configuration space representation the expressions shown in 
Eqs.~(\ref{DeltaNLOmassiveformal}) and (\ref{DeltaNNLOmassiveformal})
read ($x=|{\mbox{\boldmath $x$}}|$, $y=|{\mbox{\boldmath $y$}}|$):
\begin{eqnarray}
-\,2\,M_{\mbox{\tiny Q}}^{\mbox{\tiny pole}}\,
\Delta^{\mbox{\tiny NLO}}_{\mbox{\tiny massive}}
& = &
\int\! d^3{\mbox{\boldmath $x$}}\,
\phi_{\mbox{\tiny 1S}}(x)\,
\delta V_{\mbox{\tiny c,m}}^
   {\mbox{\tiny NLO}}({\mbox{\boldmath $x$}})\,
\phi_{\mbox{\tiny 1S}}(x)
\,,
\label{DeltaNLOmassivecalculation}
\\[4mm]
-\,2\,M_{\mbox{\tiny Q}}^{\mbox{\tiny pole}}\,
\Delta^{\mbox{\tiny NNLO}}_{\mbox{\tiny massive}}
& = &
\int\! d^3{\mbox{\boldmath $x$}}\,
\phi_{\mbox{\tiny 1S}}(x)\,
\delta V_{\mbox{\tiny c,m}}^
   {\mbox{\tiny NNLO}}({\mbox{\boldmath $x$}})\,
\phi_{\mbox{\tiny 1S}}(x)
\nonumber
\\[2mm] & & - \,
\int\! d^3{\mbox{\boldmath $x$}}\,
\int\! d^3{\mbox{\boldmath $y$}}\,
\phi_{\mbox{\tiny 1S}}(x)\,
\delta V_{\mbox{\tiny c,m}}^
   {\mbox{\tiny NNLO}}({\mbox{\boldmath $x$}})\,
\bar G_{\mbox{\tiny 1S}}(x,y)\,
\delta V_{\mbox{\tiny c,m}}^
   {\mbox{\tiny NNLO}}({\mbox{\boldmath $y$}})\,
\phi_{\mbox{\tiny 1S}}(y)
\nonumber
\\[2mm] & & - \,
2\,
\int\! d^3{\mbox{\boldmath $x$}}\,
\int\! d^3{\mbox{\boldmath $y$}}\,
\phi_{\mbox{\tiny 1S}}(x)\,
\delta V_{\mbox{\tiny c,m}}^
   {\mbox{\tiny NNLO}}({\mbox{\boldmath $x$}})\,
\bar G_{\mbox{\tiny 1S}}(x,y)\,
V_{\mbox{\tiny c,massless}}^
   {\mbox{\tiny NNLO}}({\mbox{\boldmath $y$}})\,
\phi_{\mbox{\tiny 1S}}(y)
\,,
\label{DeltaNNLOmassivecalculation}
\end{eqnarray}
where
$V_{\mbox{\tiny c,massless}}^{\mbox{\tiny
NLO}}({\mbox{\boldmath $r$}})$, 
$\delta V_{\mbox{\tiny c,m}}^{\mbox{\tiny NLO}}({\mbox{\boldmath $r$}})$, 
and
$\delta V_{\mbox{\tiny c,m}}^{\mbox{\tiny NNLO}}({\mbox{\boldmath $r$}})$ 
are given in Eqs.~(\ref{VcNLOmassless}), (\ref{VcNLOmassiverspace})
and (\ref{VcNNLOmassiverspace}), respectively. The term
$\phi_{\mbox{\tiny 1S}}$ is the LO ground state wave
function (see Eq.~(\ref{phinS})) and 
\begin{equation}
\bar G_{\mbox{\tiny 1S}}(x,y)
\, = \,
\sum\limits_{\mbox{\tiny S-wave},i\ne 1}\hspace{-0.9cm}
  \int\hspace{0.6cm}
\langle\,{\mbox{\boldmath $x$}}\,|\,
\frac{|\,i\,\rangle\,\langle\,i\,|}{E_i-E_{\mbox{\tiny 1S}}}\,
|\,{\mbox{\boldmath $y$}}\,\rangle
\end{equation}
is the S-wave component of the configuration space LO Coulomb Green
function $G_c({\mbox{\boldmath $x$}},{\mbox{\boldmath $y$}},E)$
%(see text after Eq.~(\ref{NNLOSchroedinger})) 
for 
$E=E_{\mbox{\tiny 1S}}=-2M_{\mbox{\tiny Q}}^{\mbox{\tiny pole}}
\Delta^{\mbox{\tiny LO}}$, where the ground state $n=1$, ${}^3S_1$ 
energy-pole is subtracted. 
For the calculation of the NNLO light quark
mass corrections only the sum over intermediate S-wave states has to
be considered, because the light quark mass corrections to the static
potential are spin- and angle-independent.
An explicit expression for 
$\bar G_{\mbox{\tiny 1S}}(x,y)$ can be easily derived from the S-wave 
($l=0$) contribution of the configuration space Coulomb Green function 
representation by Voloshin~\cite{Voloshin2} 
($a_s\equiv\alpha_s^{(n_l)}(\mu)$,
$M\equiv M_{\mbox{\tiny Q}}^{\mbox{\tiny pole}}$),
\begin{eqnarray}
G_c^{l=0}(x,y,E)
& = &
\frac{M}{4\,\pi}\,(\,2\,k\,)\,
e^{-k(x+y)}\,\sum\limits_{n=1}^\infty\,
\frac{L_{n-1}^1(2\,k\,x)\,L_{n-1}^1(2\,k\,y)}
{(n-\frac{\gamma}{k})\,n}
\,,
\end{eqnarray}
where
\begin{eqnarray}
\gamma & = & 
\frac{M_{\mbox{\tiny Q}}^{\mbox{\tiny pole}}\,C_F\,a_s}{2}
\,,
\\[4mm] 
k^2 & = & -\,M\,E
\end{eqnarray}
and $L_n^a$ are the generalized Laguerre polynomials. The result reads
\begin{eqnarray}
\lefteqn{
\bar G_{\mbox{\tiny nS}}(x,y) \, = \,
\lim\limits_{E\to E_{\rm nS}}\,\bigg[\,
G_c^{l=0}(x,y,E) \, - \, 
\frac{\phi_{\mbox{\tiny nS}}(x)\phi_{\mbox{\tiny nS}}(y)}
{E_{\rm nS}-E}
\,\bigg]
}
\nonumber
%\label{GnSfirstline}
\\[2mm] & = &
\frac{M\,\gamma}{2\,\pi\,n}\,
\exp\Big(-\frac{\gamma}{n}\,(x+y)\Big)\,
\bigg\{\,
\nonumber
\\[2mm] 
& & \quad
  -\,\frac{2\,\gamma}{n^3}\,\bigg[\,
    x\,L_{n-2}^2\Big(\frac{2\,\gamma}{n}\,x\Big)\,
    L_{n-1}^1\Big(\frac{2\,\gamma}{n}\,y\Big)\,+\,
    y\,L_{n-1}^1\Big(\frac{2\,\gamma}{n}\,x\Big)\,
    L_{n-2}^2\Big(\frac{2\,\gamma}{n}\,y\Big)
  \,\bigg]
\nonumber
\\[2mm] 
& & \qquad
  +\,\Big(\frac{5}{2\,n^2}-\frac{\gamma}{n^3}(x+y)\Big)\,
  L_{n-1}^1\Big(\frac{2\,\gamma}{n}\,x\Big)\,
  L_{n-1}^1\Big(\frac{2\,\gamma}{n}\,y\Big)\,
\nonumber
\\[2mm] 
& & \qquad
 +\,\sum\limits_{m\ne n}^\infty\,
  \frac{L_{m-1}^1(\frac{2\,\gamma}{n}\,x)\,
        L_{m-1}^1(\frac{2\,\gamma}{n}\,y)}
  {(m-n)\,m}\,
\bigg\}
\,,
\label{GnSsecondline}
\end{eqnarray}
for the subtraction of the $n\,{}^3S_1$ energy pole,
where
\begin{eqnarray}
\phi_{\mbox{\tiny nS}}(x)
& = &
\pi^{-1/2}n^{-5/2}\,\gamma^{\frac{3}{2}}\,
\exp\Big(-\frac{\gamma}{n}\,x\Big)\, 
L_{n-1}^1\Big(\frac{2\,\gamma}{n}\,x\Big)
\,,
\label{phinS}
\\[4mm]
E_{\rm nS}
& = & - \,\frac{M\,C_F^2\,a_s^2}{4\,n^2}
\,.
\end{eqnarray}
For $n=1$ the first term in the curly brackets vanishes because
$L_{-1}^2(x)=0$. 

In the following we present a number of intermediate results that
lead to Eqs.~(\ref{DeltaNLOmassiveexplicit}) and
(\ref{DeltaNNLOmassiveexplicit}). All results are
eventually expressed in terms of the function $h_i$ ($i=0,\ldots,7$),
which have been defined in Eqs.~(\ref{h0func})--(\ref{h7func}).

\par
%\vspace{0.5cm}
%
\subsubsection*{Light Quark Mass Corrections at NLO}
The light quark mass corrections at NLO have been determined before in
Ref.~\cite{Eiras2}. They consist of a single insertion of 
$\delta V_{\mbox{\tiny c,m}}^{\mbox{\tiny NLO}}$ 
(Eq.~(\ref{VcNLOmassiverspace})) in first order
Rayleigh--Schr\"odinger perturbation theory:
\begin{eqnarray}
\int\! d^3{\mbox{\boldmath $r$}}\,
\phi_{\mbox{\tiny 1S}}^2(r)\,
\delta V_{\mbox{\tiny c,m}}^{\mbox{\tiny NLO}}
({\mbox{\boldmath $r$}})
& = &
-\,\frac{C_F^2\,a_s^2}{2}\,
M_{\mbox{\tiny Q}}^{\mbox{\tiny pole}}\,
\Big(\frac{a_s}{3\,\pi}\Big)\,\bigg[\,
\int\limits_1^\infty\,d x\,\frac{f(x)}{(1+\frac{m}{\gamma}\,x)^2}
+ \ln\Big(\frac{m}{2\,\gamma}\Big) + \frac{11}{6}
\,\bigg]
\nonumber
\\[2mm]
& = & 
-\,\frac{C_F^2\,a_s^2}{2}\,
M_{\mbox{\tiny Q}}^{\mbox{\tiny pole}}\,
\Big(\frac{a_s}{3\,\pi}\Big)\,\bigg[\,
h_0(a) + \ln\Big(\frac{a}{2}\Big) + \frac{11}{6}
\,\bigg]
\,,
\end{eqnarray}
where
\begin{eqnarray}
a
& = &
\frac{m}{\gamma}
\,.
\end{eqnarray}
The function $h_0$ can be easily calculated
analytically, see Eq.~(\ref{h0func}). 

\par
%\vspace{0.5cm}
%
\subsubsection*{Light Quark Mass Corrections at NNLO -- Single Insertions}
For the light quark mass corrections at NNLO let us first consider the
contributions coming from the single insertion of the NNLO potential
at first order Rayleigh--Schr\"odinger perturbation theory, the first
term on the RHS of Eq.~(\ref{DeltaNNLOmassivecalculation}). We already
mentioned in the text after Eq.~(\ref{VcNNLOmassiverspace}) that the
light quark mass corrections to the NNLO static potential can be
divided into three different parts, of which each vanishes
individually for $m\to 0$. In the following we present the results
from these three contributions separately in terms of the functions
$h_i$:
($\tilde m=e^{\gamma_{\mbox{\tiny E}}}\,m$,
$\tilde \mu=e^{\gamma_{\mbox{\tiny E}}}\,\mu$),
\begin{eqnarray}
\lefteqn{
\int\! d^3{\mbox{\boldmath $r$}}\,
\phi_{\mbox{\tiny 1S}}^2(r)\,
\bigg(-\frac{C_F\,a_s}{r}\bigg)\,
\bigg[-\frac{3}{2}\,\int\limits_1^\infty dx\,f(x)\,e^{-2 m r x}\,
\bigg(\beta_0\,\Big(\ln\frac{4m^2x^2}{\mu^2}
 -{\rm Ei}(2\,m\,x\,r)-{\rm Ei}(-2\,m\,x\,r)\Big) - a_1
\bigg)
}
\nonumber 
\\[2mm]
\lefteqn{
\hspace{4cm} 
+\,3\,\bigg(\ln(\tilde m\,r)+\frac{5}{6}\,\bigg)\,
 \bigg(\beta_0\,\ln(\tilde \mu\,r)+\frac{a_1}{2}\bigg) 
+\beta_0\,\frac{\pi^2}{4}
\,\bigg]
}
\nonumber
\\[2mm]
& = & 
-\,\frac{C_F^2\,a_s^2}{2}\,
M_{\mbox{\tiny Q}}^{\mbox{\tiny pole}}\,
\bigg[\,
  -\frac{3}{2}\,h_0(a)\,
    \bigg(\beta_0\,\ln\Big(\frac{4\,\gamma^2}{\mu^2}\Big)-a_1\bigg)
  -3\,\beta_0\,h_1(a)
\hspace{6cm}
\nonumber
\\[2mm] 
& & \hspace{1cm}
+\,3\,\bigg(\ln\Big(\frac{a}{2}\Big)+\frac{11}{6}\bigg)\,
  \bigg(\beta_0\,\ln\Big(\frac{\mu}{2\,\gamma}\Big)
    +\beta_0+\frac{a_1}{2}\bigg)
+\frac{3}{4}\,\beta_0\,(\pi^2-4) 
\,\bigg]
\,,
\label{NNLOsingle1}
\\[4mm]
\lefteqn{
\int\! d^3{\mbox{\boldmath $r$}}\,
\phi_{\mbox{\tiny 1S}}^2(r)\,
\bigg(\,\frac{C_F\,a_s}{r}\bigg)\,
\bigg[\,
\int\limits_1^\infty dx\,f(x)\,e^{-2 m r x}\,
\bigg(\,\frac{5}{3}+\frac{1}{x^2}\bigg(
1+\frac{1}{2\,x}\,\sqrt{x^2-1}\,(1+2x^2)\,
\ln\Big(\frac{x-\sqrt{x^2-1}}{x+\sqrt{x^2-1}}\Big)
\bigg)
\,\bigg)
}
\nonumber
\\[2mm]
\lefteqn{ 
\hspace{4cm} 
+\int\limits_1^\infty dx\,f(x)\,e^{-2 m r x}\,
\bigg( \ln(4x^2)
 -{\rm Ei}(2\,m\,x\,r)-{\rm Ei}(-2\,m\,x\,r) - \frac{5}{3}
\bigg)
}
\nonumber
\\[2mm]
\lefteqn{ 
\hspace{4cm} 
+\bigg(\ln(\tilde m\,r)+\frac{5}{6}\,\bigg)^2 + \frac{\pi^2}{12}
\,\bigg]
}
\nonumber
\\[2mm]
& = &
-\,\frac{C_F^2\,a_s^2}{2}\,
M_{\mbox{\tiny Q}}^{\mbox{\tiny pole}}\,
\bigg[\,
-h_2(a)+h_0(a)\,\bigg(2\,\ln\Big(\frac{a}{2}\Big)+\frac{5}{6}\bigg)
-2\,h_1(a) 
+\bigg(\ln\Big(\frac{a}{2}\Big)+\frac{11}{6}\bigg)^2
+\frac{\pi^2}{4}-1
\,\bigg]
\,,
\nonumber
\\
\label{NNLOsingle2}
\\[4mm]
\lefteqn{
\int\! d^3{\mbox{\boldmath $r$}}\,
\phi_{\mbox{\tiny 1S}}^2(r)\,
\bigg(-\frac{C_F\,a_s}{r}\bigg)\,
\bigg[\,\frac{57}{4}\,\bigg(\,
c_1\,\Gamma(0,2\,c_2\,m\,r) + d_1\,\Gamma(0,2\,d_2\,m\,r)
+ \ln(\tilde m\,r) + \frac{161}{228} + \frac{13}{19}\,\zeta_3
\,\bigg)
\,\bigg]
}
\nonumber
\\[2mm]
& = &
-\,\frac{C_F^2\,a_s^2}{2}\,
M_{\mbox{\tiny Q}}^{\mbox{\tiny pole}}\,
\bigg[\,
  \frac{57}{4}\,
  \bigg(
    \frac{c_1\,c_2\,a}{1+c_2\,a} + \frac{d_1\,d_2\,a}{1+d_2\,a}
    + c_1\,\ln(1+c_2\,a) + d_1\,\ln(1+d_2\,a) 
  \bigg)
\,\bigg]
\,.
\label{NNLOsingle3}
\end{eqnarray} 

\par
%\vspace{0.5cm}
%
\subsubsection*{Light Quark Mass Corrections at NNLO -- Double Insertions}
For the light quark mass corrections at NNLO coming from double
insertions of the NLO static potential at second order
Rayleigh--Schr\"odinger perturbation theory, there are two different
contributions: the first contains two insertions of the NLO light quark
mass corrections to the static potential, and the second contains one
insertion of the NLO light quark mass corrections to the static
potential and one insertion of the NLO static potential for massless
light quarks. In detail, the two contributions read: 
\begin{eqnarray}
\lefteqn{
-\,\int\! d^3{\mbox{\boldmath $x$}}\,
\int\! d^3{\mbox{\boldmath $y$}}\,
\phi_{\mbox{\tiny 1S}}(x)\,
\delta V_{\mbox{\tiny c,m}}^
   {\mbox{\tiny NNLO}}({\mbox{\boldmath $x$}})\,
\bar G_{\mbox{\tiny 1S}}(x,y)\,
\delta V_{\mbox{\tiny c,m}}^
   {\mbox{\tiny NNLO}}({\mbox{\boldmath $y$}})\,
\phi_{\mbox{\tiny 1S}}(y)
}
\nonumber
\\[2mm]
& = &
\frac{C_F^2\,a_s^2}{2}\,
M_{\mbox{\tiny Q}}^{\mbox{\tiny pole}}\,
\bigg[\,
  h_0(a)\,\bigg(
    -\frac{5}{2}\,h_0(a)+2\,h_3(a)-2\,h_4(a)
  \bigg)-h_5(a)+h_6(a)
\nonumber
\\[2mm] & & \hspace{1cm}
  -\,3\,h_0(a)\,\bigg(\ln\Big(\frac{a}{2}\Big)+\frac{3}{2}\bigg)
  +2\,h_7(a)+2\,h_3(a)\,\bigg(\ln\Big(\frac{a}{2}\Big)+\frac{11}{6}\bigg)
\nonumber
\\[2mm] & & \hspace{1cm}
  -\,\frac{1}{2}\,\ln^2\Big(\frac{a}{2}\Big)
  -\frac{5}{6}\,\ln\Big(\frac{a}{2}\Big)
  -\zeta_3
  +\frac{\pi^2}{6}
  -\frac{61}{72}
\,\bigg]
\,,
\label{NNLOdouble1}
\\[4mm]
\lefteqn{
-\,2\,
\int\! d^3{\mbox{\boldmath $x$}}\,
\int\! d^3{\mbox{\boldmath $y$}}\,
\phi_{\mbox{\tiny 1S}}(x)\,
\delta V_{\mbox{\tiny c,m}}^
   {\mbox{\tiny NNLO}}({\mbox{\boldmath $x$}})\,
\bar G_{\mbox{\tiny 1S}}(x,y)\,
V_{\mbox{\tiny c,massless}}^
   {\mbox{\tiny NNLO}}({\mbox{\boldmath $y$}})\,
\phi_{\mbox{\tiny 1S}}(y)
}
\nonumber
\\[2mm]
& = &
\frac{C_F^2\,a_s^2}{2}\,
M_{\mbox{\tiny Q}}^{\mbox{\tiny pole}}\,
\bigg[\,
  \frac{3}{2}\,\beta_0\,\bigg(\,
    h_0(a)\,\bigg(3\,\ln\Big(\frac{2\,\gamma}{\mu}\Big)-2\bigg)
    +2\,h_7(a) 
    +2\,h_3(a)\,\bigg(1-\ln\Big(\frac{2\,\gamma}{\mu}\Big)\bigg)
\nonumber
\\[2mm] & & \hspace{2cm}
    +\,\bigg(\ln\Big(\frac{a}{2}\Big)+\frac{5}{6}\bigg)\,
       \ln\Big(\frac{2\,\gamma}{\mu}\Big)
    -2\,\zeta_3 + \frac{\pi^2}{3} -1
   \,\bigg)
\nonumber
\\[2mm] & & \hspace{1cm}
  +\,\frac{3}{2}\,a_1\,\bigg(\,
    -\frac{3}{2}\,h_0(a)+h_3(a)
    -\frac{1}{2}\,\ln\Big(\frac{a}{2}\Big)-\frac{5}{12}
\,\bigg)
\,\bigg]
\,.
\label{NNLOdouble2}
\end{eqnarray} 
Adding the two results we arrive at the expression displayed in
Eq.~(\ref{DeltaNNLOmassiveexplicitdouble}). 
%We note that the of the
%terms $XXXXX$ in $$
%arise from the summation over intermediate states contained in 
%$\bar G$.

%
%
\par
\vspace{0.5cm}
\section[Heavy $Q\bar Q$ Ground State Mass at
  N$^3$LO in the Large-$\beta_0$ Approximation]
  {Perturbative Heavy-Quark--Antiquark Ground State Mass at
  N$^3$LO in the Large-$\beta_0$ Approximation}
\label{appendix1Smasslargeb0}
In this appendix we calculate the mass of the ground state of the
heavy-quark--antiquark system at N$^3$LO in the non-relativistic
expansion (or order $\epsilon^4$ in the upsilon expansion) in the
large-$\beta_0$ approximation, in terms of the heavy quark pole
mass. It is assumed that the dynamics of the 
heavy quark pair is perturbative, i.e. that 
$M_{\mbox{\tiny Q}}\gg M_{\mbox{\tiny Q}}\alpha_s\gg 
M_{\mbox{\tiny Q}}\alpha_s^2\gg\Lambda_{\rm QCD}$. 
All light quarks are treated as massless. Up to NNLO in the
non-relativistic expansion, the large-$\beta_0$ corrections can be
easily extracted from the corrections shown in
Eq.~(\ref{1Spolegeneric}) (for zero light quark masses) by taking the
highest power of $n_l$ 
%($=$ number of light quark species) 
and supplementing the result by the replacement 
$n_l\to -\frac{3}{2}\beta_0$.

Up to N$^3$LO the heavy-quark--antiquark ground state mass in the
large-$\beta_0$ approximation can be parametrized in the following
way:
\begin{eqnarray}
\Big[\,
M_{Q\bar Q}^{n=1}
\,\Big]_{\beta_0}
& = &
\Big[\,2 M_{\mbox{\tiny Q}}^{\mbox{\tiny 1S}}
\,\Big]_{\beta_0}
\nonumber
\\[2mm]
& = &  
2 M_{\mbox{\tiny Q}}^{\mbox{\tiny pole}}\,\bigg[\,
1
\, - \, \epsilon\, \Delta^{\mbox{\tiny LO}} 
\, - \, \epsilon\, \Delta^{\mbox{\tiny NLO}}_{\beta_0} 
\, - \, \epsilon\, \Delta^{\mbox{\tiny NNLO}}_{\beta_0} 
\, - \, \epsilon\, \Delta^{\mbox{\tiny NNNLO}}_{\beta_0} 
\,\bigg]
\,,
\label{MQQbarN3LO}
\end{eqnarray}
where ($a_s=\alpha_s^{(n_l)}(\mu)$)
\begin{eqnarray}
\Delta^{\mbox{\tiny LO}} 
& = &
\frac{C_F^2\,a_s^2}{8}
\,,
\label{DeltaLOb0} 
\\[4mm]
\Delta^{\mbox{\tiny NLO}}_{\beta_0} 
& = &
\frac{C_F^2\,a_s^2}{8}
\Big(\frac{a_s}{\pi}\Big)\,\beta_0\,
\bigg[\,
L + \frac{11}{6}
\,\bigg]
\,,
\label{DeltaNLOb0} 
\\[4mm]
\Delta^{\mbox{\tiny NNLO}}_{\beta_0} 
& = &
\frac{C_F^2\,a_s^2}{8}\,
\Big(\frac{a_s}{\pi}\Big)^2\,\beta_0^2\,
\bigg[\,
\frac{3}{4}\,L^2 + \frac{9}{4}\,L 
+ \frac{1}{2}\,\zeta_3 + \frac{\pi^2}{24} + \frac{77}{48}
\,\bigg]
\,,
\label{DeltaNNLOb0} 
\\[4mm]
L & = & \ln\Big(\frac{\mu}
{C_F\,a_s\,M_{\mbox{\tiny Q}}^{\mbox{\tiny pole}}}\Big)
\,.
\end{eqnarray}
To calculate $\Delta^{\mbox{\tiny NNNLO}}_{\beta_0}$ we first need the
N$^3$LO heavy-quark--antiquark static potential in the large-$\beta_0$
approximation. In momentum space representation the large-$\beta_0$
corrections are particularly simple because they only involve
insertions of massless quark one-loop vacuum polarizations into the
gluon line defining the LO static potential, supplemented by the
replacement $n_l\to -\frac{3}{2}\beta_0$,
\begin{eqnarray}
\tilde V_{\beta_0}({\mbox{\boldmath $p$}})
& = &
-\,\frac{4\,\pi\,C_F\,a_s}{{\mbox{\boldmath $p$}}^2}\,
\bigg\{\,
1 
- \Big(\frac{a_s}{4\,\pi}\Big)\,\beta_0\,
\bigg[\,\ln\Big(\frac{{\mbox{\boldmath $p$}}^2}{\mu^2}\Big)
-\frac{5}{3}\,\bigg]
+ \Big(\frac{a_s}{4\,\pi}\Big)^2\,\beta_0^2\,
\bigg[\,\ln\Big(\frac{{\mbox{\boldmath $p$}}^2}{\mu^2}\Big)
-\frac{5}{3}
\,\bigg]^2
\nonumber
\\[2mm]
& & \hspace{2.5cm}
- \Big(\frac{a_s}{4\,\pi}\Big)^3\,\beta_0^3\,
\bigg[\,\ln\Big(\frac{{\mbox{\boldmath $p$}}^2}{\mu^2}\Big)
-\frac{5}{3}
\,\bigg]^3
+ \dots
\,\bigg\}
\,.
\label{momVb0NNNLO}
\end{eqnarray}
In configuration space representation, the large-$\beta_0$ potential
reads
\begin{eqnarray}
V_{\beta_0}({\mbox{\boldmath $r$}})
& = &
\int\!\frac{d^3{\mbox{\boldmath $p$}}}{(2\pi)^3}\,\,
\tilde V_{\beta_0}({\mbox{\boldmath $p$}})
\,\exp( i {\mbox{\boldmath $p$}} {\mbox{\boldmath $r$}} )
\nonumber\\[2mm]
& = &
V^{\mbox{\tiny LO}}({\mbox{\boldmath $r$}}) \, + \,
V_{\beta_0}^{\mbox{\tiny NLO}}({\mbox{\boldmath $r$}}) \, + \,
V_{\beta_0}^{\mbox{\tiny NNLO}}({\mbox{\boldmath $r$}}) \, + \,
V_{\beta_0}^{\mbox{\tiny NNNLO}}({\mbox{\boldmath $r$}})
\, + \, \ldots
\,,
\end{eqnarray}
where ($\tilde\mu \, \equiv \, e^{\gamma_{\mbox{\tiny E}}}\,\mu$,
$r=|{\mbox{\boldmath $r$}}|$)
\begin{eqnarray}
V^{\mbox{\tiny LO}}({\mbox{\boldmath $r$}})
& = &
-\,\frac{C_F\,a_s}{r}
\,,
\label{konfigVb0LO}
\\[4mm]
V_{\beta_0}^{\mbox{\tiny NLO}}({\mbox{\boldmath $r$}})
& = &
-\,\frac{C_F\,a_s}{r}\,
\Big(\frac{a_s}{2\,\pi}\Big)\,\beta_0\,
\bigg[\,
\ln(\tilde\mu\,r) + \frac{5}{6}
\,\bigg]
\,,
\label{konfigVb0NLO}
\\[4mm]
V_{\beta_0}^{\mbox{\tiny NNLO}}({\mbox{\boldmath $r$}})
& = &
-\,\frac{C_F\,a_s}{r}\,
\Big(\frac{a_s}{2\,\pi}\Big)^2\,\beta_0^2\,
\bigg[\,
\bigg(\ln(\tilde\mu\,r) + \frac{5}{6}\bigg)^2 + \frac{\pi^2}{12}
\,\bigg]
\,,
\label{konfigVb0NNLO}
\\[4mm]
V_{\beta_0}^{\mbox{\tiny NNNLO}}({\mbox{\boldmath $r$}})
& = &
-\,\frac{C_F\,a_s}{r}\,
\Big(\frac{a_s}{2\,\pi}\Big)^3\,\beta_0^3\,
\bigg[\,
\bigg(\ln(\tilde\mu\,r) + \frac{5}{6}\bigg)^3
+ \frac{\pi^2}{4}\,\bigg(\ln(\tilde\mu\,r) + \frac{5}{6}\bigg)
+ 2\,\zeta_3
\,\bigg]
\,.
\label{konfigVb0NNNLO}
\end{eqnarray}
It is now straightforward to derive the formal result for
$\Delta^{\mbox{\tiny NNNLO}}_{\beta_0}$ using Rayleigh--Schr\"odinger
time-independent perturbation theory up to third order:
\begin{eqnarray}
-2\,M_{\mbox{\tiny Q}}^{\mbox{\tiny pole}}\,
\Delta_{\beta_0}^{\mbox{\tiny NNNLO}}
& = &
\langle\,1S\,|\,
V_{\beta_0}^{\mbox{\tiny NNNLO}}
\,|\,1S\,\rangle 
\, + \,
2\,\sum\limits_{i\ne\mbox{\tiny 1S}}\hspace{-0.55cm}\int\,\,\,
\langle\,1S\,|\,
V_{\beta_0}^{\mbox{\tiny NNLO}}\,
\frac{|\,i\,\rangle\,\langle\,i\,|}{E_{\mbox{\tiny 1S}}-E_i}\,
V_{\beta_0}^{\mbox{\tiny NLO}}\,
\,|\,1S\,\rangle 
\nonumber
\\[2mm] & & + \,
\sum_{m\ne\mbox{\tiny 1S}}\hspace{-0.65cm}\int\,\,\,
\sum_{n\ne\mbox{\tiny 1S}}\hspace{-0.6cm}\int\,\,\,
\langle\,1S\,|\,
V_{\beta_0}^{\mbox{\tiny NLO}}\,
\frac{|\,m\,\rangle\,\langle\,m\,|}{E_{\mbox{\tiny 1S}}-E_m}
\Big(\,
V_{\beta_0}^{\mbox{\tiny NLO}} 
+2 M_{\mbox{\tiny Q}}^{\mbox{\tiny pole}}\,
\Delta_{\beta_0}^{\mbox{\tiny NLO}}\,
\Big)
\frac{|\,n\,\rangle\,\langle\,n\,|}{E_{\mbox{\tiny 1S}}-E_n}\,
V_{\beta_0}^{\mbox{\tiny NLO}}\,
\,|\,1S\,\rangle 
\,,
\nonumber
%\label{Deltab0NNNLOformal}
\\[2mm] 
& = &
\int\! d^3{\mbox{\boldmath $x$}}\,
\phi_{\mbox{\tiny 1S}}(x)\,
V_{\beta_0}^
   {\mbox{\tiny NNNLO}}({\mbox{\boldmath $x$}})\,
\phi_{\mbox{\tiny 1S}}(x)
\nonumber
\\[2mm] & & 
- \, 2 \,
\int\! d^3{\mbox{\boldmath $x$}}\,
\int\! d^3{\mbox{\boldmath $y$}}\,
\phi_{\mbox{\tiny 1S}}(x)\,
V_{\beta_0}^{\mbox{\tiny NNLO}}({\mbox{\boldmath $x$}})\,
\bar G_{\mbox{\tiny 1S}}(x,y)\,
V_{\beta_0}^{\mbox{\tiny NLO}}({\mbox{\boldmath $y$}})\,
\phi_{\mbox{\tiny 1S}}(y)
\nonumber
\\[2mm] & & 
+ \,
\int\! d^3{\mbox{\boldmath $x$}}\,
\int\! d^3{\mbox{\boldmath $y$}}\,
\int\! d^3{\mbox{\boldmath $z$}}\,
\phi_{\mbox{\tiny 1S}}(x)\,
V_{\beta_0}^{\mbox{\tiny NLO}}({\mbox{\boldmath $x$}})\,
\bar G_{\mbox{\tiny 1S}}(x,y)\,
\Big(\,
V_{\beta_0}^{\mbox{\tiny NLO}}({\mbox{\boldmath $y$}}) + 
2 M_{\mbox{\tiny Q}}^{\mbox{\tiny pole}}\,
\Delta_{\beta_0}^{\mbox{\tiny NLO}}\,
\Big)\,
\nonumber
\\[2mm] & & \hspace{3.5cm} \times\,
\bar G_{\mbox{\tiny 1S}}(y,z)\,
V_{\beta_0}^{\mbox{\tiny NLO}}({\mbox{\boldmath $z$}})\,
\phi_{\mbox{\tiny 1S}}(z)
\,,
\label{Deltab0NNNLOrspace}
\end{eqnarray}
where 
\begin{eqnarray}
\phi_{1S}({\mbox{\boldmath $r$}}) & = &
\langle\,{\mbox{\boldmath $r$}}\,|\,\mbox{1S}\,\rangle 
\, = \,
\langle\,\mbox{1S}\,|\,\mbox{\boldmath $r$}\,\rangle 
\, = \,
\pi^{-\frac{1}{2}}\,
\gamma^{\frac{3}{2}}\,
e^{-\gamma\,r}
\,,
\\[4mm]
\gamma & = & 
\frac{M_{\mbox{\tiny Q}}^{\mbox{\tiny pole}}\,C_F\,a_s}{2}
\,,
\label{gammadef}
\end{eqnarray} 
is the LO ground state Coulomb wave function
and $\bar G_{\mbox{\tiny 1S}}(x,y)$ the S-wave component of the Coulomb
Green function, where the ground state energy pole is subtracted. The
explicit form of $\bar G$ is given in Eq.~(\ref{GnSsecondline}).
The reader should note the appearance of the NLO binding energy in the
third term on the RHS of Eq.~(\ref{Deltab0NNNLOrspace}), which arises
for the first time at third order Rayleigh--Schr\"odinger perturbation
theory (see e.g. Ref.~\cite{Galindo1}). 
For the calculation of the integrals in Eq.~(\ref{Deltab0NNNLOrspace})
the following formulae are helpful 
($a_c\equiv C_F\,\alpha_s^{(n_l)}(\mu)$,
$x=|{\mbox{\boldmath $x$}}|$,$y=|{\mbox{\boldmath $y$}}|$,
$z=|{\mbox{\boldmath $z$}}|$):
\begin{eqnarray}
{\rm A}_1^0 & = & 
\int\! d^3{\mbox{\boldmath $x$}}\,
\phi_{\mbox{\tiny 1S}}^2(x)\,\bigg[
-\frac{a_c}{x}
\,\bigg]
\, = \,
-\,a_c\,\gamma
\,,
\label{A10}
\\[4mm]
{\rm A}_1^1 & = & 
\int\! d^3{\mbox{\boldmath $x$}}\,
\phi_{\mbox{\tiny 1S}}^2(x)\,
\bigg[-\frac{a_c}{x}\,\ln(\tilde\mu\,x)\,\bigg]
\, = \,
-\,a_c\,\gamma\,\bigg[\,
\ln\Big(\frac{\mu}{2\,\gamma}\Big) + 1
\,\bigg]
\,,
\label{A11}
\\[4mm]
{\rm A}_1^2 & = & 
\int\! d^3{\mbox{\boldmath $x$}}\,
\phi_{\mbox{\tiny 1S}}^2(x)\,
\bigg[-\frac{a_c}{x}\,\ln^2(\tilde\mu\,x)\,\bigg]
\, = \,
-\,a_c\,\gamma\,\bigg[\,
\ln^2\Big(\frac{\mu}{2\,\gamma}\Big) 
+ 2 \, \ln\Big(\frac{\mu}{2\,\gamma}\Big)
+ \frac{\pi^2}{6}
\,\bigg]
\,,
\label{A12}
\\[4mm]
{\rm A}_1^3 & = & 
\int\! d^3{\mbox{\boldmath $x$}}\,
\phi_{\mbox{\tiny 1S}}^2(x)\,
\bigg[-\frac{a_c}{x}\,\ln^3(\tilde\mu\,x)\,\bigg]
\nonumber
\\[4mm]
& = &
-\,a_c\,\gamma\,\bigg[\,
\ln^3\Big(\frac{\mu}{2\,\gamma}\Big)
+ 3 \, \ln^2\Big(\frac{\mu}{2\,\gamma}\Big) 
+ \frac{\pi^2}{2} \, \ln\Big(\frac{\mu}{2\,\gamma}\Big)
+ \frac{\pi^2}{2}
- 2\,\zeta_3
\,\bigg]
\,,
\label{A13}
\\[4mm]
{\rm A}_2^1 & = & -\, 
\int\! d^3{\mbox{\boldmath $x$}}\,
\int\! d^3{\mbox{\boldmath $y$}}\,
\phi_{\mbox{\tiny 1S}}(x)\,\bigg[-\frac{a_c}{x}\,\bigg]\,
\bar G_{\mbox{\tiny 1S}}(x,y)\,
\bigg[-\frac{a_c}{y}\,\ln(\tilde\mu\,y)\,\bigg]\,
\phi_{\mbox{\tiny 1S}}(y)
\nonumber
\\[2mm] & = &
-\,a_c\,\gamma\,\bigg[\,
\frac{1}{2}\,\ln\Big(\frac{\mu}{2\,\gamma}\Big)
\,\bigg]
\,,
\label{A21}
\\[4mm]
{\rm A}_2^2 & = & -\,
\int\! d^3{\mbox{\boldmath $x$}}\,
\int\! d^3{\mbox{\boldmath $y$}}\,
\phi_{\mbox{\tiny 1S}}(x)\,
\bigg[-\frac{a_c}{x}\,\ln(\tilde\mu\,x)\,\bigg]\,
\bar G_{\mbox{\tiny 1S}}(x,y)\,
\bigg[-\frac{a_c}{y}\,\ln(\tilde\mu\,y)\,\bigg]\,
\phi_{\mbox{\tiny 1S}}(y)
\nonumber
\\[2mm] & = &
-\,a_c\,\gamma\,\bigg[\,
\frac{1}{2} \, \ln^2\Big(\frac{\mu}{2\,\gamma}\Big) 
+ \frac{1}{2} 
- \frac{\pi^2}{6}
+ \zeta_3
\,\bigg]
\,,
\label{A22}
\\[4mm]
{\rm A}_2^3 & = & -\,
\int\! d^3{\mbox{\boldmath $x$}}\,
\int\! d^3{\mbox{\boldmath $y$}}\,
\phi_{\mbox{\tiny 1S}}(x)\,
\bigg[-\frac{a_c}{x}\,\ln^2(\tilde\mu\,x)\,\bigg]\,
\bar G_{\mbox{\tiny 1S}}(x,y)\,
\bigg[-\frac{a_c}{y}\,\ln(\tilde\mu\,y)\,\bigg]\,
\phi_{\mbox{\tiny 1S}}(y)
\nonumber
\\[2mm] & = &
-\,a_c\,\gamma\,\bigg[\,
\frac{1}{2}\,\ln^3\Big(\frac{\mu}{2\,\gamma}\Big)
+ \bigg(2\,\zeta_3 - \frac{\pi^2}{4}\bigg)\, 
    \ln\Big(\frac{\mu}{2\,\gamma}\Big) 
- \frac{\pi^4}{180}
- \frac{\pi^2}{3}
+ 4\,\zeta_3
- 1
\,\bigg]
\,,
\label{A23}
\\[4mm]
{\rm A}_3^2 & = & 
\int\! d^3{\mbox{\boldmath $x$}}\,
\int\! d^3{\mbox{\boldmath $y$}}\,
\int\! d^3{\mbox{\boldmath $z$}}\,
\phi_{\mbox{\tiny 1S}}(x)\,
\bigg[-\frac{a_c}{x}\,\ln(\tilde\mu\,x)\,\bigg]\,
\bar G_{\mbox{\tiny 1S}}(x,y)\,
\bar G_{\mbox{\tiny 1S}}(y,z)\,
\bigg[-\frac{a_c}{z}\,\ln(\tilde\mu\,z)\,\bigg]\,
\phi_{\mbox{\tiny 1S}}(z)
\nonumber
\\[2mm] & = &
-\,\frac{1}{2}\,\bigg[\,
\frac{3}{2}\,\ln^2\Big(\frac{\mu}{2\,\gamma}\Big)
+ \frac{1}{2}\,\ln\Big(\frac{\mu}{2\,\gamma}\Big) 
+ \frac{\pi^4}{45}
- \frac{2\,\pi^2}{3}
+ \frac{9}{2}
\,\bigg]
\,,
\label{A32}
\\[4mm]
{\rm A}_3^3 & = & 
\int\! d^3{\mbox{\boldmath $x$}}\,
\int\! d^3{\mbox{\boldmath $y$}}\,
\int\! d^3{\mbox{\boldmath $z$}}\,
\phi_{\mbox{\tiny 1S}}(x)\,
\bigg[-\frac{a_c}{x}\,\ln(\tilde\mu\,x)\,\bigg]\,
\bar G_{\mbox{\tiny 1S}}(x,y)\,
\bigg[-\frac{a_c}{y}\,\ln(\tilde\mu\,y)\,\bigg]\,
\nonumber
\\[2mm] 
& & \hspace{4cm} \times\,
\bar G_{\mbox{\tiny 1S}}(y,z)\,
\bigg[-\frac{a_c}{z}\,\ln(\tilde\mu\,z)\,\bigg]\,
\phi_{\mbox{\tiny 1S}}(z)
\nonumber
\\[2mm] & = &
-\,a_c\,\gamma\,\bigg[\,
\frac{3}{4}\,\ln^3\Big(\frac{\mu}{2\,\gamma}\Big)
+ \frac{1}{2}\,\ln^2\Big(\frac{\mu}{2\,\gamma}\Big)
+ \bigg( \frac{\pi^4}{90} - \frac{\pi^2}{3} 
  + \frac{5}{2} \bigg)\, 
  \ln\Big(\frac{\mu}{2\,\gamma}\Big)
\nonumber
\\[4mm] 
& & \hspace{1.5cm}
+ \,\frac{\pi^4}{40}
- \frac{\pi^2}{6}
+ 6\,\zeta_5
- \bigg(\frac{\pi^2}{2} + 4\bigg)\,\zeta_3
+ \frac{15}{4}
\,\bigg]
\,.
\label{A33}
\end{eqnarray}
Using Eqs.~(\ref{A10})--(\ref{A33}) it is straightforward to derive
the result for $\Delta^{\mbox{\tiny NNNLO}}_{\beta_0}$:
\begin{eqnarray}
\Delta^{\mbox{\tiny NNNLO}}_{\beta_0} 
& = &
\frac{C_F^2\,a_s^2}{8}\,
\Big(\frac{a_s}{\pi}\Big)^3\,\beta_0^3\,
\bigg[\,
\frac{1}{2}\,L^3 +
\frac{15}{8}\,L^2 + 
\Big(\frac{\pi^2}{12} + \zeta_3 + \frac{25}{12}\bigg)\,L
\nonumber
\\[2mm] & & \hspace{3cm}
+ \frac{\pi^4}{1440}
+ \frac{19\,\pi^2}{144}
+ \frac{3}{2}\,\zeta_5
- \bigg(\frac{\pi^2}{8}-\frac{11}{6}\bigg)\,\zeta_3 
+ \frac{517}{864}
\,\bigg]
\,,
\label{DeltaNNNLOb0}
\end{eqnarray}
where
\begin{equation}
L \, = \, \ln\Big(\frac{\mu}{2\,\gamma}\Big)
\,.
\end{equation}

\par
\vspace{0.5cm}
\section{Order $\alpha_s^4$ $\overline{\mbox{MS}}$--1S Mass Relation in the
  Large-$\beta_0$ Approximation} 
\label{appendixmsbar1Smasslargeb0}
In this appendix we determine the order $\alpha_s^4$ (or order
$\epsilon^4$ in the upsilon expansion) relation between the heavy
quark $\overline{\mbox{MS}}$ and 1S masses in the large-$\beta_0$
approximation. We assume that there are $n_l$ light quarks that are
all massless. 

The large-$\beta_0$ approximation of the heavy quark
$\overline{\mbox{MS}}$--pole mass relation is known to all orders of
perturbation theory. The all order formula based on the resummation of
massless quark one-loop vacuum polarization insertions into the
one-loop gluon line reads~\cite{Beneke5}
($\bar a_s=\alpha_s^{(n_l)}(M_{\mbox{\tiny Q}})$)
\begin{eqnarray}
\lefteqn{
\bigg[\,
\frac{\overline M_{\mbox{\tiny Q}}(M_{\mbox{\tiny Q}})
-M_{\mbox{\tiny Q}}^{\mbox{\tiny pole}}}
{M_{\mbox{\tiny Q}}^{\mbox{\tiny pole}}}
\,\bigg]_{\beta_0}
}
\nonumber
\\[2mm]
& = & -\,
\sum\limits_{n=0}^\infty\,
\bar a_s\,
\bigg[\,\Big(\frac{\bar a_s}{4\,\pi}\Big)\,\beta_0\,\bigg]^n\,
\Big(\frac{d}{d u}\Big)^n\,
\bigg\{\,
\frac{1}{3\,\pi}\,\bigg(\,
e^{\frac{5}{3}u}\,6\,(1-u)\,
  \frac{\Gamma(u)\,\Gamma(1-2\,u)}{\Gamma(3-u)}
 + \frac{\tilde B(u)}{u}\,
\bigg)
\,\bigg\}\,\bigg|_{u=0}
\,,
\nonumber
\\
\label{msbarpoleb0}
\end{eqnarray}
where
\begin{eqnarray}
\tilde B(u) 
& = & 
\sum\limits_{n=0}^\infty\,\frac{u^n}{(n!)^2}\,
\Big(\frac{d}{d u}\Big)^n\,B(u)\,\bigg|_{u=0}
\,,
\\[4mm]
B(u) & = &
-\frac{1}{3}\,(3+2\,u)\,
\frac{\Gamma(4+2\,u)}{\Gamma(1-u)\,\Gamma^2(2+u)\,\Gamma(3+u)}
\,.
\end{eqnarray}
We note that in the large-$\beta_0$ approximation the actual
definition of the heavy quark mass as the renormalization scale 
in Eq.~(\ref{msbarpoleb0}) is
irrelevant, because this only affects subleading corrections in the
large-$\beta_0$ approximation.
Evaluating Eq.~(\ref{msbarpoleb0}) up to order $\alpha_s^4$ we arrive
at
\begin{eqnarray}
\Big[\,\overline M_{\mbox{\tiny Q}}(M_{\mbox{\tiny Q}})
\,\Big]_{\beta_0}
& = & 
M_{\mbox{\tiny Q}}^{\mbox{\tiny pole}}\,
\Big[\, 1 \, - \,
\epsilon\,\delta^{(1)} \, - \,
\epsilon^2\,\delta^{(2)}_{\beta_0} \, - \,
\epsilon^3\,\delta^{(3)}_{\beta_0} \, - \,
\epsilon^4\,\delta^{(4)}_{\beta_0} \, - \, \dots
\,\Big]
\,,
\label{msbarpoleb0fourloops}
\end{eqnarray}
where
\begin{eqnarray}
\delta^{(1)} 
& = &
\frac{4}{3}\,\Big(\frac{\bar a_s}{\pi}\Big)
\,,
\\[4mm]
\delta^{(2)}_{\beta_0} 
& = &
\Big(\frac{\bar a_s}{\pi}\Big)^2\,\beta_0\,
\bigg(\frac{\pi^2}{12} +\frac{71}{96}\bigg)
\,,
\\[4mm]
\delta^{(3)}_{\beta_0} 
& = &
\Big(\frac{\bar a_s}{\pi}\Big)^3\,\beta_0^2\,
\bigg( \frac{7}{24}\,\zeta_3 + \frac{13\,\pi^2}{144} + 
  \frac{2353}{10368}
\bigg)
\,,
\\[4mm]
\delta^{(4)}_{\beta_0} 
& = &
\Big(\frac{\bar a_s}{\pi}\Big)^4\,\beta_0^3
\bigg(\frac{71\,\pi^4}{7680} + \frac{317}{768}\,\zeta_3
  + \frac{89\,\pi^2}{1152} + \frac{42979}{331776}
\bigg)
\,.
\end{eqnarray}
The large-$\beta_0$ approximation for the $\overline{\mbox{MS}}$--1S
mass relation up to order $\alpha_s^4$ ($\epsilon^4$) using the
upsilon expansion (with $\epsilon=1$) then reads
\begin{eqnarray}
\Big[\,\overline M_{\mbox{\tiny Q}}(M_{\mbox{\tiny Q}})
\,\Big]_{\beta_0}
& = & 
M_{\mbox{\tiny Q}}^{\mbox{\tiny 1S}}\,\Big[\,
1 + 
\epsilon\,\Big(\,\Delta^{\mbox{\tiny LO}} - \delta^{(1)}\,\Big) 
\, + \,
\epsilon^2\,\Big(\,\Delta^{\mbox{\tiny NLO}}_{\beta_0}-
  \delta^{(2)}_{\beta_0}\,\Big)
\nonumber
\\[2mm] & & \hspace{1.3cm}
\, + \,
\epsilon^3\,\Big(\,\Delta^{\mbox{\tiny NNLO}}_{\beta_0}-
  \delta^{(3)}_{\beta_0}\,\Big)
\, + \,
\epsilon^4\,\Big(\,\Delta^{\mbox{\tiny NNNLO}}_{\beta_0}-
  \delta^{(4)}_{\beta_0}\,\Big) \, + \, \dots
\,,
\label{msbar1Sb0fourloops}
\end{eqnarray}
where the $\Delta$'s are given in
Eqs.~(\ref{DeltaLOb0})--(\ref{DeltaNNLOb0}) and (\ref{DeltaNNNLOb0}).
We note that, in contrast to the complete result, the inversion of
Eq.~(\ref{MQQbarN3LO}) is particularly simple in the large-$\beta_0$
approximation because higher powers of low order contributions 
can be dropped. This is because they do not contain the
highest power of $n_l$ in any order of perturbation theory.

\par
\vspace{0.5cm}
\section{Three-Loop Light Quark Mass Corrections to the Heavy Quark
  Pole--$\overline{\mbox{MS}}$ Mass Relation from Vacuum Polarization
  Insertions } 
\label{appendixmsbarmassvacuum}
In this appendix we present analytic results for the functions
${\rm f}_{\rm P}^{(0)}$, ${\rm f}_{\rm P}^{(1)}$ and 
${\rm f}_{\rm PP}^{(0)}$ (see Sec.~\ref{sectionpolemsbarmass}), 
which occur in the order $\alpha_s^2$ and
$\alpha_s^3$ corrections in the relation between the heavy quark 
$\overline{\mbox{MS}}$ and the pole mass, and which come from
insertions of massive quark one-loop vacuum polarizations into the
gluon line ($r\equiv m/M_{\mbox{\tiny Q}}$):
\begin{eqnarray}
{\rm f}_{\rm P}^{(0)}\Big(\frac{m}{M_{\mbox{\tiny Q}}}\Big)
& = &
\frac{1}{4}\, 
\int\limits_0^\infty\!\frac{d q^2}{M_{\mbox{\tiny Q}}^2}\,
\bigg[\,\frac{1}{2}\,\frac{q^2}{M_{\mbox{\tiny Q}}^2}
+\bigg(1-\frac{1}{2}\frac{q^2}{M_{\mbox{\tiny Q}}^2}\bigg)\,
\bigg(1+4\,\frac{M_{\mbox{\tiny Q}}^2}{q^2}\bigg)^{\frac{1}{2}}
\,\bigg]\,\rm P\Big(\frac{m^2}{q^2}\Big)
%\nonumber
%\\[2mm] & = &
%\frac{1}{24}\,
%\int\limits_0^\infty\! d y\,\bigg(\frac{4-y^2}{1-y}\bigg)\,
%\rm P\Big(r^2 \,\frac{1-y}{y^2}\Big)
\nonumber
\\[2mm] & = &
\frac{3}{2}\,\ln^2(r) 
+ \frac{\pi^2}{4} 
- \frac{3}{2}\,\bigg(\ln(r) + \frac{3}{2}\,\bigg)\,r^2 
\nonumber
\\[2mm] & & 
+ \,\frac{3}{2}\,(1 + r)\,(1 + r^3)\,
   \bigg(\,\Li2(-r) - \frac{1}{2}\,\ln(r)^2 + 
      \ln(r)\ln(1 + r) + \frac{\pi^2}{6}\,\bigg) 
\nonumber
\\[2mm] & & 
+ \,\frac{3}{2}\,(1 - r)\,(1 - r^3)\,
   \bigg(\,\Li2(r) - \frac{1}{2}\,\ln(r)^2 + 
      \ln(r)\,\ln(1 - r) - \frac{\pi^2}{3}\,\bigg)
\,,
\\[4mm]
{\rm f}_{\rm P}^{(1)}\Big(\frac{m}{M_{\mbox{\tiny Q}}}\Big)
& = &
\frac{1}{4}\, 
\int\limits_0^\infty\!\frac{d q^2}{M_{\mbox{\tiny Q}}^2}\,
\bigg[\,\frac{1}{2}\,\frac{q^2}{M_{\mbox{\tiny Q}}^2}
+\bigg(1-\frac{1}{2}\frac{q^2}{M_{\mbox{\tiny Q}}^2}\bigg)\,
\bigg(1+4\,\frac{M_{\mbox{\tiny Q}}^2}{q^2}\bigg)^{\frac{1}{2}}
\,\bigg]\,\rm P\Big(\frac{m^2}{q^2}\Big)
\,\ln\Big(\frac{q^2}{M_{\mbox{\tiny Q}}^2}\Big)
\nonumber
\\[2mm] & = &
  \frac{3}{4}\,r^2\,\bigg(\,\frac{1}{2} - \ln(r)\,\bigg) 
+ r^4\,\ln^2(r)\,\bigg(\,\frac{3}{4} - \ln(r)\,\bigg) 
- \frac{1}{2}(1 - 5\,r - 5\,r^3 + r^4)\,\ln^3(1 + r) 
\nonumber
\\[2mm] & & 
- \,\frac{\pi^2}{2}\bigg(\frac{r}{4}(6 - 4\,r + 6\,r^2 - r^3) - 
        r(5 + 5\,r^2 + r^3)\ln(r) 
  + (1 + 7\,r + 7\,r^3 + r^4)\,\ln(1 + r)\bigg)
\nonumber
\\[2mm] & & 
- \,\frac{3}{4}\,(1 + r)^2\,(1 + r^2)\,
     \bigg(\,\ln(1 - r^2)\,\ln(r) + \frac{1}{2} \Li2(r^2)\,\bigg)
\nonumber
\\[2mm] & & 
- \,3\,r\,(1 + r^2)\,\bigg(\, - \frac{7\,\pi^2}{6}\,\ln(2) 
       + \,\frac{1}{3}\, \ln^3(2) 
       - \ln(1 - r)\,\ln(r)\,\Big(1 - 2\,\ln(2r)\Big) 
\nonumber
\\[2mm] & & \qquad
       + \,\ln(2)\,\ln(1 + r)\,\ln\Big(\frac{1 + r}{2}\Big) 
       - 2\,\ln(r)\,\ln(1 + r)\,\ln\Big(\frac{2r(1 - r)}{1 + r}\Big) 
\nonumber
\\[2mm] & & \qquad
       - \,\ln(r)\,\Li2(-r) 
       - \Big(1 - \ln(r)\Big)\,\Li2(r) 
       + 2\,\ln(r)\,\Li2\Big(\frac{1 - r}{1 + r}\Big) 
\nonumber
\\[2mm] & & \qquad
       + \,2\,\Litri\Big(\frac{1}{1 + r}\Big) 
       + 2\,\Litri\Big(\frac{2r}{1 + r}\Big) 
       - 2\,\Litri\Big(\frac{1 + r}{2}\Big)
    \,\bigg) 
\nonumber
\\[2mm] & & 
+ \,3\,(1 - r)^2\,(1 + r + r^2)\,\bigg(\,
        \Litri(1 - r) + \Litri\Big(\frac{1}{1 + r}\Big) - 
        \frac{1}{2}\,\Litri(1 - r^2)
    \,\bigg) 
\nonumber
\\[2mm] & & 
- \,\frac{3}{4}\,(6 - 7\,r - 7\,r^3)\,\zeta_3
\,,
\\[4mm]
{\rm f}_{\rm PP}^{(0)}\Big(\frac{m}{M_{\mbox{\tiny Q}}}\Big)
& = &
\frac{1}{4}\, 
\int\limits_0^\infty\!\frac{d q^2}{M_{\mbox{\tiny Q}}^2}\,
\bigg[\,\frac{1}{2}\,\frac{q^2}{M_{\mbox{\tiny Q}}^2}
+\bigg(1-\frac{1}{2}\frac{q^2}{M_{\mbox{\tiny Q}}^2}\bigg)\,
\bigg(1+4\,\frac{M_{\mbox{\tiny Q}}^2}{q^2}\bigg)^{\frac{1}{2}}
\,\bigg]\,\bigg[\,\rm P\Big(\frac{m^2}{q^2}\Big)\,\bigg]^2
\nonumber
\\[2mm] & = &
\frac{4}{5}\,(1 - r^2)\,r^2 
+ \,\frac{3}{2}(6 - 7\,r - 7\,r^3)\,\zeta_3 
- \frac{4}{5}\, (1 + 2\,r^2)\,r^2\,\ln(r)  
\nonumber
\\[2mm] & & 
+ \,\frac{\pi^2}{30}\,r\,(96 - 30\,r + 120\,r^2 - 65\,r^3 + 8\,r^5) 
+ \pi^2\,(1 + 7\,r + 7\,r^3 + r^4)\,\ln(1 + r) 
\nonumber
\\[2mm] & & 
- \,\frac{1}{5}\,r\,(1 - r)\,(32 + 17\,r + 57\,r^2 - 8\,r^3 - 8\,r^4)\,
    \ln(1 - r)\,\ln(r) 
\nonumber
\\[2mm] & & 
+ \,\frac{1}{5}\,r\,(1 + r)\,(32 - 17\,r + 57\,r^2 + 8\,r^3 - 8\,r^4)\,
    \ln(1 + r)\,\ln(r)  
\nonumber
\\[2mm] & & 
- \,\frac{1}{5}\,r^2\,(30 + 65\,r^2 - 8\,r^4)\,\ln^2(r)
+ 6\,(1 + r)^2\,(1 - r + r^2)\,\ln(1-r)\,\ln^2(r)  
\nonumber
\\[2mm] & & 
+ \,(1 - 5\,r - 5\,r^3 + r^4)\,\ln^3(1 + r) 
+ \frac{8}{5}\, r\,(4 + 5\,r^2)\,\bigg(\,\Li2(-r) - \Li2(r)\,\bigg)  
\nonumber
\\[2mm] & & 
+ \,\bigg(\, 3\,(2 + 3\,r^4)\,\ln(r) 
     + \frac{1}{10}\,r^2\,(15 + 65\,r^2 - 8\,r^4) \,\bigg)\,\Li2(r^2) 
- 3\,(1 + 2\,r^4)\,\Litri(r^2) 
\nonumber
\\[2mm] & & 
+ \,6\,r\,(1 + r^2)\,\bigg(\,
    -\frac{\pi^2}{6}\,\Big(7\,\ln(2) + 2\,\ln(r) \Big) 
    + \frac{1}{3}\,\ln^3(2)  
\nonumber
\\[2mm] & & \qquad 
    + \,\ln\Big(\frac{1 + r}{2}\Big)\,\Big(\,
         2\,\ln(r)\,\ln\Big(\frac{1 + r}{1 - r}\Big) 
         + \ln(2)\,\ln(1 + r)\,\Big) 
    + 2\,\ln(r)\,\Li2\Big(\frac{1 - r}{1 + r}\Big)  
\nonumber
\\[2mm] & & \qquad
    + \,2\,\Litri\Big(\frac{1}{1 + r}\Big) 
    + 2\,\Litri\Big(\frac{2r}{1 + r}\Big) 
    - 2\,\Litri\Big(\frac{1 + r}{2}\Big)  
   \,\bigg)
\nonumber
\\[2mm] & & 
+ \,6\,(1 - r)^2\,(1 + r + r^2)\,\bigg(\, 
    \ln(1 + r)\,\ln^2(r) 
    - \Litri(1 - r)  
\nonumber
\\[2mm] & & \qquad
    - \,\Litri\Big(\frac{1}{1 + r}\Big) 
    + \frac{1}{2}\,\Litri(1 - r^2)  
\,\bigg)
\,.
\end{eqnarray}
The function ${\rm P}$ has been defined in Eq.~(\ref{Pdef}).
The result for ${\rm f}_{\rm P}^{(0)}$ has already been obtained
in Ref.~\cite{Broadhurst1}. The results for 
${\rm f}_{\rm P}^{(1)}$ and  
${\rm f}_{\rm PP}^{(0)}$ are new.

\par
\vspace{0.5cm}
\section{Some Formulae for Moments with Massless Light Quarks} 
\label{appendixmassless}
In this appendix the expressions for the constants $w^{0,1,2}_p$
are given. They have been taken from Ref.~\cite{Hoang1}. They arise in
the charm quark mass corrections to the moments of the $\Upsilon$ sum
rules if the charm quark is included in the evolution of the strong
coupling. The constants read ($p=1,2,3,\ldots$):
\begin{eqnarray}
w^0_p & = &
-\frac{1}{p!\,\Gamma(\frac{p}{2})}\,
\int\limits_0^\infty dt \int\limits_0^\infty du \,
\frac{1}{(1+t+u)^2}\,\ln^p\Big(\frac{(1+t)\,(1+u)}{t\,u}\Big)
\, = \,
-\,\frac{(p+1)\,\zeta_{p+1}}{\Gamma(\frac{p}{2})}
\,,
\\[4mm]
w^1_p & = &
\frac{1}{p!\,\Gamma(\frac{p}{2})}\,
\int\limits_0^\infty dt \int\limits_0^\infty du \,
\frac{1-\ln(1+t+u)}{(1+t+u)^2}\,\ln^p\Big(\frac{(1+t)\,(1+u)}{t\,u}\Big)
\nonumber\\[2mm] & = &
-\,\bigg\{\,
\frac{(1+p)}{\Gamma(\frac{p}{2})}\,\bigg[\,\gamma_{\mbox{\tiny E}}\,
  \zeta_{p+1} +
  \sum\limits_{m=0}^\infty\,\frac{\Psi(2+m)}{(1+m)^{p+1}}\,\bigg]
+ \frac{2}{\Gamma(\frac{p}{2})}\,
  \sum\limits_{l=0}^{p-1}\,\sum\limits_{m=0}^\infty\,
  (-1)^{p-l}\,\frac{(1+l)\,\Psi^{(p-l)}(2+m)}{(p-l)!\,(1+m)^{1+l}}
\,\bigg\}
\,,
\nonumber\\&&
\\[4mm]
w^2_p & = &
\frac{1}{p!\,\Gamma(\frac{p}{2})}\,
\int\limits_0^\infty dt \int\limits_0^\infty du \,
\frac{\zeta_2-2\,\ln(1+t+u)+\ln^2(1+t+u)}{(1+t+u)^2}\,
\ln^p\Big(\frac{(1+t)\,(1+u)}{t\,u}\Big)
\nonumber\\[2mm] & = &
\frac{(1+p)}{\Gamma(\frac{p}{2})}\,\bigg\{\,
\Big(\,\gamma_{\mbox{\tiny E}}^2+2\,\zeta_2\,\Big)\,\zeta_{1+p}
\nonumber\\[2mm] & & \quad
 +\, \sum\limits_{m=0}^\infty\,\frac{1}{(1+m)^{1+p}}\,\bigg[\,
2\,\gamma_{\mbox{\tiny E}}\,\Psi(2+m) - 
\Psi^\prime(2+m) + \Big(\Psi(2+m)\Big)^2
\,\bigg]
\,\bigg\}
\nonumber\\[2mm] & &
+ \,\frac{2}{\Gamma(\frac{p}{2})}\,\sum\limits_{m=0}^\infty\,
\sum\limits_{l=0}^{p-1}\,\frac{(-1)^{p-l}\,(1+l)}{(p-l)!\,(1+m)^{1+l}}\,
\bigg[\,
2\,\gamma_{\mbox{\tiny E}}\,\Psi^{(p-l)}(2+m) - \Psi^{(p-l+1)}(2+m) 
\nonumber\\[2mm] & & \mbox{\hspace{6cm}}
+\,2\,\Psi^{(p-l)}(2+m)\,\Psi(2+m)
\,\bigg]
\nonumber\\[2mm] & &
+\,\frac{4}{\Gamma(\frac{p}{2})}\,\sum\limits_{m=0}^\infty\,
\sum\limits_{l=0}^{p-2}\,\sum\limits_{k=1}^{p-l-1}\,
(-1)^{p-l}\,\frac{(1+l)\,\Psi^{(p-l-k)}(2+m)\,\Psi^{(k)}(2+m)}
{(p-l-k)!\,k!\,(1+m)^{1+l}}
\,.
\end{eqnarray}

\end{appendix}

\vspace{1.0cm}
%
%\newpage
%%%%%%%%%%%%%%%%%%%%%%%%%%%%%%%%%%%%%%%%%%%%%%%%%%%%%%%%%%%%%%%%%%%%%%%%
\sloppy
\raggedright
\def\app#1#2#3{{\it Act. Phys. Pol. }{\bf B #1} (#2) #3}
\def\apa#1#2#3{{\it Act. Phys. Austr.}{\bf #1} (#2) #3}
\def\lhc{Proc. LHC Workshop, CERN 90-10}
\def\npb#1#2#3{{\it Nucl. Phys. }{\bf B #1} (#2) #3}
\def\nP#1#2#3{{\it Nucl. Phys. }{\bf #1} (#2) #3}
\def\plb#1#2#3{{\it Phys. Lett. }{\bf B #1} (#2) #3}
\def\prd#1#2#3{{\it Phys. Rev. }{\bf D #1} (#2) #3}
\def\pra#1#2#3{{\it Phys. Rev. }{\bf A #1} (#2) #3}
\def\pR#1#2#3{{\it Phys. Rev. }{\bf #1} (#2) #3}
\def\prl#1#2#3{{\it Phys. Rev. Lett. }{\bf #1} (#2) #3}
\def\prc#1#2#3{{\it Phys. Reports }{\bf #1} (#2) #3}
\def\cpc#1#2#3{{\it Comp. Phys. Commun. }{\bf #1} (#2) #3}
\def\nim#1#2#3{{\it Nucl. Inst. Meth. }{\bf #1} (#2) #3}
\def\pr#1#2#3{{\it Phys. Reports }{\bf #1} (#2) #3}
\def\sovnp#1#2#3{{\it Sov. J. Nucl. Phys. }{\bf #1} (#2) #3}
\def\sovpJ#1#2#3{{\it Sov. Phys. LETP }{\bf #1} (#2) #3}
\def\jl#1#2#3{{\it JETP Lett. }{\bf #1} (#2) #3}
\def\jet#1#2#3{{\it JETP Lett. }{\bf #1} (#2) #3}
\def\zpc#1#2#3{{\it Z. Phys. }{\bf C #1} (#2) #3}
\def\ptp#1#2#3{{\it Prog.~Theor.~Phys.~}{\bf #1} (#2) #3}
\def\nca#1#2#3{{\it Nuovo~Cim.~}{\bf #1A} (#2) #3}
\def\ap#1#2#3{{\it Ann. Phys. }{\bf #1} (#2) #3}
\def\hpa#1#2#3{{\it Helv. Phys. Acta }{\bf #1} (#2) #3}
\def\ijmpA#1#2#3{{\it Int. J. Mod. Phys. }{\bf A #1} (#2) #3}
\def\ZETF#1#2#3{{\it Zh. Eksp. Teor. Fiz. }{\bf #1} (#2) #3}
\def\jmp#1#2#3{{\it J. Math. Phys. }{\bf #1} (#2) #3}
\def\yf#1#2#3{{\it Yad. Fiz. }{\bf #1} (#2) #3}
\def\ufn#1#2#3{{\it Usp. Fiz. Nauk }{\bf #1} (#2) #3}
\def\spu#1#2#3{{\it Sov. Phys. Usp.}{\bf #1} (#2) #3}
\def\epjc#1#2#3{{\it Eur. Phys. J. C }{\bf #1} (#2) #3}
%%%%%%%%%%%%%%%%%%%%%%%%%%%%%%%%%%%%%%%%%%%%%%%%%%%%%%%%%%%%%%%%%%%%%%%%

\end{document}